\documentclass[ngerman,english,official]{IISthesis}

\usepackage{graphicx}           %
\usepackage[usenames]{color}
\usepackage[toc, section=section, automake]{glossaries}
\usepackage[automake]{glossaries-extra}

\usepackage{amsmath}
\RequirePackage[hidelinks]{hyperref}
\usepackage{xspace}
\usepackage{textcomp,lmodern}
\usepackage{multirow}
\usepackage{rotating}
\usepackage{lscape}
\usepackage{array} 
\renewcommand{\arraystretch}{1.10}
\usepackage[nottoc]{tocbibind}

\usepackage{amsmath,amssymb,amsfonts}
\usepackage{wasysym}
\RequirePackage{algorithmicx}
\RequirePackage{algorithm}
\RequirePackage{algpseudocode}
\usepackage{booktabs}
\usepackage{threeparttable}
\usepackage[table,xcdraw]{xcolor}
\usepackage{acronym}
\usepackage[detect-all, per-mode=symbol, range-units=single, range-phrase=~to~]{siunitx}
\usepackage[shortcuts]{extdash}
\usepackage{fixfoot}
\usepackage{url}
\usepackage{cite}
\usepackage{textcomp}
\usepackage{textgreek}
\usepackage{multicol}
\hypersetup{breaklinks=true}
\usepackage{soul}
\usepackage{makecell}
\usepackage[shortlabels,inline]{enumitem}
\usepackage{wrapfig}
\usepackage[export]{adjustbox}
\usepackage{tabularx}
\usepackage{float} 

\usepackage{bibentry} 

\usepackage{pifont}
\usepackage{makecell}
\usepackage[multiple]{footmisc}
\usepackage{refcount}
\usepackage[noabbrev,capitalise]{cleveref}
\usepackage{microtype}
\makeatletter
\def\@esphack{%
  \relax
  \ifhmode
    \spacefactor\@savsf
    \ifdim\@savsk>\z@
      \ifdim\lastskip=\z@
        \nobreak \hskip\z@skip
      \fi
      \ignorespaces
    \fi
  \else
    \ifvmode
      \if@nobreak\nobreak\else\if@noskipsec\nobreak\fi\fi
    \fi
  \fi}%
\makeatother
\usepackage{multirow}

\usepackage{tikz}
\usepackage{calc}
\usepackage{eso-pic}
\usepackage{ragged2e}
\usepackage[utf8]{inputenc}
\usepackage[T1]{fontenc}
\usepackage{orcidlink}

\usepackage[font=footnotesize,labelfont=rm,textfont=rm, position=b]{subcaption}
\usepackage{nicefrac}

\usepackage[frozencache]{minted}
\makeatletter
\newcounter{exportlisting}
\patchcmd{\minted@highlight@i}
  {\edef\minted@highlightfilename{\pydatabuffermdfivesum\detokenize{.highlight.minted}}}
  {\stepcounter{exportlisting}%
   \edef\minted@highlightfilename{listing\arabic{exportlisting}\detokenize{.highlight.minted}}}
  {}
  {\PackageError{export}{Cannot patch minted cache file names}{}}
\makeatother
\usepackage{framed}
\usepackage{hyphenat}
\usepackage{circledsteps}

\usepackage{inconsolata}

\usepackage{arydshln}
\expandafter\let\csname endtabular*\endcsname\endtabular
\makeatletter
\def\thesis@LTmarks{{%
  \setbox\@outputbox\vbox{\unvcopy\@cclv}%
  \@expl@@@mark@update@singlecol@structures@@}}
\def\thesis@LTdraw{\thesis@LTmarks\adl@LTdraw\LT@foot\vss}
\patchcmd{\LT@output}{\adl@LTdraw\LT@foot\vss}{\thesis@LTdraw}{}
  {\PackageWarning{thesis}{Cannot patch arydshln's \string\LT@output}}
\patchcmd{\LT@output}{\adl@LTdraw\LT@foot\vss}{\thesis@LTdraw}{}
  {\PackageWarning{thesis}{Cannot patch arydshln's \string\LT@output}}
\makeatother
\usepackage{afterpage}

\usepackage{pdflscape}
\usepackage{zref-user}
\usepackage[strict]{changepage}

\usepackage{comment}

\definecolor{ieee-bright-dblue-100}{rgb}{0.0, 0.3828, 0.6055}
\definecolor{ieee-bright-dblue-80}{rgb}{0.0, 0.4883, 0.6797}
\definecolor{ieee-bright-dblue-60}{rgb}{0.3633, 0.6094, 0.7617}
\definecolor{ieee-bright-dblue-40}{rgb}{0.5898, 0.7383, 0.8398}
\definecolor{ieee-bright-dblue-20}{rgb}{0.8906, 0.8984, 0.9219}
\definecolor{ieee-bright-red-100}{rgb}{0.7266, 0.0469, 0.1836}
\definecolor{ieee-bright-red-80}{rgb}{0.832, 0.3164, 0.3281}
\definecolor{ieee-bright-red-60}{rgb}{0.8906, 0.4922, 0.4805}
\definecolor{ieee-bright-red-40}{rgb}{0.9336, 0.6562, 0.6406}
\definecolor{ieee-bright-red-20}{rgb}{0.9688, 0.8203, 0.8125}
\definecolor{ieee-bright-orange-100}{rgb}{0.9961, 0.6367, 0.0}
\definecolor{ieee-bright-orange-80}{rgb}{0.9844, 0.6953, 0.3125}
\definecolor{ieee-bright-orange-60}{rgb}{0.9883, 0.7695, 0.4844}
\definecolor{ieee-bright-orange-40}{rgb}{0.9922, 0.8359, 0.6562}
\definecolor{ieee-bright-orange-20}{rgb}{0.9961, 0.9219, 0.8164}
\definecolor{ieee-bright-yellow-100}{rgb}{0.9961, 0.8164, 0.0}
\definecolor{ieee-bright-yellow-80}{rgb}{0.9961, 0.8477, 0.2148}
\definecolor{ieee-bright-yellow-60}{rgb}{0.9961, 0.875, 0.4492}
\definecolor{ieee-bright-yellow-40}{rgb}{0.9961, 0.9062, 0.6328}
\definecolor{ieee-bright-yellow-20}{rgb}{0.9961, 0.9531, 0.8125}
\definecolor{ieee-bright-lgreen-100}{rgb}{0.4688, 0.7422, 0.125}
\definecolor{ieee-bright-lgreen-80}{rgb}{0.5742, 0.7852, 0.332}
\definecolor{ieee-bright-lgreen-60}{rgb}{0.6875, 0.8398, 0.5039}
\definecolor{ieee-bright-lgreen-40}{rgb}{0.793, 0.8906, 0.6641}
\definecolor{ieee-bright-lgreen-20}{rgb}{0.8945, 0.9414, 0.8281}
\definecolor{ieee-bright-dgreen-100}{rgb}{0.0, 0.5156, 0.2383}
\definecolor{ieee-bright-dgreen-80}{rgb}{0.1641, 0.6055, 0.3867}
\definecolor{ieee-bright-dgreen-60}{rgb}{0.3906, 0.6953, 0.5234}
\definecolor{ieee-bright-dgreen-40}{rgb}{0.6094, 0.8008, 0.6719}
\definecolor{ieee-bright-dgreen-20}{rgb}{0.8047, 0.8945, 0.8359}
\definecolor{ieee-bright-purple-100}{rgb}{0.5938, 0.1133, 0.5898}
\definecolor{ieee-bright-purple-80}{rgb}{0.6992, 0.3281, 0.668}
\definecolor{ieee-bright-purple-60}{rgb}{0.7812, 0.4961, 0.7461}
\definecolor{ieee-bright-purple-40}{rgb}{0.8555, 0.6602, 0.8281}
\definecolor{ieee-bright-purple-20}{rgb}{0.9219, 0.8281, 0.9023}
\definecolor{ieee-bright-lblue-100}{rgb}{0.0, 0.6094, 0.6484}
\definecolor{ieee-bright-lblue-80}{rgb}{0.0, 0.6797, 0.7188}
\definecolor{ieee-bright-lblue-60}{rgb}{0.2109, 0.75, 0.7812}
\definecolor{ieee-bright-lblue-40}{rgb}{0.5469, 0.8242, 0.8438}
\definecolor{ieee-bright-lblue-20}{rgb}{0.7695, 0.918, 0.9219}
\definecolor{ieee-bright-cyan-100}{rgb}{0.0, 0.707, 0.8828}
\definecolor{ieee-bright-cyan-80}{rgb}{0.0, 0.7227, 0.9453}
\definecolor{ieee-bright-cyan-60}{rgb}{0.2656, 0.7812, 0.957}
\definecolor{ieee-bright-cyan-40}{rgb}{0.5547, 0.8438, 0.9688}
\definecolor{ieee-bright-cyan-20}{rgb}{0.7773, 0.9141, 0.9805}
\definecolor{ieee-bright-white-100}{rgb}{0.9961, 0.9961, 0.9961}
\definecolor{ieee-bright-white-80}{rgb}{0.9961, 0.9961, 0.9961}
\definecolor{ieee-bright-white-60}{rgb}{0.9961, 0.9961, 0.9961}
\definecolor{ieee-bright-white-40}{rgb}{0.9961, 0.9961, 0.9961}
\definecolor{ieee-bright-white-20}{rgb}{0.9961, 0.9961, 0.9961}
\definecolor{ieee-dark-red-100}{rgb}{0.5234, 0.1211, 0.2539}
\definecolor{ieee-dark-red-80}{rgb}{0.6445, 0.2812, 0.3828}
\definecolor{ieee-dark-red-60}{rgb}{0.7422, 0.4727, 0.5234}
\definecolor{ieee-dark-red-40}{rgb}{0.832, 0.6445, 0.6758}
\definecolor{ieee-dark-red-20}{rgb}{0.918, 0.8203, 0.832}
\definecolor{ieee-dark-orange-100}{rgb}{0.9062, 0.4648, 0.1328}
\definecolor{ieee-dark-orange-80}{rgb}{0.9648, 0.5664, 0.3164}
\definecolor{ieee-dark-orange-60}{rgb}{0.9766, 0.6758, 0.4805}
\definecolor{ieee-dark-orange-40}{rgb}{0.9844, 0.7773, 0.6523}
\definecolor{ieee-dark-orange-20}{rgb}{0.9922, 0.8789, 0.8125}
\definecolor{ieee-dark-yellow-100}{rgb}{0.9961, 0.7773, 0.1719}
\definecolor{ieee-dark-yellow-80}{rgb}{0.9961, 0.8086, 0.375}
\definecolor{ieee-dark-yellow-60}{rgb}{0.9961, 0.875, 0.4492}
\definecolor{ieee-dark-yellow-40}{rgb}{0.9961, 0.8984, 0.6875}
\definecolor{ieee-dark-yellow-20}{rgb}{0.9961, 0.9453, 0.8438}
\definecolor{ieee-dark-lgreen-100}{rgb}{0.3945, 0.5508, 0.0938}
\definecolor{ieee-dark-lgreen-80}{rgb}{0.5078, 0.6289, 0.293}
\definecolor{ieee-dark-lgreen-60}{rgb}{0.6367, 0.7188, 0.4688}
\definecolor{ieee-dark-lgreen-40}{rgb}{0.7539, 0.8047, 0.6367}
\definecolor{ieee-dark-lgreen-20}{rgb}{0.875, 0.9023, 0.8125}
\definecolor{ieee-dark-dgreen-100}{rgb}{0.0, 0.3867, 0.2539}
\definecolor{ieee-dark-dgreen-80}{rgb}{0.1836, 0.5, 0.3906}
\definecolor{ieee-dark-dgreen-60}{rgb}{0.3984, 0.6172, 0.5273}
\definecolor{ieee-dark-dgreen-40}{rgb}{0.5938, 0.7422, 0.6758}
\definecolor{ieee-dark-dgreen-20}{rgb}{0.793, 0.8711, 0.8359}
\definecolor{ieee-dark-purple-100}{rgb}{0.4648, 0.1445, 0.5117}
\definecolor{ieee-dark-purple-80}{rgb}{0.5898, 0.3242, 0.6016}
\definecolor{ieee-dark-purple-60}{rgb}{0.6914, 0.4883, 0.6953}
\definecolor{ieee-dark-purple-40}{rgb}{0.7969, 0.6523, 0.793}
\definecolor{ieee-dark-purple-20}{rgb}{0.8945, 0.8203, 0.8945}
\definecolor{ieee-dark-cyan-100}{rgb}{0.0, 0.4492, 0.4648}
\definecolor{ieee-dark-cyan-80}{rgb}{0.0, 0.5469, 0.5664}
\definecolor{ieee-dark-cyan-60}{rgb}{0.3047, 0.6602, 0.668}
\definecolor{ieee-dark-cyan-40}{rgb}{0.5586, 0.7695, 0.7734}
\definecolor{ieee-dark-cyan-20}{rgb}{0.7734, 0.8789, 0.8789}
\definecolor{ieee-dark-dblue-100}{rgb}{0.0, 0.1562, 0.332}
\definecolor{ieee-dark-dblue-80}{rgb}{0.1797, 0.3008, 0.4609}
\definecolor{ieee-dark-dblue-60}{rgb}{0.3828, 0.4609, 0.5859}
\definecolor{ieee-dark-dblue-40}{rgb}{0.5781, 0.6289, 0.7188}
\definecolor{ieee-dark-dblue-20}{rgb}{0.7852, 0.8047, 0.8555}
\definecolor{ieee-dark-grey-100}{rgb}{0.457, 0.4688, 0.4805}
\definecolor{ieee-dark-grey-80}{rgb}{0.5625, 0.5625, 0.5742}
\definecolor{ieee-dark-grey-60}{rgb}{0.6641, 0.6641, 0.6758}
\definecolor{ieee-dark-grey-40}{rgb}{0.7734, 0.7695, 0.7773}
\definecolor{ieee-dark-grey-20}{rgb}{0.8789, 0.8828, 0.8828}
\definecolor{ieee-dark-black-100}{rgb}{0.0, 0.0, 0.0}
\definecolor{ieee-dark-black-80}{rgb}{0.3438, 0.3477, 0.3555}
\definecolor{ieee-dark-black-60}{rgb}{0.5, 0.5078, 0.5195}
\definecolor{ieee-dark-black-40}{rgb}{0.6523, 0.6602, 0.6719}
\definecolor{ieee-dark-black-20}{rgb}{0.8164, 0.8242, 0.8281}

\newcommand{\kGE}{\ensuremath{\,}k\GE}  %

\newcommand{\cmark}{\ding{51}}%
\newcommand{\xmark}{\ding{55}}%

\newcommand{\gf}{{GlobalFoundries}}
\newcommand{\gfs}{{GlobalFoundries'}}

\newcommand{\riscv}{\mbox{RISC-V}}

\newcommand{\occamy}{{Occamy}}

\newcommand{\x}{$\times$}
\newcommand{\topar}[1]{\textbf{#1:}\,}

\newcommand{\cn}[1]{{#1}}
\setabbreviationstyle[acronym]{long-short}
\newacronym{fft}{FFT}{fast Fourier transform}
\newacronym{dnn}{DNN}{deep neural network}
\newacronym{gpu}{GPU}{graphics processing unit}

\newacronym{hpc}{HPC}{high-performance computing}
\newacronym{isa}{ISA}{instruction set architecture}
\newacronym{fpu}{FPU}{floating-point unit}
\newacronym{fp}{FP}{floating-point}
\newacronym{mac}{MAC}{multiply-accumulate}
\newacronym{muldiv}{MULDIV}{integer multiply-divide unit}
\newacronym{alu}{ALU}{arithmetic logic unit}
\newacronym{fu}{FU}{functional unit}
\newacronym{fifo}{FIFO}{first-in, first-out}
\newacronym{cgra}{CGRA}{coarse-grained reconfigurable array}
\newacronym{pim}{PIM}{processing in memory}
\newacronym{dae}{DAE}{decoupled access-execute}
\newacronym{dram}{DRAM}{dynamic random access memory}
\newacronym{sell}{SELL}{sliced ELLPACK}

\newacronym[longplural={core complexes}]{cc}{CC}{core complex}
\newacronym{dmcc}{DMCC}{data movement CC}
\newacronym{fpss}{FPSS}{floating-point subsystem}
\newacronym{ssr}{SSR}{stream semantic register}
\newacronym{issr}{ISSR}{indirection SSR}
\newacronym{essr}{ESSR}{egress SSR}
\newacronym{sssr}{SSSR}{sparse stream semantic register}
\newacronym{tcdm}{TCDM}{tightly coupled data memory}
\newacronym{frep}{FREP}{floating-point repetition}
\newacronym{dma}{DMA}{direct memory access}
\newacronym{ipu}{IPU}{integer processing unit}

\newacronym{csf}{CSF}{compressed sparse fiber}
\newacronym{csr}{CSR}{compressed sparse rows}
\newacronym{csc}{CSC}{compressed sparse columns}
\newacronym{dcsr}{DCSR}{doubly-compressed sparse rows}
\newacronym{bcsr}{BCSR}{blocked compressed sparse rows}
\newacronym{sdcsr}{sDCSR}{sliced doubly-compressed sparse rows}
\newacronym{spmv}{SpMV}{sparse-dense matrix-vector multiply}
\newacronym{sdmv}{SDMV}{sDCSR matrix-vector multiply}
\newacronym{spvv}{SpVV}{sparse vector-vector multiply}
\newacronym{csrmv}{CsrMV}{CSR matrix-vector multiply}
\newacronym{csrmm}{CsrMM}{CSR matrix-matrix multiply}
\newacronym{cvr}{CVR}{compressed vectorization-oriented sparse row}
\newacronym{dl}{DL}{deep learning}
\newacronym{la}{LA}{linear algebra}

\newacronym{ml}{ML}{machine learning}
\newacronym{simd}{SIMD}{single-instruction, multiple-data}
\newacronym{sm}{SM}{streaming multiprocessor}
\newacronym{rtl}{RTL}{register transfer level}
\newacronym{knl}{KNL}{Knights Landing}
\newacronym{uve}{UVE}{Unlimited Vector Extension}
\newacronym{fem}{FEM}{finite element analysis}
\newacronym{loess}{LOESS}{locally estimated scatterplot smoothing}
\newacronym{sve}{SVE}{Scalable Vector Extensions}

\newacronym{llc}{LLC}{low-level controller}

\newacronym{ai}{AI}{Artificial Intelligence}
\newacronym{asic}{ASIC}{application-specific integrated circuit}
\newacronym{axi4}{AXI4}{Advanced eXtensible Interface 4}
\newacronym{bmc}{BMC}{baseboard management controller}
\newacronym{cpu}{CPU}{central processing unit}
\newacronym{dm}{DM}{data mover}
\newacronym{dmac}{DMAC}{direct memory access controller}
\newacronym{dmae}{DMAE}{direct memory access engine}
\newacronym{dsp}{DSP}{digital signal processing}
\newacronym{dvfs}{DVFS}{dynamic voltage and frequency scaling}
\newacronym{e2e}{E2E}{end-to-end}
\newacronym{fpga}{FPGA}{field-programmable gate array}
\newacronym{gemm}{GEMM}{general matrix multiply}
\newacronym{gp}{GP}{general-purpose}
\newacronym{hbm}{HBM}{high-bandwidth memory}
\newacronym{hil}{HIL}{hardware-in-the-loop}
\newacronym{hlc}{HLC}{high-level controller}
\newacronym{idma}{iDMA}{intelligent DMA}
\newacronym{ihls}{iHLS}{IP-based high-level synthesis}
\newacronym{ip}{IP}{intellectual property}
\newacronym{l1}{L1}{level-one}
\newacronym{l2}{L2}{level-two}
\newacronym{l3}{L3}{level-three}
\newacronym{mcu}{MCU}{microcontroller unit}
\newacronym{noc}{NoC}{network-on-chip}
\newacronym{ooc}{OOC}{out-of-context}
\newacronym{obi}{OBI}{open bus protocol}
\newacronym{os}{OS}{operating system}
\newacronym{pcf}{PCF}{power control firmware}
\newacronym{pcs}{PCS}{power controller system}
\newacronym{pdk}{PDK}{process design kit}
\newacronym{pe}{PE}{processing element}
\newacronym{pfct}{PFCT}{periodic frequency control task}
\newacronym{pulp}{PULP}{parallel ultra-low power}
\newacronym{pvct}{PVCT}{periodic voltage control task}
\newacronym{pvt}{PVT}{process-voltage-temperature}
\newacronym{rpc}{RPC}{reduced pin count}
\newacronym{sdma}{sDMAE}{sensor DMAE}
\newacronym{soa}{SoA}{state of the art}
\newacronym[longplural={scratchpad memories}]{spm}{SPM}{scratchpad memory}
\newacronym{spmm}{SpMM}{sparse-dense matrix multiply}
\newacronym{spmspm}{SpMSpM}{sparse-sparse matrix multiply}
\newacronym{sram}{SRAM}{static random-access memory}
\newacronym{ulp}{ULP}{ultra-low-power}
\newacronym{vrm}{VRM}{voltage regulator module}
\newacronym{mpc}{MPC}{model predictive control}
\newacronym{nn}{NN}{neural network}
\newacronym{llm}{LLM}{large language model}
\newacronym{d2d}{D2D}{die-to-die}
\newacronym{c2c}{C2C}{chip-to-chip}
\newacronym{grli}{GRLI}{globally regular, locally irregular}
\newacronym{gpt}{GPT}{generative pretrained transformer}
\newacronym{gcn}{GCN}{graph convolutional network}
\newacronym{gsm}{GSM}{Graduate Student Member}
\newacronym{fma}{FMA}{fused multiply-accumulate}
\newacronym{iotlb}{IOTLB}{IO translation lookaside buffer}
\newacronym{phy}{PHY}{physical layer}
\newacronym{ddr}{DDR}{double-data-rate}
\newacronym{ate}{ATE}{automatic test equipment}
\newacronym{fll}{FLL}{frequency-locked loop}
\newacronym{flop}{FLOP}{floating-point operation}
\newacronym{dsa}{DSA}{domain-specific accelerator}

\newacronym{s7ps}{S7PS}{symmetric 7-point star}
\newacronym{sr}{SR}{stream register}
\newacronym{ipc}{IPC}{instructions per cycle}
\newacronym[longplural={systems-on-chip}]{soc}{SoC}{system-on-chip}
\newacronym{wse}{WSE}{wafer-scale engine}
\newacronym{dag}{DAG}{directed acyclic graph}
\newacronym{dlt}{DLT}{data layout transform}
\newacronym{tlb}{TLB}{translation lookaside buffer}
\newacronym{cmtr}{CMTR}{compute-to-memory time ratio}

\newacronym{snn}{SNN}{spiking neural network}
\newacronym{lif}{LIF}{leaky integrate-and-fire}
\newacronym{fm}{FM}{feature map}
\newacronym{rf}{RF}{receptive field}
\newacronym{spwr}{SpWR}{sparse weight reduction}
\newacronym{aer}{AER}{address event representation}
\newacronym{gals}{GALS}{globally asynchronous, locally synchronous}
\newacronym{lsm}{LSM}{liquid state machine}

\newacronym{etm}{ETM}{energy and thermal management}
\newacronym{qp}{QP}{quadratic program}
\newacronym{lti}{LTI}{linear time-invariant}
\newacronym{admm}{ADMM}{alternating direction method of multipliers}
\newacronym{kkt}{KKT}{Karush-Kuhn-Tucker}
\newacronym{fe}{FE}{forward elimination}
\newacronym{bs}{BS}{backward substitution}
\newacronym{amd}{ApMD}{approximate minimum degree}
\newacronym{pmca}{PMCA}{programmable multicore accelerator}
\newacronym{alap}{ALAP}{as late as possible}
\newacronym{pcg}{PCG}{preconditioned conjugate gradient}
\newacronym{lqr}{LQR}{linear quadratic regulator}
\newacronym{pid}{PID}{proportional-integral-derivative}

\newacronym{vlsu}{VLSU}{vector load-store unit}

\newacronym{ir}{IR}{intermediate representation}
\newacronym{ssa}{SSA}{single static assignment}
\newacronym{thr}{THR}{tree height reduction}

\newacronym{rvv}{RVV}{RISC-V vector extension}
\newacronym{tmu}{TMU}{Tensor Marshaling Unit}
\newacronym{sam}{SAM}{Sparse Abstract Machine}
\newacronym{cam}{CAM}{content-addressable memory}
\newacronym{ta}{TA}{tensor algebra}
\newacronym{chi}{CHI}{Coherent Hub Interface}

\DeclareSIUnit[quantity-product= ]\percent{\%}
\DeclareSIUnit{\x}{\!\ensuremath{\times}}
\DeclareSIUnit\bit{b}
\DeclareSIUnit\lbit{bit}
\DeclareSIUnit\GE{GE}
\DeclareSIUnit\kGE{\kilo\GE}
\DeclareSIUnit\MGE{\mega\GE}
\DeclareSIUnit\comp{COMP}
\DeclareSIUnit\sop{SOP}
\DeclareSIUnit\tok{tok}
\newlength\myheight
\newlength\mydepth
\settototalheight\myheight{Xygp}
\newenvironment{publications}{\begin{itemize}[labelwidth=2.4em]\footnotesize}{\end{itemize}}
\newcommand{\publication}[1]{\item[\textbf{\cite{#1}\hspace{2pt}}] \begin{NoHyper}\bibentry{#1}\end{NoHyper}}

\nobibliography* %

\DeclareSIUnit\GE{GE}
\DeclareSIUnit\kGE{\kilo\GE}
\DeclareSIUnit\MGE{\mega\GE}
\DeclareSIUnit\flop{FLOP}
\DeclareSIUnit\flops{\flop\per\second}
\DeclareSIUnit\Gflops{\giga\flops}
\DeclareSIUnit\Tflops{\tera\flops}
\DeclareSIUnit\dash{\text{-}}

\newfloat{listing}{htbp}{lst}
\floatname{listing}{Listing}
\newcounter{sublisting}
\newcounter{sublisting@save}
\crefname{sublisting}{listing}{listings}
\Crefname{sublisting}{Listing}{Listings}

\makeatletter
\crefalias{ALG@line}{line}
\AddToHook{env/algorithmic/begin}{\def\@currentcounter{ALG@line}}
\makeatother

\counterwithin{listing}{chapter}
\counterwithin{algorithm}{chapter}

\newcommand{\impval}[1]{#1}

\def\showtodo{}
\IfFileExists{identifier.tex}{
    \def\\reviewpass{-c348afeb4cf37cc3-dirty-(bb0f528084f6e0e994999672ad40a697609fef36)}

}{
    \def\reviewpass{Overleaf Version - Do not distribute!}
}

\ifx\showrevision\undefined
\else
    \AddToShipoutPictureFG{%
        \put(%
            8mm,%
            \paperheight-1.2cm%
            ){\vtop{{\null}\makebox[0pt][c]{%
                \rotatebox[origin=c]{90}{%
                    \large\textcolor{red!75}{\reviewpass}%
                }%
            }}%
        }%
    }
    \AddToShipoutPictureFG{%
        \put(%
            \paperwidth-6mm,%
            \paperheight-1.2cm%
            ){\vtop{{\null}\makebox[0pt][c]{%
                \rotatebox[origin=c]{90}{%
                    \large\textcolor{red!30}{ETH Zurich - Unpublished - Confidential - Draft - Copyright Paul Scheffler 2024-2025}%
                }%
            }}%
        }%
    }
\fi

\ifx\showtodo\undefined
    
    \newcommand{\tdval}[1]{{#1}}
\else
    
    \newcommand{\tdval}[1]{{\textcolor{red}{#1}}}
\fi

\ifx\shownumchk\undefined
    \newcommand{\nchk}[1]{#1}
\else
    \newcommand{\nchk}[1]{{\textcolor{ieee-bright-dblue-100}{#1}}}
\fi

\ifx\showrebuttal\undefined
    \newcommand{\rev}[1]{#1}
\else
    \newcommand{\rev}[1]{{\textcolor{ieee-bright-lblue-100}{#1}}}
\fi

\ifx\showrebuttal\undefined
    \newcommand{\revdel}[1]{}
\else
    \newcommand{\revdel}[1]{\textcolor{ieee-bright-red-100}{\st{#1}}}
\fi

\ifx\showrebuttal\undefined
    
\else
    
\fi

\ifx\showrebuttal\undefined
    \newcommand{\revprg}[1]{}
\else
    \newcommand{\revprg}[1]{\hspace{-0.5ex}\textcolor{ieee-bright-red-100}{\scalebox{.2}[1.5]{$\blacksquare$}}\hspace{-0.5ex}}
\fi

\makeatletter
\AtBeginEnvironment{minted}{\vspace{-\FrameSep}}
\AfterEndEnvironment{minted}{\vspace{-\FrameSep}}
\makeatother

\setminted{fontsize=\fontsize{9.6}{12.18}\selectfont}
\setminted[gas]{baselinestretch=0.94}
\fvset{backgroundcolorboxoverlap=0pt}

\newcounter{cntsideways}
\makeatletter
\AddToShipoutPictureBG{%
 \ifnum\zref@extractdefault{rotate\number\value{page}}{page}{0}=0
  \PLS@RemoveRotate
 \else
    \checkoddpage\ifoddpage
        \PLS@AddRotate{90}%
    \else
        \PLS@AddRotate{-90}%
    \fi
 \fi}

\newcommand\rotatesidewayslabel{\stepcounter{cntsideways}%
 \zlabel{tmp\thecntsideways}\zlabel{rotate\zref@extractdefault{tmp\thecntsideways}{page}{0}}}

\ifx\showlandscape\undefined
    \newenvironment{sidetab}{\begin{sidewaystable}}{\end{sidewaystable}}
    \newenvironment{sidefig}{\begin{sidewaysfigure}}{\end{sidewaysfigure}}
\else
    
    \newenvironment{sidefig}{\begin{sidewaysfigure}\rotatesidewayslabel}{\end{sidewaysfigure}}
\fi

\setlist[enumerate]{itemsep=0em}
\setlist[itemize]{itemsep=0em}

\DeclareCaptionLabelSeparator{contcols}{ (columns continue on next page): }

\makeglossaries
\renewcommand{\glossarysection}[2][]{}

\setglossarystyle{long}
\renewcommand{\glsnamefont}[1]{\textbf{#1}}
\renewcommand{\glspostdescription}{}                %
\renewcommand{\glsnumberformat}[1]{\hspace{1.5mm}(p.$\,$\glshypernumber{#1})}  %

\hypersetup{
  pdftitle = {Acceleration of Sparse and Irregular Workloads through Energy-Efficient Stream Extensions},
  pdfauthor = {Paul Scheffler},
  pdfsubject = {Ph.D. dissertation, ETH Zurich, Zurich, Switzerland},
  pdfkeywords = {...},
  pdfstartview = {Fit},     %
  bookmarksopen = {true},
  bookmarksnumbered = {true},
  bookmarksopenlevel = {0}, %
  colorlinks = {false},  %
  pdfstartpage = 1          %
}

\makeatletter
\def\bstctlcite{\@ifnextchar[{\@bstctlcite}{\@bstctlcite[@auxout]}}
\def\bstctlcite[#1]#2{\@bsphack
  \@for\@citeb:=#2\do{%
    \edef\@citeb{\expandafter\@firstofone\@citeb}%
    \if@filesw\immediate\write\csname #1\endcsname{\string\citation{\@citeb}}\fi}%
  \@esphack}
\makeatother

\ifx\showfeedback\undefined
    \newcommand{\mmp}[1]{#1}
    \newcommand{\mmpcut}[1]{}
\else
    \newcommand{\mmp}[1]{{\textcolor{ieee-bright-lblue-100}{#1}}}
    \newcommand{\mmpcut}[1]{\hspace{-0.3ex}\textcolor{ieee-bright-lblue-100}{\scalebox{.3}[1.4]{$\blacksquare$}}\hspace{-0.3ex}}
\fi

\ifx\showfinal\undefined
    \newcommand{\fin}[1]{#1}
\else
    \newcommand{\fin}[1]{{\textcolor{ieee-bright-lgreen-100}{#1}}}
\fi

\makeatletter
  \def\theHALC@line{\thealgorithm-\theALC@line}
  \def\theHALC@rem{\thealgorithm-\theALC@rem}
\makeatother

\newcommand{\thesauthor}{\mmp{\textbf{P.~Scheffler}}}

\begin{document}
\bstctlcite[@auxout]{IEEEexample:BSTcontrol}  %

\newcolumntype{R}{>{\raggedleft\arraybackslash}X}

\title{Acceleration of Sparse and Irregular Workloads through Energy-Efficient Stream Extensions}
\author{Paul Scheffler}

\dissnum{31794}
\acdegree{DOCTOR OF SCIENCES \\ (Dr. sc. ETH Zurich)}
\dateofbirth{18.10.1996}
\citizen{Luxembourg}
\examiner{Prof.\ Dr.\ Luca\ Benini}
\coexaminer{Prof.\ Dr.\ Miquel Moretó Planas}
\thesisyear{2026}

\volume{\tdval{-}}
\isbn{\tdval{?-?????-???-?}}         %
\isbnLong{\tdval{???-?-?????-???-?}} %
\published{2026}

\frontmatter

\maketitle       %

\selectlanguage{english}
\clearpage
\chapter{Acknowledgments}
\label{chap:acknowledgments}

Working toward my Ph.D. was the greatest adventure I've embarked on so far.
In a mere five years, it taught me more about engineering, research, and life itself than I ever thought I could learn, constantly keeping me at the edge of my seat by presenting me with new and exciting challenges to overcome.
However, I was never alone on this transformative journey;
in fact, none of it would have been possible without the support of my mentors, collaborators, friends, and family who accompanied me each step of the way.

First and foremost, I want to thank my supervisor, Professor Luca Benini, for expertly guiding me through even the toughest of challenges and giving me the chance to be a small part of an incredible research group.
You granted me the exceptional opportunity to contribute not only to cutting-edge research, large multinational projects, and multiple silicon demonstrators, but also to some of the greatest open-source hardware in existence.
Above all else, it's thanks to your tremendous expertise, experience, and patience that I made it this far.

I am grateful to Professor Torsten Hoefler for sharing his profound insights on sparse workloads and their acceleration, as well as his contributions to and feedback on my very first publication.
I would like to thank Professor Miquel Moretó Planas for co-refereeing my thesis and providing his thorough and insightful feedback on it.

I am deeply thankful to my friends and office mates Thomas, Nils, Matteo, and Philippe for making every day at work as pleasant, productive, and enriching as possible.
You made J85 a truly magical place, and not just because it was a veritable hardware forge equipped for any eventuality: together, we created and achieved things that I will forever be proud of.
Thank you for our daily support, both on our main and numerous side quests, as well as inside and outside of work.

In particular, none of this would have been possible without my good friend Thomas.
Throughout the eleven years we've known each other and spent together at ETH, you have supported me in more endeavors than I can even remember, working with me on countless assignments, projects, student supervisions, papers, tapeouts, and even teaching me how to drive.
You saved me many times when I was at my wit's end, and I cannot thank you enough.

I want to thank all of my numerous colleagues and co-authors over the years in Zurich and Bologna for our fruitful collaborations.
I owe thanks to Thomas, Gianna, Chi, Alessandro, Simone, Luca, Bowen, Florian, and Fabian for partaking, to some extent, in my strange obsession of accelerating sparse and irregular workloads.
I thank Cristian, Michael, Robert, Sergio, Christopher, and Enrico for always having an open ear for technical and personal worries alike.
I am grateful for Philippe's pioneering work on pushing end-to-end open-source silicon to new heights, giving renewed meaning to the many hours I spent maintaining and improving our open-source IPs.

I am tremendously thankful to Dr.~Frank Kagan G\"urkaynak for leading the Microelectronics Design Center (DZ), serving as my co-supervisor, and teaching me so much of what I know about digital VLSI design.
Thank you for not only making every last one of our tapeouts possible, but also for your tireless efforts on managing projects, coordinating with external partners, and teaching this wonderful craft to entire generations of students.
I want to thank the entire DZ crew, Frank, Beat, Zerun, Alfonso, and Arianna, for their continued work on maintaining a state-of-the-art EDA environment and enabling chip and system design on a scale unmatched by other universities.
Thank you to Christoph, Adam, Mateo, Hansj\"org, and Thomas for your excellent technical support, from IT issues to getting PCBs manufactured.

Last, but certainly not least, I want to thank my family for fervently supporting me in everything I do and never giving up on me.
I want to thank my mother and brother for always being there for me whenever I was sick, sad, in trouble, or entirely lost in life, and stopping at nothing to pull me out of whatever holes I dug myself into.
I want to thank my father for inspiring me to pursue a Ph.D., and for fundamentally shaping my platonic ideal of what a good academic should be.

\begin{comment}

- Luca Benini
- Prof. Dr. Miquel Moretó Planas

- Torsten Hoefler

- Thomas
- J85
- Various Colleagues

- DZ and IT, and Frank, Hansjoerg and Thomas

- Partners of EPI SGA2

- Family
\end{comment}

\clearpage
\chapter{Abstract}
\label{chap:abstract}

The slowdown of Moore's law and Dennard scaling has motivated computer architects to uphold performance scaling through architectural specialization.
So far, most of their efforts have focused on \emph{highly regular} workloads, which operate on contiguous data and exhibit good data locality, ample reuse, and a data-independent execution flow.
Driven by an ever-growing demand, hardware designers have tuned their architectures to efficiently handle large volumes of highly regular compute, resulting in the wide and heavily vectorized datapaths of today's CPUs, GPUs, and domain-specific accelerators.

As this specialization is hitting diminishing returns and compute problems continue to grow in size and complexity, attention is shifting towards \emph{sparse} and \emph{irregular} workloads.
These workloads are essential to large-dataset applications like sparse neural networks, multiphysics simulations, and graph analytics.
However, they lack the convenient properties of highly regular compute that modern architectures are optimized for, resulting in very low (<\SI{10}{\percent}) functional unit utilization and poor performance on traditional CPUs and GPUs.

Many recent proposals tackle this inefficiency in software or hardware.
However, most are specialized to a specific application or domain, and few aim for the \emph{general} and \emph{flexible} acceleration of irregular workloads while maintaining high efficiency on regular workloads.
The most promising approaches to general hardware acceleration leverage \emph{memory streams}, which decouple data movement from execution and enable both high compute and bandwidth utilization.

In this thesis, we explore the design and programming of energy-efficient hardware stream extensions that significantly accelerate sparse and irregular workloads in general-purpose systems \emph{without} specializing to specific application domains or compromising on regular-workload performance.
We follow a bottom-up approach.

We begin by designing a flexible, core-coupled stream extension that supports both affine and \emph{indirect} streams, as well as the \emph{intersection} and \emph{union} between sparse tensors.
We then integrate this extension into an existing energy-efficient multicore RISC-V cluster and evaluate it on general sparse linear algebra, achieving significant performance and energy efficiency benefits.
To demonstrate the generality of our solution, we further use it to accelerate stencil computations, spiking neural network inference, and model-predictive control.
We evaluate our extension at scale \fin{in} a 432-core dual-chiplet, dual-HBM2E 2.5D \mbox{RISC-V} system fabricated in a \SI{12}{\nano\meter} FinFET process, achieving leading compute utilization and energy efficiency on sparse \fin{and irregular} workloads while remaining competitive on regular workloads.
Finally, we extend chip-level interconnects with \emph{end-to-end} irregular streams, improving performance, energy efficiency, and bus utilization through bus packing and near-memory stream handling.

\selectlanguage{ngerman}
\glsresetall
\clearpage
\chapter{Zusammenfassung}
\label{chap:zusammenfassung}

Die Entschleunigung des Mooreschen Gesetzes und der Dennard-Skalie\- rung motiviert Computerarchitekten dazu, die bisherige Leistungsentwicklung durch architektonische Spezialisierung aufrechtzuerhalten.
Bislang konzentrierten sich diese Bemühungen auf \emph{hochgradig regelmässige} Rechenlasten, die zusammenhängende Daten verarbeiten und eine gute Lokalität und Wiederverwendung von Daten sowie einen datenunabhängigen Programmfluss aufweisen.
Aufgrund stetig wachsender Nachfrage haben Hardwareentwickler ihre Architekturen für die effiziente Verarbeitung regelmässiger Rechenlasten in grossen Mengen optimiert, was zu den breiten und stark vektorisierten Datenpfaden heutiger CPUs, GPUs und Hardwarebeschleuniger geführt hat.

Da diese Spezialisierung an ihre Grenzen stösst und Rechenprobleme immer grösser und komplexer werden, verlagert sich die Aufmerksamkeit auf \emph{unregelmässige} und \emph{dünnbesetzte} Rechenlasten.
Diese sind für Anwendungen mit grossen Datensätzen wie dünnbesetzte neuronale Netze, Multiphysik-Simulationen und Graphenanalytik unerlässlich.
Jedoch fehlen ihnen die günstigen Eigenschaften hochgradig regelmässiger Berechnungen, für die moderne Architekturen optimiert sind, was zu einer sehr geringen (<\SI{10}{\percent}) Auslastung von Funktionseinheiten und niedriger Rechenleistung auf herkömmlichen CPUs und GPUs führt.

Viele neuartige Ansätze begegnen dieser Ineffizienz in Software oder Hardware;
die meisten sind jedoch auf bestimmte Anwendungen oder Bereiche spezialisiert, und nur wenige bestreben die \emph{allgemeine} und \emph{flexible} Beschleunigung unregelmässiger Rechenlasten unter Beibehaltung hoher Effizienz für regelmässige Rechenlasten.
Die vielversprechendsten Ansätze zur allgemeinen Hardwarebeschleunigung nutzen \emph{Speicherströme}, welche die Datenbewegung von der Berechnung entkoppeln und sowohl eine hohe Rechenauslastung als auch eine hohe Bandbreitennutzung ermöglichen.

In dieser Dissertation erforschen wir den Entwurf und die Programmierung energieeffizienter Speicherstrom-Hardwareerweiterungen, die unregelmässige und dünnbesetzte Rechenlasten in Allzwecksystemen erheblich beschleunigen, ohne sich auf bestimmte Anwendungsbereiche zu spezialisieren oder Kompromisse bei der Rechenleistung regelmässiger Rechenlasten einzugehen.
Wir verfolgen einen Bottom-up-Ansatz.

Zunächst entwerfen wir eine flexible, kerngekoppelte Erweiterung, die sowohl affine und indirekte Speicherströme als auch die laufende Schnittmenge und Vereinigung dünnbesetzter Tensoren beschleunigt.
Anschliessend integrieren wir unsere Erweiterung in ein existierendes energieeffizientes Mehrkern-RISC-V-Cluster und evaluieren sie in der Beschleunigung allgemeiner dünnbesetzter linearer Algebra, wobei wir erhebliche Leistungs- und Energieeffizienzsteigerungen erzielen.
Um die Flexibilität unserer Lösung zu demonstrieren, nutzen wir sie ausserdem zur Beschleunigung von Stencil-Berechnungen, der Inferenz von gepulsten neuronalen Netzen sowie modellprädiktiver Regelung.
Wir evaluieren unsere Erweiterung im Ausbau auf ein 2.5D-RISC-V-System mit 432 Kernen auf zwei Chiplets und zwei HBM2E-Stacks, welches in einem \SI{12}{\nano\meter}-FinFET-Prozess hergestellt wurde. Dabei erreichen wir eine führende Rechenauslastung und Energieeffizienz bei unregelmässigen Rechenlasten, während regelmässige Rechenlasten effizient bleiben.
Schliesslich erweitern wir auch Chip-weite Interconnects mit unregelmässigen Speicherströmen, was den Rechendurchsatz, die Energieeffizienz und die Bus-Auslastung durch Bus-füllende und speicherseitige Ansätze verbessert.

\selectlanguage{english}

\tableofcontents  %

\mainmatter
\setlength{\baselineskip}{12pt plus 0.33pt minus 0.12pt}
\setlength{\parskip}{0pt plus 3pt}

\glsresetall

\clearpage
\chapter{Introduction}
\label{chap:introduction}

\section{Motivation}
\label{chap:introduction:motivation}

The slowdown of Moore's law and Dennard scaling has motivated computer architects to uphold performance scaling trends through architectural specialization \cite{Fuchs2019TheAW}. %
\emph{Data-parallel compute-intensive workloads} in domains such as data analytics, \gls{ml}, and computational physics are an increasingly ubiquitous specialization target, finding deployment at any scale from edge nodes \cite{Rossi202144A1} and mobile devices to data centers and supercomputers \cite{sato2020a64fx}.

So far, \mmp{many} of these workloads \mmp{were} dominated by \emph{highly regular} computation, wherein large chunks of processed data are kept in homogeneous, randomly addressable arrays iterated over in predictable, affine patterns. %
Common application domains include \glspl{dnn}, image processing, and Monte Carlo methods, which mostly involve dense \gls{la}, convolutions, and embarrassingly parallel (e.g. element-wise) operations.
These workloads exhibit good temporal and spatial data locality, enabling ample and trivial data reuse. Their execution flow is usually data-oblivious and therefore amenable to static optimization, and the processed data can easily be partitioned for parallelization. %

Driven by an ever-growing demand \cite{openai2018aicompute}, hardware designers have continuously tuned their architectures to increasing volumes of highly regular compute.
This resulted in widening \gls{simd} datapaths \cite{stephens2017sve, sodani2016knl}, expanding vector capabilities in high-performance \cite{stephens2017sve, sodani2016knl} and embedded \cite{Gobieski2019MANICAV, Johns2020AMR} processors alike, as well as the emergence of \glspl{gpu} as universal massively parallel compute devices \cite{Navarro2014ASO}.
It also led to a notable increase in the development of highly specialized accelerators that aim to minimize overall compute \mmp{energy~\cite{Fuchs2019TheAW}}.

As the specialization to such workloads is hitting diminishing returns~\cite{Fuchs2019TheAW}, and compute problems continue to grow in size and complexity \cite{openai2018aicompute, hoefler2021sparsity}, attention is shifting towards \emph{sparse and irregular workloads}~\cite{hoefler2021sparsity, Zhang2025Survey, Dave2021HardwareAO}.
Example applications include graph analytics, sparse and graph neural networks, and computational physics.
While still data-parallel and compute-intensive, these workloads lack the convenient properties of regular compute:
they use heterogeneous multi-level data structures to encode dimensionally vast, sparse data, resulting in dynamic, non-contiguous memory access patterns during data traversal.
Accordingly, their data locality is nontrivial and limited, their memory access patterns are data-dependent, and their heterogeneous data is hard to partition for efficient parallelization.

These workloads pose a major challenge to general-purpose \glspl{cpu} \cite{Xie2018CVREV, Talati2021ProdigyIT, Yang2021SpZipAS, liu2013knc} and \glspl{gpu} \cite{Shi2020EfficientSM, Greathouse2014EfficientSM} optimized for highly regular compute:
to achieve decent performance and energy efficiency, their \gls{simd} datapaths require contiguous data chunks absent from most irregular workloads, and their functional units must be highly utilized despite the significant control overhead of traversing sparse data structures.
Additionally, the deep, tiered-storage memory hierarchies \mmp{that modern \gls{cpu} and \gls{gpu}} architectures rely on are ill-equipped to handle irregular access patterns \cite{Talati2021ProdigyIT, Yang2021SpZipAS}, resulting in long latencies, poor bandwidth utilization, and cache thrashing.

Motivated by these inefficiencies in established architectures, the goal of this thesis is to explore the design and application of \emph{energy-efficient hardware stream extensions} to accelerate sparse and irregular workloads on general-purpose systems.
The remainder of our introduction will be structured as follows.
We will first provide the reader with an overview of sparse and irregular workloads (\cref{chap:introduction:workloads}), existing hardware and software acceleration approaches for these workloads (\cref{chap:introduction:acceleration}), and hardware stream extensions (\cref{chap:introduction:hwsext})%
.
We will then outline the chapters of this thesis  (\cref{chap:introduction:outline}), providing a summary for each and explaining how they relate to each other, and present our contributions (\cref{chap:introduction:contributions}).
Finally, we will provide a list of papers the author has published while working toward \mmp{his} degree and explain how they relate to this thesis (\cref{chap:introduction:publications}).

\section{Sparse and Irregular Workloads}
\label{chap:introduction:workloads}

Before providing an overview of existing acceleration approaches, we will first define and discuss the concept of sparsity, sparse and irregular workloads, and the representation and storage of sparse data structures.

\topar{Sparsity} An $n$-dimensional linear tensor $T \in \mathcal{X}^{k_1\times ... \times k_n}$ with elements in $\mathcal{X}$ is called \emph{sparse} if most of its elements are \emph{zero}~\cite{Kjolstad2017taco}.
Sparse tensors are commonly used to express linear transforms or relational models;
depending on the problem domain, they can vary significantly in their dimensions, density, and nonzero distribution.
For instance, weight tensors in sparse \gls{ml} usually exhibit \emph{low} and \emph{structured} sparsity (\SIrange{50}{90}{\percent}, organized in small blocks)~\cite{hoefler2021sparsity}, while problem matrices in computational physics and graph analytics exhibit \emph{high} and \emph{unstructured} sparsity (\SIrange{95}{99.99}{\percent})~\cite{Davis2011TheUO}.
We can generalize the notion of sparsity to generic \emph{sparse data structures}, in which only some elements in an indexable space are explicitly defined and all others have an implicit default value.

\topar{Sparse and irregular workloads} A \emph{sparse workload} processes sparse data structures as defined above.
An \emph{irregular workload} involves irregular memory access sequences or irregular control flow~\cite{Burtscher2012irreg,Dadu2019TowardsGP}.
Irregular \emph{memory access sequences}, unlike regular ones, are not piecewise contiguous and cannot be generated using an affine expression of nested linear iterators (i.e. those of regular \texttt{for} loops).
An irregular \emph{control flow} performs compute operations in an irregular order.
If a workload involves \emph{data-dependent} memory access sequences or a data-dependent control flow, it is inherently irregular~\cite{Dadu2019TowardsGP}.

Sparse workloads are also generally irregular.
For instance, when computing the dot product of a sparse and a dense vector, it is most efficient to read the dense vector only where the sparse vector is \fin{nonzero}, forming an irregular and data-dependent memory access sequence~\cite{Dadu2019TowardsGP}.
When both vectors are sparse, computation is necessary only where both vectors are nonzero, also making the control flow irregular~\cite{Dadu2019TowardsGP,Hegde2019ExTensorAA}.
However, the inverse is not true: not all irregular workloads involve sparse data or sparse \gls{la}.
For example, stencil computations with irregular stencil shapes~\cite{Yount2015VectorFI} operate on dense data grids, but access grid elements using irregular memory access sequences.

\mmp{In this thesis, we sometimes use \emph{general sparse \gls{la}} as a shorthand expression for \gls{la} computations that may involve addition, multiplication, element-wise multiplication, or scaling among sparse and dense vectors or matrices with any shape, density, or nonzero distribution.}
\mmp{By extension, \emph{general sparse \glsunset{ta}tensor algebra}~\fin{(TA)} may involve any of these operations on multidimensional dense and sparse tensors of any kind.
For a system to \fin{efficiently handle} general sparse \gls{ta}, it must fulfill the criteria proposed by Hsu et al. for general and efficient sparse \gls{ta} computing systems~\cite{hsu2023sam}.}
\mmp{Finally, by \emph{general irregular workloads}, we mean irregular workloads of any kind; in accordance with our definitions above, this includes workloads that \emph{cannot} be mapped directly to sparse \gls{la} or \gls{ta} operations.}

\topar{Sparse storage formats} Sparse tensors and other \fin{sparse} data structures can be stored in various formats.
The choice of format can significantly affect the underlying algorithms, performance, and memory footprint of the workloads processing them~\cite{formats_chou}.

\emph{Coordinate-based formats} like COO~\cite{formats_chou} store nonzeros and their full $n$-dimensional coordinates as tuples in an array.
They scale well to extreme sparsity, such as in large graphs.
However, their metadata overhead is high: depending on the dimension count, size, and data precision, the coordinates likely take up more space than the nonzero values.
\emph{Index-based formats} like CSR~\cite{eisenstat1977csr}, \mmp{CSC~\cite{duff1989csc},} and CSF~\cite{Smith2015TensormatrixPW} store nonzeros and their coordinates on a primary axis in two separate arrays;
higher-level axes are delimited by additional pointer arrays, \mmp{enabling random accesses}.
While somewhat less flexible, these formats significantly lower metadata overheads for medium- to high-sparsity data, as the coordinates of higher-level
axes are compressed.
Finally, \emph{Bitmap-based formats}~\cite{Gondimalla2019SparTenAS,Wang2021DualsideST,Chen2024Bitmap,kanellopoulos2019smash} store the positions of nonzeros using bitmaps.
These cannot scale to medium or high sparsities as their metadata scales with their dense size, which is often prohibitively large;
however, they provide low overhead for low-sparsity use cases like \gls{ml} weight storage.
Overall, indexed formats are the most scalable, albeit with limited efficiency in low or extreme sparsity regimes.

Numerous proposed variations of these format types aim to improve workload performance or storage efficiency.
\emph{Slicing}~\cite{dang2012scoo,Sadi2019EfficientSO} and \emph{tiling}~\cite{Niu2021TileSpMVAT} formats partition large sparse tensors into a set of smaller ones, trading efficiency for memory locality and potentially lower metadata overhead.
\emph{Block-sparse} formats~\cite{Pinar1999ImprovingPO,Zachariadis2020AcceleratingSM} encode nonzero blocks of elements, accepting the inclusion of redundant zero elements to better utilize wide datapaths.
Recently, $N$:$M$ \emph{structured sparse} formats emerged as an efficient approach to \mmpcut{efficiently }handle the low, highly structured sparsity of \gls{ml} tensors~\cite{Castro2023VENOMAV,nvidia2020a100tc}: in each block of $M$ elements, exactly $N$ are nonzero, and higher sparsities are supported through redundant zero elements. Nonzero positions can be stored in few-bit indices as $M$ is typically small. In recent Nvidia GPUs, $N=2$ and $M=4$ is supported in hardware~\cite{nvidia2020a100tc}.

\section{Existing Acceleration Approaches}
\label{chap:introduction:acceleration}

Existing acceleration approaches for sparse and irregular workloads tackle their inefficient compute through software, hardware, and co-design solutions.
In this section, we provide a high-level overview of \gls{soa} acceleration methods.
Individual solutions will be discussed in further detail and compared to our contributions in the related work sections of \Cref{chap:sssrs,chap:workloads,chap:occamy,chap:chiplevel}.

\topar{Software Libraries}
Software acceleration on commercial CPUs and GPUs is usually provided through specialized libraries.
For generic sparse \gls{la}, \mmp{in addition to portable solutions like \fin{\emph{BLIS}}~\cite{vzee2015blis}}, some hardware vendors provide highly optimized, hardware-specific subroutine libraries like Intel's \emph{oneMKL}~\cite{intel2020onemkl}, AMD's \emph{AOCL-Sparse}~\cite{Amd2020aoclsparse}, or Nvidia's \emph{cuSPARSE}~\cite{nvidia2018cuda}.
These can be used directly or leveraged through higher-level math software like \emph{MATLAB}~\cite{mathworks1984matlab}, \emph{SciPy}~\cite{Virtanen2020SciPy}, or \emph{Eigen}~\cite{guennebaud2010eigen}.
Beyond generic \gls{la}, there are also specialized solutions for specific domains like high-dimensional \gls{ta}~\cite{Kjolstad2017taco,ahrens2025finch}, graph processing~\cite{davis2019ssgblas}, or stencil codes~\cite{Matsumura2020AN5DAS,You2021DRStencilED,Zhang2023RevisitingTB}.
Sparse \gls{ml} is usually handled through existing \gls{ml} frameworks; these, in turn, rely on both generic vendor \gls{la} libraries and specialized libraries like \emph{cuSPARSELt}~\cite{nvidia2020cusparselt} that target \gls{ml}-specific structured sparsity.

\Gls{soa} research on software acceleration focuses primarily on optimizing the underlying algorithms and sparse storage formats for existing CPU and GPU architectures to increase performance.
Most works target generic sparse \gls{la} kernels~\cite{Xie2018CVREV,Zhang2018VectorizedPS,Fei2021Regu2DAV,Niu2021TileSpMVAT,Merrill2016MergeBasedPS,Sgherzi2024spchar}, sparse \gls{ml}~\cite{Sun2022SpDNN,gnn_spmm_sc20}, or specific workload types like stencil codes~\cite{Yount2015VectorFI,You2021DRStencilED,Zhang2023RevisitingTB}.
Common optimization approaches include tuning access patterns for tiered memory hierarchies~\cite{Xie2018CVREV,Karimi2022VCSRAE,Shi2020EfficientSM}, leveraging architectural \gls{simd} and vector features~\cite{Zhang2018VectorizedPS,Xie2018CVREV,Fei2021Regu2DAV,Yount2015VectorFI}, and effective parallelization~\cite{Merrill2016MergeBasedPS,Niu2021TileSpMVAT}.
Thus, most \gls{soa} software solutions forego portability to improve performance for specific architectures and applications, making it difficult for them to build on each other.

\topar{Hardware Accelerators} While targeted software optimizations can yield important performance benefits, disruptive improvements require tackling the underlying inefficiencies in hardware.

Numerous \glspl{dsa} for sparse and irregular workloads have already been proposed.
Sparse \gls{ml} accelerators are particularly common, culminating in vast design~\cite{Dave2021HardwareAO} and training methodology surveys~\cite{hoefler2021sparsity}.
These accelerators usually aim to maximize energy efficiency and per-area throughput for a specific inference application, with some having already been demonstrated in silicon~\cite{huang2025hybr, yue2024cimblock, feng2024pntcloud, zhang2021snap}.
However, they are highly specialized for inference on particular network types: they target specific operations, number formats, and domain-characteristic (low or structured) sparsity regimes.
\mmp{Another common target for \glspl{dsa} are sparse system solvers~\cite{hao2025telos,feldman2024azul,feldmann2023spatula}, which exhibit a similarly high degree of specialization: they target particular problems with their own domain-specific compute flow and structural sparsity assumptions.}
\mmp{As a result, these designs cannot accelerate general sparse and irregular workloads.}

More flexible designs targeting generic sparse~\gls{la}~\cite{Srivastava2020MatRaptorAS, zhang2021gamma, Pal2018OuterSPACEAO, park2020output, Sadi2019EfficientSO,kanellopoulos2019smash,yang2024trapezoid,he2020sptpu} or \gls{ta}~\cite{Hegde2019ExTensorAA,zhong2024feasta} have also been proposed.
These accelerators are not bound to a specific workload or application domain.
However, most are still hardwired for specific operations~\cite{Srivastava2020MatRaptorAS, Pal2018OuterSPACEAO, park2020output, Sadi2019EfficientSO} or dataflows~\cite{Srivastava2020MatRaptorAS, zhang2021gamma, Pal2018OuterSPACEAO}.
They also usually accelerate either one-sided~\cite{Sadi2019EfficientSO,Song2022Serpens} or two-sided sparse computations~\cite{Srivastava2020MatRaptorAS,Pal2018OuterSPACEAO,Hegde2019ExTensorAA,zhong2024feasta}, but rarely both.
Thus, these designs can still only accelerate a subset of sparse workloads.

Recently, highly flexible accelerators based on \glspl{cgra} have been proposed~\cite{Dadu2019TowardsGP,koul2024onyx,chen2025opal}.
These provide simple, interconnected, programmable \glspl{pe} operating on exchanged data streams.
Their dataflow-oriented architecture makes them almost as flexible as general-purpose processors in handling sparse and irregular workloads while enabling higher efficiency.
However, \glspl{cgra} still cannot replace general-purpose processors as they provide limited compute capabilities~\cite{koul2024onyx,chen2025opal} and are challenging to program productively~\cite{liu2019cgrasurvey}.

\topar{ISA Extensions} Accelerators provide excellent performance and energy efficiency, but cannot replace general-purpose processors. A promising idea is therefore to extend existing processors and their \glspl{isa} to improve performance and energy efficiency on sparse and irregular workloads.

In CPUs, one-sided sparsity and static irregular memory access sequences can be accelerated with vector \emph{scatter-gather} instructions, which are also found in recent application-class cores.
For instance, Intel's \emph{Knights Landing} \glsunset{knl}(\gls{knl})~\cite{sodani2016knl} and Arm's \emph{Scalable Vector Extensions} \glsunset{sve}(\gls{sve})~\cite{stephens2017sve} extend the x86-64 and AArch64 \glspl{isa} with support for \gls{simd} scatter-gather, respectively.
\mmp{The RISC-V \emph{vector extension}~\glsunset{rvv}(\gls{rvv})~\cite{riscv2021rvv} \fin{adopts} a vector-length-agnostic programming model and relaxes data alignment requirements.}
\mmp{However, these scatter-gather instructions are ultimately \fin{still} constrained in their vector length,} cannot accelerate two-sided sparsity or dynamic access sequences, and only improve performance when implemented in a performant fashion.
CPU extensions proposed in \gls{soa} research improve on scatter-gather by tackling its limitations~\cite{Domingos2021UnlimitedVE} or more directly targeting irregular workloads~\cite{Gong2020SAVESV,Rao2022SparseCoreSI,Yang2021SpZipAS,pavon2021via,lee2024terminus}.
In both cases, there is a trend toward leveraging \emph{memory streams} to accelerate sparse and irregular workloads, as we do in this thesis. We will discuss this in more detail in \Cref{chap:introduction:hwsext}.

GPU extensions focus on sparse \gls{ml}. With its A100 GPU~\cite{nvidia2020a100tc}, Nvidia has introduced one-sided $2$:$4$ structured sparsity support, accelerating low-precision \gls{ml} inference tasks by up to 2\x~over dense computation.
AMD has \mmpcut{since }adopted $2$:$4$ sparsity support in MI300X~\cite{Smith2024MI300X}.
Unfortunately, $2$:$4$ encoding is highly specialized for the low, structured sparsity of sparse \gls{ml}; it cannot meaningfully accelerate the unstructured medium-to-high sparsity of many other sparse workloads, which are usually tackled through specialized software \fin{on GPUs}~\cite{Tsai2020SparseLA,Niu2021TileSpMVAT,Merrill2016MergeBasedPS}.
Thus, some recent works on GPU hardware extensions aim to accelerate sparsity in a more general fashion~\cite{Liu2023TSTC,zhu2019sparsetensorcore,Wang2021DualsideST}.

\section{Hardware Stream Extensions}
\label{chap:introduction:hwsext}

As indicated by the trends observed in \Cref{chap:introduction:acceleration}, the most promising approach to effectively accelerating sparse and irregular workloads may be \emph{stream extensions} \cite{Domingos2021UnlimitedVE, Wang2019StreambasedMA, Talati2021ProdigyIT, Yang2021SpZipAS, Schuiki2021StreamSR}. These move data between a general-purpose processor and memory in \emph{streams}, using programmable hardware units to generate addresses in a decoupled fashion.
They maintain all the functionality and flexibility of the extended processor, but also free it from micromanaging memory accesses and hide memory access latencies, significantly increasing compute throughput and bandwidth efficiency.

\emph{Affine} and \emph{indirect} streams, which cover a majority of streaming access patterns in common benchmark workloads \cite{Wang2019StreambasedMA}, can accelerate not only dense and one-sided sparse \gls{la}, but also some nonlinear workloads with irregular memory access sequences.
By generating streams from each other~\cite{Yang2021SpZipAS} or \emph{joining} them using ordered set operations (e.g. intersection or union)~\cite{Rao2022SparseCoreSI}, they can also accelerate two-sided sparse computations and complex irregular control flows.

Streams can be set up at runtime by programming a streaming unit with their \emph{semantics} (e.g. base address, strides, or data width).
Stream elements are either prefetched into a cache~\cite{Talati2021ProdigyIT} or provided directly to the processor through push/pop instructions or mapped registers~\cite{Wang2019StreambasedMA, Schuiki2021StreamSR}.
They can also be packed into vector registers for efficient use with \gls{simd} or vector processors~\cite{Domingos2021UnlimitedVE}.

What makes stream extensions unique is their high flexibility and \fin{programmability. They} %
are not bound to a particular storage format, algorithm, or iteration order, making them \emph{general-purpose} extensions~\cite{Domingos2021UnlimitedVE, Wang2019StreambasedMA, Talati2021ProdigyIT}.
They can be tightly coupled to cores, sharing compute and memory system resources and minimizing their latency and area footprint~\cite{Schuiki2021StreamSR}, or implemented at the interconnect level~\cite{Talati2021ProdigyIT,Yang2021SpZipAS}.
The manipulation of streams may be freely \fin{combined and interleaved} with regular instruction execution \cite{Wang2019StreambasedMA,Schuiki2021StreamSR}.
\mmp{Stream semantics can be explicitly programmed or inferred by a compiler \cite{Talati2021ProdigyIT};}
\mmp{in fact, some recent compiler works are dedicated to the efficient mapping of irregular workloads to abstract hardware streams~\cite{hsu2023sam,siracusa2026ember}.}

Therefore, in this thesis, we aim to design \emph{energy-efficient} hardware stream extensions and leverage them for the \emph{general} acceleration of sparse and irregular workloads on general-purpose systems.

\section{Thesis Outline}
\label{chap:introduction:outline}

\begin{figure}
    \centering
    \includegraphics[width=0.85\linewidth]{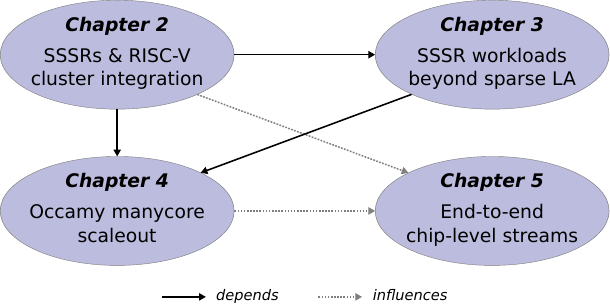}
    \caption{Overview of the thesis chapters and how they relate.}
    \label{fig:introduction:overview}
\end{figure}

In this section, we briefly outline the following chapters of this thesis and how they relate to each other; a high-level overview is depicted in \cref{fig:introduction:overview}.
Most of the discussed material has previously been published in peer-reviewed conference proceedings and journals. We provide a full list of related publications in \cref{chap:introduction:publications}.

\paragraph*{\textbf{Chapter 2}} We first design a lightweight core-coupled irregular stream extension called \emph{sparse stream semantic registers} (\glsunset{sssr}\glspl{sssr}).
In addition to dense compute, \glspl{sssr} can accelerate sparse-dense and sparse-sparse workloads through streaming indirection, intersection, and union of indexed tensors.
We integrate and evaluate our extensions in the existing eight-core RISC-V \emph{Snitch} cluster~\cite{Zaruba2020SnitchAT}, incurring minimal area overhead.
Comparing the \gls{sssr}-extended cluster to one using the baseline RV32G instruction set, we observe significant speedups and increases in functional unit utilization and energy efficiency across general sparse-dense and sparse-sparse \gls{la} kernels.

\paragraph*{\textbf{Chapter 3}} To demonstrate the \mmp{flexibility} of \glspl{sssr}, we use them to accelerate and improve the energy efficiency of irregular workloads whose performance-critical compute goes \emph{beyond} sparse \gls{la}.
We first describe our work on extending the Snitch cluster's LLVM toolchain and runtime to better support \glspl{sssr}.
We then present a generic methodology to accelerate stencil computations with register-mapped indirect stream extensions, and apply it to significantly accelerate various stencil codes using \glspl{sssr}.
Next, we leverage \glspl{sssr} to accelerate the sparse reduction of input currents in spiking neural networks.
Finally, we use \glspl{sssr} to accelerate the sparse compute of model-predictive control, enabling the real-time energy and thermal management of large application-class manycores using a single memory-extended cluster.

\paragraph*{\textbf{Chapter 4}} Having demonstrated the performance and energy efficiency benefits of \glspl{sssr}, we scale out our extended compute cluster to a hierarchical 2.5D manycore accelerator called \emph{Occamy}.
Fabricated in GlobalFoundries' \SI{12}{\nano\meter} FinFET technology, Occamy is a dual-chiplet 2.5D system with 432 cores, \SI{32}{\gibi\byte} of HBM2E \mmp{\gls{dram}}, and a peak performance of \SI{768}{DP\text{-}\giga\flop\per\second}.
It efficiently and flexibly \mmp{handles both regular and irregular compute}, avoiding the inefficiencies of traditional cache hierarchies through explicit  data movement using \gls{dma} engines.
We scale up our existing sparse \gls{la} and stencil workloads and implement further \gls{ml} applications to show that Occamy achieves leading \gls{fpu} utilization and energy efficiency on irregular compute while maintaining competitive efficiency on regular compute.

\paragraph*{\textbf{Chapter 5}} On Occamy, irregular streams are confined to each cluster's shared on-chip memory; this is highly efficient, but presents a hurdle for workloads with very large and sparse datasets that cannot efficiently be partitioned.
In this chapter, we design \emph{AXI Pack}, an extension for wide-bus chip-level interconnects that enables efficient \emph{end-to-end} irregular streams of narrow data elements between processing elements and memory endpoints.
\emph{AXI Pack} enhances Arm's widespread \gls{axi4} protocol with bus-packed strided and indirect streams handled at memory endpoints, enabling significant speedups and energy efficiency benefits on a RISC-V vector processor system with wide buses.

\paragraph*{\textbf{Chapter 6}} We summarize the core results and achievements of this thesis and provide an outlook, identifying possible directions for future work.

\section{Contributions}
\label{chap:introduction:contributions}

The main contributions of this thesis and its related publications can be summarized as follows:

\begin{enumerate}

\item Design and implementation of \glspl{sssr}, a flexible core-coupled stream extension accelerating irregular workloads by decoupling their memory accesses from execution and eliminating associated overheads. \glspl{sssr} can stream both affine and indirect memory accesses and perform the streaming \emph{intersection} and \emph{union} of indexed sparse tensors.

\item Integration of \glspl{sssr} into an existing energy-efficient multicore {RISC-V} compute cluster and their evaluation on general (both one- and two-sided, varying operand density) sparse \gls{la}, enabling significant performance and energy efficiency benefits in exchange for minimal area overhead.

\item Acceleration of irregular workloads beyond sparse \gls{la} on the \gls{sssr}-extended cluster, demonstrating the generality of \glspl{sssr} and achieving significant performance and energy efficiency benefits on generic stencil computations, spiking neural network inference, and online model-predictive control.

\item Scale-up of \gls{sssr} acceleration to \emph{Occamy}, a 432-core dual-chiplet, dual-HBM2E 2.5D RISC-V system with chiplets fabricated in a \SI{12}{\nano\meter} FinFET node. Occamy achieves leading compute utilization and energy efficiency on sparse and irregular workloads without compromising on regular compute performance.

\item Design and evaluation of \emph{AXI Pack}, an extension to the wide\-spread \gls{axi4} protocol providing efficient \emph{end-to-end}~strided and indirect streams.
\emph{AXI Pack} improves performance~and energy efficiency by maximizing memory throughput and bandwidth efficiency through bus packing and memory-side indirection.

\end{enumerate}

Some of the work discussed in this thesis was developed in collaboration with Thomas Benz, Gianna Paulin, Chi Zhang, Alessandro Ottaviano, Simone Manoni, and Luca Colagrande.
Their respective theses may, in part, cover the same material from a different perspective and with a different focus.

\section{List of Publications}
\label{chap:introduction:publications}

Most of the work presented in this thesis is covered in the following conference and journal publications:

\begin{publications}
    \publication{Scheffler2021IndirectionSS}
    \publication{scheffler2023sssr}
    \publication{zhang2023axipack}
    \publication{scheffler2024saris}
    \publication{scheffler2025occamy}
    \publication{manoni2025spikestream}
    \publication{ottaviano2025mpc}
\end{publications}

The following publications are covered in part or may provide additional insight on the topics discussed in this thesis:

\begin{publications}
    \publication{paulin2022softtiles}
    \publication{ottaviano2023cheshire}
    \publication{benz2024idma}
    \publication{zhang2024coal}
\end{publications}

The following publications with contributions from the author are not covered in this thesis:

\begin{publications}
    \publication{riedel2021banshee}
    \publication{sauter2024insbas}
    \publication{garofalo2025carfield}
\end{publications}

\clearpage
\chapter{Cluster-level Extensions}
\label{chap:sssrs}

In this chapter, we \mmpcut{will }introduce and evaluate the design of our cluster-level sparse streaming extensions, \emph{sparse stream semantic registers} (\glspl{sssr}), and demonstrate them through integration into the existing multi-core RISC-V \emph{Snitch} compute cluster~\cite{Zaruba2020SnitchAT}.

We \mmpcut{will }initially motivate and evaluate our design using \emph{general} sparse \gls{la}. As discussed in \Cref{chap:introduction}, sparse \gls{la} is performance-critical to numerous applications, but challenging to accelerate uniformly across different operations, formats, and sparsity regimes.
In the following \cref{chap:workloads}, we will then demonstrate that \glspl{sssr} are flexible enough to accelerate irregular workloads beyond those based on \gls{la}.

\section{Introduction}
\label{chap:sssrs:introduction}
We recall from \Cref{chap:introduction} that sparse \gls{la} poses a formidable challenge for contemporary CPU and GPU architectures, which are optimized for highly regular and vectorizable workloads.
To reduce memory footprint and eliminate redundant computation, sparse tensors are stored in compressed formats that hold only the values and positions of nonzero elements.
This results in nested indirect data structures that cannot be randomly accessed, irregular memory access patterns, and significant control overheads.
As a consequence, even the most fundamental sparse tensor operations, \emph{addition} and \emph{multiplication}, are inherently irregular. Worse, their optimal execution strategy depends on operand sparsity: while one sparse operand can be handled through indirect memory accesses, two sparse operands require comparing nonzero positions in both tensors.

\glsunset{csr}%
\begin{listing}[t]
  \includegraphics[keepaspectratio, scale=1.215]{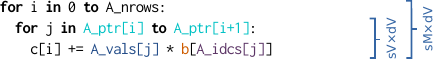}
  \caption{Multiplication of a sparse matrix \texttt{A} in \gls{csr} format with a dense vector \texttt{b} (\texttt{sM×dV}), iterating sparse-dense dot products (\texttt{sV×dV}).}
  \label{lst:sssrs:intro_a}
\end{listing}

Consider the multiplication of a sparse matrix \verb|A|, stored in the widespread \gls{csr} format, with a dense vector \verb|b|, which can be implemented as shown in \Cref{lst:sssrs:intro_a}.
Here, \verb|A_vals| and \verb|A_idcs| store the nonzero values and their column indices, and \verb|A_ptrs| delimits the rows of \verb|A|. For each element of the result vector \verb|c|, we perform a \emph{sparse-dense} dot product \verb|sV×dV|\footnote{We henceforth use the prefixes \texttt{s} and \texttt{d} to denote \emph{sparse} and \emph{dense} tensors, respectively. Analogously, \texttt{M} denotes a \emph{matrix} and \texttt{V} a \emph{vector}. \mmp{While some operations, like \texttt{sV×dV} (SpVV) and \texttt{sM×dV} (SpMV), already have established abbreviations, we use a systematic naming scheme in this chapter for both clarity and convenience.}} as shown.
Compiling this \verb|sV×dV| for the \riscv{} instruction set yields a loop of nine instructions, of which only one, a fused \gls{mac}, performs necessary compute \cite{Scheffler2021IndirectionSS}. 
Thus, an in-order core can achieve at most \SI{11}{\percent} \gls{fpu} utilization%
\footnote{\mmp{Loop unrolling could further eliminate branches and pointer increments, theoretically enabling up to \SI{16.7}{\percent} FPU utilization for infinitely many nonzeros. In practice, this marginal gain is strongly counteracted by unroll setup overheads (especially for few nonzeros) and an increase in instruction cache pressure.}}
, while a superscalar core would need at least nine issue slots to keep a single \gls{fpu} busy, even when ignoring dependencies and control bottlenecks.
In both cases, the main efficiency bottleneck is the \emph{indirection} \verb|b[A_idcs[j]]|, which requires half of the non-compute instructions and is also needed for sparse-dense addition.

\glsunset{csf}%
\begin{listing}[t]
  \includegraphics[keepaspectratio, scale=1.215]{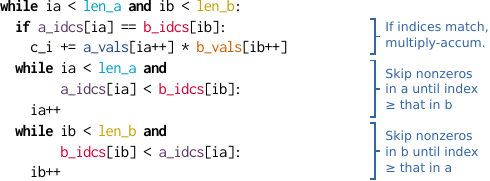}
  \hfill
  \caption{Dot product of two sparse vectors \texttt{a} and \texttt{b} stored in the \gls{csf} format (\texttt{sV×sV}).}
  \label{lst:sssrs:intro_c}
\end{listing}

Now, consider the multiplication of \verb|A| with a sparse vector \verb|b| stored in the common \gls{csf} format. The \emph{sparse-sparse} dot product \verb|sV×sV| computed for each result element can be implemented as shown in \Cref{lst:sssrs:intro_c}. %
We \emph{intersect} the nonzero indices in the two vectors, since only positions where both vectors are nonzero contribute to the result.
Compared to \verb|sV×dV|, the control overhead is even higher and dynamically depends on the sparsity patterns of both operands. Sparse-sparse tensor addition needs yet another approach: we must form a \emph{union} of indices, since a nonzero in either operand produces a  nonzero in the result.

Sparse \gls{la} is essential in many applications~\cite{Zhang2025Survey}. The SuiteSparse matrix collection~\cite{Davis2011TheUO} hosts sparse matrices from numerous real-world problems in domains including computational physics, mathematics, economics, and biology.
In \gls{ml}, model sparsification can significantly reduce the operations and memory required for a given inference accuracy \cite{hoefler2021sparsity}; while some approaches target sparse-dense compute by sparsifying only weights, others also exploit activation sparsity.
Graphs, usually represented as highly sparse matrices, are common operands in both sparse-dense and sparse-sparse workloads, such as PageRank \cite{Page1999ThePC} and triangle counting \cite{Rao2022SparseCoreSI}, respectively.

Established superscalar out-of-order architectures struggle with the high control overhead of sparse \gls{la}.
\mmp{For instance,} a Fujitsu A64FX running \texttt{sM×dV} with a highly optimized sparse matrix format achieves at most \SI{130.9}{\Gflops}, or \SI{4.7}{\percent} of its peak \mmp{compute~\cite{alappat2020a64fx}}.
\mmp{As we will explore in \cref{chap:sssrs:relwork}, this inefficiency is not unique to A64FX, but can be observed across superscalar out-of-order \glspl{cpu}}.
\mmp{Some recent hardware proposals,} %
\mmp{like those we discussed in \cref{chap:introduction:acceleration},} %
aim to improve sparse \gls{la} performance, but most of them are limited by their lack of generality or large microarchitectural impact.
Most general-purpose \gls{isa} extensions consider only one-sided \fin{sparsity~\cite{Talati2021ProdigyIT,Yang2021SpZipAS,Wang2019StreambasedMA},} and those targeting sparse-sparse workloads incur large area \fin{impacts~\cite{pavon2021via,siracusa2023tmu} or} cannot efficiently handle the former~\cite{Rao2022SparseCoreSI}.
Sparse \gls{ml} solutions often rely on custom formats and the domain's unique low and structured sparsity~\cite{Wang2021DualsideST, Gondimalla2019SparTenAS}, while more general sparse \fin{\gls{la}~\cite{Srivastava2020MatRaptorAS,zhang2021gamma,Pal2018OuterSPACEAO} and \gls{ta}~\cite{Hegde2019ExTensorAA, Dadu2019TowardsGP} accelerators} incur significant silicon area impacts compared to their host systems and other solutions.

\begin{listing}[t]
\includegraphics[keepaspectratio, scale=1.215]{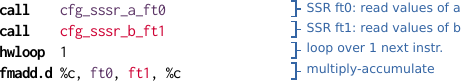}
\caption{\texttt{sV×dV} and \texttt{sV×sV} in RISC-V assembly using \glspl{sssr} and a hardware loop. The kernels differ only in the used \gls{sssr} configuration.}
\label{lst:sssrs:intro_d}
\end{listing}

\glsunset{sssr}
In this chapter, we present \glspl{sssr}, a modular, tightly core-coupled \gls{isa} extension tackling these inefficiencies by significantly accelerating and improving \gls{fpu} utilization on both sparse-dense \emph{and} sparse-sparse \gls{la} \emph{without} major microarchitectural impact.
\Glspl{sssr} %
build on the existing architecture of \glspl{ssr}~\cite{Schuiki2021StreamSR}, which enable near-complete \gls{fpu} utilization in in-order cores for dense linear algebra and other data-oblivious workloads\footnote{A workload whose memory access patterns are input-invariant.}~\cite{Ramachandran2020DataOA} by mapping memory streams with regular address patterns directly to architectural registers.
In addition to affine streams, \glspl{sssr} also accelerate the \emph{indirection}, \emph{intersection}, and \emph{union} operations necessary for general sparse \gls{la}, incurring only a small increase in overall hardware complexity while significantly improving sparse performance. For example, \glspl{sssr} allow us to write both \verb|sV×dV| and \verb|sV×sV| in \riscv{} assembly as in \Cref{lst:sssrs:intro_d}, where the registers \verb|ft0| and \verb|ft1| have stream semantics.
Since all index processing and comparison is handled in hardware, the loop body consists only of the useful \verb|fmadd.d| instruction, which we iterate using a hardware loop.

We evaluate \glspl{sssr} by integrating them into the RISC-V Snitch \cite{Zaruba2020SnitchAT} core and compute cluster, in which we evaluate their benefits on various sparse \gls{la} workloads using generated vectors and real-world sparse matrices.
On a single core, \gls{sssr} indirection, intersection, and union enable speedups of up to 7.0\x, 7.7\x, and 9.8\x~over an optimized RISC-V baseline on sparse-dense multiply, sparse-sparse multiply, and sparse-sparse addition, respectively, with peak indirection \gls{fpu} utilizations of up to \SI{80}{\percent}\footnote{Full \gls{fpu} utilization would require separate index memory ports, which we forgo to minimize our architectural impact (see  \cref{chap:sssrs:arch:integr}).}. On an eight-core cluster, sparse-dense \verb|sM×dV| using \glspl{sssr} is up to 4.9\x~faster and 2.9\x~more energy efficient, while sparse-sparse \verb|sM×sV| is up to 5.9\x~faster and 3.0\x~more energy efficient. \glspl{sssr} incur only 1.8\% in additional area in an eight-core cluster (\SI{128}{\kibi\byte} data memory and \SI{8}{\kibi\byte} instruction cache) over regular \glspl{ssr}.
The synthesizable \gls{rtl} description for \glspl{sssr} and their integration into Snitch are available free and open-source\footnote{\url{https://github.com/pulp-platform/snitch_cluster}: \gls{sssr} IP in {hw/}\\{snitch\_ssr} and integration in {hw/snitch\_cluster}.}.

Our contributions in this chapter are:

\begin{enumerate}
    \item A lightweight, modular \gls{isa} extension handling streaming indirection, intersection, and union in hardware to efficiently handle sparse general \gls{la} in single-issue cores, with a complete open-source hardware implementation in an open RISC-V core and multi-core cluster.

    \item A programming model for \glspl{sssr}, efficient sparse-dense and sparse-sparse \gls{la} kernels operating on vectors and matrices, and an exploration of further accelerable workloads.

    \item Significant performance and energy efficiency benefits on single cores (\mmp{up to $7.0$\x, $7.7$\x, and $9.8$\x~faster \texttt{sM×dV}, \texttt{sV×sV}, and \texttt{sV+sV}}) \mmp{as well as} eight-core clusters with realistic \gls{dram} models (\mmp{up to $4.9$\x, $5.9$\x~faster and $2.9$\x, $3.0$\x~more energy efficient \texttt{sM×dV} and \texttt{sM×sV}}), incurring a minimal \SI{1.8}{\percent} impact on cluster area.

    \item A comparison of \glspl{sssr} to \gls{soa} CPU, GPU, and accelerator approaches, achieving higher generality, a lower area impact, and {1.7}\x~higher parallel \verb|sM×dV| \gls{fpu} utilization than GPUs.
\end{enumerate}

The remainder of this chapter is organized as follows: \Cref{chap:sssrs:arch} describes our \gls{sssr} architecture and its integration into the Snitch core and cluster. \Cref{chap:sssrs:progmodel} presents our \gls{sssr} programming model and sparse \gls{la} kernels. \Cref{chap:sssrs:eval} evaluates the performance, timing, area, and energy efficiency benefits of \glspl{sssr}. Finally, \Cref{chap:sssrs:relwork} discusses related CPU, GPU, and accelerator work and compares it to our approach.

\section{Architecture}
\label{chap:sssrs:arch}

To accelerate sparse workloads, we start from and extend the existing affine address generator of \glspl{ssr}. We present two new generator designs, the \emph{indirection} and \emph{egress} address generators, as well as two extended, backward-compatible \gls{ssr} types using them. We will then present the sparse \gls{ssr} streamer, which combines and connects the index datapaths of three extended \glspl{ssr} to enable index intersection and union. Finally, we will discuss the streamer's integration into the Snitch core and \mmp{cluster~\cite{Zaruba2020SnitchAT},} in which we evaluate our extensions. We note that \glspl{sssr} are not strictly dependent on a specific core or instruction set and can be used with any core, \gls{isa}, or hardware loop.

We designed two extended \gls{ssr} address generators to handle indexed streams. The \emph{indirection} generator (\cref{chap:sssrs:arch:iag}) can \emph{read} an index stream from memory to either generate indirect addresses or match read indices against those of another such generator for intersection or union. The \emph{egress} generator (\cref{chap:sssrs:arch:eag}) can \emph{write} an externally provided index stream to memory while emitting corresponding data addresses.
\mmp{\Cref{chap:sssrs:arch:sssrs} presents the new sparse \gls{ssr} types using these generators, and} \Cref{chap:sssrs:arch:streamer} explains how \mmp{their features combine} to enable end-to-end sparse-sparse index processing.

\begin{figure}%
\captionsetup[subfigure]{margin=0ex}%
\setkeys{Gin}{height=7.2cm}%
\subfloat[Indirection address generator, showing the per-mode datapaths.]{%
\label{fig:sssrs:arch_addr_gen}%
\includegraphics{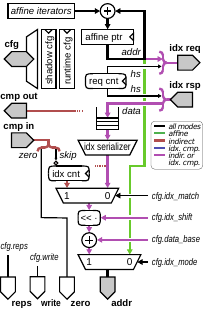}%
}\hfill\addtocounter{subfigure}{1}
\subfloat[\gls{sssr} streamer with two ISSRs, one ESSR, and a shared index comparator.]{%
\label{fig:sssrs:arch_sssr_streamer}%
\includegraphics{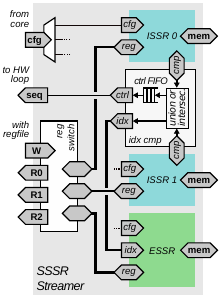}%
}\addtocounter{subfigure}{-2}%
\par%
\centering%
\vspace{1em}%
\subfloat[Indirection \gls{ssr} (left) and egress \gls{ssr} (right) with their corresponding new address generators and existing data mover architectures (grayed out).]{%
\label{fig:sssrs:arch_ssrs}%
\includegraphics[width=\linewidth, height=3.5cm]{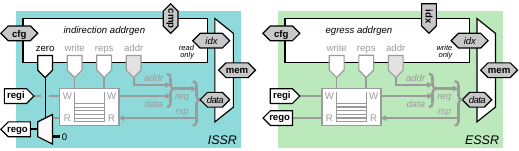}%
}\hfill%
\caption{Architecture of our \gls{sssr} \gls{isa} extension, showcasing address generation, data movers, the per-core \gls{sssr} streamer, and the inter-\gls{ssr} index comparator.}%
\label{fig:sssrs:arch_overview}%
\end{figure}

\subsection{\mmp{Indirection} Address Generator}
\label{chap:sssrs:arch:iag}

\cref{fig:sssrs:arch_addr_gen} shows the indirection address generator, highlighting the datapaths used by different generation modes. Like the original generator design, it exposes shadowed configuration registers to the host core, enabling the setup of a new stream while another is still running, and a control interface to the surrounding \gls{ssr} logic. The index stream is fetched through an additional read-only memory port.

All generation modes reuse the existing affine address generator with up to four nested levels, but for different purposes. During \emph{affine iteration}, it directly provides the address generator's output as shown. During \emph{indirection} or \emph{index matching}, %
it generates addresses for the words in memory containing the indices to be read; unlike fetching individual indices, this fully utilizes the memory bus. These words are fetched into a decoupling \gls{fifo} queue. To prevent downstream blocking or a queue overflow, an outstanding request counter limits inflight requests.

\glspl{sssr} can handle indices of any unsigned $2^n$\,\si{\byte} integer type that fits onto the used memory bus (i.e., $8$, $16$, $32$, and \mmp{\SI{64}{\lbit}} on Snitch). An \emph{index serializer} extracts indices of the configured size from the buffered index words. To increase flexibility and programmability, we support arbitrary alignment of the index array base in memory. In indirection mode, the serialized indices are then left-shifted by a small programmable offset, enabling optional power-of-two striding; this enables indirection into the upper axes of power-of-two-strided tensors without incurring a costly hardware multiplier. Finally, the shifted indices are added to the configured data base address, resulting in the desired indirect address stream.

During index matching, the indices are instead passed to an external matching unit, detailed in \cref{chap:sssrs:arch:streamer}, which compares the read indices with those in another \gls{ssr}'s generator streaming the other operand. Unlike in indirection, the emitted addresses simply count up from the configured data base address with a unit stride, since there is exactly one nonzero value per index.
However, depending on which index stream is ahead, the matching unit may instruct either generator to \emph{skip} addresses or \emph{insert} null data elements into the \gls{ssr} data stream, forming the \emph{intersection} or \emph{union} of the index streams.

\subsection{Egress Address Generator}
\label{chap:sssrs:arch:eag}

The \emph{egress} address generator is much simpler. It addresses the need to write back the indices of a joint stream alongside its data. For this purpose, it reads joint indices generated by an index matching unit through an additional port, passes them to an index \emph{coalescer}, and writes them back through a \emph{write-only} memory port, effectively reversing the indirection generator's index fetch datapath. It writes exactly one index per emitted address, but allows index writing to lead by a parameterized number of elements, decoupling the index and data streams and preventing unnecessary blocking.

Both extended generators are fully parameterized. The address and data widths, affine loop count, and index queue depth can all be adapted. Reduced internal index and address widths may be used for area-efficient designs operating on smaller memories.

\subsection{Sparse SSRs}
\label{chap:sssrs:arch:sssrs}

\cref{fig:sssrs:arch_ssrs} shows our extended sparse \glspl{ssr} built around the proposed address generators, the \gls{issr} and \gls{essr}. Like the original \gls{ssr}, both provide a data queue decoupling the register and memory ports \mmp{that} is used in both read and write directions.
Both expose their address generator's index matching ports, and the \gls{issr} adds some multiplexing logic to enable the indirection generator to inject zero data elements into read streams for stream union. %

In both the \gls{issr} and \gls{essr}, we chose to combine the index and data memory ports using a round-robin arbiter instead of requiring two memory ports. %
During affine streaming, this does not affect data throughput as the index port is unused. For both indirect and egress streaming, this tradeoff is motivated as follows: only one index word must be transmitted each $n$ data words, where $n$ is the number of indices per memory word. Thus, the peak achievable data mover utilization is $\nicefrac{n}{n+1}$; on our evaluation platform with $64$-bit data, this corresponds to $67\%$, $80\%$, and $88\%$ for $32$-, $16$-, and $8$-bit indices, respectively. %
Alternatively, we could arbitrate memory accesses in the streamer or expose two core memory ports per \gls{ssr}, trading higher peak utilization on sparse streams for a significantly larger cluster interconnect. In this work, we consider area-efficient \gls{sssr} variants with one port, enabling the drop-in replacement of and direct comparisons to the existing single-port \glspl{ssr}.

\subsection{SSSR \mmp{Streamer}}
\label{chap:sssrs:arch:streamer}

\cref{fig:sssrs:arch_sssr_streamer} shows the \gls{sssr} \emph{streamer} in its default configuration. Like the original \gls{ssr} streamer, it combines all available \glspl{ssr} and provides shared configuration and register file interfaces to the host processor. A \emph{register switch} maps each \gls{ssr}'s data channel to a predefined register when enabled.

The \gls{sssr} streamer is fully parameterizable and may be configured to map any number of \glspl{ssr}, \glspl{issr}, or \glspl{essr} to any register indices. However, its optional inter-\gls{ssr} index stream join capabilities, if present, are subject to some configuration constraints. A single streamer provides at most \emph{one} index comparator; thus, only \emph{two} \glspl{issr} in a streamer may support index comparison between each other, and one optional third \gls{essr} may consume the joined index stream for writeback. Thus, the default configuration comprises two index-comparing \glspl{issr} and one \gls{essr}.

\begin{figure}
    \centering
    \includegraphics[width=\linewidth]{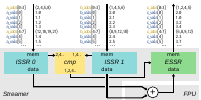}
    \caption{Example \gls{sssr} memory request sequences and index union dataflow for end-to-end \texttt{sV+sV} with 64-bit FP data and 16-bit indices.}
    \label{fig:sssrs:arch_svPsv_example}
\end{figure}

The index comparator %
determines which, if either, of the \gls{issr} index streams is ahead, and advances both as necessary to emit the union or intersection of indices. %
\Cref{fig:sssrs:arch_svPsv_example} demonstrates on an \verb|sV+sV| example how the index comparator generates a union of \gls{issr} indices and forwards them to an \gls{essr} for writeback, while the host core's FPU performs the required additions.
The desired comparison mode is configured in and forwarded from the \glspl{issr}. %
Since the number of index-data pairs in a joint stream is unknown ahead of time, the comparator also exposes a generic \emph{stream control} queue, which provides a single bit indicating either that another element pair is available or that the joint data stream is complete. This queue can be read by a host core's branch unit or hardware loop unit in a decoupled fashion to determine the end of a joint stream.

\subsection{Snitch Core and Cluster Integration}
\label{chap:sssrs:arch:integr}

\begin{figure}
    \centering
    \includegraphics[width=\linewidth]{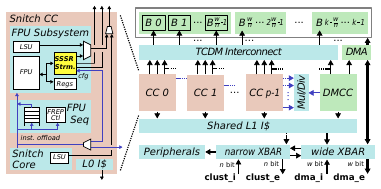}%
    \caption{Snitch Cluster used in evaluation with SSSR streamer.}
    \label{fig:sssrs:arch_cluster}
\end{figure}

\cref{fig:sssrs:arch_cluster} shows the integration of the default \gls{sssr} streamer into the energy-efficient Snitch \gls{cc} and Snitch cluster~\cite{Zaruba2020SnitchAT}, which we will use in our evaluation.
Each worker \gls{cc} (red) supports the base \mmp{32-bit} RISC-V \gls{isa} (RV32G) and consists of a single-issue, in-order integer processor offloading instructions to a double-precision \gls{fpu}.
The \gls{fpu}'s utilization can be maximized through \glspl{ssr} and the \gls{frep} hardware loop, \mmp{which we explain in more detail below.}
The \gls{fpu} also supports \gls{simd} operation down to FP8 precision~\cite{bertaccini2024fpu}, which we will leverage in future chapters.

\mmp{\gls{frep}~\cite{Zaruba2020SnitchAT} is a hardware loop instruction that repeats a subsequent block of \gls{fpu} instructions a specified number of times, avoiding the need for an explicit branch instruction or loop counter.}
\mmp{It operates on a shallow circular buffer of offloaded \gls{fpu} instructions and can issue instructions to the \gls{fpu} without any involvement from the offloading integer processor.}
\mmp{When combined} with \gls{ssr} streams, this allows the integer processor and \gls{fpu} susbystem to \mmp{concurrently issue and execute useful instructions} in a \emph{pseudo-dual-issue} fashion:
\mmp{while the \gls{fpu} subsystem is busy executing repeated floating-point instructions, the integer core is free to issue following instructions unless and until synchronization with the \gls{fpu} datapath is logically enforced, for instance through a load-store dependency or a register move.}

Like the previously integrated \gls{ssr} streamer, the \gls{sssr} streamer is multiplexed with the \gls{fpu}'s register file.
The \glspl{cc} preserve their existing memory topology, combining the integer core, FPU, and first \gls{issr} into one memory port and providing exclusive ports to the second \gls{issr} and \gls{essr}.
We also extend the existing \gls{frep} hardware loop with a new mode, \mmp{which uses our stream control interface instead of a fixed iteration count} to issue one iteration per joint stream element.

\begin{table}
\setlength{\tabcolsep}{0.6em}
\centering
\begin{tabular}{lllr}
\toprule
 & Parameter                         & Possible values                               & Used \\
 \midrule
$p$ & Worker core count              & $\mathbb{N}^{> 0}$                            & 8    \\
$n$ & Narrow (core, bank) width      & 32, 64                                        & 64   \\
$w$ & Wide (L1 I\$, DMA) width       & $\{n \leq 2^j \leq 1024~|~j \in \mathbb{N}\}$ & 512  \\
$k$ & Memory bank count              & multiples of~$\nicefrac{w}{n}$                & 32   \\
$D$ & TCDM size in \si{\kibi\byte}   & $\{2^j~|~j \in \mathbb{N}\}$                  & 128  \\
$I$ & L1 I\$ size in \si{\kibi\byte} & $\{2^j~|~j \in \mathbb{N}\}$                  & 8    \\
\bottomrule
\end{tabular}
\caption{Snitch cluster parameters and values used in evaluation.}
\label{tbl:sssrs:arch_cluster}
\end{table}

\Cref{tbl:sssrs:arch_cluster} lists the essential design parameters and their possible values for the fully configurable cluster.
All worker \glspl{cc} (by default 8) share an L1 instruction cache and an integer multiply-divide unit.
A \gls{dma} engine~\cite{benz2024idma} connected to a wide memory bus enables high-bandwidth data transfers between the \gls{tcdm} (by default \SI{128}{\kibi\byte} across 32 banks) and main memory; it is controlled by a lightweight \mmp{\gls{dmcc},} which can also be used for cluster management and coordination.

\section{Programming Model}
\label{chap:sssrs:progmodel}

The \glspl{sssr} are configured by the host core using a register interface accessed through custom instructions. This enables fast job setups: in our kernels, it takes at most 10 cycles to set up and launch new jobs for all three \glspl{sssr} from a cold state. In our configuration, \glspl{issr} 0 and 1 map to temporary \gls{fpu} registers \verb|ft0| and \verb|ft1| and the \gls{essr} to \verb|ft2|. The redirection of register accesses to \glspl{ssr} is toggled through a common control register~\cite{Schuiki2021StreamSR}.
In this \mmp{chapter}, we focus on accelerating sparse \gls{la} at the instruction level
by creating a library of hand-optimized operators, as is commonly done in high-performance \gls{la} packages.
In the next \mmp{chapter} (\cref{chap:workloads:rtllvm}), we will shortly discuss our extensions to the Snitch cluster toolchain and runtime to better support \glspl{sssr}, simplifying the implementation of further workloads.

\subsection{Accelerable Formats and Operations}
\label{chap:sssrs:progmodel:formops}

While we will demonstrate the generality of \glspl{sssr} in \Cref{chap:workloads} and explore further applications in \Cref{chap:sssrs:progmodel:further}, our main motivation in this \mmp{chapter} is accelerating general sparse \gls{la} on a set of established tensor formats.
Unlike approaches focused on low sparsities \cite{nvidia2020a100tc, Wang2021DualsideST, Gondimalla2019SparTenAS}, \glspl{sssr} target flexibility and scalability to achieve notable speedups across a wider sparsity range (see \cref{chap:sssrs:eval:sperf}).
\glspl{sssr} support iterating along the major axis of any tensor format where said axis is given by two arrays: a \emph{value} array storing nonzero values and an \emph{index} array storing their positions. We call this pair of arrays a sparse \emph{fiber}.
Fiber-based tensor formats include the common \gls{csr}~\cite{eisenstat1977csr} and \glsunset{csc}\gls{csc}~\cite{duff1989csc} matrix formats, the generalized \gls{csf}~\cite{Smith2015TensormatrixPW} tensor format, and many format variations that slice data, pack it into blocks, or further compress upper tensor dimensions.
The index shifter discussed in \Cref{chap:sssrs:arch:iag} also enables iterating on higher-level axes if all axes beneath it have power-of-two dimensions, which can be ensured through data chunking in practice. %
Indices can be of any unsigned $2^n$\,\si{\byte} integer type that fits onto the memory bus.%

The data width of \glspl{sssr} is fixed to that of the host processor, but \gls{simd} computations on blocked formats like BCSR \cite{Pinar1999ImprovingPO} are trivially supported through zero padding.
For example, the 64-bit \glspl{fpu} in Snitch can leverage their 2- and 4-way \gls{simd} capabilities on 32- and 16-bit formats when computing on blocks of two or four elements, respectively.
We can operate on nonzero \gls{simd} blocks the same way as on single double-precision elements; this substitution is orthogonal to the use of \glspl{sssr} and simply consolidates narrow nonzero elements into 64-bit nonzero \gls{simd} blocks. Therefore, \glspl{sssr} should enable similar peak speedups and somewhat reduced speedups for a given nonzero pattern, as proportionally more nonzero elements must pack together to overcome \gls{sssr} setup costs.

\glspl{sssr} are not bound to specific sparsity regimes. They enable significant efficiency gains on sparse-sparse, sparse-dense, and even dense problems through \gls{ssr} backward-compatibility. The supported formats have no structural sparsity requirements and scale well to extreme sparsities. Furthermore, \glspl{sssr} do not impose a specific higher-order dataflow: for example, sparse-sparse matrix multiplication (\verb|sM×sM|) may be done in an inner-product, outer-product, or row-wise fashion to best suit the application and data.

To optimize area and leakage, the \gls{sssr} streamer parameterization can be tailored to specific application requirements.
\mmp{For instance, in \cref{chap:sssrs:progmodel:sdla}, we will explain how one \gls{issr} and one \gls{ssr} are sufficient to accelerate sparse-dense multiplication}, with an additional \gls{issr} needed for addition. \mmp{Likewise, in \cref{chap:sssrs:progmodel:ssla}, we will show how two \glspl{issr} suffice to accelerate sparse-sparse multiplication,} with an additional \gls{essr} required for addition.

\subsection{Sparse Linear Algebra Kernels}
\label{chap:sssrs:progmodel:slakern}

To demonstrate the versatility of our extensions and evaluate their benefits, we create a set of sparse-dense and sparse-sparse linear algebra kernels for the \riscv{} Snitch cluster extended with \glspl{sssr}. The kernels operate on \gls{csr} matrices and \gls{csf} vectors. Each is implemented for 8-, 16-, and \mmp{32-bit} index types and, except for intersection-based kernels, in three variants:

\begin{itemize}
    \item \textsc{base}: Stock \riscv{} optimized baseline
    \item \textsc{ssr}: \riscv{} extended with \gls{frep} loop and \glspl{ssr}
    \item \textsc{sssr}: \riscv{} extended with \gls{frep} loop and \glspl{sssr}
\end{itemize}

To obtain the \textsc{base} kernels, we compile high-level C code with \verb|clang -Ofast| and hand-optimize the resulting assembly for Snitch \glspl{cc} through instruction reordering and loop unrolling to minimize memory or dependency stalls.
The \textsc{ssr} kernels further improve performance by mapping regular, affine value streams such as sparse value arrays to \glspl{ssr}, but without using our sparsity extensions. We note that this excludes kernels using intersection, since regular \glspl{ssr} cannot further accelerate conditional stream loads.
Finally, we implemented all kernels as \textsc{sssr} variants fully leveraging our extensions, which we will discuss in detail.

\subsection{Sparse-Dense Kernels}
\label{chap:sssrs:progmodel:sdla}

\begin{listing}[t]
\includegraphics[keepaspectratio, scale=1.215]{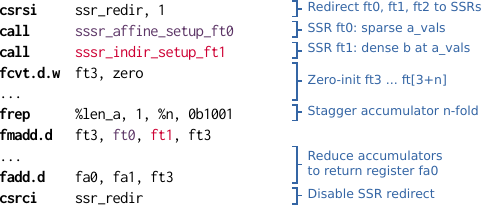}
\caption{\textsc{sssr} variant of \texttt{sV×dV} kernel: \texttt{ft0} streams the values of sparse vector \emph{a} and \texttt{ft1} streams indirected values from dense vector \emph{b}.}
\label{lst:sssrs:prog_svXdv_kernel}
\end{listing}

\topar{sV×dV} \cref{lst:sssrs:prog_svXdv_kernel} shows our \textsc{sssr} \verb|sV×dV| kernel. Versions for different index sizes differ only in the indirection configuration for \texttt{ft1}. Our goal is to continuously issue the \gls{mac} instruction \verb|fmadd.d| with minimal overhead to maximize \gls{fpu} utilization. We achieve this in three phases:

\begin{enumerate}
    \item \noprotrusion\emph{Setup}: We first enable register redirection to \glspl{ssr}. We then set up \gls{issr} \verb|ft0| to linearly stream the sparse vector \verb|a|'s values in affine mode and \gls{issr} \verb|ft1| to stream the dense vector \verb|b| indirected at \verb|a|'s indices. Finally, we zero-initialize a contiguous block of accumulator registers starting at \verb|ft3|.
    \item \noprotrusion\emph{Compute}: We use \gls{frep} to iterate the useful \verb|fmadd.d| instruction, streaming both multiplicands from the previously configured \glspl{issr}. To prevent stalls caused by register dependencies on the accumulated sum, we use \gls{frep}'s existing register staggering feature to increment the accumulator register index on each iteration, maintaining several partial sums in our contiguous block of registers; this is described in detail by Zaruba et al.~\cite{Zaruba2020SnitchAT}. Due to the limited peak throughput our \glspl{issr} with one memory port impose, the larger the index type, the fewer accumulators are needed to hide these  stalls.
    \item \noprotrusion\emph{Teardown}: We reduce our accumulators into the return register \verb|fa0| and disable \gls{ssr} redirection.
\end{enumerate}

Since the compute and teardown phases contain only \gls{fpu} instructions, the core is free to continue execution while the \gls{fpu} is kept busy by \gls{frep}; this also enables the seamless setup of new shadowed \gls{ssr} jobs. Synchronization between the core and \gls{fpu} can be enforced whenever necessary.

\topar{sM×dV} Instead of simply iterating our \verb|sV×dV| kernel for each matrix row, we further optimize \verb|sM×dV| to reduce setup and reduction overheads and maximize FPU utilization. We stream the entire matrix fiber in single \gls{ssr} and \gls{issr} jobs, significantly reducing our setup overhead. Furthermore, we unroll the first few \verb|fmadd| in each row, adding branches to shorter reductions, and issue an \gls{frep} loop and a full reduction only when necessary to accelerate short rows.

We use 32-bit row pointers in all variants to maximize row scaling. While we default on multiplying a \gls{csr} matrix from the left, our kernels provide runtime parameters to use different power-of-two vector and arbitrary result strides; this enables multiplying a \gls{csr} or \gls{csc} matrix with any power-of-two-strided dense tensor axis from either side.

\topar{sM×dM} We multiply a \gls{csr} matrix with a power-of-two-column row-major matrix. We simply iterate our \verb|sM×dV| kernel here, as the additional overhead of iterating over dense data in a third-order loop is negligible (see \cref{chap:sssrs:eval:sperf:sd}). We again provide parameters for custom dense matrix and result strides, enabling multiplication of  \gls{csr} or \gls{csc} matrices with row- and column-major matrices from either side.
The only restriction is that our index shifter, \mmp{which we chose as an alternative to a costly hardware multiplier (recall \cref{chap:sssrs:arch:iag}),} requires the dense axis we indirect into to be power-of-two-strided.
In practice, this requirement can be accommodated by tiling matrices using the cluster \gls{dma}'s strided transfers.

\topar{sV+dV} We assume that the result is accumulated onto the dense vector; if necessary, a copy can be created using \glspl{ssr} beforehand. We configure \verb|ft2| to stream the sparse vector values, \verb|ft0| to gather the dense addends, and \verb|ft2| to scatter sum elements back into the dense vector, and use \gls{frep} to execute \verb|fadd.d ft1, ft0, ft2| once for each sparse value. Either vector could also be scaled by a constant with the same performance by iterating an \verb|fmadd| instead. This element-wise addition can be extended to fiber-based tensors of \emph{any} shape through higher-level iteration.

\topar{sV$\odot$dV} Since one operand is dense, the result indices are the same as the sparse operand indices; they can optionally be copied with \glspl{ssr} if needed. We multiply each sparse value with a gathered dense co-operand as for \verb|sV+dV|, but instead linearly write out the results to yield the corresponding result value array. Again, this element-wise multiplication can be iterated to accommodate tensors of any shape.

\subsection{Sparse-Sparse Kernels}
\label{chap:sssrs:progmodel:ssla}

\topar{sV×sV} Since both indirection and intersection are fully handled inside \glspl{sssr}, the \textsc{sssr} \verb|sV×sV| kernel is identical to its \verb|sV×dV| counterpart except for the \gls{sssr} and \gls{frep} configurations used. We configure \verb|ft0| and \verb|ft1| to stream the fibers of each operand and activate intersection such that only value pairs with matching indices are streamed to the \gls{fpu}. Since the number of emitted pairs is unknown ahead of time in intersection, we use our stream-controlled \gls{frep} variant (\verb|frep.s|) to issue one \verb|fmadd.d| for each.

\topar{sM×sV} We iterate our \verb|sV×sV| code for each result element. Since our intersection approach requires increasing indices in fibers, we launch new \gls{sssr} jobs for each row unlike for \verb|sM×sV|. However, we can hide some of this configuration overhead through the decoupling of core and \gls{fpu} subsystem and the shadowed \gls{sssr} configuration interface.

\topar{sM×sM} Three sparse-sparse matrix multiply dataflows, all accelerable with \glspl{sssr}, are commonly differentiated:

\begin{itemize}
    \item \noprotrusion\emph{inner}: compute output element-wise by computing inner products (\verb|sV×sV|) for each row-column pair.
    \item \noprotrusion\emph{row-wise}: compute output row-wise by accumulating  sparse rows of left matrix (\verb|sV+sV|) scaled with nonzeros in each row of right matrix.
    \item \noprotrusion\emph{outer}: compute outer product for each row-column pair (dense vector scaling), then accumulate all partial sparse result matrices (\verb|sM+sM)|.
\end{itemize}

Each method has a different iteration order, memory footprint, communication overhead, and data reuse, making each attractive for different applications and sparsity regimes. \glspl{sssr} can accelerate all three methods by accelerating the subkernels pointed out above. Here, we limit ourselves to implementing a \gls{csr}×\gls{csc} \emph{inner} kernel iterating our \verb|sM×sV| kernel.

\begin{listing}[t]
\includegraphics[keepaspectratio, scale=1.215]{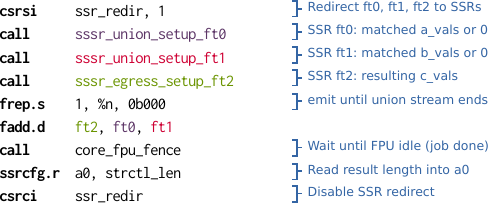}
\caption{\textsc{sssr} variant of \texttt{sV+sV} kernel: \texttt{ft0} and \texttt{ft1} stream in sparse operands \emph{a} and \emph{b} and \texttt{ft2} streams out sparse result \emph{c}.}
\label{lst:sssrs:prog_svPsv_kernel}
\end{listing}

\topar{sV+sV} As shown in \cref{lst:sssrs:prog_svPsv_kernel}, we read two sparse vectors from \glspl{issr} \verb|ft0| and \verb|ft1| and write their sparse sum to \gls{essr} \verb|ft2|. The index comparator performs index \emph{union} so that \verb|ft0| and \verb|ft1| emit either an element value or \emph{zero} if an index only occurs in the co-operand. This allows us to simply iterate \verb|fadd.d| as shown, with our stream-controlled \verb|frep.s| ensuring the correct iteration count. Since \verb|ft2| is configured as an egress, one compared index will be written to memory for each computed sum value, resulting in the sum index vector. We read the result length from an \gls{essr} configuration register after the job is done, which we ensure by synchronizing the core and \gls{fpu} pipelines. When iterating this kernel to add higher-dimensional sparse tensors (e.g., \verb|sM+sM|), we can defer this read to overlap computation with configuration as for \verb|sM×sV|.

\topar{sV$\odot$sV} This kernel is almost identical to \verb|sV+sV|; we instead configure the index comparator for \emph{intersection} and iterate \verb|fmul.d|. The same insights on higher-dimensional tensors and control-compute overlap also apply here.

\subsection{Further Applications}
\label{chap:sssrs:progmodel:further}

\glspl{sssr} accelerate general sparse \gls{la} and therefore the many applications built on it, including \gls{fem} simulations, sparse and graph neural networks, and graph workloads like PageRank \cite{Page1999ThePC}. However, the indirection, intersection, and union they accelerate are \emph{general-purpose} operations that underlie many more irregular workloads. We will demonstrate \gls{sssr} acceleration on some of these in \cref{chap:workloads}, but want to shortly introduce a few more here:

\topar{Codebook decoding} \glspl{issr} can stream compressed data stored as index arrays pointing into smaller arrays of repeated values. Such \emph{codebooks} could be used to efficiently store quantized deep learning parameters~\cite{Dave2021HardwareAO}, image channels~\cite{marcellin2002jpeg}, and nonzero values of sparse tensors. Each \gls{issr} in a streamer can independently stream codebooked vectors. Two \glspl{issr} can handle sparse-dense \gls{la} with codebooked nonzeros (one gathering dense operands, the other nonzeros) with similar performance as our sparse-dense kernels, significantly reducing the size of compressible tensors.

\topar{Graph pattern matching} Intersection between the sparse adjacency vectors of graph nodes can be leveraged to identify and count subgraph embeddings and assess the similarity of graphs \cite{Rao2022SparseCoreSI}. Thus, \glspl{sssr} may be used to accelerate graph pattern matching workloads in fields like computer vision and drug discovery.

\topar{Scatter-gather data transforms} \glspl{issr} enable a streaming variant of \emph{scatter-gather}. Thus, they can accelerate any existing scatter-gather-based transforms rearranging data, including sparse matrix transpose~\cite{Stathis2004SparseMT}, parallel radix sort\cite{Zagha1991RadixSF}, or densifying sparse tensors by scattering their nonzeros.

\section{Evaluation}
\label{chap:sssrs:eval}

We will first evaluate the performance benefits of \glspl{sssr} by using our \gls{la} kernels on a single Snitch \gls{cc} and in parallel cluster \verb|sM×dV| and \verb|sM×sV| implementations backed by \mmp{\emph{DRAMSys}~\cite{Jung2015DRAMSysAF},} a realistic \gls{dram} simulator \mmp{(details provided in \cref{chap:sssrs:eval:cperf})}.
Next, we \mmpcut{will }discuss the area and timing impacts of different hardware configurations of our streamer. Finally, we \mmpcut{will }investigate the energy efficiency benefits of our extensions.

Like the baseline Snitch platform we integrate them into, our hardware extensions are designed and modeled as parametric, fully synthesizable SystemVerilog \gls{rtl} hardware descriptions. This allows us to simulate both the baseline and our extended systems with full cycle accuracy and visibility, enabling the precise tracing and utilization tracking of all components. While we could use any SystemVerilog-capable simulator, we use \emph{Questa Advanced Simulator} here. Our implementations used to evaluate area, timing, and power use the same \gls{rtl} descriptions we simulate as inputs. Hardware resources external to the simulated \gls{cc} or cluster are modeled behaviorally. Unless noted otherwise, the evaluated cluster and \gls{cc} use default \gls{sssr} streamers and are parameterized as specified in \cref{tbl:sssrs:arch_cluster}.

For all experiments, dense test tensors are obtained by sampling normally distributed values and sparse vectors are generated for a given nonzero count and dimension with normally distributed values and uniformly distributed indices. The sparse matrices used are all real-world-problem matrices from the SuiteSparse Matrix Collection~\cite{Davis2011TheUO};
they cover various problem domains and aspect ratios and have~2\si{\kilo\relax} to 3.2\si{\kilo\relax} columns, 2.8\si{\kilo\relax} to 543\si{\kilo\relax} nonzeros\mmp{, and densities of \SI{0.032}{\percent} to \SI{10.1}{\percent} (or, equivalently, sparsities of \SI{89.9}{\percent} to \SI{99.968}{\percent}).}

To evaluate the performance benefits of \glspl{sssr} without the influence of other system components, we first evaluate our \gls{la} kernels in \gls{rtl} simulation of a single \gls{cc} by connecting it to an exclusive instruction cache and exclusive three-port data memory. These behave similarly to the shared instruction cache and \gls{tcdm} channels in a cluster, respectively, except for additional cache misses and bank conflicts whose effects we include and discuss in \cref{chap:sssrs:eval:cperf}.

\subsection{Single-Core Performance}
\label{chap:sssrs:eval:sperf}

\subsubsection{Sparse-dense Kernels}
\label{chap:sssrs:eval:sperf:sd}

\begin{figure}
\captionsetup[subfigure]{aboveskip=3.3pt, belowskip=-1.3pt, font=footnotesize, margin=0ex}%
\subfloat[\gls{cc} \texttt{sV×dV} \gls{fpu} util. vs. sparse nonzeros (dashed: without reductions).]{%
\label{fig:sssrs:res_cc_svXdv_util}\hspace{-0.29em}%
\includegraphics[height=3.985cm]{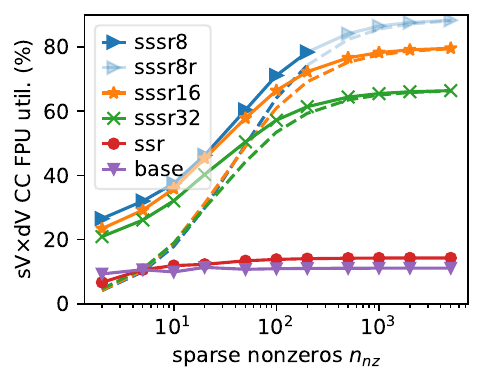}\hspace{0.7ex}%
\vspace{-0.45em}%
}\hspace{3.4ex}%
\subfloat[\gls{cc} \texttt{sV+dV} \gls{fpu} util. vs. sparse nonzeros (no reductions needed).]{%
\label{fig:sssrs:res_cc_svPdv_util}\hspace{-0.26em}%
\includegraphics[height=3.985cm]{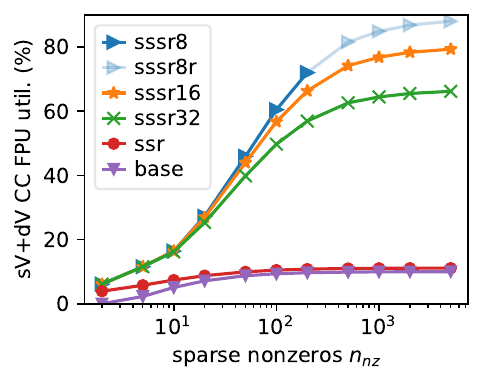}\hspace{0.7ex}%
\vspace{-0.45em}%
}\\%
\subfloat[\gls{cc} \texttt{sM×dV} speedups of \textsc{ssr} and \textsc{sssr} over \textsc{base} for selected matrices.] {%
\hspace{0.095em}%
\label{fig:sssrs:res_cc_smXdv_speedup}%
\includegraphics[height=4.09cm]{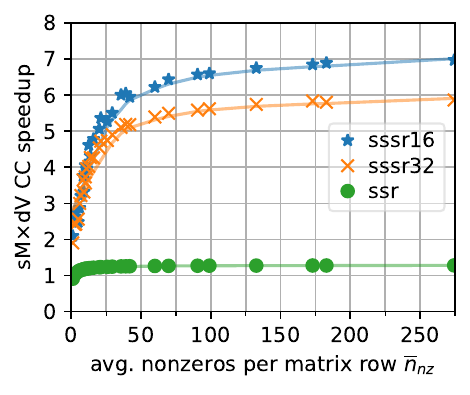}\hspace{0.7ex}%
\vspace{-0.45em}%
}\hspace{3.4ex}%
\subfloat[\gls{cc} \texttt{sV×sV} speedups of \textsc{sssr} over \textsc{base}. Both vectors have dense size 60\si{\kilo\relax}.] {%
\vspace{0.15em}%
\label{fig:sssrs:res_cc_svXsv_speedup}\hspace{-0.47em}%
\includegraphics[height=3.96cm]{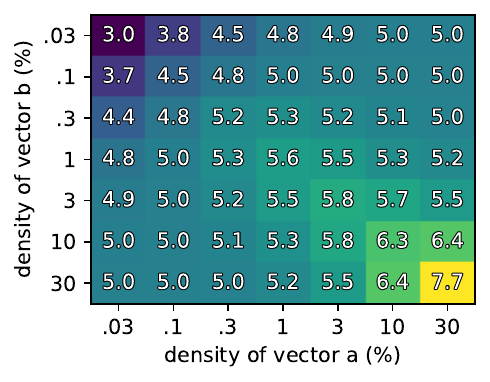}\hspace{0.7ex}%
\vspace{-0.45em}%
}\\%
\subfloat[\gls{cc} \texttt{sV+sV} speedups of \textsc{sssr} over \textsc{base}. Both vectors have dense size 60\si{\kilo\relax}.] {%
\kern-0.5em%
\vspace{0.12em}%
\label{fig:sssrs:res_cc_svPsv_speedup}%
\includegraphics[height=3.96cm]{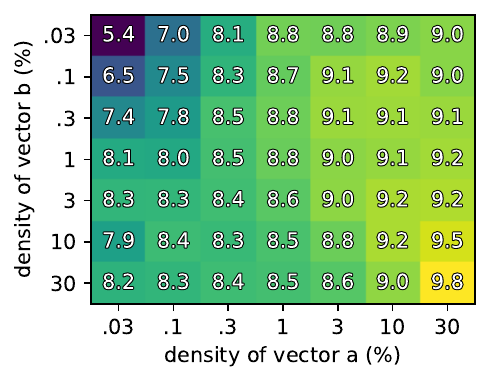}\hspace{0.7ex}%
\vspace{-0.45em}%
}\hspace{3.4ex}%
\subfloat[\gls{cc} \texttt{sM×sV} speedups of \textsc{sssr} over \textsc{base} for selected matrices.] {%
\hspace{0.12em}%
\label{fig:sssrs:res_cc_smXsv_speedup}%
\includegraphics[height=4.09cm]{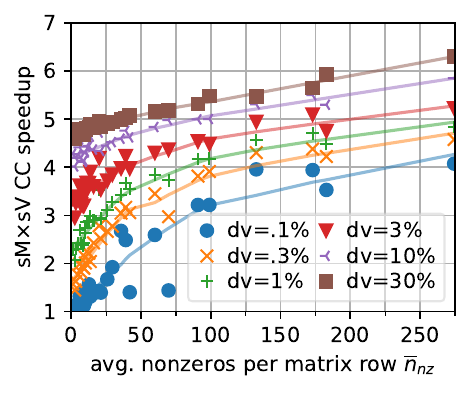}\hspace{0.7ex}%
\vspace{-0.45em}%
}\\\vspace{-2.12pt}%
\caption{\linespread{0.987}\selectfont{}Single-\gls{cc} performance results for kernels leveraging indirection, intersection, and union. \ref{fig:sssrs:res_cc_svXdv_util} and \ref{fig:sssrs:res_cc_svPdv_util} include results that assume recurring 8-bit indices (\texttt{sssr8r}).  The trend lines in \ref{fig:sssrs:res_cc_smXdv_speedup} and \ref{fig:sssrs:res_cc_smXsv_speedup} are fitted using \gls{loess}.}
  \label{fig:sssrs:res_cc}
\end{figure}

\topar{sV×dV} \cref{fig:sssrs:res_cc_svXdv_util} shows the \gls{fpu} utilization for our \verb|sV×dV| kernels for \mmp{sparse vector nonzero counts ($n_{nz}$) of 1 to 5000}.
\mmp{For the dense vector length of 8000 used here, this corresponds to densities between \SI{0.0125}{\percent} and \SI{62.5}{\percent}; }
\mmp{however, we} note that the kernel runtimes do not depend on the dense vector's length as long as it fits into the \gls{tcdm}.
Non-\textsc{sssr} variants perform identically for all index sizes, as a RISC-V load of any size incurs one instruction. \textsc{sssr} kernel \gls{fpu} utilizations are shown with and without accumulator reductions (dashed). Since an 8-bit-indexed \gls{csf} vector can hold at most 256 nonzeros, we also show results for 8-bit-index indirection with repeated indices (\texttt{sssr8r}).

All three \textsc{sssr} index variants significantly outperform the \textsc{base} and \textsc{ssr} kernels, with \fin{32- and 16-bit} variants approaching their arbitration-imposed utilization limits of  \SI{67}{\percent} and \SI{80}{\percent} and achieving 4.7\x~and 5.6\x~higher \gls{fpu} utilization than \textsc{ssr}, \fin{respectively}. The 8-bit-indexed kernel reaches up to 5.8\x~higher utilization, and 8-bit indirection with repeated indices approaches its arbitration limit of \SI{88}{\percent}. We note that \textsc{base} and \textsc{ssr} also approach their issue-bound utilization limits of one \gls{mac} every nine and seven \fin{cycles.}

Because they process nonzeros much faster, the \textsc{sssr} kernels need a significantly higher $n_{nz}$ to fully overcome their setup and reduction overheads. For few nonzeros, their utilization without reductions is even lower than for \textsc{base} and \textsc{ssr}, motivating our row unrolling in \textsc{sssr} \verb|sM×dV| kernels. Since \textsc{sssr} kernels with smaller indices need more accumulators to sustain their peak utilization, they outperform their larger-index counterparts only for higher $n_{nz}$.

\topar{sV+dV} \cref{fig:sssrs:res_cc_svPdv_util} shows the FPU utilization for our \verb|sV+dV| kernels. Here, the result is \emph{accumulated} onto the dense vector. We observe similar trends as for \verb|sV×dV| with two key differences. Since a result value is written back for each nonzero, \textsc{base} and \textsc{ssr} peak utilizations further decrease to one \gls{mac} every ten and nine cycles. For the same reason, no more reduction is necessary in \textsc{sssr} kernels as results are immediately written back by streaming indirection using the second \gls{issr}. The \texttt{sssr} kernels using 32-bit indices, 16-bit indices, and 8-bit indices with reuse still approach their arbitration-imposed limits of 67, 80, and \SI{88}{\percent}.

\topar{sM×dV} \cref{fig:sssrs:res_cc_smXdv_speedup} shows the speedups of \textsc{sssr} \texttt{sM×dV} kernels over \textsc{base} against the average nonzeros per matrix row $\overline{n}_{nz}$, reflecting the \glspl{mac} performed per row. We assume here that the \gls{tcdm} is large enough to store the full matrix. We do not evaluate 8-bit indices as they are not large enough to index the chosen matrices' columns.

As for \texttt{sV×dV}, our \fin{32- and 16-bit} \textsc{sssr} kernel speedups over \textsc{base} approach their theoretical limits and reach up to 5.9\x~and 7.0\x, requiring a higher $\overline{n}_{nz}$ than \textsc{base} and \textsc{ssr} to compensate their row iteration and reduction overheads. \gls{fpu} utilizations reach up to \SI{66}{\percent} and \SI{79}{\percent}, respectively. Again, the 16-bit \textsc{sssr} kernel outperforms the 32-bit variant only once $\overline{n}_{nz} \apprge 20$ due to its longer reduction sequence.

\topar{sM×dM} Even in edge cases, speedups and \gls{fpu} utilizations are nearly identical to the \verb|sM×dV| kernels we iterate: for the tiny sparse matrix \emph{Ragusa18} with only 64 nonzeros, \gls{fpu} utilization changes by only \SI{0.12}{\percent} when multiplying it with a two-column dense matrix instead of a dense vector.

\subsubsection{Sparse-sparse Kernels}

\topar{sV×sV} \cref{fig:sssrs:res_cc_svXsv_speedup} shows the speedup of our 16-bit \textsc{sssr} \verb|sV×sV| kernel over its \textsc{base} counterpart for vector densities of \fin{\SI{0.03}{\percent} to \SI{30}{\percent}} and \mmp{a vector length of 60\si{\kilo\relax} elements}. We observe speedups of \fin{{3.0\x} to 7.7\x}, increasing near-symmetrically with the density of both operands. Speedups are generally higher for similar densities; as densities diverge, speedups converge to 5.0\x.

To understand these trends, we first consider two steady-state edge cases. While encountering \emph{no} intersections and scanning one vector's nonzeros, \texttt{base} needs \emph{five} and \glspl{sssr} \emph{one} cycle per nonzero, explaining the 5.0\x~speedup limit for divergent densities. While processing \emph{only} intersections, \textsc{base} needs 18 and \glspl{sssr} 1.25 cycles per nonzero pair, yielding a peak achievable speedup of 14.4\x~and FPU utilization of \SI{80}{\percent} if nonzero positions were to coincide exactly. Thus, as index matching likelihood increases with both densities, so do speedups. Finally, as both vectors become very sparse and contain few elements, intersection quickly terminates and speedups are offset by \gls{sssr} setup costs. They are also modulated by the random vectors' nonzero positions and sequences, introducing slight asymmetries.

\topar{sV+sV} \cref{fig:sssrs:res_cc_svPsv_speedup} shows the speedup of our 16-bit \textsc{sssr} \verb|sV+sV| kernel over \textsc{base} for the same vectors as for \verb|sV×sV|. Speedups range from \fin{5.4\x} to 9.8\x~and increase with the density of both operands with slight operand asymmetry.

To understand these trends, we again consider edge cases. While reading only summand $a$ or $b$, \glspl{sssr} process each element up to 9.6\x~and 8.8\x~faster assuming continuous index comparison. In practice, occasional \gls{essr} index write backpressure decreases speedups to the observed 9.0\x~and 8.2\x. Speedups are asymmetric in $a$ and $b$ as ternary branching code in \textsc{base} incurs either one or two branch cycles. For continuous index matches, peak speedups are again 14.4\x.  Thus, speedups increase with increasing density in both operands and for similar densities.

\topar{sM×sV}  \cref{fig:sssrs:res_cc_smXsv_speedup} shows the speedup of our 16-bit \textsc{sssr} \texttt{sM×sV} kernel over \textsc{base}  against the average nonzeros per matrix row $\overline{n}_{nz}$ for different vector densities. We again assume that the \gls{tcdm} stores the full matrix.

We observe similar trends as for the \texttt{sV×dV} kernel we iterate: speedups increase with the density of both arguments and approach those of \texttt{sV×dV} for corresponding densities, reaching up to 6.3\x. However, the smaller dense vector dimensions result in an earlier onset of offsetting and modulating effects as densities decrease: for \SI{0.1}{\percent} vector density or $2$ to $3$ nonzeros, speedups are significantly reduced and strongly vary with nonzero patterns. However, speedups remain above 1, meaning we still amortize \gls{sssr} setup even for few nonzeros in both arguments.

\subsection{Cluster Performance}
\label{chap:sssrs:eval:cperf}

To evaluate multi-core and system-level performance, we implement parallel scaleouts of \verb|sM×dV| and \verb|sM×sV| on a Snitch cluster, \mmp{leveraging the existing baremetal cluster runtime}. We reuse our architecture-optimized single-core kernels, distribute dynamically sized chunks of rows among cores, and implement a double-buffered data movement scheme for the matrix using the cluster \gls{dma} engine.

To ensure realistic memory throughput, latency, and scheduling, we connect our cluster to one of eight channels of a Micron MT54A16G808A00AC-36 HBM2E \gls{dram} (\SI{3.6}{\giga b\per\second\per pin} or \SI{57.6}{\giga\byte\per\second} peak bandwidth, \SI{88}{\nano\second} average round-trip latency\footnote{Averaged from \gls{dma} matrix chunk reads in \gls{sssr} \texttt{sM×dV} using peak-speedup matrix \texttt{mycielskian12}. Excludes on-chip interconnect.}), which we simulate using \emph{DRAMSys}~\cite{Jung2015DRAMSysAF}. All input data and instructions initially reside in this \gls{dram} and all results are written back to it. A \SI{16}{\kibi\byte} 4-way L2 instruction cache, bypassed by \gls{dma} data reads and writes, caches instruction fetches. In addition to DRAM, \gls{phy}, and controller delays modeled by \emph{DRAMSys}, we model the delay of an on-chip interconnect with 16 additional cycles of forward and backward latency.
\mmp{In a later subsection}, we also explore the impact of interconnect latency and DRAM channel bandwidth sharing.

\begin{figure}
\setkeys{Gin}{height=5.5cm}%
\subfloat[Cluster \texttt{sM×dV} speedups of \textsc{sssr} over \textsc{base} for 16-bit indices and selected matrices.]{%
\includegraphics[left]{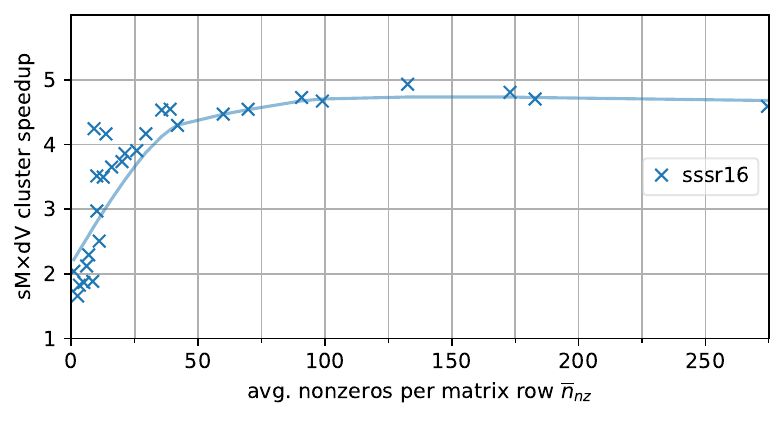}%
\label{fig:sssrs:res_cl_csrmv_speedup}%
}\hfill \vspace{0.6em}\\%
\subfloat[Cluster \texttt{sM×sV} speedups of \textsc{sssr} over \textsc{base} for 16-bit indices, varying vector densities, and selected matrices.]{%
\includegraphics[left]{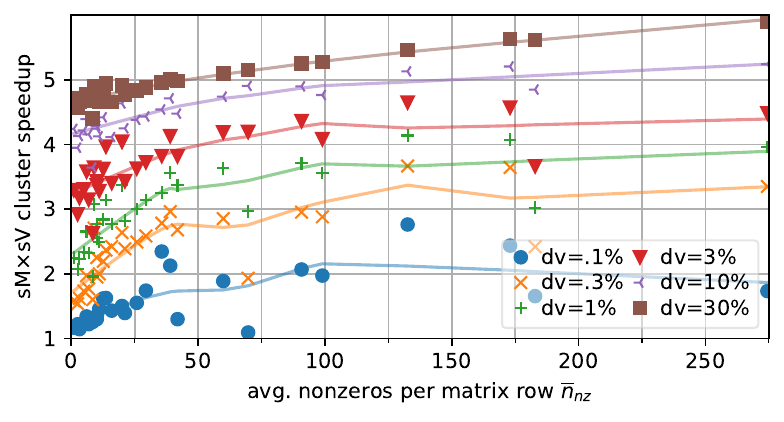}%
\label{fig:sssrs:res_cl_csrmspv_speedup}%
}\hfill\\%
\caption{Eight-core cluster performance results for selected \gls{la} kernel scaleouts. Trend lines are fitted using \gls{loess}.}%
\label{fig:sssrs:res_cl}%
\end{figure}

\topar{sM×dV} \cref{fig:sssrs:res_cl_csrmv_speedup} shows the speedups of the 16-bit \textsc{sssr} kernel over the \textsc{base} kernel on \verb|sM×dV| parallelized on a cluster. Improvements are notable even for $\overline{n}_{nz}=1$ at 1.7\x~and reach up to 4.9\x, sustaining over 4\x~for $\overline{n}_{nz}>30$.
We observe the same trends as for single-core \verb|sM×dV|, but with reduced speedups and stronger variations, which we attribute to multiple causes.
Most importantly, \emph{bank conflicts}, aggravated by the pseudorandom access patterns of indirection, lower the achievable \gls{tcdm} bandwidth and thus \gls{issr} throughput. Additionally, the \emph{initial vector transfer} cannot be overlapped with useful computation, modulating speedups with the vector's length. Our matrix \emph{double-buffering} and \emph{row distribution} schemes also incur some computational imbalance and overhead. %
The high latency, limited throughput, and scheduling of our \emph{\gls{dram} model}, though mostly hidden through the double-buffered \gls{dma} transfer of long, contiguous matrix chunks, also slightly decrease speedups by a median of \SI{1.8}{\percent} and up to \SI{6.9}{\percent}. Finally, we observe occasional stalls due to instruction cache misses. %
Despite these inherent parallelization and scaleout overheads, \glspl{sssr} enable up to \SI{46.8}{\percent} overall \gls{fpu} utilization and extreme efficiency gains at minimal area cost: \emph{eight} cores with \glspl{sssr} achieve the same computational throughput as \emph{39} cores running \textsc{base}.

\topar{sM×sV} \cref{fig:sssrs:res_cl_csrmspv_speedup} shows 16-bit \textsc{sssr} \verb|sM×sV| speedups for different vector densities, which peak at 5.9\x. As with \verb|sM×dV|, we see similar trends as in the used single-core kernel with diminished speedups due to initial vector copy, instruction cache and \gls{tcdm} stalls, work-sharing imbalances, and \gls{dram} inefficiencies.
For high vector densities, performance is mainly affected by \gls{tcdm} stalls and setup overheads: the peak speedup is only \SI{6.9}{\percent} lower than in the single-core case.
Because each core processes only a subset of matrix rows and
due to variations in \gls{dram} response time, few-nonzero modulation is more pronounced overall.
As the vector density $d_v$ decreases and computational imbalance increases, \gls{dram} inefficiencies increase: for $d_v=\SI{30}{\percent}$, speedups are only \SI{0.40}{\percent} lower on average than with an ideal memory system, while for $d_v=\SI{0.1}{\percent}$, they are \SI{16}{\percent} lower.
The minimum observed speedup of 1.1\x~is \emph{higher} than in the single-core case, which we attribute to the notably smaller code size and thus fewer instruction cache stalls the \textsc{sssr} kernel incurs.

\subsubsection{Bandwidth and Latency Sensitivity}
\label{chap:sssrs:eval:cperf:bwlat}

\begin{figure}%
\subfloat[Cluster \texttt{sM×dV} and \texttt{sM×sV} speedups vs. limited DRAM channel bandwidth. The red dots indicate where speedups have degraded by \SI{10}{\percent}.]{%
\includegraphics[width=\linewidth]{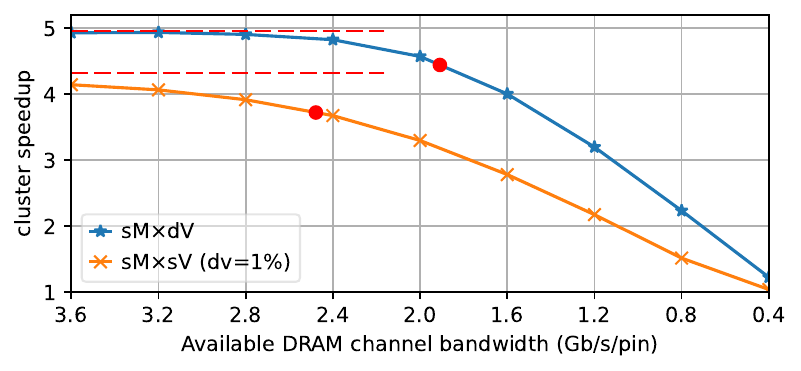}%
\label{fig:sssrs:res_cl_csrmv_bw}%
}\vspace{0.6em}\\%
\subfloat[Cluster \texttt{sM×dV} and \texttt{sM×sV} speedups vs. on-chip interconnect latency.]{%
\includegraphics[width=\linewidth]{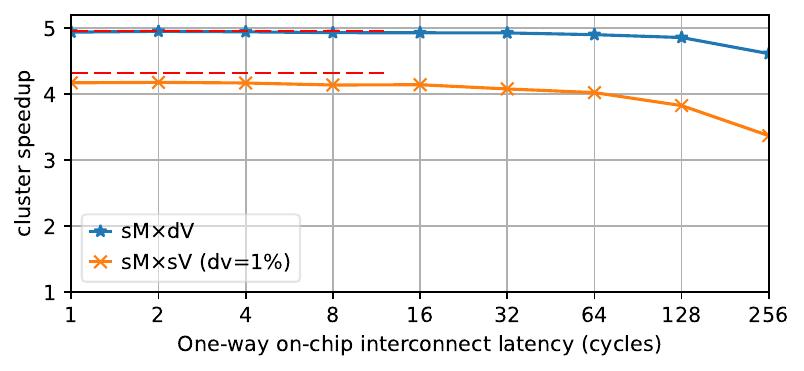}%
\label{fig:sssrs:res_cl_csrmv_lat}%
}%
\caption{Sensitivity of \gls{sssr} Snitch cluster speedups to limited DRAM channel bandwidth and varying on-chip interconnect latency. The red dashed lines show speedups for unlimited bandwidth and zero latency.}%
\label{fig:sssrs:res_cl_bwlat}%
\end{figure}

To evaluate the sensitivity of \gls{sssr} speedups to bandwidth sharing and interconnect scaling, we rerun cluster \texttt{sM×dV} and \texttt{sM×sV} with limited \gls{dram} throughput and varying interconnect latencies. We use the peak-\texttt{sM×dV}-speedup matrix \texttt{mycielskian12} ($\overline{n}_{nz} = 133$, \SI{4.3}{\percent} density) to ensure high \gls{dram} pressure and use \SI{1}{\percent} vector density for \texttt{sM×sV}. In both cases, we also indicate ideal speedups assuming a memory system with unlimited bandwidth and zero latency.

\Cref{fig:sssrs:res_cl_csrmv_bw} shows how speedups decrease as the available \gls{dram} channel bandwidth is restricted from its full throughput of $3.6$ to \SI{0.4}{\giga b\per\second\per pin} \mmp{(9\x~less)}, simulating channel sharing with other bus agents. We maintain the one-way interconnect latency of 16 cycles. %
For \gls{sssr}-accelerated \texttt{sM×dV} with full \gls{dram} bandwidth, the cluster incurs an average throughput of $R_T = \SI{1.6}{\giga b\per\second\per pin}$ \mmp{(\SI{44}{\percent} of peak)}; we thus expect our accelerated \texttt{sM×dV} to be compute-bound above $R_T$ and memory-bound below it. %
Indeed, speedups start decreasing linearly below $R_T$ until they reach 1\x~just below \SI{0.4}{\giga b\per\second\per pin}, past which both the accelerated and baseline variants are memory-bound and perform the same. %
However, we also observe some speedup degradation \emph{above} $R_T$; this is because \gls{dram} bandwidths higher than $R_T$ can hide minor throughput variations due to computational imbalances and \gls{dram} scheduling. %
For \texttt{sM×sV}, we observe similar trends: in this case, $R_T$ is higher due to faster processing of the matrix fiber, and speedups in the compute-bound region degrade slightly faster due to increased computational imbalance%
.

\Cref{fig:sssrs:res_cl_csrmv_lat} shows how speedups decrease as on-chip interconnect latency increases, assuming full \gls{dram} channel throughput. For \texttt{sM×dV}, speedups remain near-constant until one-way latency exceeds 64 cycles, comparable to that of the downstream \gls{dram} subsystem. This is because the cluster's \gls{dma} engine moves matrix data in double-buffered chunks tens of \si{\kibi\byte} in length, which saturate the bus for hundreds of cycles; performance is negatively impacted only once the total access latency becomes comparable to the durations of these transfers. %
For \texttt{sM×sV}, the same trends hold as matrix data is still moved in large, double-buffered chunks. However, greater variations in computation rate and imbalance result in slightly more pronounced speedup losses as latency increases.

Overall, the efficiency of \glspl{sssr} in removing instruction bottlenecks moves sparse workloads  %
closer to the memory bound, and they still provide notable speedups well into the memory-bound regime and only fully lose their benefits when  \gls{dram} throughput is reduced to an order of magnitude less than regular channel bandwidth. %
This is despite \texttt{sM×dV} and \texttt{sM×sV} being very memory-intensive; sparse workloads allowing for matrix data reuse, such as matrix-matrix multiplications and iterative solvers, are significantly less memory-bound. %
Furthermore, our double-buffered data movement approach proves highly latency-resilient, compensating for hundreds of cycles in interconnect latency in addition to \gls{dram} delays with minimal speedup losses.

\subsection{Area and Timing}
\label{chap:sssrs:eval:at}

\begin{figure}
\centering%
\subfloat[Area breakdown of the default streamer and major subcomponents
\\(areas in \si{\kilo\GE}). The residual area is shown rightmost in light gray.]{%
\label{fig:sssrs:res_synth_areahier}%
\includegraphics[width=\linewidth]{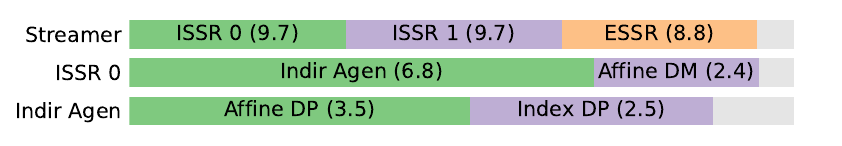}%
}\vspace{0.5em}\hfill%
\subfloat[Area and min. clock period for different streamer configurations.]{%
\label{fig:sssrs:res_synth_atcmp}%
\includegraphics[height=4.5cm]{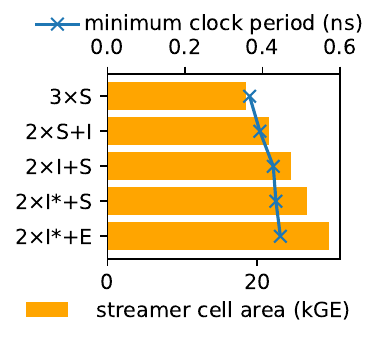}%
}\hspace{1.2em}%
\subfloat[Streamer area vs. clock period for different target clocks.]{%
\label{fig:sssrs:res_synth_at}%
\includegraphics[height=4.5cm]{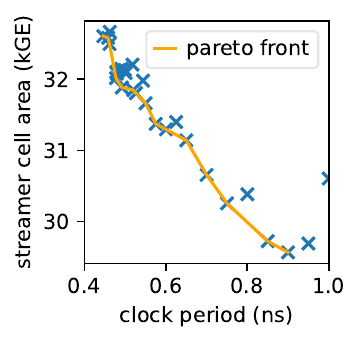}%
}%
\caption{GF12LP+ synthesis area and timing results for \gls{sssr} streamer. S: \gls{ssr}, I: \gls{issr}, E: \gls{essr}, I*: \gls{issr} with shared comparator.}%
\label{fig:sssrs:res_synth}%
\end{figure}

We synthesize the \gls{sssr} streamer in various configurations using Synopsys \emph{Design Compiler} for GlobalFoundries’ 12LP+ FinFET technology. We target the typical (TT) process corner at \SI{25}{\celsius} with \SI{0.8}{\volt} supply voltage. Unless stated otherwise, a \SI{1}{\giga\hertz} clock and IO delays of \SI{35}{\percent} and \SI{60}{\percent} on the core and \gls{tcdm} interfaces are constrained. In our default streamer configuration, each of the two  \glspl{issr} and single \gls{essr} has four data FIFO stages, 17-bit addresses to cover our \SI{128}{\kibi\byte} \gls{tcdm}, and four affine loop levels.

\mmp{\cref{fig:sssrs:res_synth_areahier}} shows the default streamer's hierarchical area distribution. Each \gls{issr} contributes \SI{9.7}{\kGE} and the \gls{essr} \SI{8.8}{\kGE} to the \SI{30}{\kGE} streamer. \gls{issr} area is dominated by its address generator, only \SI{37}{\percent} of which is fully dedicated to index streaming. \cref{fig:sssrs:res_synth_atcmp} compares the area and minimum achievable clock period for different streamer configurations. If only indirection capabilities are needed, only \SI{3.0}{\kGE} (\SI{16}{\percent}) in additional area is incurred per \gls{issr}. Intersection between two \fin{\glspl{issr}} incurs another \SI{2.1}{\kGE}.
The full \gls{sssr} streamer with added union capability incurs \SI{11}{\kGE} (\SI{60}{\percent}) \fin{in} area overhead over \fin{a regular \gls{ssr} streamer} and only moderately increases the minimum clock period from \SI{367}{\pico\second} to \SI{446}{\pico\second}, \mmp{still well below Snitch's \SI{1}{\nano\second} target}.
\mmp{\cref{fig:sssrs:res_synth_at}} shows the achieved minimum period and area for different period goals, demonstrating that \fin{the \gls{sssr} streamer's} area scales gracefully as timing pressure increases. At the cluster level, adding \gls{sssr} streamers to all eight worker cores incurs only a minimal \SI{1.8}{\percent} cell area overhead compared to providing only regular \glspl{ssr}.

\subsection{Energy and Power}
\label{chap:sssrs:eval:ep}

We estimate the power and total energy consumption of a Snitch cluster running our workloads from \cref{chap:sssrs:eval:cperf} with 16-bit-index \textsc{base} and \textsc{sssr} kernels. We use the same interconnect and main memory models as in \cref{chap:sssrs:eval:cperf}. We target the TT process corner of  GlobalFoundries’ 12LP+ FinFET technology at \SI{1}{\giga\hertz}.
We \mmp{fully} implement \mmp{(synthesize, place, and route)} the cluster using Synopsys \emph{Fusion Compiler} and estimate power for the low- and high-efficiency matrices \verb|cryg2500| and \verb|cavity12| \mmp{on the routed netlist} using Synopsys \emph{PrimeTime}, then scale dynamic power with component utilizations measured in RTL simulation.

\begin{figure}%
\centering%
\subfloat[\texttt{sM×dV}]{%
\centering%
\includegraphics[width=0.48\linewidth]{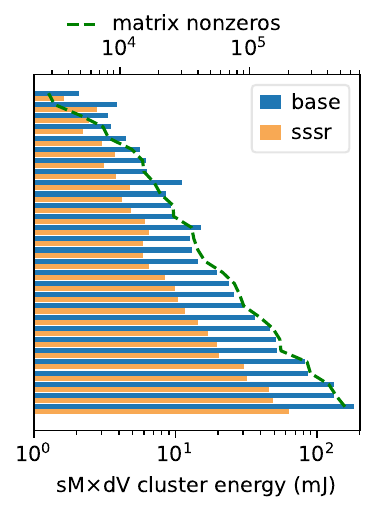}%
\label{fig:sssrs:res_pow_csrmv}%
}\hspace{1.0em}%
\subfloat[\texttt{sM×sV} (\SI{1}{\percent} vector density)]{%
\centering%
\includegraphics[width=0.48\linewidth]{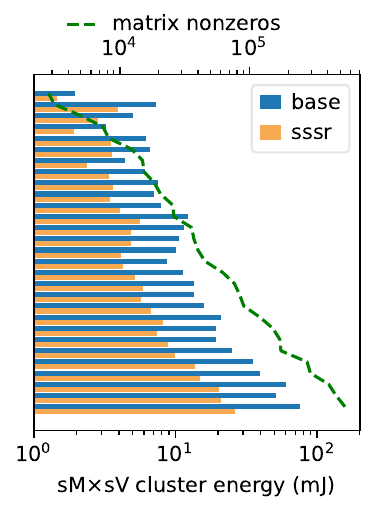}%
\label{fig:sssrs:res_pow_csrmspv}%
}\hspace{0.75ex}%
\caption{Snitch Cluster workload energy estimates in GF12LP+. The benchmarks and data are the same as in \Cref{chap:sssrs:eval:cperf}.}%
\label{fig:sssrs:res_cl_pow}%
\end{figure}

\cref{fig:sssrs:res_pow_csrmv} shows the total energies for \verb|sM×dV| using both kernel variants for each matrix. While the median cluster power is predictably lower for the \textsc{base} kernel (\SI{195}{\milli\watt} vs. \SI{285}{\milli\watt}), \glspl{sssr} reduce the minimum energy per \texttt{fmadd} from \SI{282}{\pico\joule} to \SI{103}{\pico\joule} and achieve energy efficiency improvements of up to 2.9\x.
\cref{fig:sssrs:res_pow_csrmspv} shows the total energies for \texttt{sM×sV} for a vector density of \SI{1}{\percent}. Again, \glspl{sssr} incur higher median power, but lower the minimum energy per matrix nonzero from \SI{107}{\pico\joule} to \SI{43}{\pico\joule} and achieve energy efficiency improvements of up to 3.0\x.

\section{Related Work}
\label{chap:sssrs:relwork}

In this section, we compare our approach to numerous \mmp{\gls{soa}} CPU, GPU, and accelerator solutions, covering both hardware and software proposals. We summarize our findings in two tables. \cref{tbl:sssr:relwork_comp_sw} collects the evaluation platform, matrix format, and highest fraction of peak compute achieved on all discussed FP64 \texttt{sM×dV} software implementations. \cref{tbl:sssr:relwork_comp} compares the features, flexibility, and architectural cost of all discussed hardware designs.

\subsection{CPU Software and Hardware}
\label{chap:sssrs:relwork:cpu}

\topar{Scalar processor architectures} %
A processor with \glspl{sssr} can be considered an extended and specialized \gls{dae} architecture~\cite{Smith1984DecoupledAC}.
\gls{dae} architectures combine two decoupled processors, one handling \emph{memory accesses} and another \emph{execution}, which communicate through register-mapped \glspl{fifo}.
\glspl{sssr} %
provide a lightweight access processor design with multiple performance enhancements.
They leverage multi-element \emph{streams} as abstractions, which %
improves %
decoupling, minimizes %
offloads from the main instruction stream, and attains near-ideal compute and memory throughputs.
Their job types define a compact, orthogonal instruction set that minimizes access code complexity and enables lightweight implementation.
Unlike the original \gls{dae} concept, \glspl{sssr} support %
multiple concurrent streams on different registers.
Our low-cost sparsity extensions maintain high throughput on indirect and joint streams without the hardware and bookkeeping overheads of a more conventional access processor architecture, enabling efficient general sparse compute without major execute processor modifications.
\mmp{We note that like \glspl{sssr}, many of the streaming \gls{isa} extensions we later discuss in this section can, to some extent, be considered specialized \gls{dae} access processors}.

\begin{table}
\setlength{\tabcolsep}{0.58em}
\centering
\ifx\showrebuttal\undefined
\else
\color{blue}
\fi
\begin{tabular}{cp{7.6em}llr}
\toprule
 & Work &%
 Platform &%
 Format &%
 \hspace{-0.5em}Pk. FP Ut. \\
 \midrule
 \multirow{4}{*}{\adjustbox{angle=90}{CPU}}%
 & CVR~\cite{Xie2018CVREV}                        & Xeon Phi 7250  & CVR               & \SI{0.69}{\percent} \\
 & Zhang et al.~\cite{Zhang2018VectorizedPS}      & Xeon Phi 7230  & SELL-like         & \SI{1.5}{\percent}  \\
 & Regu2D~\cite{Fei2021Regu2DAV}                  & Xeon Gold 6132 & Regu2D            & \SI{3.1}{\percent}  \\
 & \scalebox{.89}[1.0]{Alappat et al.~\cite{alappat2020a64fx}} & A64FX          & SELL-C-\textsigma & \SI{4.7}{\percent}  \\
 \midrule
 \multirow{6}{*}{\adjustbox{angle=90}{GPU}}%
 & Tsai et al.~\cite{Tsai2020SparseLA}            & V100           & CSR             & \SI{1.6}{\percent}  \\
 & Merrill et al.~\cite{Merrill2016MergeBasedPS}  & K40            & CSR             & \SI{2.0}{\percent}  \\
 & TileSpMV~\cite{Niu2021TileSpMVAT}              & A100           & tile-adapt.     & \SI{2.9}{\percent}  \\
 & Tsai et al.~\cite{Tsai2020SparseLA}            & Radeon VII     & CSR             & \SI{3.2}{\percent}  \\
 & cuSPARSE~\cite{nvidia2018cuda}                 & GTX 1080 Ti    & CSR             & \SI{17}{\percent}   \\
 & TileSpMV~\cite{Niu2021TileSpMVAT}              & Titan RTX      & tile-adapt.     & \SI{27}{\percent}   \\
 \midrule
 & \textbf{SSSRs (ours)}                        & Snitch + \glspl{sssr} & CSR & \SI{47}{\percent} \\
\bottomrule
\end{tabular}
\caption{Overview of the discussed double-precision (FP64) \texttt{sM×dV} software implementations.}
\label{tbl:sssr:relwork_comp_sw}
\end{table}

\newcommand{\rot}[1]{\adjustbox{angle=-30,llap,lap=1em}{#1}\hspace{1.5em}}
\newcommand{\tm}{\checkmark}
\newcommand{\lH}{H}
\newcommand{\lM}{M}
\newcommand{\lL}{L}

\begin{table}
\setlength{\tabcolsep}{0.718em}
\centering
\fontsize{8.7pt}{10.7pt}\selectfont
\begin{tabular}{cp{10.6em}cccccr}
\toprule
 & \vspace{2.3em}\adjustbox{lap=1em}{Work} &%
 \rot{\,~~Open source} &%
 \rot{\,~1-sided accel.} &%
 \rot{\,~2-sided accel.} &%
 \rot{~~Usage flexib.\textsuperscript{*}} &%
 \rot{Sparsity flexib.\textsuperscript{\textdagger}} &%
 \rot{~~\si{\kGE} per core\textsuperscript{\textdaggerdbl}} \\
 \midrule
 \multirow{11}{*}{\adjustbox{angle=90}{CPU}}%
 & SVE S/G~\cite{stephens2017sve}               &     & \tm &     & \lM & \lH & -  \\
 & KNL S/G~\cite{sodani2016knl}                 &     & \tm &     & \lM & \lH & -  \\
 & \mmp{RVV S/G~\cite{riscv2021rvv}}            &     & \tm &     & \lH & \lH & -  \\
 & UVE \cite{Domingos2021UnlimitedVE}    & \tm\textsuperscript{**} & \tm &    & \lH & \lH & 72\textsuperscript{\P} \\
 & Gong et al. \cite{Gong2020SAVESV}            &     & \tm & \tm & \lM & \lL & 31\textsuperscript{\P} \\
 & \mmp{VIA~\cite{pavon2021via}}                &     & \tm & \tm & \lM & \lH & 593\textsuperscript{\S} \\
 & Prodigy \cite{Talati2021ProdigyIT}           &     & \tm &     & \lM & \lH & 10\textsuperscript{\P} \\
 & SpZip \cite{Yang2021SpZipAS}                 &     & \tm &     & \lM & \lH & 116 \\
 & Z. Wang et al. \cite{Wang2019StreambasedMA}  &     & \tm &     & \lH & \lH & -   \\
 & SparseCore \cite{Rao2022SparseCoreSI}        &     &     & \tm & \lH & \lH & 619 \\
 & \mmp{TMU~\cite{siracusa2023tmu}}             &     & \tm & \tm & \lM & \lH & 354 \\
 \midrule
 \multirow{3}{*}{\adjustbox{angle=90}{GPU}}%
 & A100 \cite{nvidia2020a100tc}                 &     & \tm &     & \lM & \lL & -   \\
 & Zhu et al. \cite{zhu2019sparsetensorcore}    &     & \tm &     & \lL & \lM & 12\textsuperscript{\S}  \\
 & Y. Wang et al. \cite{Wang2021DualsideST}     &     &     & \tm & \lL & \lM & 157\textsuperscript{\S} \\
 \midrule
 \multirow{10}{*}{\adjustbox{angle=90}{Accelerators}}%
 & MatRaptor \cite{Srivastava2020MatRaptorAS}   &     &     & \tm & \lL & \lH & - \\
 & \mmp{GAMMA \cite{zhang2021gamma}}            &     &     & \tm & \lL & \lH & - \\
 & OuterSPACE \cite{Pal2018OuterSPACEAO}        &     &     & \tm & \lL & \lH & - \\
 & \mmp{Trapezoid~\cite{yang2024trapezoid}}     &     & \tm & \tm & \lL & \lH & - \\
 & \mmp{Sparse-TPU~\cite{he2020sptpu}}          &     & \tm &     & \lL & \lM & - \\
 & Sadi et al. \cite{Sadi2019EfficientSO}       &     & \tm &     & \lL & \lH & - \\
 & EIE \cite{Han2016EIEEI}                      &     & \tm & \tm & \lL & \lH & - \\
 & ExTensor \cite{Hegde2019ExTensorAA}          &     &     & \tm & \lM & \lH & - \\
 & \mmp{FEASTA~\cite{zhong2024feasta}}          &     &     & \tm & \lM & \lH & - \\
 & Dadu et al. \cite{Dadu2019TowardsGP}         &     & \tm & \tm & \lM & \lH & - \\
 & SISA \cite{Besta2021SISASI}                  &     &     & \tm & \lM & \lH & - \\
 \midrule
 & \textbf{SSSRs (ours)}                        & \tm & \tm & \tm & \lH & \lH & 30 \\
\bottomrule
\end{tabular}
\\[0.5em]
\justifying\noindent
L/M/H = Low/Medium/High. *\,Qualitative, based on operand granularity, use constraints, overheads. \textdagger\,Qualitative, based on efficiently handled range, structure. \textdaggerdbl\,Based on authors' data, different cores used. **\,gem5 model only, no RTL. \P\,Architectural state only at 6T=1.5GE per bit. \S\,Only node given; we assume GlobalFoundries' 22nm and 12nm, respectively.
\caption{Overview of the discussed hardware designs.}
\label{tbl:sssr:relwork_comp}
\end{table}

\topar{Vector processor architectures} %
Indirection in \glspl{issr} is similar to vector processor \emph{scatter-gather}, which finds recent adoption in superscalar out-of-order processors.
Arm's \gls{sve}~\cite{stephens2017sve} and Intel's \gls{knl}~\cite{sodani2016knl} add support for \gls{simd} scatter-gather; however, the short lengths and alignment requirements of \gls{simd} vectors limit scatter-gather benefits on sparse workloads compared to \glspl{issr}.
\mmp{The RISC-V vector extension (\gls{rvv})~\cite{riscv2021rvv} is vector length agnostic and relaxes some data alignment requirements, but it does not address the overhead of issuing explicit vector load and store instructions.}
\mmp{The \gls{uve}~\cite{Domingos2021UnlimitedVE} introduces long indirect streams coupled directly to vector register accesses,} similar to \glspl{issr}; however, it requires vectorizable workloads and cannot operate efficiently at scalar granularity.
Gong et al.~\cite{Gong2020SAVESV} extend a vector multiply-add unit to skip ineffectual and coalesce useful computations. While this is orthogonal to \glspl{sssr}, it is only effective on low, regular sparsity in uncompressed data \fin{that explicitly stores zeros}.
\mmp{VIA~\cite{pavon2021via} extends a vector processor's functional units with a smart scratchpad that can either be mapped directly or used as a \gls{cam} to accelerate sparse-dense or sparse-sparse \gls{la}, respectively. Although VIA accelerates both one- and two-sided sparse \gls{la}, even the smallest configuration evaluated by the authors comes at a significant area cost of \SI{593}{\kGE}, or about two full Snitch \glspl{cc} with \glspl{sssr}.}

\topar{Software optimizations for scalar and vector processors} Numerous software solutions aim to accelerate sparse \gls{la} on scalar and vector processors.
The CVR matrix format~\cite{Xie2018CVREV} aims to improve \texttt{sM×dV} \gls{simd} lane and cache efficiency on x86 CPUs. Still, it utilizes at most \SI{0.69}{\percent} of peak FP64 compute on Xeon Phi 7250.
Zhang et al.~\cite{Zhang2018VectorizedPS} optimize \texttt{sM×dV} %
for Intel's AVX-512 extensions, proposing a format based on \gls{sell} that enables peak FP64 compute utilizations of up to \SI{1.5}{\percent} on Xeon Phi 7230.
Regu2D~\cite{Fei2021Regu2DAV} leverages an adaptive 2D partitioning scheme and arranges rows as similar-sized groups with zero-padding as necessary to fill \gls{simd} lanes, improving their platform's peak FP64 utilization to \SI{3.1}{\percent}.
Alappat et al.~\cite{alappat2020a64fx} model the performance of \texttt{sM×dV} on Fujitsu's A64FX using the \gls{sell}-C-\textsigma~matrix format and report up to \SI{130.9}{\Gflops}, or \SI{4.7}{\percent} of peak FP64 compute. Nevertheless, this is 9.9\x~less than what we achieve on a Snitch cluster with \glspl{sssr}.

\mmp{Instead of creating new algorithms and formats to improve performance, some works aim to characterize the execution of existing proposals, often to choose the optimal method for a given scenario.}
\mbox{Z. Xie} et al.~\cite{Xie2022APS} present a \gls{csr}-interfaced \verb|sM×sM| library for multi- and manycores that selects the best internal format and algorithm with a deep learning model based on nonzero patterns.
\mmp{WISE~\cite{yesil2023wise} is a framework that predicts the speedups of various \texttt{sM×dV} methods over a baseline and chooses the best method for each specific matrix.}
\mmp{Sgherzi et al. present SpChar~\cite{Sgherzi2024spchar}, a characterization methodology for general sparse computation, and use it to explore what hardware and software characteristics most impact \texttt{sM×dV} performance on Arm-based systems.}
\mmp{These works are useful to identify the best existing approach for a given problem and platform, but they inherently cannot outperform the leading solution in each scenario.}
\mmp{That said, they are orthogonal to our work on \glspl{sssr}, which may also benefit from case-by-case performance characterization and careful software method selection.}

\subsection{Streaming ISA Extensions}
\label{chap:sssrs:relwork:sisaext}

General-purpose processor extensions increasingly leverage streams as abstractions to accelerate compute.
Prodigy~\cite{Talati2021ProdigyIT} uses compile-time-inferred metadata to prefetch nested indirect streams. However, it does not eliminate load-store or bookkeeping instructions.
SpZip~\cite{Yang2021SpZipAS} reads and writes compressed irregular data using decoupled streaming units. This is similar to \glspl{sssr} reading and writing fiber-based formats, but SpZip does not support intersection or union, %
and a fetcher-compressor pair incurs \SI{116}{\kGE} or 3.9\x~more area than an \gls{sssr} streamer.
Z. Wang et al.~\cite{Wang2019StreambasedMA} propose a decoupled register-mapped extension accelerating affine and indirect memory streams similar to \glspl{issr}. However, streams need to be stepped explicitly, severely limiting possible speedups on in-order cores, and intersection and union are not supported.
SparseCore~\cite{Rao2022SparseCoreSI} accelerates intersection, union, and subtraction of indexed streams and provides dedicated functional units. However, indirection is not accelerated, making the handling of one-sided sparsity inherently inefficient. Additionally, SparseCore incurs \SI{0.183}{\milli\meter^2}  per stream unit in Open-Cell \SI{15}{\nano\meter}, which is comparable to two entire Snitch \glspl{cc} with \glspl{sssr}.
\mmp{Finally, the \gls{tmu} combines multiple lanes of programmable units to accelerate various traversal and merging operations on multi-dimensional sparse tensors, including indirection, intersection, and union. While the \gls{tmu} is \gls{ta}-complete, its incurs a significant area of \SI{354}{\kGE}, or about one entire Snitch \gls{cc} with \glspl{sssr}, and its callback-based communication is less agile than the direct register mapping of \glspl{sssr}.}

\subsection{GPU Software and Hardware}
\label{chap:sssrs:relwork:gpu}

\topar{GPU software} %
Moving from general-purpose processors to more specialized architectures, GPUs have been the subject of intensive investigations aimed at improving their efficiency on sparse data computations.
One-sided sparsity on GPUs is often accelerated in \emph{software} through specialized algorithms and tensor formats.
Merrill et al.~\cite{Merrill2016MergeBasedPS} improve the performance consistency of \gls{csr} \verb|sM×dV| on GPUs using a merge-based decomposition, reaching up to \SI{2.0}{\percent} of peak FP64 compute on a K40 GPU.
Nvidia provides optimized sparse \gls{la} kernels for their GPUs through \emph{cuSPARSE}~\cite{nvidia2018cuda}. We evaluate their \gls{csr} \verb|sM×dV| kernels from CUDA Toolkit 10.0 on a Jetson AGX Xavier (FP32 only) and a GTX 1080 Ti GPU (FP32 and FP64), profiling 100 invocations with \emph{nvprof} on the same matrices as for \glspl{sssr}. For FP32, both GPUs exhibit high peak \gls{sm} occupancies of \SI{96}{\percent} and \SI{87}{\percent}, but low peak floating-point utilizations of and \SI{2.1}{\percent} and \SI{0.75}{\percent} on active \glspl{sm}, suggesting low thread parallelism among warps. FP64 on the GTX 1080 Ti shows similar \gls{sm} occupancies, but a notably higher peak floating-point utilization of \SI{17}{\percent}, likely due to the 32\x~fewer FP64 cores per \gls{sm}. %
Tsai et al.~\cite{Tsai2020SparseLA} create optimized \texttt{sM×dV} library kernels for both Nvidia and AMD GPUs and various formats. Their \gls{csr} kernels regularly outperform vendor code and reach up to \SI{1.6}{\percent} and \SI{3.2}{\percent} of peak FP64 utilization on Nvidia V100 \pagebreak[4] and AMD Radeon VII GPUs, respectively.
TileSpMV~\cite{Niu2021TileSpMVAT} uses warp-optimized kernels on sparse tiles in a given or adaptively selected format. It achieves peak FP64 utilizations of \SI{2.9}{\percent} on A100 and \SI{27}{\percent} on Titan RTX, which again has 32\x~fewer FP64 than FP32 cores per \gls{sm}.
Still, this is 1.7\x~lower than in a Snitch cluster with \glspl{sssr}.
Shi et al.~\cite{Shi2020EfficientSM} accelerate \verb|sM×dM| by encoding matrices as arrays of coordinate-encoded tiles, improving access contiguity and data reuse. However, the average FP32 speedups of 1.7\x~over cuSPARSE on SuiteSparse matrices are insufficient to reach FP utilizations comparable to those of Snitch with \glspl{sssr}.
Karimi et al.~\cite{Karimi2022VCSRAE} design a set of memory-aware matrix formats accelerating \verb|sM×dV| by improving cache usage and thread-level parallelism. While improving performance, their best format achieves median FP32 speedups of only 1.8\x~(K40) and 3.5\x~(V100) over cuSPARSE \gls{csr} on Suite- Sparse matrices, still not enough to match our utilizations.

Some sparse-sparse software approaches for GPUs also exist: Zacha\-riadis et al.~\cite{Zachariadis2020AcceleratingSM} accelerate \verb|sM×sV| using a tensor-core-based outer-product approach with dense 16\x16 tiles, accepting low flexibility and unnecessary computations within blocks.
Li et al.~\cite{Li2020AdaptiveSO} optimize general sparse matrix-vector multiply by dynamically choosing one of eight \verb|sM×dV|/\verb|sM×sV| approaches based on tensor properties using a \gls{ml} model. This solution inherently cannot exceed the peak performance of the existing approaches chosen from, but can be adapted to improve overall performance on other systems, including ones with \glspl{sssr}.

\topar{GPU hardware} %
Extensions to GPU \emph{hardware} accelerating sparsity have also been proposed.
Nvidia's A100 architecture~\cite{nvidia2020a100tc} introduces support for one-sided \emph{structured} sparsity by efficiently handling up to two zeros in every block of four values.
\glspl{sssr} efficiently handle a much wider range of sparsities ($\gg\;$\SI{50}{\percent}) without any structural requirements and enable random accesses into a full \gls{tcdm}.
Zhu et al.~\cite{zhu2019sparsetensorcore} present an algorithm and co-optimized tensor core modifications for efficient sparse-dense neural network inference.
Y. Wang et al.~\cite{Wang2021DualsideST} extend V100 tensor cores to efficiently handle sparse-sparse matrix multiply and convolution with an outer product approach. However, the bitmap encoding used hinders efficient performance and memory scaling beyond low sparsities, and the dense output allows for the overall approach to be replicated with scattering \glspl{issr}.

\subsection{Hardware Accelerators}
\label{chap:sssrs:relwork:hwaccel}

\mmp{Most sparse workload accelerators target} a single application domain like \gls{dl}~\cite{Dave2021HardwareAO} or specific \gls{la} operations and dataflows.
\mmp{These designs often} use custom data formats and precisions tailored to their use case and are incapable of general-purpose computation. Notably, a general-purpose processor with \glspl{sssr} can implement many of the employed specialized computation schemes.
\mmp{MatRaptor~\cite{Srivastava2020MatRaptorAS} and GAMMA~\cite{zhang2021gamma} accelerate \texttt{sM×sM} using row-wise dataflows, whereas OuterSPACE~\cite{Pal2018OuterSPACEAO} uses an outer-product approach; \glspl{sssr} can efficiently handle both dataflows.}
\mmp{Trapezoid~\cite{yang2024trapezoid} accelerates \texttt{dM×dM}, \texttt{sM×dM}, and \texttt{sM×sM} with a 2D array of \glspl{pe} that supports different dataflows to best suit the operand sparsities; \glspl{sssr} can achieve the same through different software kernels.}
\mmp{\mbox{Sparse-TPU}~\cite{he2020sptpu} is a sytolic array for \texttt{sM×dV} that packs nonzeros into array columns and serially streams the required vector elements through them; a core with \glspl{sssr} could adopt its effective packing approach without being limited by its spatial architecture, which is ill-suited to highly sparse matrices.}
Sadi et al.~\cite{Sadi2019EfficientSO} accelerate large-scale \verb|sM×dV| by column-slicing matrices and then accumulating the sparse results, improving vector locality and reuse; iterating over slices could be done on a host core while \glspl{sssr} handle multiplication and result accumulation.
Han et al.'s EIE~\cite{Han2016EIEEI} minimizes DRAM access energy in \gls{dl} inference by keeping full, weight-shared models in on-chip SRAM; \glspl{sssr} also enable high utilization on sparse data with codebooked values, but are not limited to one application, schedule, or dataflow.

\mmp{Unlike the above accelerators, however, some designs tackle sparse \gls{ta} in a more general fashion, forgoing peak efficiency in a specific domain for higher algorithmic flexibility}.
\mmp{The \gls{sam}~\cite{hsu2023sam} proposes a stream-based machine model for sparse \gls{ta} accelerators, including various portable primitives. While \gls{sam} is merely an abstraction and not an actual design, it defines a set of formal requirements for general and efficient sparse \gls{ta} computing systems, all of which a core with \glspl{sssr} fulfills.}
ExTensor~\cite{Hegde2019ExTensorAA} accelerates sparse \gls{ta} by hierarchically intersecting \gls{csf} tensors to avoid redundant computation and transfers; a processor with \glspl{sssr} can replicate this by handling the upper axes in software while the streamer intersects the performance-critical primary axes.
\mmp{FEASTA~\cite{zhong2024feasta} is a programmable accelerator for sparse \gls{ta} that intersects and reductively merges multi-dimensional sparse fibers in hardware. While FEASTA supports multiple dataflows, it does not provide efficient sparse-dense processing through indirection.}
Dadu et al.~\cite{Dadu2019TowardsGP} present a compute fabric accelerating both stream joins and streaming indirection. However, the loosely-coupled spatial compute approach incurs notable costs in area and programmability.
SISA~\cite{Besta2021SISASI} introduces a set-centric \gls{isa} with the goal of accelerating graph mining algorithms. In addition to set intersection and union, it also accelerates set difference, membership and size queries, and element insertion or removal. It supports both indexed and bitmap sets and a binary-search-based %
\emph{galloping} mode for index joins between sets of divergent sizes. While SISA does not explicitly accelerate one-sided sparsity, many of its additional capabilities could be added to SSSRs. However, SISA's architectural cost is discussed only in loose analogy to other designs, making such tradeoffs hard to assess.

\mmp{Finally, instead of designing new accelerators, some recent proposals investigate how to more efficiently map sparse and irregular workloads to existing ones.}
\mmp{COGNATE~\cite{sudusinghe2025cognate} uses transfer learning to apply effective optimizations to sparse tensor programs before deploying them to accelerators, collecting inexpensive data samples from general-purpose \glspl{cpu} to train cost models.}
\mmp{HotTiles~\cite{gerogiannis2024hottiles} tackles \texttt{sM×dM} using a heterogeneous team of accelerator \glspl{pe} by partitioning the sparse matrix into denser and sparser tiles and handling each with the \gls{pe} type better suited to it.}
\mmp{These methods are orthogonal to the use of \glspl{sssr}, whose performance could also benefit from targeted offline software optimizations and use in a heterogeneous scenario.}

\section{Conclusion}
\label{chap:sssrs:conclusion}

In this chapter, we presented \emph{\glspl{sssr}}, a modular, tightly core-coupled \gls{isa} extension that enables efficient general sparse linear algebra on sparse-fiber-based formats including the widespread \gls{csr} and \gls{csf} while also providing acceleration of affine streams.
We evaluated our extension in the RISC-V Snitch cluster by presenting efficient single-core sparse-dense and sparse-sparse \gls{la} kernels and parallelizing matrix-vector products on an eight-core cluster.

In a single core using 16-bit indices, the streaming indirection of \glspl{sssr} enables peak \verb|sV×dV| \gls{fpu} utilizations of \SI{80}{\percent} and \verb|sM×dV| speedups of up to 7.0\x~over our optimized baseline.
The intersection and union of \glspl{sssr} enable speedups of $3.0$\,--\,$7.7$\x~and $5.4$\,--\,$9.8$\x~on sparse-sparse dot product and vector addition across a wide sparsity range, enabling \verb|sM×sV| speedups of up to 6.3\x.
Scaled out to a Snitch cluster with interconnect and DRAM models exhibiting realistic latency, throughput, and scheduling, \verb|sM×dV| using \glspl{sssr} is up to 4.9\x~faster and 2.9\x~more energy efficient, while \verb|sM×sV| is up to 5.9\x~faster overall and 3.0\x~more energy efficient for a vector density of \SI{1}{\percent}.
These benefits incur only \SI{11}{\kGE} per core over affine streamers, or \SI{1.8}{\percent} across a cluster, which may further be reduced thanks to the streamer's modular design.

Unlike \mmp{many} recent CPU, GPU, and accelerator proposals, \glspl{sssr} accelerate one- \emph{and} two-sided sparse compute while remaining flexible in dataflow and operand sparsity and incurring a minimal area impact. A Snitch cluster with \glspl{sssr} running \verb|sM×dV| achieves 9.9\x~and 1.7\x~higher peak floating-point \fin{compute} utilizations than recent CPU and GPU software \fin{proposals}.

\clearpage
\chapter[Workloads Beyond Sparse LA]{Workloads Beyond Sparse Linear Algebra}
\label{chap:workloads}

In this chapter, we \mmpcut{will }leverage our cluster-level extensions designed in \cref{chap:sssrs}, \emph{\glspl{sssr}}, to significantly accelerate three irregular workloads whose performance-critical compute does not directly map to sparse \gls{la}:
\emph{stencil computations}%
, \emph{spiking neural networks}%
, and \emph{model predictive control}%
.
In the process, we \mmpcut{will }demonstrate that \glspl{sssr} are flexible enough to accelerate diverse irregular applications beyond sparse \gls{la}.

Our introduction will briefly outline each of the targeted workloads, their irregular nature, and our work on them.
We will then shortly present the extensions made to the Snitch cluster's LLVM toolchain and runtime %
to improve \gls{sssr} support and workload performance and enable our work.
After this, we will present our acceleration work on each of the three workloads in detail.

\section{Introduction}
\label{chap:workloads:introduction}

As discussed in \cref{chap:introduction:workloads}, while many irregular workloads are accelerable in large part through \gls{la} kernels, this is not always the case:
some workloads involve performance-critical irregular compute that goes beyond sparse \gls{la}.
This chapter will explore the acceleration of three such workloads in particular using \glspl{sssr},
showing that our cluster-level extensions do not just accelerate sparse \gls{la}, but the irregular memory accesses and control flows underlying general irregular workloads.
As an evaluation system, we will continue using the Snitch cluster with and without \glspl{sssr}.

In \cref{chap:workloads:stc}, we consider \emph{stencil computations}, which iteratively update points in $n$D data grids in fixed, but generally irregular patterns called \emph{stencils}.
Although stencil codes do not involve sparse \gls{la}, their irregular memory accesses result in limited \gls{fpu} utilization on conventional CPUs~\cite{Zhang2020DataLT,Yount2015VectorFI,Zhao2018DeliveringPS} and GPUs~\cite{Rawat2019OnOptimizingCS,You2021DRStencilED,Matsumura2020AN5DAS,Zhang2023RevisitingTB}.
Since stencil shapes are constant across grid point iterations, they can be described by fixed offset arrays and \emph{streamed} as indices, advancing the base address on each point iteration;
we formalize this idea into a generic acceleration methodology called \emph{SARIS}, which is applicable to any \gls{sr} extension with indirection support.
Leveraging SARIS in our \glspl{sssr}-enhanced Snitch cluster on various 2D and 3D stencils, we achieve geomean speedups and energy efficiency gains of 2.7\x~and 1.6\x~over an RV32G cluster, respectively, while increasing geomean \gls{fpu} utilization from \SI{35}{\percent} to \SI{81}{\percent}.
On the Occamy manycore we present in \cref{chap:occamy}, we achieve up to \SI{83}{\percent} of peak FP64 performance, which is \SI{20}{\percent} higher than a leading GPU code generator.

In \cref{chap:workloads:snn}, we investigate \glsunset{snn}\emph{spiking neural network} (SNN) inference using the \gls{lif}~\cite{Abbott1999lif} neuron model.
Updating neuron potentials requires \glspl{spwr} of dynamic length interspersed with nonlinear activations, which incurs irregular memory accesses and cannot directly be mapped to a sparse \gls{la} kernel.
Fortunately, \glspl{sssr} can accelerate any indirect stream with low overhead  and their accesses can freely be interleaved with unrelated compute;
they can even leverage the Snitch cluster's FP8 and FP16 \gls{simd} support by packing together output channels.
This allows us to speed up and improve the energy efficiency of FP16 inference on S-VGG11 by \nchk{5.5\x} and \nchk{3.3\x}, respectively, with FP8 inference providing a further \nchk{1.7\x} improvement on both.
Overall FP16 \gls{fpu} utilization is increased from \nchk{\SI{9.3}{\percent}} to \nchk{\SI{52}{\percent}}.
Compared to the leading neuromorphic processor \emph{LSMCore}~\cite{wang2022lsmcore}, \gls{snn} inference on the \gls{sssr}-extended Snitch cluster in FP8 is \nchk{4.4\x} more energy efficient.

In \cref{chap:workloads:mpc}, we accelerate online \glsunset{mpc}\emph{model predictive control} (MPC),
an advanced control method that keeps dynamic systems within specified constraints by predicting their state trajectory~\cite{jeres2014embed2,wang2010fast}.
When formulated as a \gls{qp} and tackled with the \gls{admm}~\cite{stellato2020osqp}, \gls{mpc} is dominated by a \emph{direct sparse solver} step
that incurs complex internal data dependencies and cannot be mapped to regular sparse \gls{la}.
We first accelerate the existing \emph{QDLDL}~\cite{stellato2020osqp} direct solver with \glspl{sssr}, making it up to 2.0\x~faster than on an RV32G cluster.
Finding that performance is limited by the high number of data dependencies, we design an optimized solver called \emph{ParSPL} that drastically reduces them, accelerating both baseline and \gls{sssr}-accelerated \gls{mpc}.
On \gls{etm} problems in application-class manycores, \glspl{sssr} now enable solver speedups and energy efficiency gains of up to 4.6\x~and 2.2\x, respectively, and raise the peak \gls{fpu} utilization from \SI{11}{\percent} to \SI{57}{\percent}.
On an end-to-end \gls{etm} scenario with 9\x9 cores, \glspl{sssr} and ParSPL together reduce the \gls{mpc} iteration time from \SI{8.74}{\milli\second} to \SI{0.69}{\milli\second}, enabling online control with ample slack.
The resulting \gls{etm} solution is lightweight enough for co-integration, outperforms \gls{soa} \gls{mpc}-based \gls{etm} approaches in terms of area and energy efficiency, and is competitive with existing fast \gls{mpc} methods on various platforms.

Before going into detail on the above irregular workloads, we will shortly discuss our extensions to Snitch's custom LLVM toolchain and its runtime, which facilitate the use of \glspl{sssr} and improve performance in some scenarios.
Both the toolchain and runtime improvements are leveraged in our accelerated implementations.

\section{LLVM Toolchain and Runtime}
\label{chap:workloads:rtllvm}

The Snitch cluster provides a custom LLVM 15 toolchain, which we extend to facilitate and improve the implementation of \gls{sssr}-accelerated workloads\footnote{\url{https://github.com/pulp-platform/llvm-project/tree/15.0.0-sssr-0.1.0}.}.
We also extend the cluster runtime to leverage these features and further simplify \gls{sssr} programming.

\topar{LLVM Toolchain} We first add RISC-V assembler support for the \texttt{scfg(r|w)*} instructions that read and write \gls{sssr} configuration registers.
We then add intrinsic compiler functions, usable from C/C++ code, to read and write configuration registers as well as \emph{push} and \emph{pop} \gls{sssr} streams.
The push and pop intrinsics allow programmers to access \glspl{sssr} from high-level code.
They are initially translated to \gls{ir} \fin{pseudo-instructions} that read or write a \gls{ssa} variable with stream contents.
This form is maintained into the RISC-V backend, where a dedicated pass \emph{merges} them into RISC-V instruction arguments whenever possible and maps them to register moves otherwise.
The emitted merges are guarded against reordering until after register allocation to prevent unexpected behavior.
In blocks where \glspl{sssr} are used, the backend cannot allocate them as regular registers to prevent collisions.

In addition to intrinsics, we also add first support for automatic \gls{sssr} \emph{stream inference} in the frontend, which internally emits push and pop \fin{pseudo-instructions}.
However, since this initial implementation only supports affine streams and we want to ensure optimal performance for the workloads we accelerate, we will stick to manually implementing \gls{sssr} streams in this thesis.

We \fin{further add} some compiler enhancements to improve workload performance.
First, we tune the in-order machine model of the Snitch core, defining appropriate resource groups and latencies to more closely model its microarchitecture\footnote{Provided in the file \texttt{llvm/lib/Target/RISCV/RISCVSchedSnitch.td.}}.
We then add a proposed \glsunset{thr}\emph{tree height reduction} (THR) pass~\cite{ueno2022thr} which reassociates arithmetic operation trees to be as shallow and balanced as possible, maximizing pipeline parallelism and hiding dependency stalls especially in the \gls{fpu}.
Finally, we tune the \gls{thr} pass to the Snitch microarchitecture by adapting its weighting and adding a maximum tree height feature, ensuring high pipeline parallelism (and thus \gls{fpu} utilization) in compiled code while avoiding register spilling in the backend.

\topar{Snitch Runtime} We add runtime helper functions using our compiler intrinsics to easily set up indirect, affine, and joining streams, as well as enable and disable \gls{sssr} redirection.
We also add defines for configuration register offsets to keep stream setup code readable.

To simplify the use of inline assembly for performance-critical sections within C/C++ code, we define macros to create blocks in which \glspl{sssr} are manually controlled.
Within these blocks, \glspl{sssr} are pre-allocated as register variables and can be added in inline assembly blocks as inputs and outputs to prevent misoptimization or reordering.

\section{Stencil Computations}
\label{chap:workloads:stc}

In this section, we will present the acceleration of \emph{stencil computations} using \glspl{sssr} in the Snitch cluster. We will begin by providing some background and summarizing our contributions.

\subsection{Background and Contributions}
\label{chap:workloads:stc:bgcontr}

Stencil computations are performance-critical to numerous data-paral\-lel, compute-intensive applications in domains like physical simulation, signal processing, and machine learning.
They iterate over points in multi-dimensional data grids and update them based on their neighbors in a fixed pattern called a \emph{stencil}.

Stencil codes share a common structure, but are highly diverse in their computational profiles, ranging from simple image filters~\cite{Pouchet2012Polybench} to complex seismic propagation operators~\cite{Jacquelin2022ScalableDH}.
The number and dimensions of I/O arrays, operations per point, and stencil shape all vary significantly between codes~\cite{Rawat2019OnOptimizingCS}, making their acceleration on modern parallel processors challenging.
Existing software proposals accelerating stencils present both generic solutions, such as code generators~\cite{Rawat2019OnOptimizingCS, You2021DRStencilED, Zhang2023RevisitingTB, Matsumura2020AN5DAS} or platform optimizations~\cite{Zhang2020DataLT, Yount2015VectorFI, Zhao2018DeliveringPS}, and tuned implementations of specific codes~\cite{Jacquelin2022ScalableDH, Rocki2020FastSC,Armejach2020svestencil}.
In both cases, the target platform's architectural features are heavily leveraged.

One major source of remaining inefficiencies are \emph{memory accesses}.
For instance, small stencils tend to result in low operational intensities and \emph{memory-bound} execution~\cite{Singh2020NEROAN, Rawat2019OnOptimizingCS, Zhao2018DeliveringPS}.
Codes with irregular stencil shapes or many I/O arrays suffer from \emph{irregular access patterns} inefficiently handled by tiered memory hierarchies~\cite{Denzler2021CasperAS, Singh2020NEROAN, Yount2015VectorFI}.
Finally, both small and irregular stencils introduce significant address calculation and load-store overheads, degrading performance particularly in energy-efficient single-issue, in-order processors.

In this section, we directly tackle these memory access challenges by accelerating generic stencil codes using \glspl{sssr}.
To this end, we develop \emph{SARIS}, a portable methodology for \textbf{S}tencil \textbf{A}cceleration using \textbf{R}egister-mapped \textbf{I}ndirect \textbf{S}treams.
SARIS itself is not specific to \glspl{sssr}: it can be used with any indirection-capable stream extension that directly maps to register accesses~\cite{Domingos2021UnlimitedVE,Wang2019StreambasedMA}, which we will generically refer to as \emph{stream registers} (SRs).
It encodes the offsets of grid elements accessed in the loop body of stencil codes in index arrays; it then reuses these indices on each point update, using the point's coordinates as an indirection base.
SARIS is highly flexible, applicable to \emph{any} stencil shape, and amenable to parallelization.
It leverages concurrent streams through multiple indirect \glspl{sr} and additional affine \glspl{sr} where beneficial.
It is orthogonal to existing code optimizations including arithmetic reassociation, loop unrolling, and the use of hardware loops.

We use SARIS to optimize parallel stencil codes for our \gls{sssr}-extended Snitch compute cluster.
We implement both platform-optimized RISC-V baseline codes and variants using SARIS to leverage the \gls{sssr} and \gls{frep} extensions,
which we provide free and open-source\footnote{\url{https://github.com/pulp-platform/snitch_cluster/tree/main/sw/saris}}.
Comparing our optimized baseline and SARIS-accelerated codes in cycle-accurate simulation of a cluster, we find significant speedups of \impval{2.72\x}, near-ideal \gls{fpu} utilizations of \impval{\SI{81}{\percent}}, and energy efficiency improvements of \impval{1.58\x} on average.
We also compare the relative performance of \gls{sssr}-accelerated SARIS codes on our 432-core scaleout \emph{Occamy}, presented in \Cref{chap:occamy}, to leading \gls{cpu}, \gls{gpu}, and \gls{wse} solutions;
despite SARIS' generality and high orthogonality to \gls{soa} optimizations, we achieve a leading peak floating-point compute utilization of \SI{83}{\percent}, which is \SI{20}{\percent} higher than that achieved by the leading \gls{gpu} code generator \emph{AN5D}~\cite{Matsumura2020AN5DAS}.

To summarize, our contributions are as follows:

\begin{itemize}
\item We present SARIS, a generic, portable, and highly flexible methodology to accelerate stencil codes with register-mapped indirect streams that is orthogonal to existing code optimizations.

\item We create optimized RV32G baseline and SARIS-accelerated parallel implementations of various 2D and 3D stencil codes for the Snitch cluster with \glspl{sssr}.

\item We evaluate our SARIS-accelerated codes in cycle-accurate RTL simulation of an \gls{sssr}-extended Snitch cluster, achieving significant speedups of \impval{2.72\x}~over \fin{RV32G} baseline codes, near-ideal \gls{fpu} utilizations of \impval{\SI{81}{\percent}}, and energy efficiency improvements of \impval{1.58\x}~ on average.

\item We compare the performance achieved by SARIS on our Occamy manycore scaleout, described in \cref{chap:occamy}, to state-of-the-art CPU, GPU, and \gls{wse} solutions, achieving up to \SI{20}{\percent} higher fractions of peak compute than a leading GPU code generator.
\end{itemize}

\subsection{The SARIS Method}
\label{chap:workloads:stc:saris}

We first describe our generic SARIS method and how it can be combined with existing code optimizations. We then present our implementation of optimized baseline and SARIS-accelerated stencil codes on the Snitch cluster without and with \glspl{sssr} in \Cref{chap:workloads:stc:map}.

\begin{figure}
    \centering
    \includegraphics[width=0.53\linewidth]{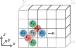}%
    \caption{Visualization of a \gls{s7ps} stencil, its constant parameters, and its 3D grid iteration. Padding cells are shown in light grey.}
    \label{fig:workloads:stc:s7ps}
\end{figure}

We describe SARIS using the example of a \glsreset{s7ps}\gls{s7ps} stencil iterating over a single data array as shown in \Cref{fig:workloads:stc:s7ps}.
We assume double-precision floating-point data, alternating buffers, and a grid halo initialized to sensible boundary conditions.

\Cref{lst:workloads:stc:s7ps_base_pseudo} shows the baseline pseudocode for one time iteration, where \texttt{inp} represents the current and \texttt{out} the next iteration buffer.
Compiling the innermost point loop for the RV32G architecture%
\footnote{We assume our grid tile is small enough for $y$ neighbors to be addressed using \SI{12}{\bit} immediate offsets, but too big to do this for $z$ neighbors, i.e. $16\leq N\leq 128$.}
without further optimizations yields the assembly code in \Cref{lst:workloads:stc:s7ps_base_asm}.
Out of \impval{20} loop instructions, only \impval{7} (\impval{\SI{35}{\percent}}) do useful compute, while \impval{12} (\impval{\SI{60}{\percent}}) are dedicated to memory accesses and address calculation.
\begin{listing}[t]%
\caption{\gls{s7ps} stencil baseline time iteration pseudocode.}%
\label{lst:workloads:stc:s7ps_base_pseudo}
\begin{minted}{python}
for z in 1 to N-1:
  for y in 1 to N-1:
    for x in 1 to N-1:
      out[z][y][x] = c0 * inp[z][y][x]
        + cx * (inp[z][y][x-1] + inp[z][y][x+1])
        + cy * (inp[z][y-1][x] + inp[z][y+1][x])
        + cz * (inp[z-1][y][x] + inp[z+1][y][x]);
\end{minted}
\end{listing}
As a result, \gls{fpu} utilization is limited to \impval{\SI{35}{\percent}} of its peak on energy-efficient single-issue in-order cores\footnote{While multi-issue out-of-order cores could further increase utilization through instruction-level parallelism, the associated energy overhead would be significant.}, even when ignoring stalls due to data dependencies and irregular memory access patterns.
While unrolling and register-tiling data along the $x$ axis could eliminate up to \impval{two} loads per output point, this would still yield at most \impval{\SI{39}{\percent}} \gls{fpu} utilization.
As we will show in \Cref{chap:workloads:stc:eval}, this limited compute efficiency manifests across all considered stencil codes.

\renewcommand\fcolorbox[4][]{\textcolor{cyan}{\strut#4}}
\begin{listing}[t]%
\caption{\gls{s7ps} stencil baseline point loop RISC-V assembly.}%
\label{lst:workloads:stc:s7ps_base_asm}
\begin{minted}{gas}
x:  fld     ft0, 0(t0)              # inp[z][y][x]
    fmul.d  ft0, %[c0], ft0
    fld     ft1, -8(t0)             # inp[z][y][x-1]
    fld     ft2,  8(t0)             # inp[z][y][x+1]
    fadd.d  ft1, ft1, ft2
    fmadd.d ft0, %[cx], ft1, ft0
    fld     ft1, -YOFFS(t0)         # inp[z][y-1][x]
    fld     ft2,  YOFFS(t0)         # inp[z][y+1][x]
    fadd.d  ft1, ft1, ft2
    fmadd.d ft0, %[cy], ft1, ft0
    fld     ft1, 0(t1)              # inp[z-1][y][x]
    fld     ft2, 0(t2)              # inp[z+1][y][x]
    fadd.d  ft1, ft1, ft2
    fmadd.d ft0, %[cz], ft1, ft0
    fsd     ft0, 0(t3)
    addi    t0, 8
    addi    t1, 8
    addi    t2, 8
    addi    t3, 8
    bne     t0, a0, x
\end{minted}
\end{listing}

The SARIS method accelerates stencil computation by minimizing the above memory access overheads through the use of \glspl{sr}. It involves the following steps on the point loop:

\begin{enumerate}
    \item Map all \emph{grid data loads} to indirect stream reads.
    \item \noprotrusion\emph{Partition} these reads among available indirect \glspl{sr}, maximizing their concurrent use and balancing their utilization.
    \item Map \emph{grid data stores} or loads of \emph{constant stencil coefficients} that cannot be kept in the register file to remaining \glspl{sr}.
    \item Determine a \emph{point loop schedule} specifying in which order the computations in the point loop access streams; this will determine the index arrays for each indirect \gls{sr}.
\end{enumerate}

\begin{figure}
    \centering
    \includegraphics[width=0.76\linewidth]{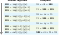}%
    \caption{SARIS point loop schedule for the \gls{s7ps} stencil.}
    \label{fig:workloads:stc:s7ps_pls}
\end{figure}

The method may be iterated on for best performance, especially when combined with orthogonal optimizations like those \fin{later discussed in this section}.
We demonstrate its steps on our \fin{\gls{s7ps} stencil} code example;
as in the \gls{sssr}-extended Snitch cluster, we assume the two indirect \glspl{sr} (\texttt{SR0} and \texttt{SR1}) and one affine \gls{sr} (\texttt{SR2}) to be available:
\begin{enumerate}
    \item We map all seven grid data loads to indirect stream reads.
    \item For each axis, we map the two opposing grid point loads to \texttt{SR0} and \texttt{SR1}, respectively, so they can concurrently be read by an addition operation. The remaining center point load is mapped to \texttt{SR0}, which already results in minimal utilization imbalance between \texttt{SR0} and \texttt{SR1}.
    \item Since all four stencil coefficients ($c_0$, $c_x$, $c_y$, $c_z$) can be kept in the register file, we map our single grid data store to \texttt{SR2} using an affine 3D pattern along the grid.
    \item We use the same computation order as our baseline. The resulting point loop schedule, shown in \Cref{fig:workloads:stc:s7ps_pls}, lists each compute operation and its stream accesses in order.
\end{enumerate}

\begin{listing}%
\caption{\gls{s7ps} stencil SARIS time iteration pseudocode.}%
\label{lst:workloads:stc:s7ps_saris_pseudo}
\begin{minted}{python}
# Configure input offsets as indices (SR 0,1)
sr_set_idcs(SR0, {&inp[1][1][1], &inp[1][1][2],
                  &inp[1][2][1], &inp[2][1][1]});
sr_set_idcs(SR1, {&inp[1][1][0], &inp[1][0][1],
                  &inp[0][1][1] });
# Affine write stream stores outputs (SR 2)
sr_affine_write_3d(SR2, &out[1][1][1], ...);

for z in 0 to N-2:
  for y in 0 to N-2:
    for x in 0 to N-2:
      # Launch indirect input reads with
      # point offset as array base (SR 0,1).
      sr_indir_read(SR0|SR1, OFFS(x,y,z), ...);
      # Computation in order of point loop schedule
      acc =  c0 * sr_read(SR0);
      acc += cx * (sr_read(SR0) + sr_read(SR1));
      acc += cy * (sr_read(SR0) + sr_read(SR1));
      acc += cz * (sr_read(SR0) + sr_read(SR1));
      sr_write(SR2, acc);
\end{minted}
\end{listing}

\begin{listing}%
\caption{\gls{s7ps} stencil SARIS point loop RISC-V assembly.}%
\label{lst:workloads:stc:s7ps_saris_asm}
\begin{minted}{gas}
x:  SRIR    SR0|SR1, t0           # SSSRs: 3 instructions
    fmul.d  ft0, %[c0], SR0
    fadd.d  ft1, SR0, SR1
    fmadd.d ft0, %[cx], ft1, ft0
    fadd.d  ft1, SR0, SR1
    fmadd.d ft0, %[cy], ft1, ft0
    fadd.d  ft1, SR0, SR1
    fmadd.d SR2, %[cz], ft1, ft0
    addi    t0, 8
    bne     t0, a0, x
\end{minted}
\end{listing}

The SARIS-accelerated pseudocode using our new point loop schedule is shown in \Cref{lst:workloads:stc:s7ps_saris_pseudo}.
We first configure the static index arrays for \texttt{SR0} and \texttt{SR1}; for simplicity, we keep all indices positive by defining offsets around the iteration origin $(1,1,1)$ and changing the axis iteration ranges to $0\;..\;N-2$.
We then configure \texttt{SR2} with the 3D iteration needed to write our output data.
Inside the point loop, we launch \texttt{SR0} and \texttt{SR1} with the grid point's address offset as the base; the indices remain the same for each grid point iteration.
Finally, we perform our computation, reading and writing \glspl{sr} in the order defined by our point loop schedule.

\Cref{lst:workloads:stc:s7ps_saris_asm} shows the resulting RISC-V assembly for the SARIS-accelerated point loop without further optimizations. %
Except for launching \texttt{SR0} and \texttt{SR1} (\impval{3} instructions on Snitch with \glspl{sssr}), incrementing the grid point, and the loop branch, \emph{all} loop instructions now perform useful compute, almost doubling the ratio of useful compute instructions from \impval{\SI{35}{\percent}} to \impval{\SI{58}{\percent}}. %
Unlike for the baseline, however, the non-compute overhead in the point loop is \emph{static} and can significantly be reduced through complementary optimizations, enabling our near-ideal mean \gls{fpu} utilizations of \impval{\SI{81}{\percent}} in \Cref{chap:workloads:stc:eval}.

Going beyond our example, SARIS can accelerate \emph{any} stencil code as it makes very few assumptions. %
It supports any sequence of computations on grids of any dimensionality and size. %
The indirect streams enable arbitrarily shaped stencils, and since the indices include array bases, any number of I/O arrays may be streamed. %
Indices can even dynamically be rewritten if necessary.

In principle, SARIS could automatically be performed by the compiler whenever a fixed sequence of address offsets is repeatedly accessed with changing bases. This generalized formulation could also enable its application beyond stencil codes.
Here, we focus on the manual application of the SARIS methodology to accelerate stencil codes as is routinely done when optimizing low-level library code;
compiler inference is left as future work.

\subsubsection{Complementary Optimizations}
\label{chap:workloads:stc:saris:opt}

SARIS is highly flexible; to further increase \gls{fpu} utilization, it can be combined with existing code optimizations, \mmp{which can either be implemented manually or \fin{inferred} automatically by a compiler}: %

\topar{Unrolling} \fin{By} unrolling the point loop and processing blocks of points at once, the impact of SARIS' static overheads can significantly be reduced. To this end, \gls{sr} streams can be extended to access grid points for multiple iterations. This also enables multi-dimensional unrolls, which can reduce parallelization imbalance.

\topar{Reordering and reassociation} \fin{By} reordering indirect stream indices, point loop computations can freely be reordered and reassociated. This can improve performance by avoiding memory and dependency stalls, especially when combined with unrolling.

\topar{Hardware loops} SARIS is orthogonal to hardware loops, which reduce point iteration overheads. If they provide a separate instruction buffer, they also reduce instruction cache pressure and misses.

\topar{Parallelization} SARIS is compatible with parallelization at the grid point level. As every point iteration launches independent indirect streams, (unrolled blocks of) points can freely be distributed among multiple cores, for example in an interleaved fashion. Additionally, grids can be subdivided into tiles as usual.

\subsection{SSSR Cluster Implementation}
\label{chap:workloads:stc:map}

\begin{table}[t!]
\centering
\begin{tabular}{p{7.2em}crrrr}
\toprule
  Code%
& Dims.%
& Rad.%
& \#Loads%
& \#Coeffs.%
& \#FLOPs%
\\ \midrule
\texttt{jacobi\_2d}~\cite{Pouchet2012Polybench}     & 2D & 1  & 5   & 1   & 5   \\
\texttt{j2d5pt}~\cite{Matsumura2020AN5DAS}          & 2D & 1  & 5   & 6   & 10  \\
\texttt{box2d1r}~\cite{Matsumura2020AN5DAS}         & 2D & 1  & 9   & 9   & 17  \\
\texttt{j2d9pt}~\cite{Matsumura2020AN5DAS}          & 2D & 2  & 9   & 10  & 18  \\
\texttt{j2d9pt\_gol}~\cite{Matsumura2020AN5DAS}     & 2D & 1  & 9   & 10  & 18  \\
\texttt{star2d3r}~\cite{Matsumura2020AN5DAS}        & 2D & 3  & 13  & 13  & 25  \\
\texttt{star3d2r}~\cite{Matsumura2020AN5DAS}        & 3D & 2  & 13  & 13  & 25  \\
\texttt{ac\_iso\_cd}~\cite{Jacquelin2022ScalableDH} & 3D & 4  & 26  & 13  & 38  \\
\texttt{box3d1r}~\cite{Matsumura2020AN5DAS}         & 3D & 1  & 27  & 27  & 53  \\
\texttt{j3d27pt}~\cite{Matsumura2020AN5DAS}         & 3D & 1  & 27  & 28  & 54  \\
\bottomrule
\end{tabular}
\caption{Overview of implemented stencil codes sorted by FLOPs per grid point. The grid loads and coefficients are also per grid point.}
\label{tbl:workloads:stc:codes}
\end{table}
We evaluate SARIS on our default \gls{sssr}-enhanced Snitch cluster introduced in \cref{chap:sssrs:arch:integr}, which provides eight single-issue, in-order RV32G worker cores extended with \glspl{sssr} and \gls{frep}.
The default \gls{sssr} streamer in each core provides two indirection-capable \glspl{issr} (\texttt{ft0} and \texttt{ft1}) and another \gls{essr} (\texttt{ft2}) capable of affine streams.
We recall that the \gls{frep} hardware loop allows the integer core and \gls{fpu} to concurrently execute instructions in a \emph{pseudo-dual-issue} fashion, enabling \glsunset{ipc}\glspl{ipc} (instructions per cycle) of above one in principle.
The default cluster features \SI{128}{\kibi\byte} of \fin{low-latency} \gls{tcdm} across 32 banks and a \fin{512-bit} programmable \gls{dma} engine that enables high-bandwidth bulk data transfers between \gls{tcdm} and main memory.

\Cref{tbl:workloads:stc:codes} lists the stencil codes we implement on the Snitch cluster and their performance-relevant characteristics, sorted by the number of \glspl{flop} per grid point.
All codes use double-precision data.
For each, we implement one time iteration on a $64^2$ (2D) or $16^3$ (3D) grid tile including halos.
We double-buffer the grid tiles in \gls{tcdm} and program the \gls{dma} engine to concurrently transfer them from and to main memory in 2D or 3D transfers as is required for continuous computation on a larger grid.
Like our target platform, our codes are available open-source.

Each code is implemented in two parallelized variants, which are both compiled using the extended Snitch LLVM 15 toolchain introduced in~\Cref{chap:workloads:rtllvm}.
The \emph{optimized baseline} (\textsc{base}) variants target the RV32G architecture without extensions; they are compiled using the \texttt{-Ofast} optimization level and the improved \gls{thr} pass discussed in~\Cref{chap:workloads:rtllvm} to maximize \gls{fpu} utilization.
The \emph{SARIS-accelerated} (\textsc{saris}) variants target the RV32G architecture with the \gls{sssr} and \gls{frep} extensions; they also use \texttt{-Ofast} \fin{with} \gls{thr} enabled, but have manually scheduled point loops using SARIS and use FREP where possible.
Both \textsc{base} and \textsc{saris} variants parallelize their point loops among the eight cluster cores using four-fold $x$-axis and two-fold $y$-axis iteration interleaving, and further unroll their point loops up to four-fold \fin{if and \emph{only} if} beneficial to performance.

\subsection{Performance and Power Evaluation}
\label{chap:workloads:stc:eval}

We analyze the performance (\Cref{chap:workloads:stc:eval:perf}) and energy efficiency benefits (\Cref{chap:workloads:stc:eval:enpow}) of SARIS-based \gls{sssr} acceleration in our eight-core Snitch cluster.

\subsubsection{Performance}
\label{chap:workloads:stc:eval:perf}

\begin{figure}
\captionsetup[subfigure]{justification=centering}%
\centering
\subfloat[Execution speedup of \textsc{saris} over \textsc{base} code variants.]{%
\label{fig:workloads:stc:perf_su}%
\includegraphics[width=\linewidth]{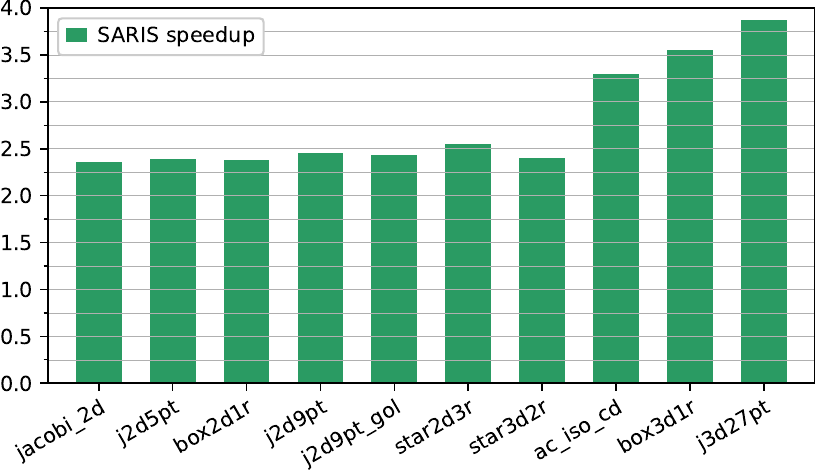}%
}%
\\[1em]%
\subfloat[\gls{fpu} utilization (as a fraction) and per-core \glsunset{ipc}\gls{ipc} for both variants.]{%
\label{fig:workloads:stc:perf_fpipc}%
\includegraphics[width=\linewidth]{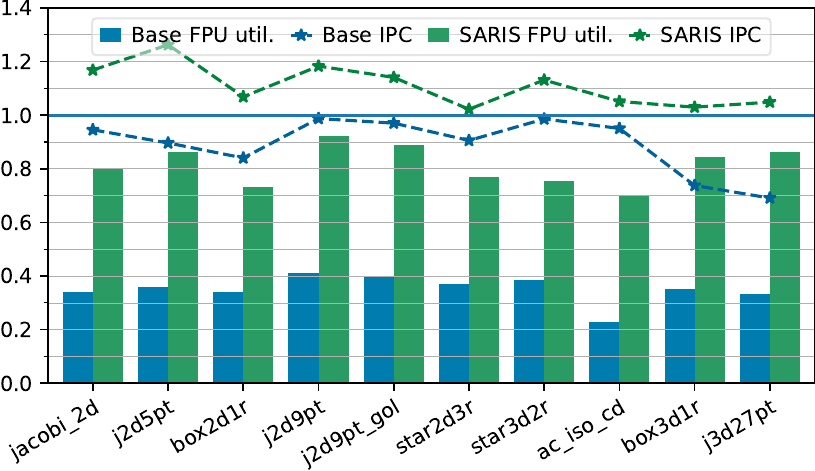}%
}
\caption{Performance comparison of \textsc{base} and \textsc{saris} stencil code variants, with codes sorted by \glspl{flop} per grid point.}%
\label{fig:workloads:stc:perf}%
\end{figure}

We execute our stencil codes in cycle-accurate simulations of the Snitch cluster's \gls{rtl} description. We then extract exact runtimes and utilization metrics from simulation traces.

\Cref{fig:workloads:stc:perf_su} shows the speedups of \textsc{saris} over \textsc{base} code variants.
SARIS achieves a significant geomean speedup of \impval{2.72\x}, with a clear increasing trend as \glspl{flop} per grid point increase;
the codes with the fewest (\texttt{jacobi\_2d}) and most (\texttt{j3d27pt}) \glspl{flop} per grid point also exhibit lowest and highest speedups of \impval{2.36\x}~and \impval{3.87\x}, respectively.
In between, speedups remain approximately stable up to \texttt{star3d2r} (geomean \impval{2.42\x}), then increase further.

To explain this trend, we observe that codes past this point involve many \emph{grid loads} and \emph{coefficients} (see \Cref{tbl:workloads:stc:codes}), increasing the register pressure in \textsc{base} variants.
This reduces the benefits of unrolling, which may exhaust architectural registers and require inefficient stack accesses.
However, reducing unrolling increases dependency stalls, also negatively impacting performance.
SARIS avoids this register bottleneck by streaming grid points and register-exhausting coefficients directly from \gls{tcdm}, maintaining the full benefits of unrolling and further increasing speedups.

\Cref{fig:workloads:stc:perf_fpipc} depicts the \gls{fpu} utilization and per-core \glsunset{ipc}\gls{ipc} for both variants.
As expected from our example in \Cref{chap:workloads:stc:saris}, SARIS-based \gls{sssr} acceleration significantly improves the geomean \gls{fpu} utilization from \impval{\SI{35}{\percent}} to a near-ideal \impval{\SI{81}{\percent}}.
It also enables the effective use of Snitch's pseudo-dual-issue capabilities, increasing geomean \gls{ipc} from \impval{0.89} to \impval{1.11}.
SARIS performance remains consistently high, with \gls{fpu} utilization and \gls{ipc} never dropping below \impval{\SI{70}{\percent}} and \impval{1.0}, respectively.
Variations in \gls{fpu} utilization and \gls{ipc} can be attributed to multiple effects.
For non-register-bound codes (up to \texttt{star3d2r}), \gls{ipc} variations in both variants resemble each other as both are impacted by \gls{tcdm} \emph{access contention}; codes with many \emph{grid loads} or large \emph{stencil overlaps} between cores exhibit more stalls due to bank conflicts.
For register-bound codes (past \texttt{star3d2r}), the \textsc{base} \gls{ipc} drops down to a minimum of \impval{0.69} due to dependency stalls, while \textsc{saris} variants avoid the register bottleneck through streaming as previously explained.
Finally, as the number of \emph{grid loads} and the \emph{stencil radius} increase, so does the setup overhead of SARIS: more indices must be stored for fewer point iterations doing useful compute. This is why \texttt{ac\_iso\_cd}, having the largest radius and many point loads, exhibits the minimum \textsc{saris} \gls{fpu} utilization of \impval{\SI{70}{\percent}}.

Overall, the inefficiencies preventing full \gls{fpu} utilization in \textsc{saris} codes are index and \gls{sssr} initialization overheads, \gls{tcdm} access contention, instruction cache misses, and core runtime imbalances.

\subsubsection{Energy and Power}
\label{chap:workloads:stc:eval:enpow}

We implement the Snitch cluster in  GlobalFoundries’ 12LP+ FinFET technology using \emph{Fusion Compiler} and execute all considered stencil codes in post-layout gate-level simulation at the cluster's target clock speed of \SI{1}{\giga\hertz}.
We record the resulting switching activity and use it for cluster power estimation in \emph{PrimeTime}, assuming typical operating conditions of \SI{25}{\celsius} and \SI{0.8}{\volt} core supply voltage.

\begin{figure}[t]
  \centering
  \includegraphics[width=\linewidth]{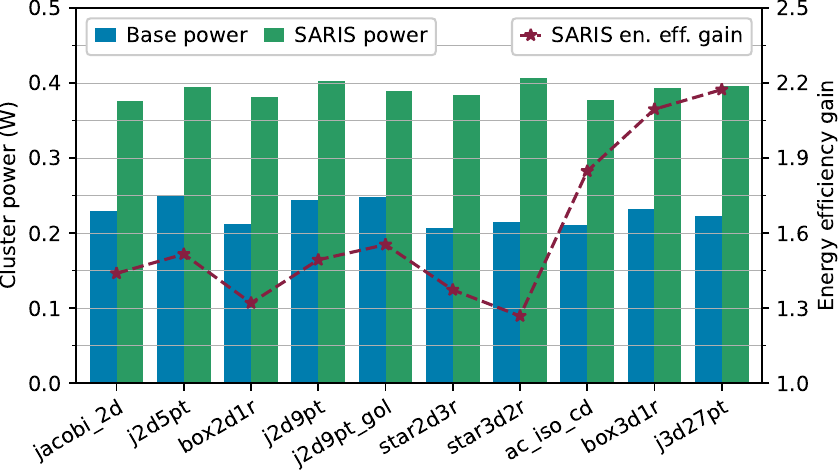}%
  \caption{Cluster power consumption for \textsc{base} and \textsc{saris} variants and \textsc{saris} energy efficiency improvement over \textsc{base}.}
  \label{fig:workloads:stc:eval:enpow}
\end{figure}

\Cref{fig:workloads:stc:eval:enpow} shows the resulting power consumption for each code, as well as the energy efficiency benefits of SARIS-based \gls{sssr} acceleration.
The power consumption for both variants is fairly stable across codes with geomeans of \impval{\SI{227}{\milli\watt}} and \impval{\SI{390}{\milli\watt}} for \textsc{base} and \textsc{saris}, respectively.
As would be expected, the slight variations in power consumption among either variant resemble those in \gls{fpu} utilization.
While the increased \gls{fpu} utilization of \textsc{saris} variants results in \impval{1.72\x}~higher mean power consumption, their significant speedups result in notable energy efficiency gains for all codes, ranging from \impval{1.27\x}~to \impval{2.17\x}~with a geomean of \impval{1.58\x}.
Accordingly, we observe the same increase in energy efficiency gains for register-bound codes as \fin{we did} for speedups.

\subsection{Related Work}
\label{chap:workloads:stc:relwork}

While some \fin{existing} \gls{sr} proposals like %
the \fin{\glsreset{uve}\gls{uve}}~\cite{Domingos2021UnlimitedVE} accelerate selected stencil benchmarks using affine streams,
\fin{to our knowledge}, SARIS is the first generic stencil acceleration method leveraging indirect \glspl{sr} to enable near-ideal FPU utilizations.
Beyond \glspl{sr}, numerous existing works accelerate stencil codes with dedicated hardware or through optimized CPU, GPU, and \gls{wse} software:

\topar{Hardware Accelerators} \fin{Dedicated} stencil accelerators often target \glspl{fpga} or operate near memory.
\emph{Casper}~\cite{Denzler2021CasperAS} accelerates stencils near a CPU's last-level cache by leveraging its high bandwidth and data locality, incurring minimal area.
\emph{Nero}~\cite{Singh2020NEROAN} is a CAPI2-interfaced \gls{fpga}+HBM solution using high-level synthesis to accelerate two compound kernels fundamental to weather prediction.
\emph{SODA}~\cite{Chi2018SODASW} automatically generates optimized dataflow architectures for stencil algorithms on \glspl{fpga} with provably optimal data reuse.

\topar{CPUs} Many CPU software approaches improve stencil performance through \glspl{dlt} and vectorization.
Zhang et al.~\cite{Zhang2020DataLT} vectorize stencil codes for \mmp{Arm's} NEON extension by packing grid loads for multiple output points contiguously in memory, reaching up to \impval{\SI{29}{\percent}} of peak compute on one core of an FT-2000+ processor.
Yount~\cite{Yount2015VectorFI} proposes \emph{vector folding}, a \gls{dlt} forming $n$D \emph{folds} in memory to vectorize stencils along multiple axes, using up to \impval{\SI{30}{\percent}} of a Xeon Phi 7120A's peak compute.
\emph{Bricks}~\cite{Zhao2018DeliveringPS} reorganizes large grids into fine-grain blocks contiguous in memory to exploit the full $n$D data locality of stencils and reduce \gls{tlb} pressure, achieving up to \impval{\SI{45}{\percent}} of peak compute on Xeon Gold 6130.
\mmp{Finally, Armejach et al.~\cite{Armejach2020svestencil} accelerate vectorized stencil codes on a simulated eight-core Armv8 system with Arm's \gls{sve}, reaching up to \SI{47}{\percent} of peak FP64 compute through textbook optimizations including loop unrolling, loop fusion, and register-level data reuse.}

\topar{GPUs} Numerous highly optimized code generators have been proposed for stencils on GPUs.
\emph{ARTEMIS}~\cite{Rawat2019OnOptimizingCS} leverages many existing and novel tiling, decomposition, fusion, and folding optimizations with parameters guided by profiling; it enables up to \impval{\SI{36}{\percent}} of peak performance on a P100 GPU.
\emph{DRStencil}~\cite{You2021DRStencilED} improves operational intensity and data reuse by fusing time steps and partitioning the resulting stencils with autotuned parameters, achieving up to \impval{\SI{48}{\percent}} of peak compute on P100.
\emph{AN5D}~\cite{Matsumura2020AN5DAS} uses a performance model to generate highly optimized CUDA code with automatic spatial and temporal blocking and \gls{sm} double-buffering from a C source, improving performance scaling and further increasing peak compute utilization to \impval{\SI{69}{\percent}} on V100 SXM2.
\emph{EBISU}~\cite{Zhang2023RevisitingTB} leverages the much larger \gls{sm} in newer GPUs by trading deep temporal blocking for lower device occupancy, achieving notable speedups over AN5D despite using only \impval{\SI{49}{\percent}} of peak compute on A100.

\topar{Wafer-Scale Engines} \fin{The} spatial architecture and high on-chip bandwidth of \glspl{wse} make them attractive for stencil computation scaleouts on large grids.
Rocki et al.~\cite{Rocki2020FastSC} solve a large linear system arising from a 7-point stencil with mixed precision on Cerebras WSE-1, \pagebreak[3] utilizing \impval{\SI{28}{\percent}} of the system's peak performance.
Jacquelin et al.~\cite{Jacquelin2022ScalableDH} scale out a \fin{25-point acoustic isotropic constant-density seismic simulation kernel (\texttt{ac\_iso\_cd} in our evaluation)} on the even larger WSE-2, mapping grid planes to tiles and streaming along the $z$ axis; they use \impval{\SI{28}{\percent}} of their engine's peak compute.

\begin{table}
\setlength{\tabcolsep}{0.6em}
\centering
\begin{tabular}{cp{8.6em}p{8.6em}cr}
\toprule
& Work &%
Platform &%
Prec. &%
\% Pk. \\
\midrule
\multirow{4}{*}{\adjustbox{angle=90}{CPU}}%
& Zhang et al.~\cite{Zhang2020DataLT}            & FT-2000+ (1 core) & FP64    & \SI{29}{\percent} \\
& Yount~\cite{Yount2015VectorFI}                 & Xeon Phi 7120A    & FP32    & \SI{30}{\percent} \\
& \emph{Bricks}~\cite{Zhao2018DeliveringPS}      & Xeon Gold 6130    & FP32    & \SI{45}{\percent} \\
& \scalebox{0.945}[1]{Armejach et al.}~\cite{Armejach2020svestencil}    &Armv8+SVE syst.  & FP64    & \SI{47}{\percent} \\
\midrule
\multirow{4}{*}{\adjustbox{angle=90}{GPU}}%
& \emph{ARTEMIS}~\cite{Rawat2019OnOptimizingCS}  & Tesla P100        & FP64    & \SI{36}{\percent} \\
& \emph{DRStencil}~\cite{You2021DRStencilED}     & Tesla P100        & FP64    & \SI{48}{\percent} \\
& \emph{AN5D}~\cite{Matsumura2020AN5DAS}         & Tesla V100 SXM2   & FP32    & \SI{69}{\percent} \\
& \emph{EBISU}~\cite{Zhang2023RevisitingTB}      & A100              & FP64    & \SI{49}{\percent} \\
\midrule
\multirow{2}{*}{\adjustbox{angle=90}{WSE}}%
& Rocki et al.~\cite{Rocki2020FastSC}            & Cerebras WSE-1    & FP16-32 & \SI{28}{\percent} \\
& \scalebox{0.95}[1]{Jacquelin et al.}~\cite{Jacquelin2022ScalableDH} & Cerebras WSE-2    & FP32    & \SI{28}{\percent} \\
\midrule
& \textbf{SARIS (Ours)}                          & Occamy            & FP64    & \textbf{\SI{83}{\percent}} \\
\bottomrule
\end{tabular}
\caption{Overview of discussed stencil software approaches. \emph{\%~Pk.} is the highest fraction of peak compute achieved.}
\label{tbl:workloads:stc:relwork_sw}
\end{table}

\Cref{tbl:workloads:stc:relwork_sw} summarizes the software works discussed above, listing the highest fraction of peak compute reported by each, and compares them to our work.
To ensure a fair system-level comparison, we compare these figures not to our cluster-level \gls{fpu} utilizations, but to the highest fraction of peak compute we achieve on the Occamy manycore scaleout we present in \cref{chap:occamy}.
We see that SARIS-based \gls{sssr} acceleration on Occamy reaches the highest peak floating-point performance utilization of \SI{83}{\percent}, which is \SI{20}{\percent} higher than the leading GPU code generator \emph{AN5D}.
Methodologically, most of the discussed software innovations are orthogonal to SARIS; while some \gls{dlt} approaches may be obviated by the fine-grain gathering abilities of indirect \glspl{sr}, further optimizations improving data reuse, arithmetic intensity, locality, and tiling could all benefit our Occamy scaleout with SARIS, potentially enabling even higher performance.

\section{Spiking Neural Networks}
\label{chap:workloads:snn}

After accelerating stencil codes with \glspl{sssr} using SARIS, we turn our attention to \emph{spiking neural networks} (SNNs). We again begin with some background on the topic and a summary of our contributions.

\subsection{Background and Contributions}
\label{chap:workloads:snn:bgcontr}

The rapid advancement of deep learning across a wide range of disciplines has led to a massive growth in computational demand~\cite{Castro2023VENOMAV}.
This surge fuels interest in \emph{neuromorphic computing}, which emulates the operating principles of biological neurons and their synaptic connections to build more compact and energy-efficient \gls{ml} models.

\glspl{snn} are at the forefront of this trend~\cite{yang2023neurorv}, achieving competitive performance on image classification and object detection tasks at a lower power envelope and memory footprint than their non-spiking counterparts~\cite{lien2022objdet,lee2022ptb,kim2023cdnn}.
To handle their event-driven nature, spike-based communication, and complex activation functions, several neuromorphic processors and accelerators have been proposed,
including both von-Neumann and non-von-Neumann architectures.
The latter feature large arrays of neuron and synapse circuits~\cite{schuman2022neuroopp} and are implemented as analog and mixed-signal circuits~\cite{moradi2018dynaps, benjamin2014neurogrid}, \gls{gals} processors \cite{davies2018loihi, frenkel2019odin}, or traditional fully synchronous digital accelerators \cite{lee2022ptb, lien2022objdet, kim2023cdnn, dimauro2022sne, wang2022lsmcore}.

In practice, analog neuromorphic processors face accuracy issues, as their continuous signals are vulnerable to voltage and temperature noise \cite{ottati2023spike}.
\gls{gals} processors are challenging to design due to the low maturity of asynchronous design automation tools \cite{yang2023neurorv}.
Both of these paradigms are also difficult to integrate with existing infrastructure, often requiring entirely new software stacks \cite{yang2023neurorv}.

Meanwhile, fully synchronous digital accelerators face fewer integration challenges, but usually only support specific network topologies, neuron models, or fixed low-precision arithmetic~\cite{dimauro2022sne, wang2022lsmcore, kim2023cdnn}.
The latter is especially problematic, as aggressive integer quantization is not as mature in \glspl{snn} as in non-spiking networks and can lead to significant accuracy degradation~\cite{lien2022objdet,zanatta2024snn}:
for instance, Zanatta et al.~\cite{zanatta2024snn}~find that quantizing a pretrained \gls{snn} from FP32 to INT4 renders it ineffective on a drone obstacle avoidance task, suggesting that floating-point compute remains beneficial to \gls{snn} inference.

To avoid the fixed compute constraints of synchronous digital accelerators and keep up with the rapid evolution of the \gls{snn} ecosystem, a convenient hardware target could be energy-efficient \emph{general-purpose} processors.
However, traditional CPU architectures face challenges with the event-driven nature of \glspl{snn}, which result in sparse data structures \cite{yang2023neurorv}, irregular memory access patterns, and thus poor utilization of functional units.

In this section, we address these challenges by accelerating \gls{snn} inference on the general-purpose Snitch cluster using \glspl{sssr}, which can efficiently handle the indirect streams underlying \gls{snn} inference.
Unlike \gls{soa} solutions that require large, highly specialized hardware units, our approach requires only our lightweight stream extension that can also be used to accelerate various other irregular workloads.
We first implement parallel FP16 and FP8 SIMD (RV32GXsf\footnote{The \emph{Xsf} (smallfloat) \gls{isa} extension~\cite{bertaccini2024fpu} adds support for the SIMD FP16 and FP8 computations we leverage in this section; it is orthogonal to \glspl{sssr}.}) baselines on the Snitch cluster by adopting a compressed representation for the sparse input \glspl{fm}.
In the process, we identify dynamic-length \glsunset{spwr}\emph{sparse weight reductions} (SpWRs), which use indirect streams to gather weights associated with input spikes, as the main source of inefficiency in neuromorphic processing on CPUs.
By accelerating these \glspl{spwr} using \glspl{sssr}, we can speed up end-to-end FP16 inference on S-VGG11 by \nchk{5.5\x} and improve its energy efficiency by \nchk{3.3\x}, increasing its \gls{fpu} utilization from \nchk{\SI{9.3}{\percent}} to \nchk{\SI{52}{\percent}}.
FP8 inference provides a further \nchk{1.7\x} improvement to both runtime and energy efficiency.
Compared to \gls{soa} neuromorphic processors, our software solution on the \gls{sssr}-enhanced Snitch cluster achieves competitive performance, outperforming Intel's \emph{Loihi}~\cite{davies2018loihi} by \nchk{1.3\x} and \nchk{2.4\x} in FP16 and FP8, respectively.
Compared to the leading neuromorphic processor \emph{LSMCore}~\cite{wang2022lsmcore}, S-VGG11 inference on the \gls{sssr}-extended Snitch cluster is \nchk{2.2\x} and \nchk{4.4\x} more energy efficient in FP16 and FP8, respectively.

To summarize, our contributions are as follows:

\begin{itemize}
    \item We implement parallel, double-buffered FP16 and FP8 \gls{snn} inference on the Snitch cluster with and without \glspl{sssr}, accelerating the performance-critical \glspl{spwr} using indirect \gls{sssr} streams.

    \item We evaluate the performance and energy efficiency of both variants on S-VGG11, finding that \glspl{sssr} accelerate FP16 inference by \nchk{5.5\x} and improve its energy efficiency by \nchk{3.3\x}, with FP8 further improving both by \nchk{1.7\x}. They also increase overall FP16 \gls{fpu} utilization from \nchk{\SI{9.3}{\percent}} to \nchk{\SI{52}{\percent}}.

    \item We qualitatively and quantitatively compare our \gls{sssr}-acceler-\linebreak ated solution to \gls{soa} neuromorphic processors, finding that we achieve competitive performance and up to \nchk{4.4\x} higher energy efficiency in FP8 than the leading \emph{LSMCore}.
\end{itemize}

Before discussing our implementation of \gls{snn} inference on the Snitch cluster in \cref{chap:workloads:snn:map}, we will shortly \fin{provide some background on} spiking neurons and the \gls{lif} model we will use for them.

\subsubsection{Spiking Neurons}

\glspl{snn} use spiking neurons as their computational primitives, which communicate through discrete spikes. As a neuron receives spikes from its neighbors, its potential rises over time until a threshold is reached, causing an output spike to be generated.

\begin{equation}\label{eq:workloads:snn:lif}
    \begin{cases}
        i_m(t) = \sum_{n=1}^N s_{i, n}(t) w_{n} \\
        v_m(t) = v_m(t-1)\alpha + r i_m(t) - v_{rst} s_{o, m}(t) \\
        s_{o,m}(t) = \begin{cases}
            1 & \text{if } v_m(t) \ge v_{th} \\
            0 & \text{otherwise}
        \end{cases}
    \end{cases}
\end{equation}

\cref{eq:workloads:snn:lif} \fin{describes} the commonly used \gls{lif}~\cite{Abbott1999lif} model for a spiking neuron $m$. Here, $s_{i,n}(t)$ and $s_{o,m}(t)$ represent the input spike from synapse $n$ and the output spike of neuron $m$ at time $t$, respectively.
 $w_{n}$ is the weight assigned to the incoming synapse $n$.
 $i_m(t)$ denotes the total incoming spike current and $v_m(t)$ the potential of neuron $m$ at time $t$.
 $v_r^{st}$, $v_{th}$, $r$, and $\alpha$ are the reset potential, membrane threshold, membrane resistance, and decay factor, respectively. We note that $r$ is usually 1.

\subsection{SSSR Cluster Implementation}
\label{chap:workloads:snn:map}

In this section, we \mmpcut{will }discuss the implementation of \gls{snn} layer inference on our extended Snitch cluster and its acceleration with \glspl{sssr}.
We \mmpcut{will }first introduce an efficient representation of spike \glspl{fm} using a \gls{csr}-like format.
We \mmpcut{will }then present our cluster-parallel layer computation and its acceleration using \glspl{sssr}.
Next, we will discuss our cluster \gls{tcdm} double-buffering and tiling scheme.
Finally, we \mmpcut{will }describe the encoding of regular input images into spikes in the first layer.

\subsubsection{Sparse \gls{fm} Format}

We efficiently store \glspl{fm} in a \gls{csr}-like format:
for each neuron, we store the indices of spiking \emph{channels} in a shared index array \verb|c_idcs|, wherein neurons are delimited by a pointer array \verb|s_ptr|.
Since spikes are binary, there is no value array; the intensity of any nonzero spike is one.
For \mmp{fully connected} layers, distinguishing channels and spatial neuron coordinates becomes unnecessary, and thus \verb|s_ptr| becomes a single value counting the number of spikes.

Unlike the commonly used \gls{aer}, which requires tracking the temporal and spatial coordinates of each spike~\cite{dimauro2022sne}, our format avoids these overheads: time steps are processed at once and the spatial representation is dense, significantly simplifying processing.

\subsubsection{Layer Computation}

For each layer, we parallelize inference among cluster cores by distributing output channels of individual neurons using a workload stealing scheme:
an atomically incremented counter indicates the next output neuron and channel to process.
At a higher level, this scheme is combined with tensor double-buffering and tiling in the cluster \gls{tcdm}, which we will describe in the next section.

\begin{listing}[t]%
\caption{Single-core pseudocode to process one output channel of a single neuron; work is distributed by a workload-stealing scheme.}
\label{lst:workloads:snn:siter}
\begin{minted}{python}
# core_oneur, core_ochan, core_w_base set by greedy schedule
# Initialize next state using state v(t)
vnext = v[core_oneur][core_ochan]
# Iterate over spatial filter dimensions
for frow in 0 to KH:
  for fcol in 0 to KW:
    # Compute index of connected neuron
    ineur = core_oneur + frow * KH + fcol;
    # SpWR: accumulate weighted spikes onto state
    for ichan in s_ptr_i[ineur] to s_ptr_i[ineur+1]:
      vnext += core_w_base[c_idcs_i[ichan]]
# Apply activation function and write back
act(v, vnext, core_oneur, core_ochan, s_idcs_tile)
\end{minted}
\end{listing}
\cref{lst:workloads:snn:siter} shows the single-core pseudocode for processing one output channel of a single neuron, which includes the fused activation.
We first initialize the next potential $v(t+1)$ using the current state $v(t)$.
We then iterate over the spatial filter dimensions $k_h$ and $k_w$ in two outer loops, and accumulate the weighted input spikes of neighboring neurons in an inner loop.

The inner loop is a \gls{spwr}, which performs a streaming indirection.
As such, we can accelerate it using a single \gls{issr} that streams the weights (\verb|core_w_base|) at the spiking input channels' indices (\verb|c_idcs_i|).
We can also accelerate the loop itself using \gls{frep}, avoiding register dependencies through its staggering feature as we did in \cref{chap:sssrs}.
Together, \glspl{sssr} and \gls{frep} reduce the performance-critical inner \gls{spwr} loop body from seven instructions to one useful floating-point addition.

Depending on the target precision, we can use the \gls{fpu}'s \gls{simd}
capabilities by mapping multiple output channels to different lanes:
for instance, FP16 processing on the default cluster's 64-bit-wide FPUs would accumulate four batched output channels in parallel.
For this purpose, the weight tiles are arranged in an $HWC$ layout, keeping channels contiguous in memory.
Upon activation, the per-channel potentials are extracted from the individual \gls{simd} lanes using bit-masking and branching operations.

For each neuron in our output \gls{fm} tile, we allocate a worst-case-sized channel index buffer in the \gls{tcdm}, assuming that every channel spikes.
When an output spike is generated for a channel, its index is atomically appended to the appropriate index array; we use atomically incremented counters for this.
The resulting index arrays are later unified to a single array by the \gls{dma} core when the output \gls{fm} tile is written back.

\subsubsection{Double-Buffering and Tiling}

We keep the full input and output tensors of each layer in main memory.
To ensure efficient processing, we double-buffer tiles of them in the cluster \gls{tcdm};
while the worker cores process the current tile, the \gls{dma} core simultaneously copies the previous tile's output out and the next tile's inputs in.

In each tile iteration, we first copy in the weight tile using a single asynchronous 2D \gls{dma} transfer.
We then copy in the input \gls{fm} by issuing multiple (row-wise) asynchronous 1D transfers.
Once the weight and input \gls{fm} transfers are issued, the \gls{dma} core generates the output \gls{fm} tile's pointers (\verb|s_ptr_o|) using the lengths of the per-neuron index buffers.
Finally, we use the \gls{dma} engine to individually copy out rows of pointers and the per-neuron index arrays to their target locations in main memory.

\subsubsection{Spike Encoding}

Visual \glspl{snn} can receive their inputs from either frame-based cameras or event cameras. In the former case, dense frames must be converted into spikes to be processed.
A common approach is to use a frame's channels as input currents ($i_m(t)$) in the network's first layer~\cite{bellec2020solrnn,xiao2022online}; this is analogous to a traditional convolutional layer up to the activation step.
We also use this method.
To maximize efficiency, we use an \emph{im2col} transform, turning the convolution into a dense matrix multiplication which we can trivially accelerate using affine \gls{sssr} streams.
We use a 2D \gls{dma} transfer from main memory to \gls{tcdm} to reorganize the $HWC$-encoded frame into the required format on the fly.

\subsection{Performance and Power Evaluation}
\label{chap:workloads:snn:eval}

In this section, we evaluate the performance, power, and energy efficiency of \gls{sssr}-accelerated \gls{snn} inference on our extended Snitch cluster.
We compare an \mmpcut{unaccelerated }RV32GXsf baseline in FP16 to our \gls{sssr}-accelerated implementation in FP16 and FP8.

We consider a low-latency, single-timestep S-VGG11 network that was trained for image recognition on the CIFAR10 dataset adapted as presented in~\cite{zhu2024snn}.
This network performs frame-to-spike encoding on the first layer.
We account for varying sparsity in the input \glspl{fm} by evaluating a batch of 128 input images and reporting standard deviations (shown as error bars) on our performance metrics.

\subsubsection{Performance}

We measure and compare the \mmp{execution time} and \gls{fpu} utilization of baseline RV32GXsf and \gls{sssr}-accelerated \gls{snn} inference through cycle-accurate \gls{rtl} simulation of the Snitch cluster with \glspl{sssr}.

\begin{figure}
  \centering
  \includegraphics[width=\linewidth]{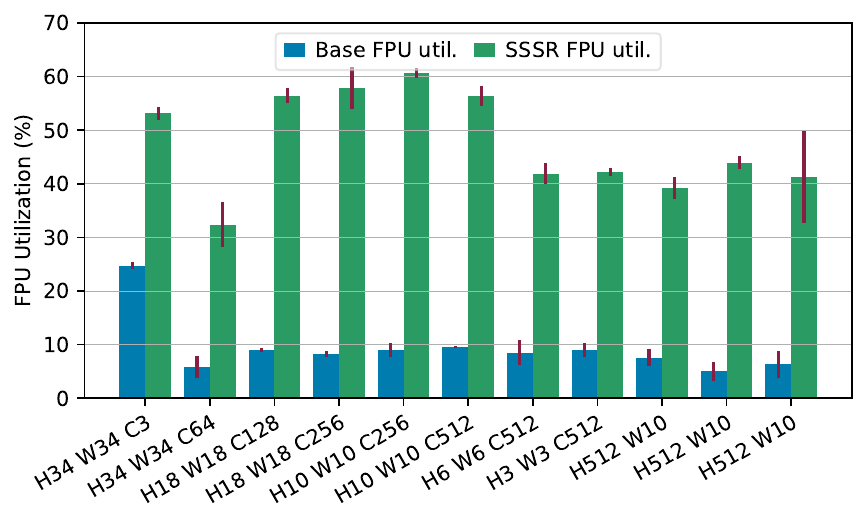}
  \vspace{-2.0em}
  \caption{Average FPU utilizations across S-VGG11 layers in FP16.}
  \label{fig:workloads:snn:eval:fpipc}
\end{figure}

\cref{fig:workloads:snn:eval:fpipc} shows the \gls{fpu} utilization achieved by the baseline and \gls{sssr}-accelerated FP16 implementations for each layer of the S-VGG11 network.
The first layer achieves the highest baseline utilization as it is dominated by dense compute: it encodes dense input frames into spikes using a matrix multiplication.
Nevertheless, affine \gls{sssr} streams enable a significant increase in \gls{fpu} utilization from \SI{25}{\percent} to \SI{53}{\percent}.
The second layer is the first spiking convolutional layer and exhibits a low \gls{fpu} utilization for both implementations;
this is primarily due to the low input channel count, which keeps streams short and increases the relative importance of scheduling and setup overheads.
In the remaining convolutional layers, larger channel counts enable \gls{sssr}-accelerated \gls{fpu} utilizations of up to \SI{61}{\percent} on the fifth layer. %
In the fully connected layers, the achieved \gls{fpu} utilization is somewhat reduced due to a high degree of activation sparsity.

Overall, \glspl{sssr} increase the \gls{fpu} utilization \mmp{of FP16 S-VGG11 inference} from \SI{9.3}{\percent} to \SI{52}{\percent}.
\mmp{As was the case for the sparse \gls{la} kernels and stencil codes we previously accelerated with \glspl{sssr}, the dominant remaining inefficiencies limiting \gls{fpu} utilization are stream setup overheads and \gls{tcdm} access contention. Additionally, some overheads are incurred by the \gls{simd} lane unpacking step required for activation.}

\begin{figure}
  \centering
  \includegraphics[width=\linewidth]{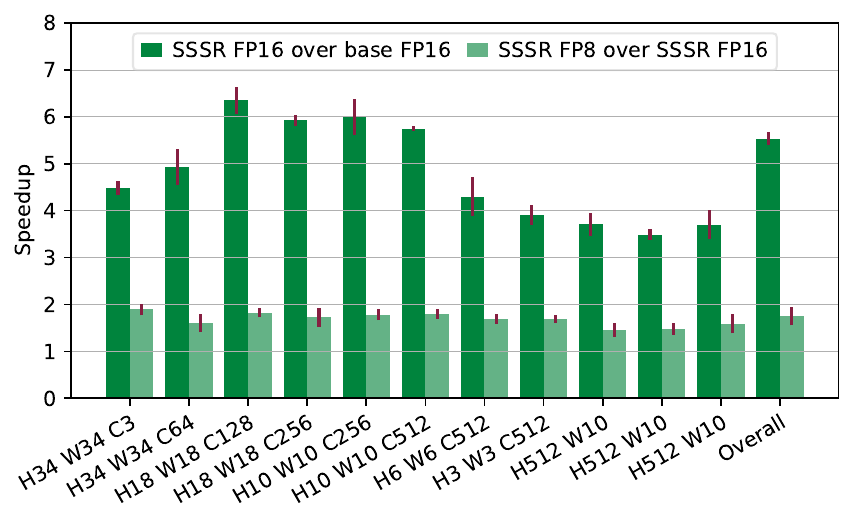}
  \vspace{-2.0em}
  \caption{Average speedups across S-VGG11 layers.}
  \label{fig:workloads:snn:eval:spdup}
\end{figure}

\cref{fig:workloads:snn:eval:spdup} shows the speedups of our \gls{sssr}-accelerated implementation over our baseline in FP16, as well as of FP8 over FP16 using \glspl{sssr}.
Overall, we achieve a 5.5\x~speedup through the use of \glspl{sssr} in FP16.
The first dense layer sees a relatively low speedup, as accelerating affine streams with \glspl{sssr} is generally less beneficial than accelerating indirect ones.
As previously observed for \gls{fpu} utilization, the first convolutional layer and the fully-connected layers exhibit lower acceleration due to short streams and high activation sparsity, respectively.
As expected, we see the highest \gls{sssr} speedups in the middle convolutional layers with high channel counts, reaching up to 6.4\x.
By using FP8 instead of FP16, we can achieve a further 1.7\x~speedup overall; this deviates from ideal 2\x~speedup mainly due to the additional effort of unpacking more lanes during activation.

\subsubsection{Power and Energy}

We implement the \gls{sssr}-extended Snitch cluster in GlobalFoundries' 12LP+ FinFET technology using Synopsys \emph{Fusion Compiler}.
We then run our inference workload in post-layout gate-level simulation and estimate the corresponding cluster power using \emph{PrimeTime}, assuming a \SI{1}{\giga\hertz} clock speed and typical operating conditions (\SI{25}{\celsius}, \SI{0.8}{\volt}).

\begin{figure}
  \centering
  \includegraphics[width=\linewidth]{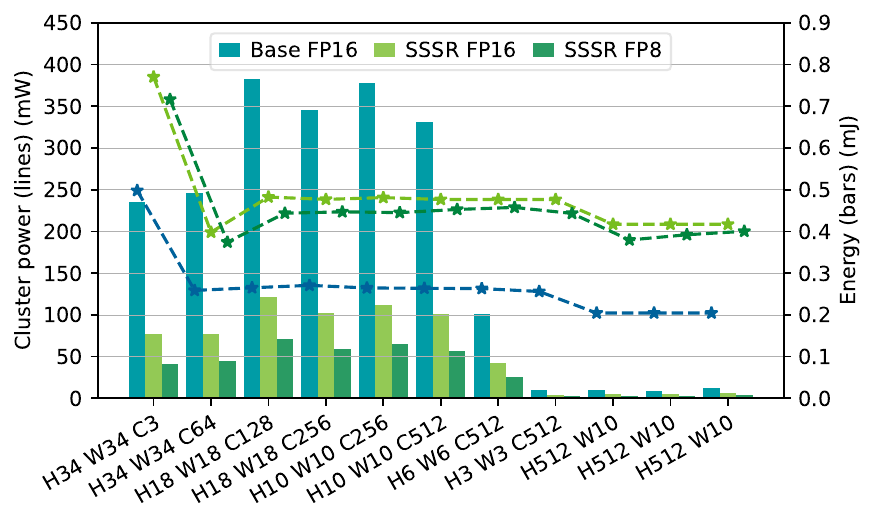}
  \vspace{-2.0em}
  \caption{\mmp{Cluster} power \mmp{(shown as lines)} and energy consumption \mmp{(shown as bars)} for each S-VGG11 layer.}
  \label{fig:workloads:snn:eval:enpow}
\end{figure}

\cref{fig:workloads:snn:eval:enpow} shows the average power and energy consumption of each S-VGG11 layer for the FP16 baseline, FP16 \gls{sssr}, and FP8 \gls{sssr} implementations.
Overall, \glspl{sssr} increase power consumption as they significantly increase compute activity.
The first encoding layer exhibits the highest power consumption as it uses fused \gls{mac} instructions and two \glspl{sssr} simultaneously, unlike the following spiking layers.
After the first spiking layer with few input channels, the power for the convolutional layers remains highly stable at \SI{132}{\milli\watt}, \SI{239}{\milli\watt}, and \SI{224}{\milli\watt} on average for the FP16 baseline, FP16 \gls{sssr}, and FP8 \gls{sssr} implementations, respectively.
The FP8 \gls{sssr} implementation consumes \SI{6.3}{\percent} less power on these layers due to the use of dedicated FP8 datapaths and a longer \gls{simd} lane unpacking process.
The power consumption of the fully connected layers is also quite stable, but lower overall due to the high activation sparsity.

We find that on average, \SI{83}{\percent} of total energy is spent on the convolutional layers.
Despite somewhat increased power consumption, \glspl{sssr} enable significant reductions in energy consumption across all layers:
in FP16, \glspl{sssr} reduce total energy consumption by 3.3\x, while using FP8 precision further reduces it by another 1.7\x.

\subsection{Related Work}
\label{chap:workloads:snn:relwork}

Existing neuromorphic processors are built using mixed-signal~\cite{moradi2018dynaps,benjamin2014neurogrid}, \gls{gals}~\cite{davies2018loihi,frenkel2019odin}, or fully synchronous digital circuits~\cite{wang2022lsmcore,yang2023neurorv}. They often use large hardware arrays to model neurons that exchange spikes through a dedicated on-chip interconnect.
Their peak performance can be expressed and compared in terms of \emph{synaptic operations per second} (\si{\sop\per\second})~\cite{frenkel2019odin}.

Among \gls{gals} processors, Intel's \emph{Loihi}~\cite{davies2018loihi} is a prominent example.
It features 128 neuromorphic cores and three x86 management cores communicating through an asynchronous \gls{noc}.
Each neuromorphic core contains 1024 spiking neural units grouped into sets of trees to model neuron connectivity.
Spikes are sent through the \gls{noc} in the \gls{aer} format.
Loihi is implemented on a \SI{60}{\milli\meter^2} die in a \SI{14}{\nano\meter} FinFET process and has a peak compute of \SI{37.5}{\giga\sop\per\second}.

Another \gls{gals} processor that has received considerable attention is \emph{ODIN}~\cite{frenkel2019odin}.
It can model up to 256 neurons and 64 \mmp{thousand} synapses, storing the neuron states in a \SI{4}{\kibi\byte} \glsunset{sram}\gls{sram} and the synapse weights in a \SI{32}{\kibi\byte} \glsunset{sram}\gls{sram}.
ODIN's 4-bit datapath is managed by a dedicated scheduler and supports both the \gls{lif} model and the biologically inspired Izhikevich activation functions;
however, overall computation must be controlled by an external host through an SPI interface.
Implemented in a \SI{28}{\nano\meter} FDSOI technology, ODIN's \SI{0.086}{\milli\meter^2} die achieves an operating frequency of \SI{75}{\mega\hertz}, a peak throughput of \SI{0.038}{\giga\sop\per\second}, and an energy efficiency of up to \SI{12.7}{\pico\joule\per\sop}.

\emph{LSMCore}~\cite{wang2022lsmcore} is a fully synchronous neuromorphic processor targeting \gls{lsm} \gls{snn} models, which incur a lower training complexity.
It features 256 input neurons and an array of 1024 4-bit liquid neuron units with \gls{lif} activation.
Sparse \glspl{fm} are stored as bitmaps, requiring a zero-skipping mechanism to select the weights of spiking neurons.
In a \SI{40}{\nano\meter} technology, LSMCore reaches a clock speed of \SI{400}{\mega\hertz} and incurs a die area of \SI{18.49}{\milli\meter^2} for a peak throughput of \SI{400}{\giga\sop\per\second}.

\emph{NeuroRVCore}~\cite{yang2023neurorv} proposes a {RISC-V} \gls{isa} extension to accelerate neuromorphic computing, which is implemented and evaluated in the existing RI5CY core.
Its 4-bit datapath features a \gls{lif}  neuron array as well as synapse- and neuron-wise current accumulator trees.
It also features a dedicated vector register file and a load-store unit for weights, neuron states, and \glspl{fm}.
Implemented in a \SI{28}{\nano\meter} technology, the full NeuroRVCore accelerator occupies \SI{2.52}{\milli\meter^2}, which is dominated by its \gls{snn} extensions.
It reaches a peak throughput of \SI{400}{\giga\sop\per\second} at an operating frequency of \SI{1}{\giga\hertz}.

Our work is closest in principle to the highly flexible approach of NeuroRVCore, as we also use a general-purpose {RISC-V} platform with \gls{isa} extensions to accelerate \gls{snn} inference.
However, \glspl{sssr} are a \emph{general-purpose} extension: unlike the domain-specific extensions of NeuroRVCore, they can accelerate numerous workloads \emph{including} \gls{snn} inference, and they implicitly support multiple data precisions through the core's existing \gls{simd} support.

\begin{figure}
    \begin{subcaptionblock}{0.522\linewidth}
        \centering%
        \includegraphics[width=\linewidth]{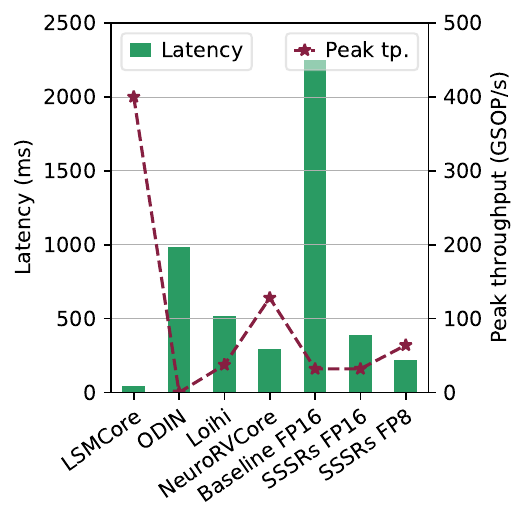}%
        \vspace{-0.1em}%
        \caption{Performance}%
        \label{fig:workloads:snn:relwork_lat}%
    \end{subcaptionblock}\hfill
    \begin{subcaptionblock}{0.478\linewidth}
        \centering%
        \includegraphics[width=\linewidth]{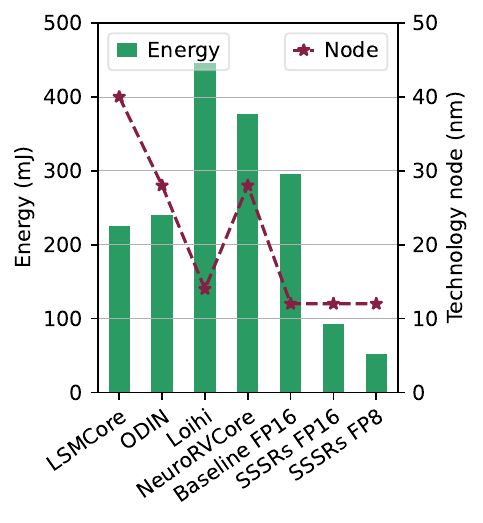}%
        \vspace{-0.1em}%
        \caption{Energy eff. and power}%
        \label{fig:workloads:snn:relwork_en}%
    \end{subcaptionblock}\hfill
\caption{Comparison with discussed neuromorphic processors on the latency and energy of the sixth S-VGG11 layer over 500 timesteps.}
\label{fig:workloads:snn:relwork}
\end{figure}

To quantitatively compare our work to the \gls{soa}, we evaluate the latency and energy consumption of the sixth layer of S-VGG11 on the CIFAR10 dataset over 500 timesteps, as was already done for the architectures discussed above in~\cite{yang2023neurorv}.
\cref{fig:workloads:snn:relwork_lat} compares the latency and peak throughput of different solutions.
Among the discussed \gls{soa} processors, LSMCore is the fastest and ODIN the slowest; this is consistent with these processors having the highest and lowest peak throughputs, respectively.
Our FP16 baseline, whose peak throughput is 12.5\x~lower than that of LSMCore,\pagebreak[3] is the slowest at \SI{2249}{\milli\second}.
However, \glspl{sssr} enable significantly lower latencies of \SI{391}{\milli\second} for FP16 and \SI{217}{\milli\second} for FP8, with the latter being the second-fastest solution and only 4.7\x~slower than the leading LSMCore.
This shows that despite being a general-purpose \gls{isa} extension, \glspl{sssr} are highly effective at accelerating \gls{snn} inference.
This is even more apparent in terms of energy consumption, which is compared in \cref{fig:workloads:snn:relwork_en}.
Our \fin{\gls{sssr}-accelerated} FP16 and FP8 implementations lead in energy efficiency, consuming 2.4\x~and 4.4\x~less energy than the \fin{\gls{soa} leader} LSMCore, respectively.
Although existing solutions are implemented using larger technology nodes, this is relativized by the higher numerical precisions our solution uses.

\section{Model Predictive Control}
\label{chap:workloads:mpc}

After stencil computations and \gls{snn} inference, the final irregular workload we will accelerate on the Snitch cluster using \glspl{sssr} is \glsunset{mpc}\emph{model predictive control} (MPC).

\subsection{Background and Contributions}
\label{chap:workloads:mpc:bgcontr}

\Gls{mpc} is an advanced control method that optimizes a dynamic system's state trajectory while respecting predefined constraints~\cite{jeres2014embed2}.
At each discrete time step, \gls{mpc} leverages a system model to solve for appropriate inputs, considering their impact on the current and predicted future system state within a given time horizon.
\gls{mpc} is better-suited to control complex dynamic systems than simpler reactive methods, but much more computationally expensive~\cite{jeres2011fpga,jeres2013embed1,sabo2018primdual}.
An emerging application for \gls{mpc} is the \glsunset{etm}\emph{energy and thermal management} (ETM) of application-class manycores~\cite{bartolini2013thermal,tilli2022twolayer,sun2020mpc}.
Classical solutions for smaller multiprocessors include closed-loop \mmp{multi-input, multi-output} algorithms and classical \gls{pid} controllers~\cite{rosedahl2017powperfop};
however, these reactive solutions are not flexible enough to handle the dynamic behavior of hundreds of \glspl{pe}~\cite{bartolini2013thermal,sun2020mpc}, leading to predictive solutions like \gls{mpc} being investigated as an alternative~\cite{bartolini2013thermal,sun2020mpc,wang2011adaptive}.
The main challenge in \gls{mpc}-based manycore \gls{etm} is its high computational cost, which grows with the number of controlled \glspl{pe}~\cite{bartolini2013thermal,tilli2022twolayer}.
Numerous works aim to reduce the large computational footprint of online \gls{mpc}, particularly those targeting embedded applications~\cite{nguyen2024tinympc,jeres2014embed2,jeres2014embed2,sabo2018primdual,vouzis2009asoc}.
They usually formulate \gls{mpc} as a convex \gls{qp}, allowing them to leverage existing \gls{qp} solution methods.
Among these, iterative \emph{first-order} methods are popular: despite their limited solution accuracy, they can efficiently be parallelized and are amenable to low-precision number formats~\cite{jeres2014embed2}.
For problems with real-time constraints, \emph{operator-splitting} methods have proven effective~\cite{stathopulos2016opsplit}; \gls{admm} in particular provides good convergence in practice~\cite{boyd2011distr}, leading to its use in various \gls{mpc} control applications~\cite{nguyen2024tinympc,jeres2013embed1,Schubiger2020Admm}.

\emph{OSQP}~\cite{stellato2020osqp} is an \gls{soa} convex \gls{qp} solver based on \gls{admm}.
The main compute-intensive step of OSQP's iterations is solving a sparse equation system using its \emph{QDLDL} direct solver.
Unfortunately, the \gls{fe} and \gls{bs} steps of QDLDL incur indirect memory accesses and numerous internal data dependencies, severely limiting its \gls{fpu} utilization on traditional \gls{cpu} architectures.
This holds back OSQP's use in \gls{mpc}-based control solutions, especially on energy-efficient embedded platforms.

In this section, we accelerate online \gls{mpc} using \glspl{sssr} by accelerating the sparse solver step of OSQP, using manycore \gls{etm} as an example control problem.
We first accelerate the indirect memory accesses of QDLDL%
, reducing its execution time by up to 2.0\x. %
\fin{In} the process, we find that average stream length, and thus performance, is limited by QDLDL's numerous internal data dependencies.
We therefore design an optimized solver called \emph{ParSPL} that reduces these dependencies by up to two orders of magnitude, accelerating both \fin{RV32G} baseline and \gls{sssr}-accelerated \gls{mpc}.
With ParSPL, \glspl{sssr} achieve solver speedups and energy efficiency gains of up to 4.6\x~and 2.2\x, respectively, \fin{raising} the peak \gls{fpu} utilization from \SI{11}{\percent} to \SI{57}{\percent}.

We propose a co-integrated \gls{etm} controller for manycores based on \emph{ControlPULP}~\cite{ottaviano2024cpulp} that uses an \gls{sssr}-extended Snitch cluster for compute.
On an end-to-end \gls{etm} scenario with 9\x9 cores, \glspl{sssr} and ParSPL together reduce the controller's \gls{mpc} iteration time from \SI{8.74}{\milli\second} to \SI{0.69}{\milli\second}, enabling online \gls{etm} with ample slack.
Our solution consumes at most \SI{435}{\milli\watt} and incurs less than \SI{1.5}{\percent} of a modern manycore's die area, outperforming \gls{soa} \gls{etm} solutions in terms of area and energy efficiency while remaining competitive with fast \gls{mpc} solutions across various platforms.

To summarize, our contributions are as follows:

\begin{itemize}

\item We accelerate the irregular memory accesses of OSQP's QDLDL solver using \glspl{sssr}, enabling speedups of up to 2.0\x~and identifying internal data dependencies as a major performance limitation.

\item We present \emph{ParSPL}, an improved solver drastically reducing internal data dependencies to accelerate both baseline and \gls{sssr}-accelerated solver steps. ParSPL enables \gls{sssr} speedups and energy efficiency gains of up to 4.6\x~and 2.2\x, respectively, raising the peak solver \gls{fpu} utilization to \SI{57}{\percent}.

\item We propose a co-integrated \emph{ControlPULP}-based \gls{etm} controller with \glspl{sssr}, consuming at most \SI{435}{\milli\watt} and occupying less than \SI{1.5}{\percent} of a modern manycore's die. On \fin{9\x9-manycore} \gls{etm}, \glspl{sssr} and ParSPL together reduce the controller's iteration time from \SI{8.74}{\milli\second} to \SI{0.69}{\milli\second}, enabling online control.

\item We compare our controller to \gls{soa} approaches, finding that we outperform \gls{mpc}-based \gls{etm} solutions in area and energy efficiency while remaining competitive with fast \gls{mpc} solutions across platforms.

\end{itemize}

In the following background subsections, we \mmpcut{will }first provide more details on the \gls{qp} formulation of \gls{mpc}, \gls{admm}, and the OSQP solver.
In \cref{chap:workloads:mpc:map}, we \mmpcut{will }then present our implementation and acceleration work.

\subsubsection{\gls{mpc} as a Quadratic Program}

For a \gls{lti} system model, \gls{mpc} seeks to minimize a quadratic function (the control \emph{cost}) while respecting a set of linear inequalities (the control \emph{constraints}). As such, it can be formulated as a convex \gls{qp} solved in each discrete time iteration~\cite{stellato2020osqp}:

\begin{equation}\label{eq:workloads:mpc:mpcqp}
    \begin{aligned}
        & \textnormal{minimize} \quad && s_{H_p}^{T}Q_{H_p}s_{H_p} + \sum_{h=0}^{H_p-1} s_{h}^TQs_h + u_{h}^TRu_h \\
        & \textnormal{subject to} \quad
        && s_{h+1} = Ds_h + Eu_h \\
        &&& s_h \in \mathcal{S}, \; u_h \in \mathcal{U} \\
        &&& s_0 = s_{init}
    \end{aligned}
\end{equation}

Here, $s_h \in \mathbb{R}^{n_s}$ represents the system state and $u_h \in \mathbb{R}^{n_u}$ the input $h$ time steps in the future.
$H_p$ is the prediction horizon length.
The cost matrices $R$, $Q$, and $Q_{H_p}$ penalize deviations from the desired input and state trajectories.
$D$ and $E$ define the system model, while the \emph{feasible sets} $\mathcal{S}$ and $\mathcal{U}$ express the linear system constraints.

For simplicity, we can also define a combined decision variable\break $x := (s_0, ..., s_{H_p}, u_0, ..., u_{H_p-1})$ of size $n := n_s(H_p+1)+ n_u H_p$, allowing us to express \cref{eq:workloads:mpc:mpcqp} in the canonical form of a convex \gls{qp}:

\begin{equation}\label{eq:workloads:mpc:canonqp}
    \begin{aligned}
        & \textnormal{minimize} \quad && (1/2)x^{T}Px + q^{T}x\\
        & \textnormal{subject to} \quad && c_l \preccurlyeq Ax \preccurlyeq c_u
    \end{aligned}
\end{equation}

Here, $P \in \mathbb{R}^{n\times n}$ is the positive-semidefinite cost matrix, $A \in \mathbb{R}^{m\times n}$ is the constraint function, and $c_l,c_u \in \mathbb{R}^m$ are the constraint vectors.
We can alternatively express the constraints through the feasible set $\mathcal{C} := \{z \in \mathbb{R}^m | c_l \preccurlyeq z \preccurlyeq c_u\}$.
In the case of \gls{mpc}, $q = \vec{0}$ as there are no linear terms in the objective function of \cref{eq:workloads:mpc:mpcqp}.

\subsubsection{First-order Methods and ADMM}

Convex \glspl{qp} can be solved using different approaches, including \emph{active-set}, \emph{interior-point}, and \emph{first-order} methods\cite{stellato2020osqp}.
First-order methods iteratively optimize a solution using only the cost function's first-order derivative (gradient).
While they provide a limited solution accuracy, they are usually precise enough for good control performance, simple enough for low-level optimization on target platforms, and amenable to parallelization~\cite{jeres2014embed2}.
Thus, they are particularly attractive for real-time control on resource-constrained embedded systems~\cite{jeres2014embed2,stathopulos2016opsplit}.

Among first-order methods, the \gls{admm} has received particular attention due to its good convergence in practice~\cite{boyd2011distr}.
To apply \gls{admm} to \cref{eq:workloads:mpc:canonqp}, a new decision variable $z \in \mathbb{R}^n$ is first added to consider our constraints as part of the minimized objective function.
Then, \emph{operator splitting} is leveraged, introducing the splitting variables $\tilde{x}$ and  $\tilde{z} \in \mathbb{R}^n$ to express our canonical \gls{qp} in the form~\cite{stellato2020osqp}:

\begin{equation}\label{eq:workloads:mpc:opsplitqp}
    \begin{aligned}
        & \textnormal{minimize} \quad (1/2)\tilde{x}^{T}P\tilde{x} + q^{T}\tilde{x} + \mathcal{I}_{Ax=z}(\tilde{x},\tilde{z}) + \mathcal{I}_{\mathcal{C}}(z) \\
        & \textnormal{subject to} \quad (\tilde{x},\tilde{z}) = (x,z)
    \end{aligned}
\end{equation}

Here, the indicator functions $\mathcal{I}_{Ax=z}(\tilde{x},\tilde{z})$ and $\mathcal{I}_{\mathcal{C}}(z)$ enforce our constraints: they are zero at $Ax=z$ and $z \in \mathcal{C}$, respectively, and infinity elsewhere.
Now, an iterative augmented Lagrangian method can be applied to solve \cref{eq:workloads:mpc:opsplitqp}; adding penalty terms to our objective function that ensure stepwise convergence ($\tilde{x} \to x$ and $\tilde{z} \to z$) yields the $k$-th iteration of \gls{admm}~\cite{stellato2020osqp}:

\begin{equation}\label{eq:workloads:mpc:admm_xztilde}
    \begin{aligned}
        (\tilde{x}^{k+1},\tilde{z}^{k+1}) \leftarrow & \operatorname {argmin}_{(\tilde{x},\tilde{z}):A\tilde{x}=\tilde{z}} (1/2)\tilde{x}^{T}P\tilde{x} + q^{T}\tilde{x} \\
        & + (\sigma/2) \lVert\tilde{x}-x^k+\sigma^{-1}w^k\rVert^{2}_{2} \\
        & + (\rho/2) \lVert\tilde{z}-z^k+\rho^{-1}y^k\rVert^{2}_{2}
    \end{aligned}
\end{equation}

\begin{equation}\label{eq:workloads:mpc:admm_x}
    x^{k+1} \leftarrow \alpha\tilde{x}^{k+1} + (1-\alpha)x^k +\sigma^{-1}w^k
\end{equation}
\vspace{-\baselineskip}
\begin{equation}\label{eq:workloads:mpc:admm_z}
    z^{k+1} \leftarrow \Pi \left( \alpha\tilde{z}^{k+1} + (1-\alpha)z^k +\rho^{-1}y^k \right)
\end{equation}
\vspace{-\baselineskip}
\begin{equation}\label{eq:workloads:mpc:admm_w}
    w^{k+1} \leftarrow w^k + \sigma \left( \alpha\tilde{x}^{k+1} + (1-\alpha)x^k -x^{k+1} \right)
\end{equation}
\vspace{-\baselineskip}
\begin{equation}\label{eq:workloads:mpc:admm_y}
    y^{k+1} \leftarrow y^k + \rho \left( \alpha\tilde{z}^{k+1} + (1-\alpha)z^k -z^{k+1} \right)
\end{equation}
\vspace{0.05em} %

Here, $\sigma > 0$ and $\rho > 0$ are the \emph{step size parameters} and $\alpha \in (0,2)$ is the \emph{relaxation parameter}.
$\Pi$ is the Euclidean projection onto the feasible set $\mathcal{C}$.
The iteration variables $w^k$ and $y^k$ are dual to the constraints $\tilde{x}=x$ and $\tilde{z}=z$.
Since we can show that $w^k=0$ for $k>0$, the $w^k$ update step in \cref{eq:workloads:mpc:admm_w} is not needed in practice.

\subsubsection{OSQP}
\label{chap:workloads:mpc:bg:osqp}

OSQP~\cite{stellato2020osqp} is an open-source convex \gls{qp} solver suite based on \gls{admm}.
Its algorithm, which is based on \cref{eq:workloads:mpc:admm_xztilde,eq:workloads:mpc:admm_x,eq:workloads:mpc:admm_z,eq:workloads:mpc:admm_w,eq:workloads:mpc:admm_y}, is shown in \cref{alg:workloads:mpc:osqp}.
A complete discussion of how \cref{alg:workloads:mpc:osqp:linsys_xnu,alg:workloads:mpc:osqp:linsys_z} are derived from \cref{eq:workloads:mpc:admm_xztilde} can be found in~\cite{stellato2020osqp}.

To solve the system on \cref{alg:workloads:mpc:osqp:linsys_xnu}, OSQP uses its QDLDL direct solver.
The constant coefficient matrix on the left-hand side, labelled $K$, is called the \gls{kkt} matrix.
$K$ is Cholesky-factorized to the form $LDL^T$ offline,  where L is a sparse lower-triangular matrix and D is a diagonal matrix.
This allows QDLDL to solve \cref{alg:workloads:mpc:osqp:linsys_xnu} in three steps: \gls{fe} using $L$, scaling by $D^{-1}$, and \gls{bs} using $L^T$.

OSQP performs further offline optimizations.
To maximize the sparsity of $L$, the rows of $K$ are reordered offline using the \gls{amd} algorithm, resulting in a permutation matrix $P$ such that $PKP{^T} = LDL^{T}$.
To avoid divisions, the rows of $L$ are normalized to its diagonal, and $D$ is stored pre-inverted and scaled to compensate for the normalization of $L$.

To decide termination, OSQP tracks two residuals that converge to zero for solvable problems.
The primal residual $r_{prim} = Ax -z$ expresses the compliance to constraints, while the dual residual $r_{dual} = Px + q + A^{T}y$ expresses the optimality of the objective function.
For both, we define respective termination tolerances $\varepsilon_{prim}$ and $\varepsilon_{dual}$; if the squared residual norms fall below these values, \cref{alg:workloads:mpc:osqp} terminates.
To avoid costly computation of residuals, OSQP also allows termination after a fixed number of iterations \texttt{max\_iter}. In this case, \texttt{max\_iter} must be carefully selected to maintain proper control performance.

\begin{algorithm}[t]
\caption{OSQP solution algorithm~\cite{stellato2020osqp}.}\label{alg:workloads:mpc:osqp}
\begin{algorithmic}[1]
\State \textbf{given} initial vals. $x^0$, $z^0$, $y^0$, params. $\rho>0$, $\sigma>0$, $\alpha \in (0,2)$
\Repeat
    \State $ \textnormal{solve}
    \left[
    \begin{array}{@{}cc@{}}
        P+\sigma I & A^T \\
        A & -\rho^{-1}I
    \end{array}
    \right]
    \left[
    \begin{array}{@{}c@{}}
        \tilde{x}^{k+1} \\
        \nu^{k+1}
    \end{array}
    \right]
    =
    \left[
    \begin{array}{@{}c@{}}
        \sigma x^k -q \\
        z^k - \rho^{-1} y^k
    \end{array}
    \right]
    $ \label{alg:workloads:mpc:osqp:linsys_xnu}
    \State $\tilde{z}^{k+1} \leftarrow z^k + \rho^{-1}(\nu^{k+1}-y^k)$ \label{alg:workloads:mpc:osqp:linsys_z}
    \State $x^{k+1} \leftarrow \alpha\tilde{x}^{k+1} + (1-\alpha)x^k$\label{alg:workloads:mpc:osqp:x}
    \State $z^{k+1} \leftarrow \Pi \left( \alpha\tilde{z}^{k+1} + (1-\alpha)z^k +\rho^{-1}y^k \right)$\label{alg:workloads:mpc:osqp:z}
    \State $y^{k+1} \leftarrow y^k + \rho \left( \alpha\tilde{z}^{k+1} + (1-\alpha)z^k -z^{k+1} \right)$\label{alg:workloads:mpc:osqp:y}
\Until termination criteria or \texttt{max\_iter} reached
\end{algorithmic}
\end{algorithm}

\subsection{SSSR Cluster Implementation}
\label{chap:workloads:mpc:map}

In this section, we \mmpcut{will }implement and accelerate the performance-critical sparse system solver of OSQP on a co-integrated controller using the Snitch cluster with \glspl{sssr} as its compute unit.
We note that the remainder of \cref{alg:workloads:mpc:osqp} consists of simple dense vector operations that can easily be accelerated with \glspl{sssr} using affine streams.

We \mmpcut{will }first introduce our co-integrated controller architecture.
We\mmpcut{ will} then describe how we implemented the performance-critical kernels of the existing QDLDL solver on the cluster, and how we accelerated them using \glspl{sssr}.
Motivated by performance challenges we find in this implementation, we\mmpcut{ will} present an improved system solver, \emph{ParSPL}, that speeds up both baseline RV32G and \gls{sssr}-accelerated \gls{mpc} by significantly reducing data dependencies.
In the following \cref{chap:workloads:mpc:eval}, we \mmpcut{will }evaluate the performance and energy efficiency benefits of our work.

\subsubsection{Co-integrated Controller}

We use \emph{ControlPULP}~\cite{ottaviano2024cpulp} as the base for our co-integrated controller. ControlPULP combines a real-time management domain with a \gls{pmca} for data processing.

ControlPULP's management domain schedules and coordinates workload execution and handles communication with the surrounding \gls{soc}.
It is built around a \emph{CV32RT}~\cite{balas2023cv32rt} manager core running a real-time  stack based on \emph{FreeRTOS}.
It also includes a small L2 \gls{spm}, a \gls{dma} engine, and an \gls{axi4} interconnect connecting the controller components to each other and to the host \gls{soc}.

The \gls{pmca} used is the Snitch cluster with \glspl{sssr} introduced in \cref{chap:sssrs:arch:integr}, with two \glspl{issr} and one \gls{essr} in each of its eight compute cores. We make one important modification: %
to enable the acceleration of large \gls{etm} problems targeting manycores, we increase the size of our \gls{tcdm} to \SI{1}{\mebi\byte}.
We do this by increasing bank depth while maintaining the bank count.
In select cases, we will project the performance for even larger \gls{etm} problems by assuming that \SI{2}{\mebi\byte} of \gls{tcdm} are available.

\subsubsection{QDLDL Implementation}

As outlined in \cref{chap:workloads:mpc:bg:osqp}, the QDLDL solver consists of three steps: \gls{fe} using $L$, scaling by $D^{-1}$, and \gls{bs} using $L^T$.
Scaling by $D^{-1}$ is a trivial dense element-wise multiplication, which we can accelerate to near-ideal performance using three affine \gls{sssr} streams.
\gls{fe} and \gls{bs} are sparse kernels with nontrivial dependencies whose implementation and acceleration we discuss in detail.

The lower-triangular sparse matrix $L$ could be stored in either a row-major (\gls{csr}) or column-major (\gls{csc}) format, which would better suit \gls{bs} or \gls{fe}, respectively.
Since QDLDL requires both of these kernels, we choose the \gls{csc} format here without loss of generality.
Furthermore, since the diagonal of $L$ is densely filled with ones by definition, we store it without its diagonal, i.e. as $L - I$.

\topar{Forward Elimination} \cref{lst:workloads:mpc:base_fe} shows the parallel baseline pseudocode for \gls{fe}, where we label the right-hand vector $b$ and store the solution back to $b$ in-place.
The outer loop iterates over the columns of $L$ and the inner loop over the values in each column.
The inner loop is parallelized among cores (\verb|par_for|), and the outer loop must synchronize between cores after its completion (\verb|par_sync|) to ensure consistent memory accesses to $\verb|b|$ between column iterations.

\definecolor{parcol}{HTML}{32a852}

\begin{listing}[t]%
\caption{Parallel baseline pseudocode for forward elimination \\($Lx=b$). The solution $x$ is written back to the input $b$ in-place.}
\label{lst:workloads:mpc:base_fe}
\begin{minted}[escapeinside=||]{python}
for i in 0 to L_ncols-1:
    b_val = b[i]
    |\textbf{\textcolor{parcol}{par\_for}}| k in L_ptr[i] to L_ptr[i+1]:
        b[L_idcs[k]] -= b_val * L_vals[k]
    |\textbf{\textcolor{parcol}{par\_sync()}}|
\end{minted}
\end{listing}

The inner loop of \gls{fe} is a streaming accumulate: the values of $L$ are read contiguously (\verb|L_vals[k]|) and the values of $b$ are both read and written indirectly using the indices of $L$ (\verb|b[L_idcs[k]]|).
We can directly accelerate this inner loop using our default \gls{sssr} streamer: the \gls{essr} streams the values of $L$ while the two \glspl{issr} perform the indirect reads and writes on $b$, respectively.
We note that the concurrent indirect read and write streams do \emph{not} violate memory consistency because
\begin{enumerate*}[label=(\roman*)]
    \item each row index of $L$ occurs at most once per column,
    \item within a column, all cores work on disjoint sets of row indices, and
    \item since each indirected value of $b$ must be read before its updated value can be computed and written, an indirect write can never interfere with its corresponding indirect read.
\end{enumerate*}

With \glspl{sssr} and the \gls{frep} hardware loop, the inner loop \mmpcut{thus }results in continuous fused multiply-decumulate (\verb|fnmsub.d|) issues.
However, the outer loop presents a performance hurdle: after each column, the memory accesses of all cores must be synchronized, significantly limiting attainable stream lengths.
Based on our performance analysis in \cref{chap:sssrs:eval}, we can predict that columns with few nonzeros will limit our \gls{sssr} speedups;
this will help motivate the design of our improved ParSPL solver.

\topar{Backward Substitution} \cref{lst:workloads:mpc:base_bs} shows the baseline pseudocode for \gls{bs}, which again operates on the right-hand vector $b$ in-place.
As for \gls{fe}, there are two nested loops, which now perform row-major iteration over the transposed $L^T$.
\begin{listing}[t]%
\caption{Parallel baseline pseudocode for backward substitution \\($L^T x=b$). The solution $x$ is written back to the input $b$ in-place.}
\label{lst:workloads:mpc:base_bs}
\begin{minted}[escapeinside=||]{python}
for i in L_ncols-1 to 0:
    ret = 0
    |\textbf{\textcolor{parcol}{par\_for}}| k in L_ptr[i] to L_ptr[i+1]:
        ret -= L_vals[k] * b[L_idcs[k]]
    core_vals[core_id] = ret
    |\textbf{\textcolor{parcol}{par\_sync()}}|
    if core_id == NUM_CORES-1:
        b[i] = sum(core_vals)
    |\textbf{\textcolor{parcol}{par\_sync()}}|
\end{minted}
\end{listing}

The outer row loop iterates in reverse to follow the correct order of substitution dependencies.
The inner loop is parallelized among cores (\verb|par_for|) and accumulates per-core product terms; the per-core sums are then reduced by the last worker core, as it is likely the least busy core.
To preserve memory consistency, we synchronize cores (\verb|par_sync|) after processing each row, as well as before per-core result reduction.

The inner loop of \gls{bs} can easily be accelerated using the default \gls{sssr} streamer: we use one \gls{issr} to indirectly stream \verb|b[L_idcs[k]]| and either remaining \gls{ssr} to continuously stream \verb|L_vals[k]|.
Together with \gls{frep}, this results in continuous \verb|fnmsub.d| issues for the inner loop.
As for \gls{fe}, the synchronizations required for each row limit the maximum length of \gls{sssr} streams and increase their relative setup and imbalance overheads.
This brings us to our improved ParSPL solver that aims to reduce synchronizations to a minimum.

\subsubsection{ParSPL}

\emph{ParSPL} aims to improve both the baseline and \gls{sssr}-accelerated performance of OSQP's sparse solver step by significantly reducing the number of synchronization steps between cores.
ParSPL preprocesses the $L$ matrix offline to avoid data dependencies: it reorders and partitions $L$ to end up with a collection of parallelizable subtasks that depend on each other.
During runtime, it then follows a predetermined schedule to process these subtasks with specialized compute kernels, which we can significantly accelerate with \glspl{sssr}.

\begin{figure}
    \setkeys{Gin}{width=4.05cm}%
    \centering
    \subfloat[L after ApMD reordering]{%
        \centering%
        \includegraphics{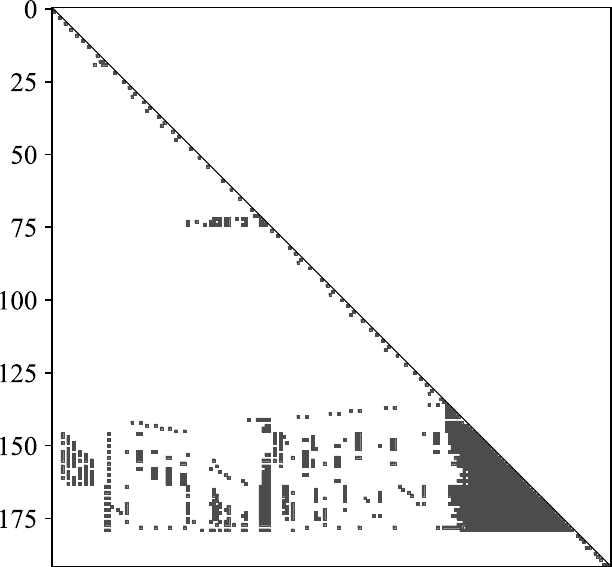}
        \label{fig:workloads:mpc:parspl:a}
        \vspace{0.45em}
    }\hspace{1.2em}%
    \subfloat[Partial level sheduling]{%
        \centering%
        \includegraphics{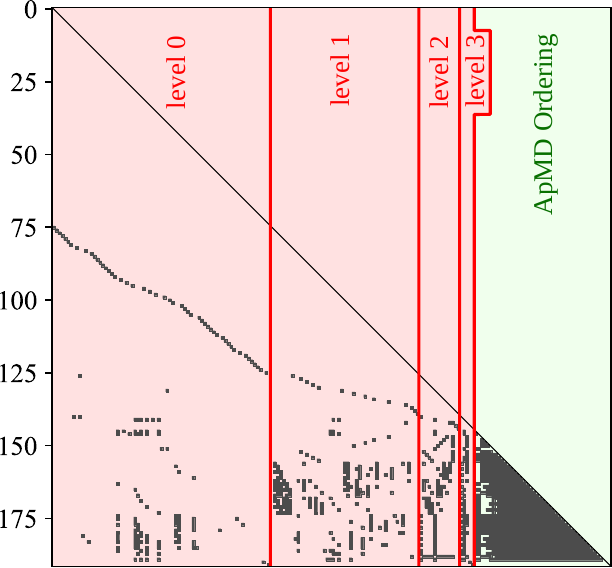}
        \label{fig:workloads:mpc:parspl:b}
        \vspace{0.45em}
    }\\\vspace{0.6em}%
    \subfloat[Graph-based tiling]{%
        \centering%
        \includegraphics{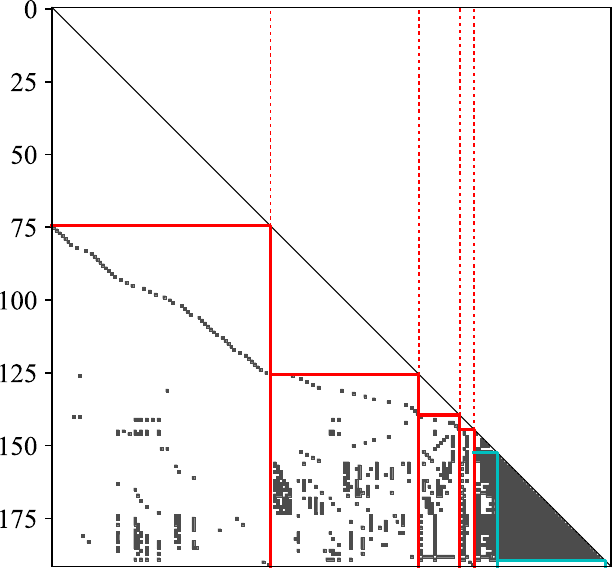}
        \label{fig:workloads:mpc:parspl:c}
        \vspace{0.91em}
    }\hspace{1.2em}%
    \subfloat[Kernel mapping]{%
        \centering%
        \includegraphics{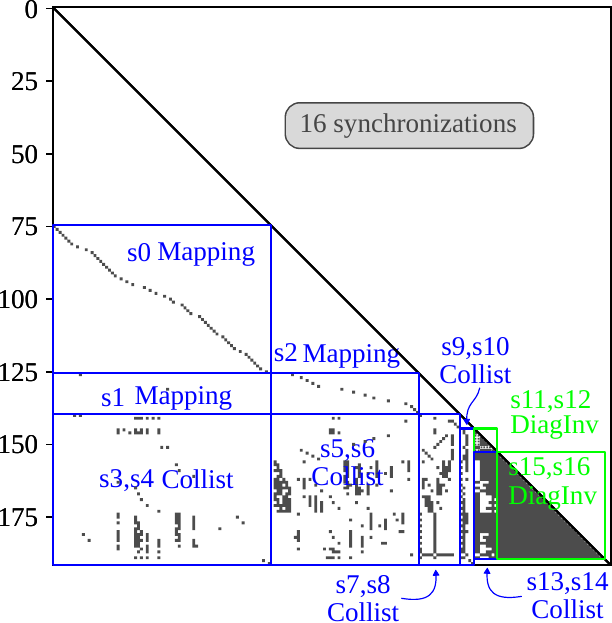}
        \label{fig:workloads:mpc:parspl:d}
        \vspace{-0.23em}
    }\\\vspace{0.6em}%
    \subfloat[ALAP sched. \& merging]{%
        \centering%
        \includegraphics{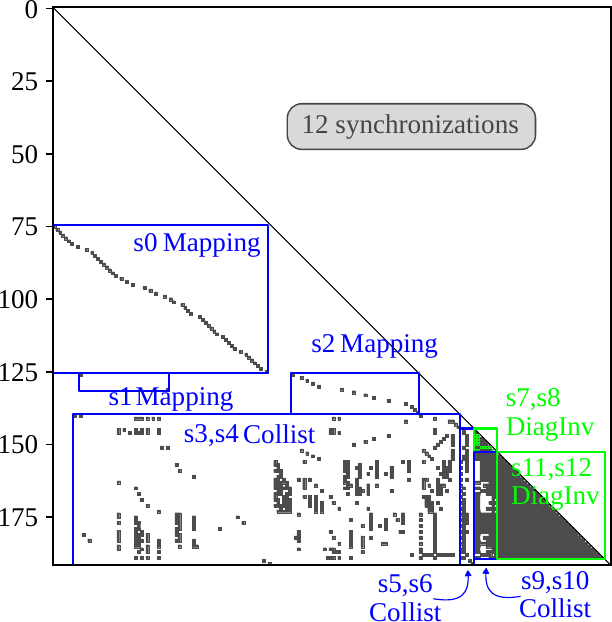}
        \label{fig:workloads:mpc:parspl:e}
        \vspace{-0.27em}
    }\hspace{1.2em}%
    \subfloat[Intra-tile parallelization]{%
        \centering%
        \includegraphics{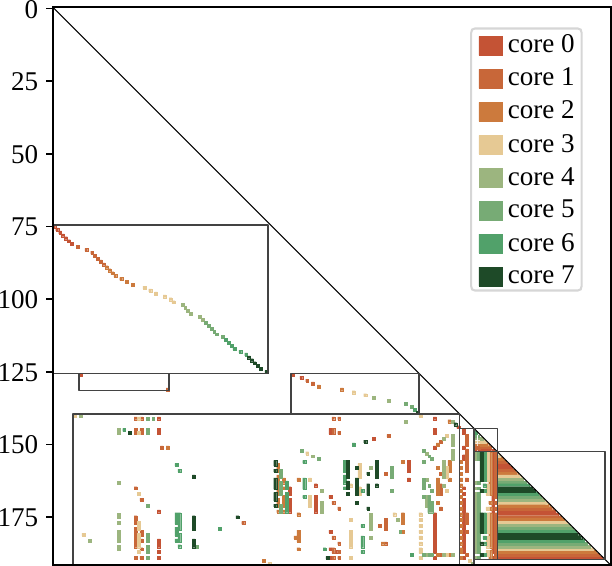}
        \label{fig:workloads:mpc:parspl:f}
        \vspace{0.91em}
    }
  \caption{ParSPL preprocessing stages for the lower-triangular sparse matrix L of the small \gls{etm} problem \texttt{P3x3\_H2}. The goal is to minimize dependencies and thus required inter-core synchronizations (\texttt{s*}).}
  \label{fig:workloads:mpc:parspl}

\end{figure}

\cref{fig:workloads:mpc:parspl} visualizes the preprocessing stages of ParSPL on an example $L$ matrix taken from a 3\x 3-core \gls{etm} problem (\verb|P3x3_H2|)~\cite{ottaviano2025mpc}.
\cref{fig:workloads:mpc:parspl:a} shows $L$ in its initial state (after \gls{amd} reordering of $K$) as processed by our baseline QDLDL implementation.
We \mmpcut{will }describe each preprocessing step, with their respective results shown in \cref{fig:workloads:mpc:parspl:b,fig:workloads:mpc:parspl:c,fig:workloads:mpc:parspl:d,fig:workloads:mpc:parspl:e,fig:workloads:mpc:parspl:f}, in sequence.

\topar{Partial level scheduling (\cref{fig:workloads:mpc:parspl:b})} Level scheduling~\cite{saad2003iterative} is a common technique to reduce dependencies in sparse triangular solvers by rearranging the matrix into a sequence of \emph{levels}.
Within a level, the input and output indices are disjoint, and synchronizations are necessary only at the end of levels.
To create levels, sparser columns with fewer dependencies on others are rearranged to be processed as soon as possible.
However, as columns become denser (sometimes forming highly dense sub-diagonal triangles as shown), the benefits of levels diminish and the original \gls{amd} ordering eventually yields better results.
Thus, we employ \emph{partial} level scheduling, discarding levels whose column counts fall below a reasonable threshold~\cite{bradley2016sptrv}.
In the manycore \gls{etm} application we target, at most four levels are typically preserved.

\topar{Graph-based tiling (\cref{fig:workloads:mpc:parspl:c})} As shown with red lines, we first use the previously scheduled levels to create parallelizable tiles that depend on each other.
We then form additional tiles, shown with cyan lines, by separating out highly dense triangle structures under the diagonal, which are more efficiently handled using a specialized dense kernel.
To do this, we view $L$ as the adjacency matrix of a graph, where any nonzero entry $l_{ij}$ presents an edge between nodes $i$ and $j$.
We assign to each nonzero a weight $1/(i-j)^2$, rating entries by their proximity to the diagonal.
We then recursively use the Ford-Fulkerson algorithm to separate out highly dense subdiagonal triangles through min-cuts until a heuristic termination criterion is reached.

\topar{Kernel mapping (\cref{fig:workloads:mpc:parspl:d})} We map each tile to one or more calls to three specialized kernels, which each come in variants for \gls{fe} and \gls{bs} as well as with and without \gls{sssr} acceleration.
Sparse tiles are mapped to the \verb|Mapping| and \verb|Collist| kernels.
The \verb|Mapping| kernels are applied to subregions of a tile where each row and each column contain at most one entry; this eliminates all dependencies between columns, enabling \gls{sssr} streams of maximal length within the subregion.
Since no reduction is required, only one synchronization per \verb|Mapping| subtile is needed.
The \verb|Collist| kernels operate on \gls{csc} subtiles and are similar to our QDLDL baseline kernels.
The main difference is that there are no input dependencies between columns, so only two synchronizations are necessary, one before and one after inter-core result reduction, enabling much longer \gls{sssr} streams.
The remaining dense sub-diagonal triangles are handled using the dedicated \verb|Diaginv| kernels.
These kernels exploit the fact that the inverse of a lower-triangular matrix is again lower-triangular, precomputing and storing the inverses of these structures in a dense format.
They then compute the desired result using regular dense \gls{la}, which can trivially be accelerated with affine \glspl{sssr} streams.
\verb|Diaginv| kernels again require only two synchronizations, one before and one after reduction.

\topar{ALAP tile scheduling and merging (\cref{fig:workloads:mpc:parspl:e})} We first merge horizontally aligned subtiles using the same kernel, which further reduces the number of synchronizations.
We then build a dependency graph between all subtiles based on individual nonzero positions, eliminating any zero padding from subtiles in the process.
We use this dependency graph to schedule all subtiles \gls{alap}, potentially eliminating some more pre-reduction synchronizations and resulting in our final tile execution sequence.

\topar{Intra-tile parallelization (\cref{fig:workloads:mpc:parspl:f})} Now that the subtiles are finalized and scheduled, we can parallelize them as outlined in the kernel mapping step:
For \verb|Mapping| kernels, each core processes a row-contiguous subset of nonzeros. For \verb|Collist| and \verb|Diaginv|, each core processes a predetermined subset of columns or rows, respectively, chosen to minimize imbalance between cores.
ParSPL automatically generates C code for the final parallelized computation schedule, providing data structures for each subtile to be passed to the corresponding kernel type.

\subsection{Performance and Power Evaluation}
\label{chap:workloads:mpc:eval}

We evaluate our acceleration work on the co-integrated controller described in \cref{chap:workloads:mpc:map}, which features a Snitch cluster with \glspl{sssr} as its main compute element.
As a target \gls{mpc} problem, we consider \gls{etm} in an application-class \gls{hpc} manycore system with different square \gls{noc} topologies, ranging from 3\x 3 (9-core) to 12\x 12 (144-core) configurations.
We use the thermal models described and optimized in \cite{ottaviano2025mpc}, which use the following naming convention: "\verb|Paxb_Hc|" models an \texttt{a}\x\texttt{b} processor grid with a prediction horizon \texttt{c}. For our analysis, we choose a prediction horizon of 2.

We first evaluate the performance and energy efficiency of the performance-critical sparse solver step by itself, demonstrating the significant benefits of both \gls{sssr} acceleration over a baseline RV32G \fin{ISA,} and ParSPL over a naive implementation of QDLDL.
We then consider an end-to-end \gls{etm} scenario and show that the joint acceleration of \glspl{sssr} and ParSPL enables the timely thermal control of a 9\x9 manycore, completing an \gls{mpc} control step in under \SI{1}{\milli\second}.

\subsubsection{Performance}

We run cycle-accurate \gls{rtl} simulations of the controller's compute cluster to evaluate the performance of both the vanilla QDLDL and ParSPL solvers on FP64 \gls{etm} problems.
We evaluate both solvers with and without \gls{sssr} acceleration.
Since the largest problem considered (\verb|P12x12_H2|) slightly exceeds the controller's default \SI{1}{\mebi\byte} \gls{tcdm} size, we consider a larger \SI{2}{\mebi\byte} configuration in this section.

\begin{figure}
  \centering
  \includegraphics[width=\linewidth]{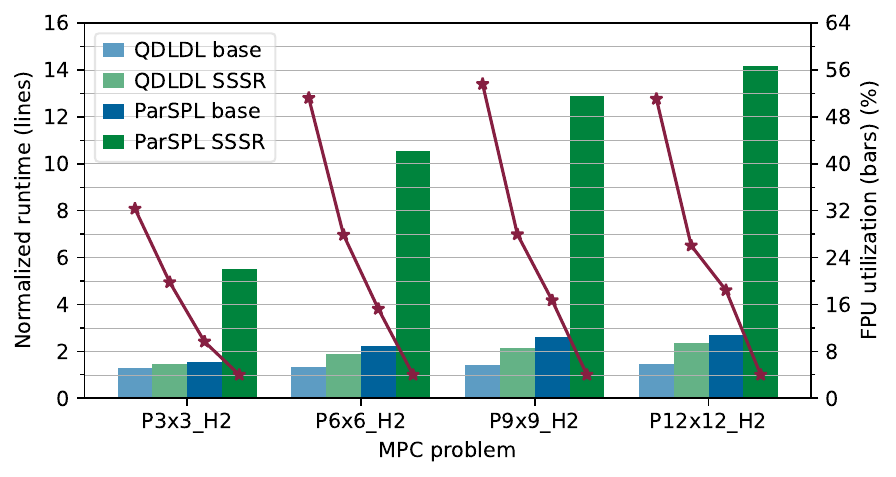}
  \caption{Normalized runtime \mmp{(lines)} and FPU utilization \mmp{(bars)} measured in \gls{rtl} simulation, with and without \glspl{sssr}, for both the vanilla QDLDL and ParSPL solver implementations.}
  \label{fig:workloads:mpc:eval:fputil_rt}
\end{figure}

\Cref{fig:workloads:mpc:eval:fputil_rt} shows the relative runtimes and \gls{fpu} utilizations for \gls{etm} problems of different sizes.
For both QDLDL and ParSPL, \glspl{sssr} accelerate all problems, showing a positive trend as problem size increases and \gls{sssr} streams become longer on average.
They also significantly raise the peak ParSPL \gls{fpu} utilization from \SI{10.8}{\percent} to \SI{56.6}{\percent}.
Further increases in \gls{fpu} utilization are limited primarily by \gls{tcdm} banking conflicts and remaining workload imbalance between cores, as well as kernel and \gls{sssr} setup overheads.

\begin{figure}
  \centering
  \includegraphics[width=\linewidth]{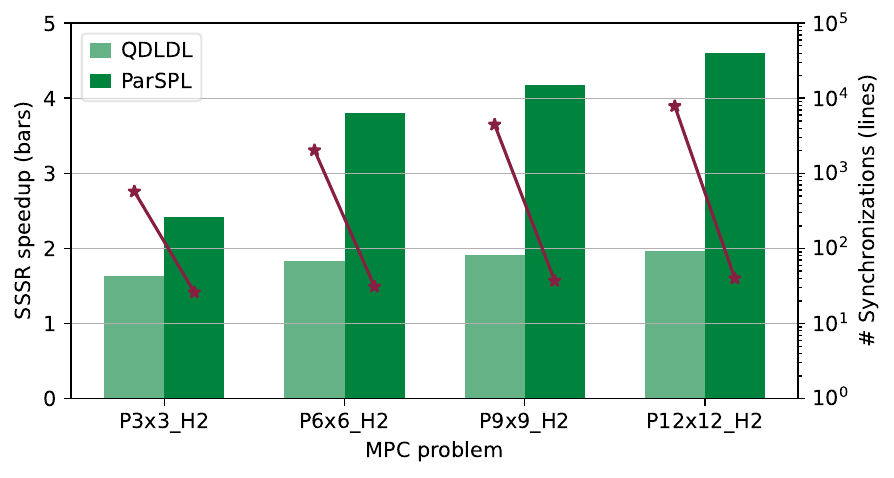}
  \caption{SSSR \fin{speedups} \mmp{(bars)} and \fin{number of necessary synchronizations} \mmp{(lines)} for \fin{both the vanilla QDLDL} and ParSPL solvers.}
  \label{fig:workloads:mpc:eval:spdup_syncs}
\end{figure}

As expected from our design considerations, ParSPL is notably faster than our vanilla QDLDL implementation and \emph{increases} the speedups provided by \glspl{sssr}.
This is because, as shown in \cref{fig:workloads:mpc:eval:spdup_syncs}, ParSPL drastically reduces required inter-core synchronizations from \emph{thousands} to \emph{tens} per solver step, enabling much longer \gls{sssr} streams on average.
As a result, ParSPL increases peak \gls{sssr} speedups from $2.0$\x~to $4.6$\x.
Combined, \glspl{sssr} and ParSPL enable speedups of up to 12.8\x~over an RV32G baseline controller running vanilla QDLDL.

\subsubsection{Energy and Power}

We implement the controller's compute cluster using Synopsys \emph{Fusion Compiler} in GlobalFoundries' 12LP+ FinFET technology, targeting a \SI{1}{\giga\hertz} clock under typical conditions (\SI{25}{\celsius}, \SI{0.8}{\volt}).
We then use Synopsys \emph{PrimeTime} to estimate the power and total energy consumption of the cluster running one iteration of the sparse system solver step for the \verb|P3x3_H2| and \verb|P9x9_H2| \gls{etm} problems in FP64.

\begin{figure}
  \centering
  \includegraphics[width=\linewidth]{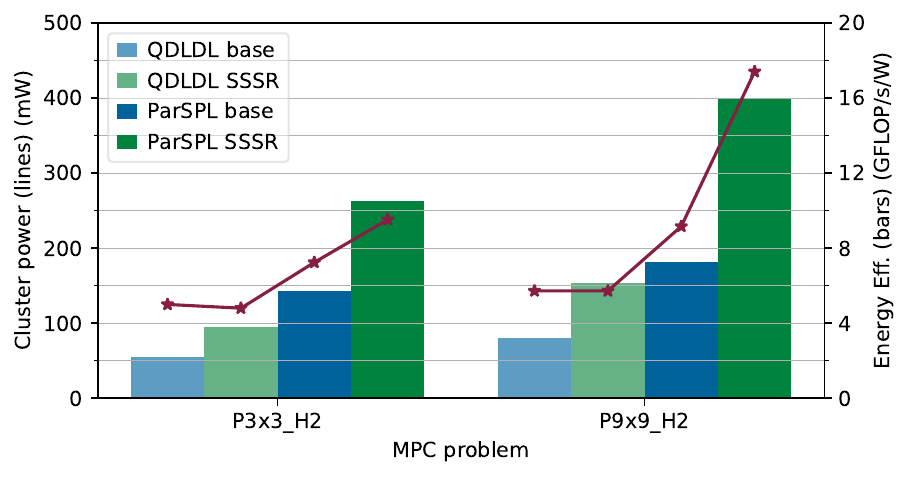}
  \caption{Cluster power \mmp{(lines)} and energy efficiency \mmp{(bars)} for vanilla OSQP and ParSPL solvers with and without SSSR acceleration.}
  \label{fig:workloads:mpc:eval:pow_eneff}
\end{figure}

\cref{fig:workloads:mpc:eval:pow_eneff} shows the results. As expected, the increased compute utilization of both ParSPL and \fin{\glspl{sssr}} increases the cluster's power consumption, with peak ParSPL power increasing from \SI{229}{\milli\watt} to \SI{435}{\milli\watt} \fin{due to \glspl{sssr}}.
However, thanks to their significant speedups, both also bring notable energy efficiency benefits: ParSPL alone improves the peak energy efficiency on an \fin{RV32G} baseline architecture by 2.3\x, while \glspl{sssr} further increase the peak efficiency by 2.2\x.
As was the case for performance, energy efficiency increases with problem size, as larger problems involve longer streams on average.
Together, ParSPL and \glspl{sssr} increase energy efficiency by up to 5.0\x, reaching up to \SI{16.0}{\giga\flop\per\second\per\watt}.

\subsubsection{End-to-end \gls{etm} Scenario}

To demonstrate the end-to-end benefits of \gls{sssr} acceleration and ParSPL on \gls{mpc}-based \gls{etm}, we consider a 9\x 9 manycore using the \verb|P9x9_H2| model in FP32 precision.
As explored in detail \mmp{by Ottaviano et al.}~\cite{ottaviano2025mpc}, a \verb|max_iter| of 15 is sufficient for convergence on this scenario, and an \gls{mpc} step period of at most \SI{3}{\milli\second} is required.
We will first show that our work enables a ControlPULP-based controller to easily meet this deadline, and then demonstrate that the incurred area overhead for the co-integrated controller is negligible.

We compare a baseline RV32G controller running our vanilla QDLDL implementation to a controller with \glspl{sssr} running our ParSPL solver.
Across all iterations, our baseline controller spends \SI{4114}{\kilo\relax} cycles on the solver step, while our enhanced controller with \glspl{sssr} running ParSPL only requires \SI{307}{\kilo\relax} cycles.
For the remaining dense \gls{la} steps in each iteration of \cref{alg:workloads:mpc:osqp}, we trivially accelerate them using affine \gls{sssr} streams, reducing their total runtime from \SI{258}{\kilo\relax} to \SI{37.5}{\kilo\relax} cycles.
Assuming a conservative controller clock period of \SI{500}{\mega\hertz}, \glspl{sssr} and ParSPL therefore reduce the time required for one \gls{mpc} iteration from \SI{8.74}{\milli\second} to \SI{0.69}{\milli\second},  enabling \gls{etm} within the \SI{3}{\milli\second} deadline with a large slack.
This slack time could be used to scale to larger systems or run other real-time tasks.

Finally, we show the feasibility of co-integrating our controller by comparing its area to that of a surrounding manycore.
We implement the ControlPULP-based controller using Synopsys \emph{Fusion Compiler} in GlobalFoundries' 22FDX technology, targeting our relaxed \SI{500}{\mega\hertz} clock under typical conditions.
We configure ControlPULP with \SI{128}{\kibi\byte} of L2 \gls{spm}, \SI{1}{\mebi\byte} of cluster \gls{tcdm}, and 32-bit-wide FPUs for the worker cores.
For this controller configuration, we find a die area of \SI{2.99}{\milli\meter^2}.
Modern \gls{hpc} manycore dies in comparable technologies usually have areas of 200 to \SI{800}{\milli\meter^2}~\cite{nvidia2023grace, intel2022alderlake}, depending on the packaging, microarchitecture, and number of cores.
Therefore, our controller incurs a negligible area overhead of \fin{\SI{0.4}{\percent}} to \SI{1.5}{\percent}.

\subsection{Related Work}
\label{chap:workloads:mpc:relwork}

We first examine \gls{soa} \gls{mpc} schemes specialized for online \gls{etm}.
We then examine online \gls{mpc} works, usually relying on \gls{admm}, for various general control scenarios across different problem scales.
\cref{tab:workloads:mpc:relwork} gives a concise overview of the works we discuss in this section and compares them to ours.

\newcommand{\mpcrot}[1]{\rotatebox[origin=c]{0}{#1}}
\newcommand{\tilt}[1]{\hspace{-1cm}\rotatebox[origin=c]{32}{#1}}
\newcommand{\noc}{\textcolor{ieee-dark-grey-40}{0}}
\newcommand{\na}{\textcolor{ieee-dark-grey-40}{n.a.}}
\newcommand{\nad}{\textcolor{ieee-dark-grey-40}{-}}
\newcommand{\dl}[2]{\makecell[cc]{#1 \\ #2}}
\newcommand{\dll}[2]{\makecell[cl]{#1 \\ #2}}
\newcommand{\dllb}[2]{\makecell[cl]{\textbf{#1} \\ \textbf{#2}}}
\newcommand{\dlb}[2]{\makecell[cc]{\textbf{#1} \\ \textbf{#2}}}
\newcommand{\tl}[3]{\makecell[cc]{#1 \\ #2 \\ #3}}
\newcommand{\tll}[3]{\makecell[cl]{#1 \\ #2 \\ #3}}
\newcommand{\tlb}[3]{\makecell[cc]{\textbf{#1} \\ \textbf{#2} \\ \textbf{#3}}}

\glsunset{pe}
\glsunset{pcg}

\newcommand{\tmark}[1]{\textsuperscript{#1}}

\newsavebox{\MpcSaveBox}
\sbox{\MpcSaveBox}{%
    \renewcommand{\arraystretch}{2.0}
    \setlength{\tabcolsep}{1.5pt}
    \begin{tabular}{lcc|cccc|cccccc|ccc}

    \arrayrulecolor{ieee-dark-black-100}\toprule &
    \multicolumn{2}{c}{\textbf{\glsentrytext{mpc} appl. profile}} &
    \multicolumn{4}{c}{\textbf{Algorithm (best referenced configuration)}} &
    \multicolumn{6}{c}{\textbf{Deployment Platform}} &
    \multicolumn{3}{c}{\textbf{\Gls{mpc} Performance}} \\

    \arrayrulecolor{ieee-dark-black-100}\midrule

    &
    \mpcrot{\textbf{Class}} &
    \mpcrot{\dlb{Referenced}{application}} &
    \mpcrot{\dlb{\glsentrytext{qp} sol.}{method}} &
    \mpcrot{\dlb{Solver}{ }} &
    \mpcrot{\dlb{Exploits}{sparsity?}} &
    \mpcrot{\dlb{Data}{\,\,precision\,\,}} &
    \mpcrot{\dlb{Pred.}{~horizon~}} &
    \mpcrot{\dlb{HW}{topology}} &
    \mpcrot{\dlb{HW}{platform}} &
    \mpcrot{\dlb{Mem.}{hierarchy}} &
    \mpcrot{\dlb{Tgt. freq.}{[\si{\mega\hertz}]}} &
    \mpcrot{\dlb{Pk. pwr.}{[\si{\watt}]}} &
    \mpcrot{\dlb{Mem. ftpr.}{[\si{\kibi\byte}]}} &
    \mpcrot{\dlb{Time to}{solve}} &
    \mpcrot{\dlb{Best}{sample time}} \\

    \arrayrulecolor{ieee-dark-black-40}\cdashline{1-16}
    \multicolumn{8}{c}{\textbf{Online \gls{mpc} for various control applications}} & \multicolumn{8}{c}{\textbf{Online \gls{mpc} for various control applications}} \\
    \arrayrulecolor{ieee-dark-black-40}\cdashline{1-16}

    Nguyen \emph{et al.}~\cite{nguyen2024tinympc} &
    Emb. & 6-DOF quadr. & ADMM & tinyMPC & \textcolor{ieee-bright-red-100}{\xmark} & FP32 & 15-20 &
    MCU & STM32F405 &
    \dl{\SI{192}{\kibi\byte} DRAM +}{\SI{1}{\mebi\byte} flash} &
    168 &
    $<$0.1 &
    $<$192 & $<$2ms-\na & 2ms \\

    Ndje \emph{et al.}~\cite{ndje2021fcnl} &
    Emb. & DC motor & FNLQDMC & \na & \textcolor{ieee-bright-red-100}{\xmark} &
    Custom & 2 & MCU & Arduino Due &
    \dl{\SI{96}{\kibi\byte} \glsentrytext{spm} +}{\SI{512}{\kibi\byte} flash} &
    84 &
    \na &
    83 & 0.8-2.9ms & 3-8ms \\

    Chaber \emph{et al.}~\cite{chaber2019fastal} &
    Emb. & Servo & \dl{FGPC}{FDMC} & Custom & \textcolor{ieee-bright-red-100}{\xmark} &
    FP32 & \dl{50}{50} & MCU & STM32F746 &
    \dl{\SI{320}{\kibi\byte} \glsentrytext{spm} +}{\SI{1}{\mebi\byte} flash} &
    216 &
    $<$0.6 &
    \na & \dl{\SI{11}{\micro\second}}{\SI{584}{\micro\second}} & \dl{\SI{250}{\micro\second}}{1ms} \\

    Jerez \emph{et al.}~\cite{jeres2013embed1,jeres2014embed2}~ &
    Emb. & AFM control & \dl{\glsentrytext{fg}}{~\glsentrytext{admm} (expl.)~} & Custom &
    \textcolor{ieee-bright-red-100}{\xmark} & Fix. pt. & 16 & \glsentrytext{dsa} &
    \dl{Virtex6 \glsentrytext{fpga}\tmark{a}}{Spartan6 \glsentrytext{fpga}\tmark{a}} & \nad &
    \dl{400}{230} &
    \na &
    \na & \dl{4.9-8.9$\mu$s}{0.5-0.9$\mu$s} & 1-1.5$\mu$s \\

    Jerez \emph{et al.}~\cite{jeres2011fpga} &
    Emb. & Spring-mass &
    \dl{Int.-point +}{Newton} & Custom & \textcolor{ieee-bright-dgreen-100}{\cmark} &
    Fix. pt. & \dl{4}{9} & \glsentrytext{dsa} & Virtex6 \glsentrytext{fpga} &
    \nad & 400 &
    \na &
    \na & 371ms & 388ms \\

    Malouche \emph{et al.}~\cite{malouche2015auto} &
    Emb. & Robot & GPC & Custom & \textcolor{ieee-bright-red-100}{\xmark} &
    FP32 & 20 & MCU & STM32F407 &
    \dl{\SI{192}{\kibi\byte} \glsentrytext{spm} +}{\SI{1}{\mebi\byte} flash} &
    168 &
    $<$0.07 &
    10.6 & \SI{4.9}{\micro\second} & 5.3ms \\

    Sabo \emph{et al.}~\cite{sabo2018primdual} &
    Emb. & Double integrator & Primal-dual & Custom & \textcolor{ieee-bright-dgreen-100}{\cmark} &
    Fix. pt. & 9 & \glsentrytext{dsa} & Digil. Nexys 4 &
    \nad & 100 &
    \na &
    \na & \na & \na \\

    Vouzis \emph{et al.}~\cite{vouzis2009asoc} &
    Emb. & \dl{Rot. antenna}{Glucose regul.} & Newton & Custom & \textcolor{ieee-bright-red-100}{\xmark} &
    \glsentrytext{lns} & \dl{20}{5} & \glsentrytext{dsa} & Virtex4 \glsentrytext{fpga} &
    2 BlockRAMs & 50 &
    \na &
    \na & \dl{0.45ms}{0.68ms} & \dl{\na}{few min} \\

    Wills \emph{et al.}~\cite{wills2012fast} &
    Emb. & ~14th-order harmon.~ & Active-set & Custom &
    \textcolor{ieee-bright-red-100}{\xmark} &
    Custom\tmark{c} & 12 & \glsentrytext{dsa} & Stratix III \glsentrytext{fpga} &
    \nad & 70 &
    \na &
    \na & 30\si{\micro\second} & 200\si{\micro\second} \\

    Wang \emph{et al.}~\cite{wang2010fast} &
    HPC & \tl{Oscill. masses}{Supply chain}{Random} & Int.-point & Custom &
    \textcolor{ieee-bright-dgreen-100}{\cmark} & \na & 10/30 &
    Desk. CPU & Athlon & \na & 3000 &
    35 (TDP) &
    \na\tmark{e} & \dl{$<$1ms}{$<$12ms\tmark{b} \\ $<$25ms} &
    \dl{5ms}{12ms \\ 25ms} \\

    Schubiger \emph{et al.}~\cite{Schubiger2020Admm}~~ &
    \glsentrytext{hpc} & Random & ADMM & \dl{PCG}{(cuOSQP)} &
    \textcolor{ieee-bright-dgreen-100}{\cmark} &
    \dl{P32}{FP64} & \na & GP-GPU &
    \dl{i9-9900K +}{RTX2080Ti} & \dl{\SI{64}{\gibi\byte} DDR4}{\SI{11}{\gibi\byte} VRAM} &
    3600 &
    \dl{250}{(RTX TDP)} &
    \na & \dl{100ms-90s}{200ms-100s} & \dl{200ms-100s}{ } \\

    \arrayrulecolor{ieee-dark-black-40}\cdashline{1-16}
    \multicolumn{8}{c}{\textbf{Online \gls{mpc} for \gls{etm} of \gls{hpc} processors}} &
    \multicolumn{8}{c}{\textbf{Online \gls{mpc} for \gls{etm} of \gls{hpc} processors}} \\
    \arrayrulecolor{ieee-dark-black-40}\cdashline{1-16}

    Bartolini \emph{et al.}~\cite{bartolini2013thermal} &
    \glsentrytext{hpc} & \dl{\glsentrytext{etm}}{distrib. 1 \gls{pe}} &
    Active-set & \na & \textcolor{ieee-bright-red-100}{\xmark} &
    \texttt{FP64} & 2 & Desk. CPU & \na & \na\tmark{d} &
    2400 &
    \na &
    \na & 4.69\si{\micro\second}\tmark{f} & 1-10ms \\

    Tilli \emph{et al.}~\cite{tilli2022twolayer} &
    \glsentrytext{hpc} & \dl{\glsentrytext{etm}}{distrib. 1 \gls{pe}} &
    \na & \na & \textcolor{ieee-bright-red-100}{\xmark} &
    \texttt{FP64} & 2 & Desk. CPU & Intel i7 8th gen. & \na\tmark{e} &
    4600 &
    95 (TDP) &
    \na & 3\si{\micro\second}\tmark{f} & 1ms \\

    Wang \emph{et al.}~\cite{wang2011adaptive} &
    \glsentrytext{hpc} & \dl{\glsentrytext{etm}}{centr. 16 \glspl{pe}} &
    Least-square & \texttt{lsqlin} & \textcolor{ieee-bright-red-100}{\xmark} &
    \na & 8 & Desk. CPU & Xeon X5365 & \na\tmark{e} & 3000 &
    150 (TDP) &
    \na & \na & 40ms \\

    Maity \emph{et al.}~\cite{maity2022future} &
    Emb. & \dl{\glsentrytext{etm}}{centr. 4 \glspl{pe}} &
    \dl{Heur.}{beam search} & \na & \textcolor{ieee-bright-red-100}{\xmark} &
    FP32 & \na & \glsentrytext{dsa} & Odroid XU4 &
    \na & 2000 &
    \na &
    \na & Few ms & \na \\

    \textbf{This work} &
    \textbf{Emb.} & \textbf{\dl{\glsentrytext{etm}}{centr. 81 \glspl{pe}\tmark{\normalfont g}}} & \textbf{ADMM} &
    \textbf{\dl{QDLDL/}{ParSPL}} & \textcolor{ieee-bright-dgreen-100}{\cmark} &
    \textbf{\dl{FP32}{FP64}} & \textbf{2} &
    \textbf{\dl{MCU with}{PMCA}} & \textbf{ControlPULP} &
    \textbf{\SI{1}{\mebi\byte}~\glsentrytext{spm}\tmark{\normalfont h}} &
    \textbf{500}\tmark{g} &
    \textbf{0.435}\tmark{i} &
    \textbf{600}\tmark{g} & \textbf{0.69ms}\tmark{g} & \textbf{$<$1ms}\tmark{g} \\

    \arrayrulecolor{ieee-dark-black-100}\bottomrule
    \end{tabular}
}

\afterpage{%
    \clearpage%
    \ifodd\value{page}
        \expandafter\afterpage%
    \fi
    {%

\begin{table}[p]

    \captionsetup{labelsep=contcols}
    \caption{\glsentrytext{qp} acceleration solutions for various applications and control problems using \glsentrytext{mpc}.}
    \label{tab:workloads:mpc:relwork}

    \begin{threeparttable}
    \centering

    \adjustbox{clip,trim=0 0 {.502\width} 0,totalheight=0.70\textheight}{\usebox{\MpcSaveBox}}

    \begin{adjustbox}{scale=0.68}
    \begin{minipage}{1.47\linewidth}
    \begin{tablenotes}[para,flushleft]
        \item[a] The papers report only potential performance and resource usage without effective measurements.%
        \item[b] The paper reports average time per iteration; no explicit sample time is given.%
        \item[c] Custom floating-point format with 7 exponent bits and a mantissa in the range [7,15].%
        \item[d] Data from Fig. 9 of~\cite{bartolini2013thermal} obtained on a dual-core desktop processor.
    \end{tablenotes}
    \end{minipage}
    \end{adjustbox}

    \end{threeparttable}
\end{table}

\clearpage

\begin{table}[p]

    \caption*{\Cref{tab:workloads:mpc:relwork} (columns continued from previous page): \glsentrytext{qp} acceleration solutions for various applications and control problems using \glsentrytext{mpc}.}

    \begin{threeparttable}
    \centering

    \adjustbox{clip,trim={.498\width} 0 0 0,totalheight=0.70\textheight}{\usebox{\MpcSaveBox}}

    \begin{adjustbox}{scale=0.68}
    \begin{minipage}{1.47\linewidth}
    \begin{tablenotes}[para,flushleft]
        \item[e] The paper lacks details on the exact memory hierarchy configuration, despite it being a commercial processor.
        \item[f] Time per single controlled \glsentrytext{pe}.
        \item[g] Refers to the 81-PE end-to-end problem evaluated in this work.\hspace{-0.2em}
        \item[h] Relies only on on-chip \glsentrytext{spm} hierarchy for manager domain and \glsentrytext{pmca}; see~\cref{chap:workloads:mpc:map}.
        \item[i] Refers to controlling 81 \glspl{pe} in FP64; power for the end-to-end problem in FP32 is lower at \SI{0.325}{\watt}.
    \end{tablenotes}
    \end{minipage}
    \end{adjustbox}

    \end{threeparttable}
\end{table}

\clearpage

    }%
}

\subsubsection{Online \gls{mpc} for \gls{etm}}

Due to the complexity of \gls{mpc}, classical approaches usually rely on the computational capabilities of out-of-order, application-class processors to execute the controller as part of a software-based \gls{hlc}.
Bartolini et al.~\cite{bartolini2013thermal} propose an \gls{mpc} solution based on thermal capping, employing a distributed scheme to avoid the challenges of centralized methods.
Exploiting the inherent parallelism of multi-core \glspl{soc}, each PE independently computes its future frequency based on predicted workload demands.
Their target platform, a \SI{3}{\giga\hertz} desktop CPU, enables online solution of the underlying \gls{qp} via active-set methods in approximately \SI{5}{\micro\second} per \gls{pe}, enabling sampling intervals of 1 to \SI{10}{\milli\second}.
Similarly, Tilli et al.~\cite{tilli2022twolayer} introduce a two-layer distributed MPC framework that combines a decentralized thermal capping layer, guaranteeing feasibility, with a distributed MPC layer, ensuring good performance.
Their software-based solution is evaluated on last-generation Intel processors.
Wang et al.~\cite{wang2011adaptive} instead adopt a centralized approach, demonstrating control of up to 16 \glspl{pe} using a \SI{3}{\giga\hertz} Intel Xeon X5 CPU and achieving \gls{mpc} sample times of under \SI{5}{\milli\second}.
In comparison, our centralized scheme controls 5\x~more \glspl{pe} (81) in a much lower execution time while operating at a 6\x~lower frequency.
It also achieves three orders of magnitude lower power consumption and at least two orders of magnitude lower memory footprint.
Maity et al.~\cite{maity2022future} propose a thermally aware scheduler for mobile \glspl{soc} like the Odroid-XU4.
Using a lightweight beam search heuristic executed on a quad-core Cortex-A7, their method aims to minimize peak temperature over a hyper-period. However, scalability is not addressed, as their \gls{etm} policy targets a fixed configuration of a quad-core Cortex-A15 CPU and a six-shader Mali GPU.

These works focus on controller design, formulation, and reliability for the \gls{etm} problem, but do not discuss details on the \gls{qp} solver or problem data structure.
Their \gls{hlc}-centric approach for thermal and power capping also incurs high overheads from power-hungry, operating-system-based software stacks as shown in~\cref{tab:workloads:mpc:relwork}.

\subsubsection{Online \gls{mpc} for Various Control Applications}

Early efforts toward online \gls{mpc} on embedded systems focus on accelerating sparse optimization methods.
For instance, the seminal work from Wang~et~al.~\cite{wang2010fast} fully exploits problem sparsity with an interior-point method, enabling linear compute effort growth per \gls{mpc} step with $H_p$, rather than cubic.
This enables large problems with hundreds to thousands of state variables, and constraints within sample times of \SIrange{5}{25}{\milli\second}, over 100\x~faster than a dense optimizer.
While this work inspired today's \gls{soa} solutions like \fin{OSQP,} it does not target embedded platforms directly and evaluates the control problems on a \SI{3}{\giga\hertz} desktop machine.

One way to tackle \gls{mpc}'s computational complexity is with a \fin{\gls{dsa}} accelerating the \gls{qp} solver pass.
Vouzis~et~al.~\cite{vouzis2009asoc} propose a coprocessor with a microcode interface to accelerate the computationally intensive steps of Newton’s algorithm.
To reduce memory footprint, a single 16-bit logarithmic number system unit iterates over the input data and stores intermediate results in local memory.
However, the evaluated problems are small with fewer than six decision variables ($n$) and constraints ($m$), making them 16\x~smaller than the smallest \texttt{P3$\times$3\_H2} problem analyzed in this work ($n=95$, $m=97$).
Wills~et~al.~\cite{wills2012fast} employ a similar approach using an active-set method, which involves computationally expensive matrix inversions.
In \cite{jeres2013embed1,jeres2014embed2}, Jerez~et~al. present hardware to accelerate both fast-gradient and \gls{admm} methods with dense dynamics, while in~\cite{jeres2011fpga}, they exploit sparsity to achieve a \SI{75}{\percent} memory reduction and a 6.5$\times$ speedup over an equivalent software implementation running on a \gls{cpu}.
The accelerators proposed in~\cite{jeres2013embed1,jeres2014embed2} show that sample intervals in the range of \SIrange{1}{1.5}{\micro\second} using fast-gradient methods, and \SIrange{5}{25}{\micro\second} using \gls{admm} methods, are feasible on \glspl{fpga} clocked at \SI{150}{\mega\hertz} with a \gls{qp} size of $n=704$ and $m=544$ (still smaller than \fin{\texttt{P12$\times$12\_H2}} with $n=1310 \; \mathrm{and} \; m=1312$).

All \glspl{dsa} discussed above have been implemented on \glspl{fpga}, reporting excellent execution times for fast \gls{mpc}. However, memory footprint is often not analyzed, performance on the deployed device is only estimated~\cite{jeres2014embed2}, and metrics like power consumption and incurred area are not included, as accurate evaluation would require an ASIC implementation.

Recent research works, notably those centered around OSQP, have renewed interest in \gls{admm} methods.
Nguyen~et~al.~\cite{nguyen2024tinympc} propose TinyMPC, a \gls{qp} solver that accelerates and compresses \gls{admm}, combined with \glspl{lqr}, for highly resource-constrained systems.
TinyMPC's key strategy is the pre-computation of the Riccati solution of the \gls{lqr} problem, which is equivalent to the primal update in \gls{admm}'s~\cref{eq:workloads:mpc:admm_xztilde}.
This way, it attains a speedup of 8.8\x~over vanilla OSQP on a single-core \gls{mcu} while adhering to stringent memory limitations ($<$~\SI{512}{\kibi\byte}) using \fin{random} small- and medium-scale problem instances (up to $n=696$ and $m=490$).
Based on its memory usage, execution time, and linear growth, TinyMPC could be a promising candidate for \gls{etm}.
\Gls{lqr} policies for thermal management of \gls{hpc} processors have been previously studied~\cite{zanini2013online}.
They prioritize maintaining a uniform thermal profile over minimizing power consumption and lack explicit threshold constraints on maximum temperature in their formulation.
Instead, constraints like frequency bounds and thermal limits are addressed externally through post-optimization adjustments.
This reliance on external correction limits the effectiveness of \gls{lqr} under heavy workloads, as it cannot dynamically account for nonlinearities or spatial and temporal thermal gradients, which is essential for efficient thermal management in modern \gls{hpc} processors.

Schubiger~et~al.~\cite{Schubiger2020Admm} focus instead on accelerating OSQP on large-scale problems \fin{($N_{nz} \geq 10^{4}$).}
They propose cuOSQP, a \fin{CUDA} solver for \glspl{gpu} based on the indirect \glsreset{pcg}\gls{pcg} method to solve the \gls{admm} system.
\fin{cuOSQP achieves speedups of up to 15\x~over \fin{single-threaded} vanilla QDLDL OSQP, and up to 5\x~over \fin{multi-threaded} MKL-Pardiso OSQP, for the largest evaluated problem ($N_{nz} = 10^{8}$) on an Nvidia RTX 2080 Ti \gls{gpu}.}
However, their evaluation shows that vanilla QDLDL OSQP is roughly $10\times$ faster than \gls{pcg}-based cuOSQP for \fin{$N_{nz} < 10^{6}$,} the problem range we evaluate in this work.
While this supports our choice of a QDLDL-based approach, indirect methods remain promising candidates for very large \gls{etm} control problems involving thousands of \glspl{pe}, a scale not yet achieved in commercial application-class manycores.

Finally, several works have addressed the mapping of fast \gls{mpc} algorithms on commercial \glspl{mcu}~\cite{chaber2019fastal,malouche2015auto,ndje2021fcnl}.
Despite achieving sub-\si{\milli\second} solution times with a memory footprint in the tens of \si{\kibi\byte}, the tackled problems are small, featuring tens of decision variables and constraints.
Furthermore, conventional \glspl{mcu} rely on memory hierarchies with an external flash alongside on-chip \glspl{spm} or caches.
While external memory is an option for embedded controllers, a self-contained on-chip memory hierarchy improves energy efficiency, reliability, and autonomy.

Our comparative analysis shows that our acceleration solution leveraging \glspl{sssr} and ParSPL outperforms \gls{soa} \gls{mpc} schemes for \gls{etm} in \gls{hpc} processors, requiring fewer compute resources and storage while delivering low-power execution and high energy efficiency at a small fraction of the  processor die area.
It also competes with fast \gls{mpc} methods deployed on various architectures including \glspl{mcu}, \glspl{dsa}, and \glspl{gpu} and validated on other control applications.
This efficiency stems from accelerating the sparse compute inherent to \gls{admm}-based \gls{mpc} with \glspl{sssr} \fin{and optimizing} the problem offline to minimize inter-core dependencies \fin{with ParSPL}.
Moreover, unlike fixed-function accelerators, our hardware is flexible and can be adapted to different algorithms and problems, helping avoid \fin{the} premature obsolescence that often affects specialized \fin{architectures}.

\section{Conclusion}
\label{chap:workloads:conclusion}

In this chapter, we demonstrated the high flexibility and generality of \glspl{sssr} by accelerating three workloads whose performance-critical compute goes beyond sparse \gls{la}: \emph{stencil computations}, \emph{spiking neural networks} (SNNs), and \emph{model predictive control} (MPC).
We first presented our \gls{sssr}-related extensions to the Snitch cluster LLVM toolchain and runtime and then investigated each workload in detail, using the Snitch cluster without \glspl{sssr} as a baseline.

For stencil codes, we created \emph{SARIS}, a generic acceleration methodology for indirection-capable \fin{\glspl{sr}} that efficiently maps stencil memory access patterns to streamable index arrays.
Applying SARIS to \glspl{sssr} in the Snitch cluster, we achieve geomean \gls{sssr} speedups and energy efficiency improvements of 2.7\x~and 1.6\x, respectively, increasing geomean \gls{fpu} utilization from \SI{35}{\percent} to \SI{81}{\percent}.
On our Occamy manycore scaleout presented in \cref{chap:occamy}, \glspl{sssr} enable up to \SI{83}{\percent} of theoretical peak floating-point performance on stencils, \SI{20}{\percent} more than the leading GPU code generator \emph{AN5D}.

To accelerate \glspl{snn}, we accelerated the performance-critical sparse reduction of spiking weights with \glspl{sssr}, leveraging the cluster's orthogonal SIMD FP16 and FP8 support to parallelize output channels.
This enables \gls{sssr} speedups and energy efficiency gains of \nchk{5.5\x} and \nchk{3.3\x} on FP16 S-VGG11 inference, respectively, increasing \gls{fpu} utilization from \nchk{\SI{9.3}{\percent}} to \nchk{\SI{52}{\percent}}.
Switching to FP8 inference further improves both runtime and energy efficiency by \nchk{1.7\x}.
As a result, FP8 \gls{snn} inference on the Snitch cluster with \glspl{sssr} is \nchk{4.4\x} more energy efficient than on the leading neuromorphic processor \emph{LSMCore}.

Finally, we accelerated online \gls{mpc} by improving the performance of OSQP's sparse solver step.
We first accelerated the existing QDLDL solver by up to 2.0\x~with \glspl{sssr}, finding that its main performance limitation is internal data dependencies.
We then designed an optimized solver called \emph{ParSPL} that reduces these dependencies, \fin{enabling} \gls{sssr} speedups and energy efficiency gains of up to 4.6\x~and 2.2\x, respectively, and raising the peak \gls{sssr} \gls{fpu} utilization to \SI{57}{\percent}.
On \gls{etm} of a 9\x9 manycore, ParSPL and \glspl{sssr} together reduce
the \gls{mpc} iteration time from \SI{8.74}{\milli\second} to \SI{0.69}{\milli\second}, enabling online control.
The resulting \gls{etm} solution is lightweight enough for co-integration, outperforming \gls{soa} \gls{mpc}-based \gls{etm} in area and energy efficiency, and is competitive with existing fast \gls{mpc} solutions.

Across all three considered workloads, \glspl{sssr} thus enable significant performance and energy efficiency improvements by \fin{increasing} \gls{fpu} utilization, showing that the benefits \fin{they} provided on sparse \gls{la} in \cref{chap:sssrs} \emph{generalize} to other irregular workloads.
\mmp{Through future work on automatic stream inference, we could further extend the benefits of \glspl{sssr} to many more irregular workloads, avoiding the manual optimization of key kernels while simultaneously strengthening the case for stream-based acceleration.}

\clearpage
\chapter{Manycore Scaleout Prototype}
\label{chap:occamy}

In this chapter, we scale out the \gls{sssr}-extended Snitch cluster we first introduced in \cref{chap:sssrs} to a fully manufactured 2.5D manycore system called \emph{Occamy}.
Occamy features 48 Snitch clusters and 432 cores across two \SI{12}{\nano\meter} compute chiplets on a passive \SI{65}{\nano\meter} interposer. Each chiplet is paired with \SI{16}{\gibi\byte} of HBM2E DRAM and features a Linux-capable 64-bit host core for management.

We motivate our work through an evolving heterogeneous mixture of regular and irregular compute in \gls{hpc} and \gls{ml} applications that is not efficiently handled by \gls{soa} \glspl{cpu}, \glspl{gpu}, and accelerators.
Our goal is therefore to create a \emph{general-purpose} manycore compute platform that leverages our work on \glspl{sssr} to efficiently and flexibly handle sparse and irregular compute \emph{without} compromising on regular compute performance.

\section{Introduction}
\label{chap:occamy:introduction}

Surging performance demands and large, sparse models in \gls{ml} and \gls{hpc} applications push data-driven workloads toward a mixture of regular and irregular computation.
In \gls{ml}, the sparsification of dense models significantly reduces both their computational and memory footprints~\cite{hoefler2021sparsity}, and the recent proliferation of \textgreater 100B-parameter transformers has reinvigorated sparsification efforts to address severe memory capacity bottlenecks~\cite{dettmers2023spqr, xia2023flash}.
In \gls{hpc}, applications like multiphysics simulation and graph analytics heavily rely on irregular compute including sparse \gls{la}, stencil computations, and graph pattern matching to handle large, high-dimensional problems~\cite{Zhang2025Survey}.

Large-scale \gls{ml} and \gls{hpc} applications traditionally target manycore CPUs and GPUs whose parallel datapaths achieve high ($\geq$\SI{75}{\percent}) FPU utilization and energy efficiency on regular workloads~\cite{nvidia_a100_dense, siegmann2024first}.
However, these general-purpose architectures struggle with the noncontiguous memory accesses, dynamic control flow, and computational imbalance inherent to irregular workloads.
Consequently, they achieve peak FPU utilizations of less than \cn{\SI{50}{\percent}} on stencil codes~\cite{cfd_amd_epyc_rome_tpds21, hirokawa2022large,Zhang2023RevisitingTB} and less than \cn{\SI{10}{\percent}} on sparse \gls{la}~\cite{amd_sparse, alappat2020a64fx, Niu2021TileSpMVAT, siegmann2024first}.

While hardware accelerators for irregular workloads have been proposed, they usually target narrow, domain-specific compute patterns:
accelerators for sparse \gls{ml}~\cite{huang2025hybr, yue2024cimblock, feng2024pntcloud, zhang2021snap} support low-precision data formats ($\leq$\SI{16}{\bit}) and low ($\leq$\SI{80}{\percent}) or structured sparsity,
while more generic sparse \gls{la}~\cite{Srivastava2020MatRaptorAS, Pal2018OuterSPACEAO, park2020output, Sadi2019EfficientSO} and \gls{ta}~\cite{Hegde2019ExTensorAA} accelerators often target specific operators~\cite{Srivastava2020MatRaptorAS, Pal2018OuterSPACEAO, park2020output, Sadi2019EfficientSO} or dataflows~\cite{Srivastava2020MatRaptorAS, Pal2018OuterSPACEAO}.
General-purpose acceleration solutions based on \glspl{cgra}~\cite{Dadu2019TowardsGP,koul2024onyx,chen2025opal} or \gls{isa} extensions~\cite{siracusa2023tmu, Domingos2021UnlimitedVE, zhu2019sparsetensorcore, Wang2019StreambasedMA, Wang2021DualsideST, Rao2022SparseCoreSI} are more flexible, but they have not yet been demonstrated on silicon at scale~\cite{Dadu2019TowardsGP, siracusa2023tmu, Domingos2021UnlimitedVE, zhu2019sparsetensorcore, Wang2019StreambasedMA, Wang2021DualsideST, Rao2022SparseCoreSI} or tackle only one-~\cite{siracusa2023tmu, Domingos2021UnlimitedVE, zhu2019sparsetensorcore, Wang2019StreambasedMA} or two-sided sparsity~\cite{Wang2021DualsideST, Rao2022SparseCoreSI}.
Hence, there is a lack of \emph{silicon-proven} processors that efficiently handle the diverse irregular compute requirements of current \gls{ml} and \gls{hpc} applications without compromising on regular compute performance.

In this chapter, we present \emph{Occamy}, a 432-core, 768-\si{DP\dash\giga\flop\per\second} dual-chiplet {\riscv} system that scales out the \gls{sssr}-extended Snitch cluster we introduced in \cref{chap:sssrs} to flexibly handle both regular and irregular \gls{hpc} and \gls{ml} workloads with high FPU utilization and energy efficiency.
Unlike existing systems targeting irregular compute, Occamy is designed for the \emph{general} acceleration of sparse and irregular workloads and avoids specialization to specific workloads, sparsity patterns, and value densities.
Each Occamy chiplet features 24 compute clusters, a Linux-capable \revprg{\riscv}~host, \SI{16}{\gibi\byte} of HBM2E DRAM,\pagebreak[4] and a fully digital \revprg{fault-tolerant \cn{\SI{1.6}{\pico\joule\per\bit}}} \gls{d2d} link.
As discussed in \cref{chap:sssrs:arch:integr}, each cluster features eight \gls{sssr}-extended worker cores with SIMD-capable FP8-to-FP64 \glspl{fpu}~\cite{bertaccini2024fpu} and an additional \emph{DMA control} core~\cite{benz2024idma} that enables asynchronous, latency-tolerant \mmp{1D and} 2D transfers of large data tiles to or from a shared \SI{128}{\kibi\byte} \gls{tcdm}.
Occamy's hierarchical chiplet-level interconnect provides a 512-bit network for \rev{bulk} data transfers and a 64-bit network for message passing and synchronization.

We implemented Occamy's \cn{\SI{73}{\milli\meter\squared}} \rev{compute} chiplets in \gfs~\SI{12}{\nano\meter} LP+ \revprg{FinFET} technology and \rev{its} passive interposer named \emph{Hedwig} in a \SI{65}{\nano\meter} node.
On regular workloads, Occamy achieves competitive FPU utilizations of up to \cn{\SI{89}{\percent}}.
Executing stencil codes, we reach FPU utilizations of up to \cn{\SI{83}{\percent}} and a technology-node-normalized compute density of up to \cn{\SI{11.1}{DP\text{-}\giga\flop\per\second\per\milli\meter^2}}, leading \fin{the} \mmp{\gls{soa}} by \cn{1.7\x}~and
\rev{\cn{1.2\x}}, respectively. \revprg{, with respect to \gls{soa} CPUs and GPUs.}
On sparse-dense \gls{la}, we achieve up to \cn{\SI{42}{\percent}} FPU utilization and \fin{a node-normalized} compute density of up to \cn{\SI{5.95}{DP\text{-}\giga\flop\per\second\per\milli\meter^2}}, substantially surpassing the \gls{soa} by \cn{5.2\x}~and \cn{11\x}, respectively.
\rev{We further} achieve a sparse-sparse \gls{la} throughput of up to \cn{\SI{187}{\giga\comp\per\second}} at an energy efficiency of \cn{\SI{17.4}{\giga\comp\per\second\per\watt}}  and a compute density of \cn{\SI{3.63}{\giga\comp\per\second\per\milli\meter^2}}.
\rev{Finally, we reach FPU utilizations of up to \SI{75}{\percent} and \SI{54}{\percent} on \glsunset{gpt}\gls{gpt}-J inference and a \gls{gcn} layer, respectively.}
\revprg{Furthermore, }~To the best of our knowledge, Occamy is the first open-source multi-chiplet RISC-V manycore demonstrated in silicon;
the \gls{rtl} description of its digital core, including the clusters, hierarchical interconnect, and \gls{d2d} link, is available under a permissive open-source license\footnote{\url{https://github.com/pulp-platform/occamy}}.

The remainder of this chapter is organized as follows.
\Cref{chap:occamy:arch} presents Occamy's hierarchical compute chiplet architecture.
In \Cref{chap:occamy:progmodel}, we discuss Occamy's programming model, which is built on the asynchronous processing and \gls{dma}-based movement of double-buffered data tiles.
\Cref{chap:occamy:impl} describes how the Occamy chiplets, the Hedwig interposer, and the carrier board were implemented.
\Cref{chap:occamy:eval} presents our experimental results on Occamy's performance.
In \Cref{chap:occamy:relwork},  we discuss how Occamy compares to \gls{soa} CPUs and GPUs as well as existing irregular workload acceleration proposals.
Finally, we provide a conclusion in \Cref{chap:occamy:conclusion}.

\newcommand{\occbsone}{0.996}
\renewcommand{\baselinestretch}{\occbsone}\selectfont

\newcommand{\secocc}[1]{\vspace{-0.1ex}\section{#1}}
\newcommand{\subsecocc}[1]{\vspace{-0.1ex}\subsection{#1}}

\secocc{Architecture}
\label{chap:occamy:arch}

\begin{figure}
    \centering
    \begin{subcaptionblock}{0.552\linewidth}%
        \centering%
        \includegraphics[width=0.9\linewidth]{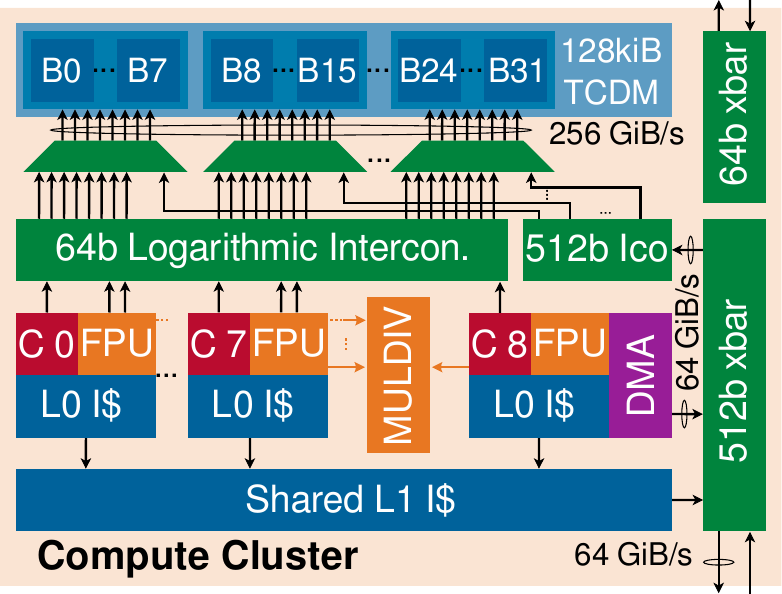}%
        \vspace{-0.5em}%
        \caption{Compute cluster}%
        \label{fig:occamy:arch:cluster}%
        \vspace{0.5em}
        \centering%
        \includegraphics[width=0.9\linewidth]{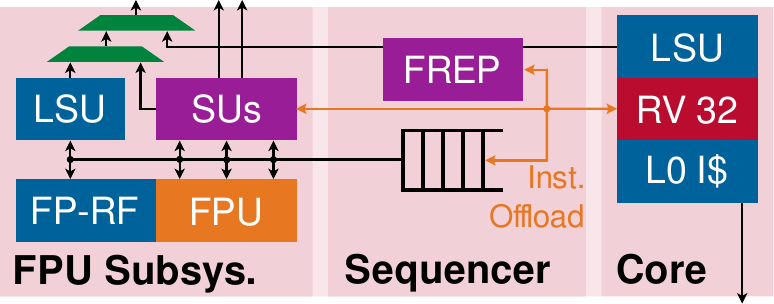}%
        \vspace{-0.4em}%
        \caption{Worker core}%
        \label{fig:occamy:arch:cc}%
    \end{subcaptionblock}%
    \begin{subcaptionblock}{0.323\linewidth}
        \centering%
        \includegraphics[width=0.913\linewidth]{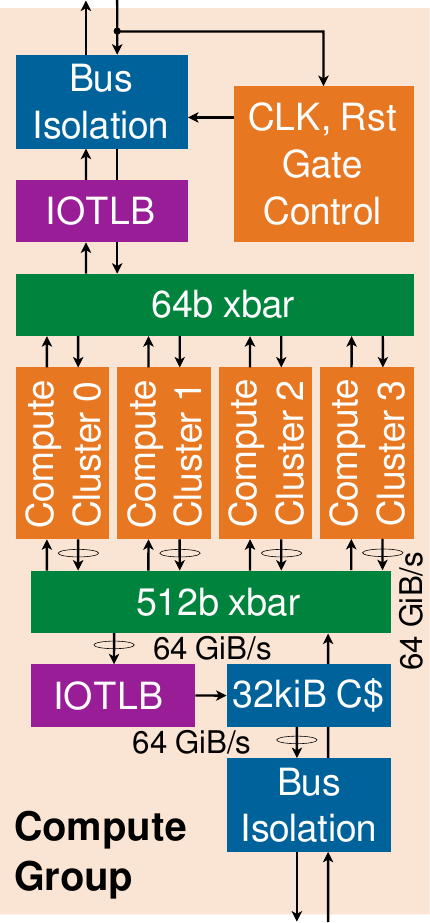}%
        \vspace{-0.4em}%
        \caption{Compute group}%
        \label{fig:occamy:arch:group}%
    \end{subcaptionblock}\vspace{0.5em}\\
    \begin{subcaptionblock}{0.968\linewidth}
        \centering%
        \includegraphics[width=0.82\linewidth]{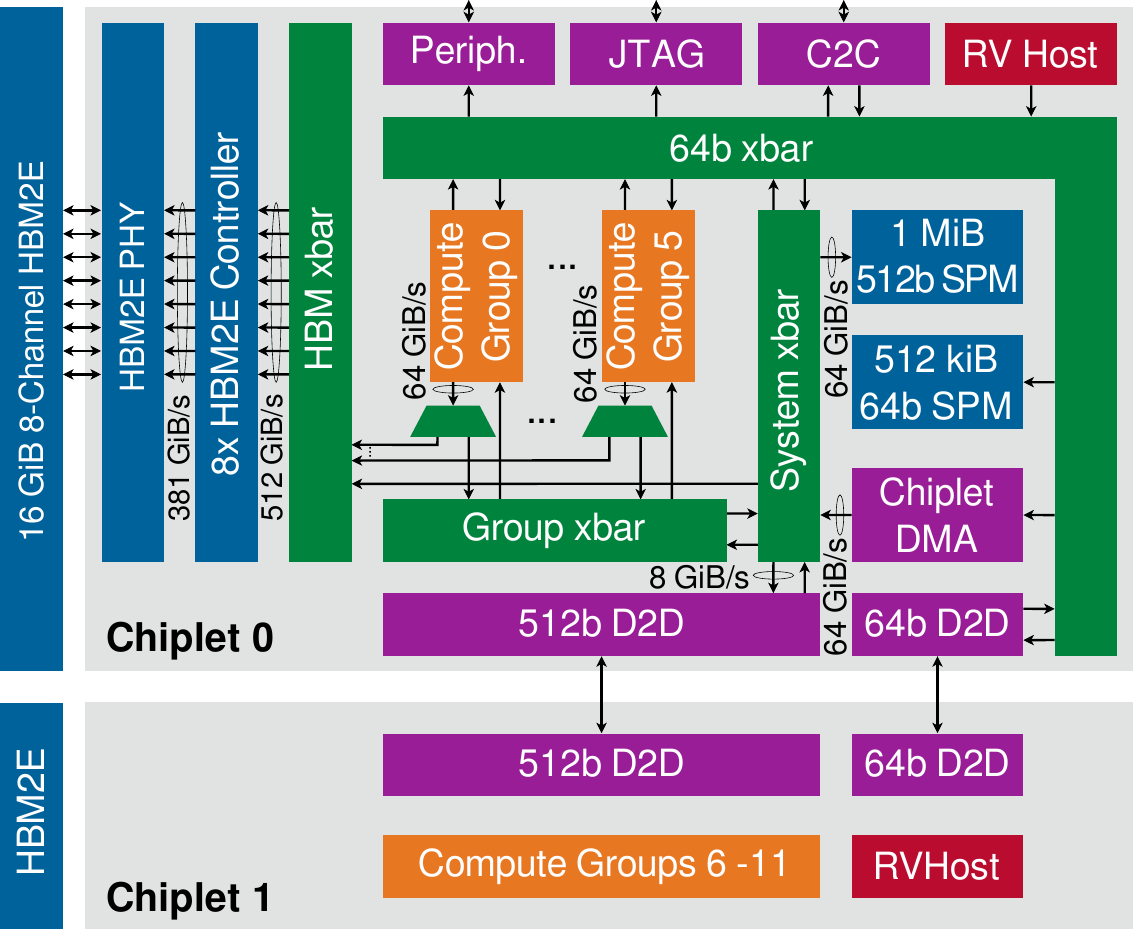}%
        \vspace{-0.2em}%
        \caption{Occamy chiplets}%
        \label{fig:arch:occamy:chiplets}%
        \vspace{-0.2em}
    \end{subcaptionblock}
    \caption{Hierarchical architecture of the two Occamy compute chiplets, including the two HBM2E interfaces and \fin{the} D2D link.}
    \label{fig:occamy:arch_overview}
\end{figure}

Occamy combines two 216-core compute chiplets on a passive interposer, each paired with an eight-device, \SI{16}{\gibi\byte} HBM2E stack and connected through a fully digital, fault-tolerant \gls{d2d} link.
\cref{fig:occamy:arch_overview} shows the hierarchical compute chiplet architecture%
, which we describe in a bottom-up fashion in the following sections.

\subsecocc{Compute Cluster}
\label{chap:occamy:arch:cluster}

Occamy's main compute element is the \gls{sssr}-extended Snitch cluster~\cite{Zaruba2020SnitchAT} introduced in \cref{chap:sssrs:arch:integr} and used in our workload evaluations in~\cref{chap:sssrs,chap:workloads}.
We quickly reintroduce the architecture of this cluster and its worker cores, focusing on our parameterization choices and features not yet discussed in detail.

Within each cluster, shown in \cref{fig:occamy:arch:cluster}, eight RV32G-compliant worker cores and one \glsunset{dma}\gls{dma} control core share a \SI{128}{\kibi\byte} 32-bank \gls{tcdm} through a single-cycle logarithmic interconnect with double-word interleaving.
The clusters also feature \SI{8}{\kibi\byte} of shared L1 instruction cache, a shared integer multiply-divide unit, a local hardware synchronization barrier, and 16 retargetable performance counters.

Each worker core, shown in \cref{fig:occamy:arch:cc}, features a 64-bit-wide \glsunset{simd}\gls{simd} \gls{fpu} supporting FP64, FP32, FP16, FP16alt \mmp{(8-bit exponent, 7-bit mantissa)}, FP8, and FP8alt \mmp{(4-bit exponent, 3-bit mantissa)} formats.
In addition to regular \gls{fma} instructions, the FPU supports widening sum-dot-product and three-addend summation instructions for all FP8 and FP16 formats~\cite{bertaccini2024fpu}.
Two \gls{isa} extensions maximize FPU utilization for both regular and irregular workloads: \glspl{sssr}, presented in \cref{chap:sssrs}, and the \gls{frep} hardware loop~\cite{Zaruba2020SnitchAT}.
The default \gls{sssr} streamer in each worker core features two \glspl{issr} and one \gls{essr}, each supporting up to 4D affine accesses.
In general, the \glspl{sssr} in each worker core enable a sustained bandwidth of up to three double-words per cycle into \gls{tcdm}.

\rev{The cluster \gls{tcdm} is dimensioned to balance throughput, interconnect complexity, and area.}
\rev{32 banks are chosen to reduce the probability of banking conflicts between \glspl{sssr} (24 per cluster) while keeping the logarithmic interconnect physically implementable.}
\rev{A capacity of \SI{128}{\kibi\byte} achieves a reasonable \glsunset{sram}\gls{sram} bit density (avoiding significant drops that would appear for $\leq$\SI{4}{\kibi\byte} banks) while keeping area prevalently allocated to compute logic.}

The \gls{dma} control core in each cluster features a tightly-coupled 512-bit \gls{dma} engine~\cite{benz2024idma}, enabling asynchronous $\leq$2D transfers between external memory (other clusters, HBM2E, or other chiplet) and the local \gls{tcdm}.
This core coordinates the computation of worker cores and their fine-grained, low-latency accesses to the \gls{tcdm} with the latency-tolerant \gls{dma} transfers of large, double-buffered data tiles to and from the \gls{tcdm}.
\rev{The \gls{dma} engine accesses} blocks of eight \gls{tcdm} banks (\emph{superbanks}) at once through a secondary interconnect\rev{, transferring} up to \SI{64}{\byte} per cycle or \SI{64}{\gibi\byte\per\second}\rev{;}
this way, it maximizes throughput without significantly slowing down ongoing memory-intensive computations, which can access up to 24 \gls{tcdm} banks at once using all \glspl{sssr}.
To reduce backpressure in the group- and chiplet-level interconnects, the \gls{dma} engine has priority access to the \gls{tcdm} over the cores and \glspl{sssr} through the secondary interconnect.
We will elaborate on the \gls{dma} engine's use in double-buffered data tiling in~\Cref{chap:occamy:progmodel:tiling}.

\subsecocc{Compute Group}
\label{chap:occamy:arch:group}

Four clusters together form the next compute hierarchy level, a \emph{group}, shown in \Cref{fig:occamy:arch:group}.
Clusters within a group have full-bandwidth access to each other through two fully connected \gls{axi4} crossbars: a 512-bit crossbar used by \gls{dma} engines and instruction caches for bulk \fin{transfers,} and an atomics-capable 64-bit crossbar used by cores for global synchronization and message passing.
\rev{Thus, groups allow their clusters to locally share data at higher bandwidths than \fin{is possible} at the global level and constitute a replicable multi-cluster design that significantly simplifies top-level chiplet implementation.}

On each crossbar, a group has one outgoing and one incoming port, providing a shared bandwidth to and from the chiplet interconnect of \SI{64}{\gibi\byte\per\second} for bulk transfers and \SI{8}{\gibi\byte\per\second} for message passing.
\rev{Providing one outgoing port per group best matches the HBM2E bandwidth available at the chiplet level and keeps the chiplet-level interconnect implementable.}
\glspl{iotlb} on the outgoing ports allow for per-group address remapping and access control at page granularity.
A remappable \SI{32}{\kibi\byte} constant cache on the outgoing 512-bit port can be used to cache program code and other immutable data.
Finally, the groups can individually be clock-gated, reset, and isolated from their interconnect ports to the chiplet level through memory-mapped registers for online power management.

\subsecocc{Occamy Chiplets}
\label{chap:occamy:arch:chiplets}

\Cref{fig:arch:occamy:chiplets} shows Occamy's top-level chiplet architecture.
Each chiplet features six groups, totaling 216 cores, and a single Linux-capable 64-bit \riscv~host processor managing the groups and all other on-chip resources\footnote{\mmp{The host processor does not need to micromanage individual clusters during computation; it merely dispatches compute tasks, similar to how they would be offloaded to a GPU. During computation, the clusters operate and communicate autonomously, with each being internally coordinated by its data movement core.}}.
Like the groups, the chiplets feature a hierarchical 512-bit \gls{axi4} network for bulk data transfers and an atomics-capable 64-bit \gls{axi4} network for synchronization, message passing, and management.
The \gls{d2d} link, which serializes cross-chiplet transactions, \fin{is comprised of} a \emph{narrow} and a \emph{wide} segment carrying 64-bit and 512-bit transactions, respectively.

The 512-bit network is composed of three fully connected crossbars.
Each group's \SI{64}{\gibi\byte\per\second} outgoing port provides access to an \emph{HBM} crossbar and a \emph{group} crossbar, interconnecting the six on-chip groups.
The \emph{HBM} crossbar provides access to the eight on-chip \rev{\cn{\SI{47.68}{\gibi\byte\per\second}}} HBM2E controllers (\rev{\cn{\SI{381.47}{\gibi\byte\per\second}}} in total).
It can be configured at runtime to interleave the HBM2E channels at page granularity, facilitating load balancing and data reuse.
A \emph{system} crossbar connects the group and HBM crossbars to the wide D2D segment and 64-bit network, providing all actors across both chiplets access to the entire memory space.
It also connects to a \SI{1}{\mebi\byte} \fin{\gls{spm},} used for low-latency on-chip storage of shared \fin{data,} and a chiplet-level \gls{dma} engine used by the host for fast, explicit data movement and memory initialization.
\rev{As a chiplet-level memory, the \SI{1}{\mebi\byte} \gls{spm} was dimensioned to be significantly larger than the \SI{128}{\kibi\byte} cluster \glspl{tcdm}, but small enough to fit into the area available to chiplet-level logic and routing.}

The 64-bit network consists of a single crossbar. It connects the Linux-capable host and the 64-bit ports of the groups to the narrow \gls{d2d} segment, the 512-bit network, and various peripherals.
It also features a \SI{512}{\kibi\byte} \gls{spm} used by the host for management tasks.
The peripherals include UART, I2C, QSPI, GPIOs, a \cn{\SI{1.33}{\giga\bit\per\second}} off-interposer \gls{c2c} link, and a JTAG test access point for live host processor debugging.
They also include RISC-V-compliant timers and platform-level interrupt controllers providing interrupts for both the host processor and all on-chip compute cores.
In essence, the 64-bit network provides an autonomous Linux-capable host platform that is tightly integrated with the remaining architecture;
we later formalized this idea into our scalable \emph{Cheshire} platform~\cite{ottaviano2023cheshire}, which has since been used on various Linux-capable \glspl{soc}~\cite{benz2023iguana,sauter2024insbas,garofalo2025carfield}.

\rev{The main benefit of our hierarchical crossbar-based interconnect over a ring or mesh topology is its \emph{symmetry}: the memory topology looks identical to all compute cores and clusters, meaning that the architectural bandwidth and latency of accesses to each hierarchy level (cluster, group, chiplet) are \emph{constant}. This greatly simplifies programming, as network performance is homogeneous and code can be written in a \emph{cluster-agnostic} way without sacrificing performance.}
\rev{However, crossbars pose challenges in physical implementation, requiring effort to provision sufficient bandwidth while managing area costs. While this was feasible for our design, larger systems may face scalability limitations. In this case, alternative \gls{noc} topologies such as mesh and torus could be more suitable.}

\rev{The \gls{d2d} link enables seamless communication across chiplets, allowing the system to scale to 432 cores and two HBM2E stacks. This approach improves overall performance while avoiding the yield challenges and high manufacturing costs associated with large monolithic dies~\cite{amd_chiplet_isscc20}. The \gls{d2d} link consists of a narrow segment, optimized for synchronization and message passing, and a wide segment, designed for high-throughput bulk data transfers.}

\begin{figure}[t]
\centering
\includegraphics[width=0.72\linewidth]{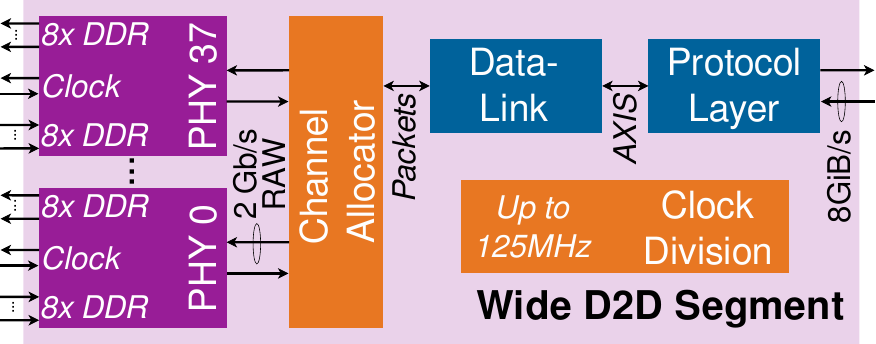}
\caption{Architecture of the wide D2D segment with its 38 source-synchronous double-data-rate PHYs, carrying up to \SI{2}{\gibi\byte\per\second} of raw data at a clock speed of \SI{125}{\mega\hertz}.}
\label{fig:arch_d2d}
\end{figure}

The wide segment is a scaled-up version of the narrow segment with \cn{38} \glspl{phy} to increase bandwidth; its architecture is shown in \Cref{fig:arch_d2d}.
The \emph{protocol layer} %
arbitrates between \gls{axi4} requests and responses and converts them to \gls{axi4}-Stream (AXIS) payloads.
\rev{It handles the five independent channels of the \gls{axi4} protocol (AW, AR, W, R, B) and applies backpressure to the \gls{axi4} interface to prevent protocol-level deadlocks. Each AXIS payload includes a header that contains information about the packet type and credits required for flow control.}
The \emph{data-link} layer further packetizes the payloads based on the available number of off-chip lanes;
it also handles credit-based \fin{flow control} to ensure that no packets are lost, as well as \fin{the} synchronization and alignment of packets if multiple \glspl{phy} are configured.
\rev{Additionally, the data-link layer features debugging capabilities, including a \emph{Raw Mode} that allows the link to operate independently of the \gls{axi4} interface by sending patterns over specific PHY channels for fault detection.}
Each \gls{phy} features an all-digital and source-synchronous interface with eight \gls{ddr} lanes in each direction.
\rev{On the transmitter side, the PHY operates with a forwarded clock derived from the system clock. On the receiver side, the transmitted packets are synchronized with the system clock and reassembled into the original payload.}
The narrow and wide segments can achieve effective duplex bandwidths of up to \cn{\SI{1.33}{\giga\bit\per\second}} and \SI{64}{\giga\bit\per\second}, respectively.
The wide segment additionally features a \emph{channel allocator} to enable fault tolerance.
An initial calibration detects faulty \glspl{phy}\footnote{\mmp{On some manufactured Occamy modules, we could successfully leverage the initial calibration to identify and avoid faulty wide-segment \glspl{phy} as intended.}}, which can individually be disabled;
the channel allocator then reshuffles packets among the functional \glspl{phy} with only linear bandwidth degradation.
\rev{This fault tolerance mechanism ensures reliable communication across chiplets, even in the presence of manufacturing defects.}

The {\occamy} chiplet relies on \glspl{fll} to generate on-chip clocks for each of its three clock domains: the \emph{compute} domain, the \emph{peripheral} domain, and the \emph{HBM2E PHY} domain.
The \emph{compute} domain includes the compute groups, the 64-bit host, the chiplet-level interconnect, the \gls{d2d} link, and the HBM2E \fin{controllers}.
Like the groups, the \gls{d2d} link and the HBM2E subsystem can be clock-gated through memory-mapped configuration registers.

\secocc{Programming Model}
\label{chap:occamy:progmodel}

Since Occamy's compute cores support the unprivileged 32-bit {\riscv} \gls{isa} (RV32G), they can be programmed using \riscv~assembly, bare-metal C/C++, or a higher-level code generator, providing the full software agility of a general-purpose system.
The extended LLVM 15 toolchain and runtime discussed in \Cref{chap:workloads:rtllvm} provide the necessary assembly, intrinsics, and library support to leverage Occamy's \gls{isa} extensions, as well as a specialized machine model and optimization passes that allow programmers to write high-performance code for regular and irregular workloads alike.

\subsecocc{Double-Buffered Tiling}
\label{chap:occamy:progmodel:tiling}

As for our cluster-level workload implementations from \cref{chap:sssrs,chap:workloads}, we use a \emph{tiled} computation model leveraging explicit \gls{dma}-based data movement to maximize compute utilization:
while the \gls{sssr}-enhanced worker cores in each cluster process local data tiles in \gls{tcdm}, the cluster's \gls{dma} engine simultaneously transfers double-buffered tiles between the \gls{tcdm} and global memory (e.g. HBM2E), using the cluster's hardware barrier and atomics in \gls{tcdm} for synchronization.
This approach combines the benefits of memory- and interconnect-efficient, latency-tolerant bulk \gls{dma} transfers and of low-latency, fine-granular access provided to worker cores by the banked \gls{tcdm}.

\begin{figure}
\centering
\includegraphics[width=0.85\linewidth]{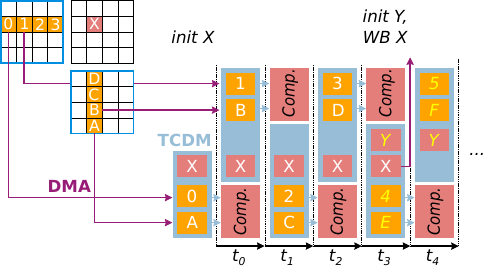}
\caption{High-level overview of our double-buffered \gls{tcdm} tiling strategy for the example workload of \fin{dense} matrix multiplication.}
\label{fig:occamy:progmodel:dm}
\end{figure}

To implement double-buffering, the \gls{dma} control core determines the necessary tile transfers for the current tiling phase and programs the \gls{dma} engine to fetch or write back the corresponding tiles.
Once both these transfers and the worker cores' computations are complete, the used tilesets are swapped and computation can immediately resume.
The \SI{128}{\kibi\byte} cluster \gls{tcdm} can support tiles tens of \si{\kibi\byte} in size, which is sufficient to hide the typical access latencies in the hundreds of cycles that accesses to HBM2E or the other chiplet may incur.

\Cref{fig:occamy:progmodel:dm} illustrates our double-buffered tiling strategy on a matrix multiplication example.
The \emph{compute} blocks represent phases where the worker cores process data, while the \emph{DMA} arrows represent \gls{dma} transfers moving tiles into or out of the cluster \gls{tcdm}.

\secocc{Silicon Implementation}
\label{chap:occamy:impl}

\begin{figure}
\centering
\hspace{-0.83em}%
\includegraphics[width=0.87\linewidth]{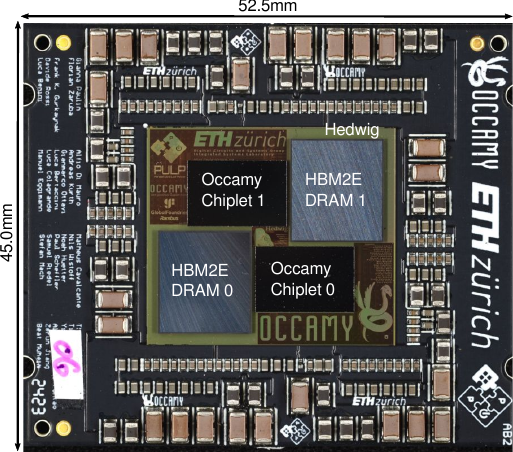}
\caption{2.5D Occamy assembly mounted on carrier PCB.}
\label{fig:occamy:impl:module}
\end{figure}

The full Occamy 2.5D system was implemented and fabricated along with a dedicated carrier board, resulting in the compute module shown in \Cref{fig:occamy:impl:module}.
The \cn{\SI{73}{\milli\meter^2}} chiplets were fabricated in \mbox{\gfs} \SI{12}{\nano\meter} LP+ FinFET node using a 13-metal stack.
The two compute dies with their respective Micron \emph{MT54A16G808A00AC-32} HBM2E stacks were mounted on \emph{Hedwig}, a passive, 4-metal-stack \SI{65}{\nano\meter} PKG-25SI interposer from \gf~we will discuss in \Cref{chap:occamy:impl:interp}.

\begin{figure}
    \includegraphics[width=\linewidth]{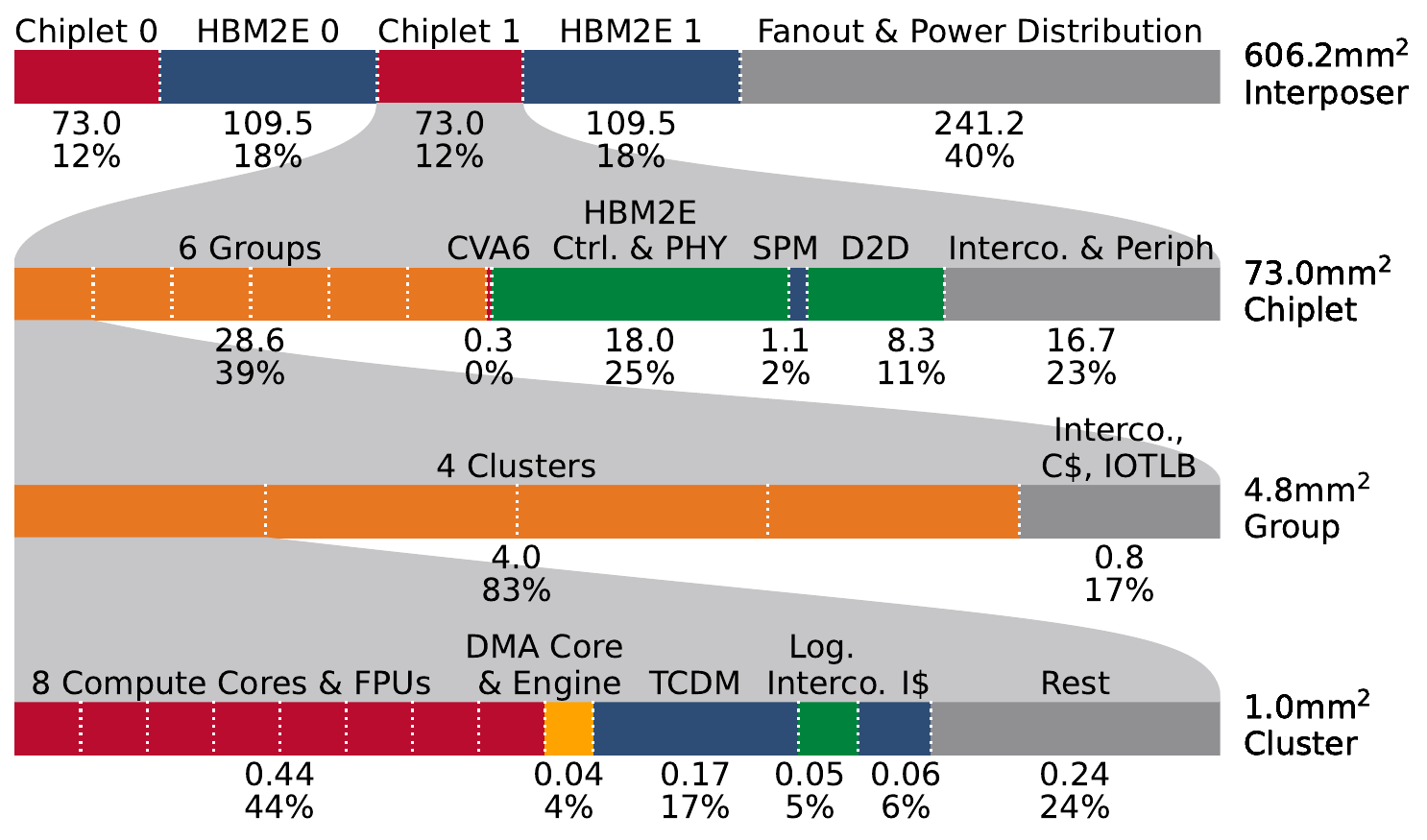}
    \caption{Hierarchical area breakdown of the Occamy system.}
    \label{fig:occamy:impl:area}
\end{figure}

\subsecocc{Occamy Chiplets}
\label{chap:occamy:impl:chiplets}

The compute chiplets were synthesized, placed, and routed hierarchically using Synopsys' \emph{Fusion Compiler} 2022.3.
We used 7.5-track standard cell libraries from Arm and HBM2E \glsunset{ip}\glspl{ip} (controller and \gls{phy}) provided by Rambus.
We targeted a nominal compute domain clock of \SI{1}{\giga\hertz} under typical conditions (\SI{0.8}{\volt}, \SI{25}{\celsius}) with a worst-case constraint of \SI{950}{\mega\hertz}.
The peripheral domain was constrained to \SI{500}{\mega\hertz} under worst-case conditions.
The HBM2E controllers and \gls{phy} were constrained as specified to match the DRAM's peak \cn{\SI{3.2}{\giga\bit\per\second\per pin}} throughput.

\Cref{fig:occamy:impl:area} presents a hierarchical area breakdown of an entire {\occamy} assembly. \Cref{fig:occamy:impl:chip} shows the resulting hierarchical chiplet layout; we describe our implementation in a bottom-up fashion.
\begin{sidefig}
    \begin{subcaptionblock}{0.277\linewidth}
        \centering%
        \includegraphics[width=\linewidth]{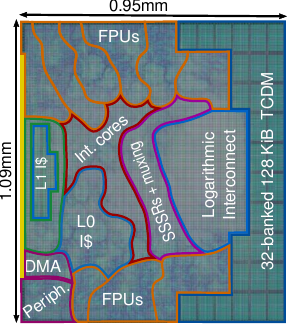}%
        \vspace{-1.3mm}%
        \caption{Compute cluster}%
        \vspace{1.6mm}%
        \label{fig:occamy:impl:cluster}%
        \vfill%
        \centering%
        \includegraphics[width=\linewidth]{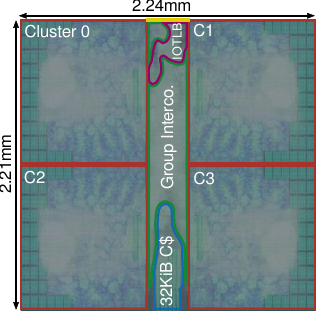}%
        \vspace{-1.3mm}%
        \caption{Compute group}%
        \label{fig:occamy:impl:group}%
    \end{subcaptionblock}\hfill
    \begin{subcaptionblock}{0.705\linewidth}
        \centering%
        \includegraphics[width=\linewidth]{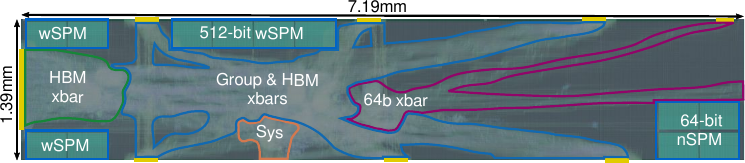}%
        \vspace{-1.3mm}%
        \caption{Chiplet-level interconnect}%
        \vspace{1.2mm}%
        \label{fig:occamy:impl:interco}%
        \vfill%
        \centering%
        \centering%
        \includegraphics[width=\linewidth]{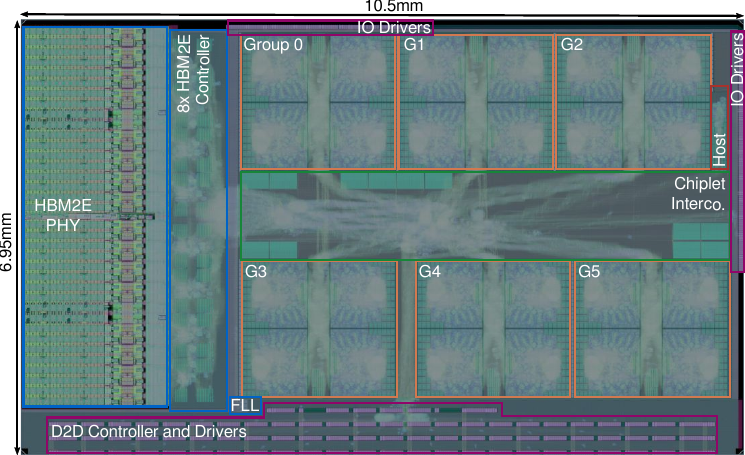}%
        \vspace{-1.3mm}%
        \caption{Chiplet}%
        \label{fig:occamy:impl:chiplet}%
    \end{subcaptionblock}\hfill
    \caption{Annotated layouts of the {\occamy} chiplet and its components (block IO shown in yellow). }
    \label{fig:occamy:impl:chip}
\end{sidefig}
In accordance with our findings in \cite{paulin2022softtiles}, we arranged the compute cluster's IO ports and L1 instruction cache on one side and its \gls{tcdm} \glsunset{sram}\glspl{sram} in a U shape on the opposite side, resulting in the \cn{\SI{1.0}{\milli\meter^2}} cluster layout shown in \Cref{fig:occamy:impl:cluster}.
The cluster is area-dominated by the worker cores with their \glspl{fpu} (\cn{\SI{44}{\percent}}) and the \gls{tcdm} (\cn{\SI{17}{\percent}}).
The group layout, shown in \Cref{fig:occamy:impl:group}, is almost entirely comprised of its four cluster macros (\cn{\SI{83}{\percent}}) and funnels its shared interconnect ports to a narrow interval on its north edge.
The chiplet interconnect is shown in \Cref{fig:occamy:impl:interco}; the global \gls{spm} \glspl{sram} were placed to avoid obstructing the shortest path for crossbar connections.
Further, we implemented the host processor, the eight HBM2E controllers, and the \gls{d2d} link as hierarchical macros to simplify integration.

The top-level chiplet layout is shown in \Cref{fig:occamy:impl:chiplet};
it is area-dominated by the six compute groups (\cn{\SI{39}{\percent}}), HBM2E interface (\cn{\SI{25}{\percent}}), and \gls{d2d} link (\cn{\SI{11}{\percent}}).
In accordance with the interposer arrangement, the west and south chiplet beachfronts are fully reserved for the HBM2E and \gls{d2d} interfaces, respectively, while the remaining off-interposer IO drivers are kept on the north and west edges.
The central chiplet bumps are reserved for power delivery.

\subsecocc{Hedwig Interposer}
\label{chap:occamy:impl:interp}

The Hedwig interposer is at the center of the 2.5D system, connecting the Occamy chiplets to each other, to their HBM2E stacks, and to the carrier board.
On its top side, it exposes \cn{\SI{45}{\micro\metre}}-wide microbump openings accepting connections from the Occamy chiplets and HBM2E stacks through \cn{\SI{15}{\micro\metre}}-wide copper micropillars.
On its bottom side, \cn{{1399}} custom-designed hexagonal C4 pads with \cn{\SI{320}{\micro\metre}} openings and a \cn{\SI{650}{\micro\metre}} pitch allow us to solder the 2.5D assembly onto the carrier PCB using low-temperature bismuth-based C4 bumps.

\begin{figure}
\centering%
    \begin{subcaptionblock}{0.75\linewidth}
        \hspace{-0.55em}%
        \includegraphics[width=\linewidth]{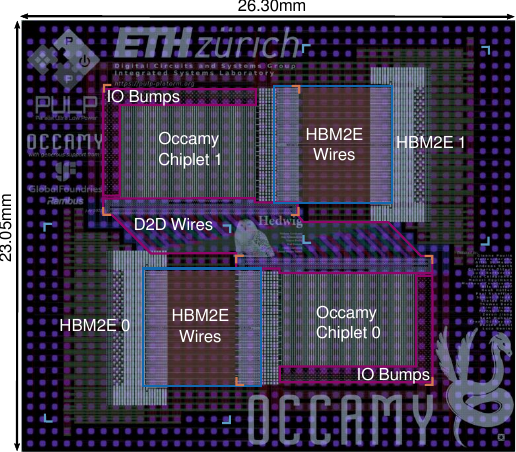}%
        \caption{Layout of the Hedwig interposer}%
        \label{fig:occamy_impl:hedwig_render}%
        \vspace{0.8em}
    \end{subcaptionblock}\hfill
    \begin{subcaptionblock}{0.72\linewidth}
        \includegraphics[width=\linewidth]{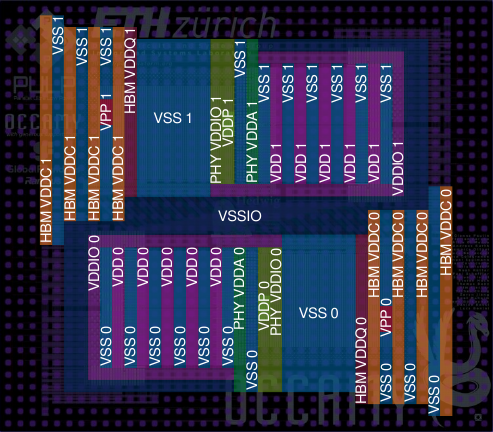}%
        \caption{C4 bump map of the Hedwig interposer}%
        \label{fig:occamy_impl:hedwig_c4}%
    \end{subcaptionblock}\hfill
    \caption{Design of the Hedwig interposer.}
    \label{fig:assembly}
\end{figure}

The connections between the Occamy chiplets and HBM2E stacks were routed on the odd metal layers with a wire width of \cn{\SI{2.5}{\micro\metre}} and a pitch of \cn{\SI{4.1}{\micro\metre}}.
The HBM2E wire length was kept below \cn{\SI{4.9}{\milli\metre}}, and ground shields were introduced on the even layers to ensure signal integrity.
\Cref{fig:occamy_impl:hedwig_render} highlights the HBM2E wires; the routing density inside the rectangles approaches \cn{\SI{100}{\percent}}.
The \cn{39} \gls{d2d} link channels were length-matched to \cn{\SI{8.8}{\milli\metre}} and combined into \cn{24} wire bundles located at the center of Hedwig as shown in \Cref{fig:occamy_impl:hedwig_render}.
The bundles were routed on the three upper metal layers in an alternating pattern, leaving the lowest layer for a shared ground connection.
A routing width of \cn{\SI{3.2}{\micro\metre}} \fin{and} a pitch of \cn{\SI{6.4}{\micro\metre}} \fin{were} chosen.

The digital IO of the Occamy chiplets and HBM2E stacks was routed to Hedwig's edge using \cn{\SI{2.5}{\micro\metre}} traces on the three uppermost layers;
placing IO-related C4 pads at the edge of Hedwig facilitates the fan-out to the carrier PCB.
\Cref{fig:occamy_impl:hedwig_c4} shows Hedwig's C4 pad map, highlighting the IO-related connections as well as the sixteen power and three ground domains.
The ground pads of IO drivers of the two chiplets are connected through the shielding of the \gls{d2d} link, resulting in a shared IO ground \emph{VSSIO}.
For the rest of the power domains, each chiplet-HBM2E pair has its own ground \emph{VSSx}.
The compute fabric, host, and HBM2E controller on each chiplet are supplied through the core power net \emph{VDDx};
the remaining power domains supply the HBM2E PHYs and memory stacks.

\subsecocc{Carrier PCB}
\label{chap:occamy:impl:pcb}

The carrier PCB completes the Occamy assembly at the lowest level.
It serves three main purposes: giving the assembly mechanical stability, implementing the fan-out of the IO signals, and stabilizing the individual power domains.
The \cn{12}-layer stack-up with \cn{\SI{70}{\micro\metre}}-thick copper foil and \emph{ROGERS RO4350B} high-stability, low-CTE laminate ensures high current delivery capabilities while providing sufficient mechanical strength.
Decoupling capacitors are placed close to Hedwig on the carrier to comply with the power delivery requirements of the HBM2E stacks.

The carrier PCB implements an industry-standard \emph{LGA 2011-3} CPU socket interface compatible with off-the-shelf mainboard sockets, facilitating the creation of application boards.
Most of the \cn{2011} pins are used for power and ground connections, with \cn{877} exposed IOs.
The carrier is designed to be strong enough to protect the fragile, \cn{\SI{120}{\micro\metre}}-thick Hedwig interposer from mechanical stress while being placed into and removed from the test socket.

\secocc{Evaluation}
\label{chap:occamy:eval}

We first create a custom testing infrastructure for {\occamy}, described in \Cref{chap:occamy:eval:meas}.
We then evaluate its performance and energy efficiency on both regular and irregular workloads in \Cref{chap:occamy:eval:dense,chap:occamy:eval:stc,chap:occamy:eval:sla}.
We further evaluate Occamy's performance and energy efficiency on dense multi-precision \gls{gemm} in \Cref{chap:occamy:eval:mulprec} and on \gls{ml} inference applications in \Cref{chap:occamy:eval:mlinf}.
Finally, we evaluate the bandwidth scaling and energy efficiency of {\occamy}'s \gls{d2d} link in \Cref{chap:occamy:eval:d2d}.

\subsecocc{Silicon Measurement Setup}
\label{chap:occamy:eval:meas}

\begin{figure}
    \begin{subcaptionblock}{0.475\linewidth}
        \centering%
        \includegraphics[width=143.26146pt,height=134.92175pt]{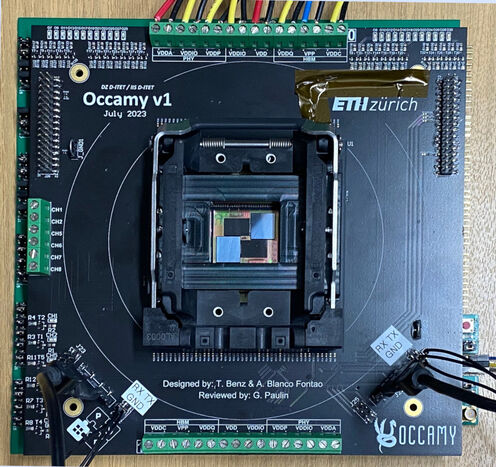}%
        \caption{Bringup board}%
        \label{fig:tester-pcb}%
    \end{subcaptionblock}\hfill
    \begin{subcaptionblock}{0.49\linewidth}
        \centering%
        \includegraphics[width=147.78526pt,height=134.37857pt]{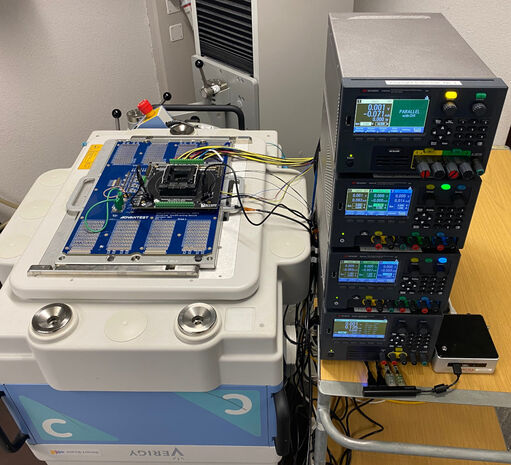}%
        \caption{ATE measurement setup}%
        \label{fig:setup}%
    \end{subcaptionblock}\hfill
\caption{Bringup board enabling both testing on an Advantest
V93000 ATE and standalone operation, and ATE measurement setup.}
\label{fig:impl_board}
\end{figure}

\Cref{fig:tester-pcb} shows {\occamy}'s \emph{bringup} board, which consists of a stack of two individual PCBs.
The upper PCB holds an {LGA 2011} zero insertion force socket for the {\occamy} system, connectors for the \cn{16} power domains, an interface to a \emph{V93000} \gls{ate} system, as well as JTAG, UART, and GPIO headers.
For standalone operation, the lower PCB provides clocking resources, reset circuitry, configuration headers, and an SD card slot for each  chiplet.

To evaluate {\occamy}, we connect the bringup board to an \gls{ate} system, ensuring a stable, low-jitter clock delivery.
In all experiments, the temperature of the {\occamy} system is kept at \cn{25°C} through an active temperature forcing system.
We use Keysight \emph{E36200} series power supplies for power delivery and current measurements.
Two Digilent \emph{JTAG-HS2} programmers are used to program the {\occamy} system through {JTAG}, and two {FTDI} chips relay {UART} data to a testing workstation.
We coordinate the power supplies, resets, programmers, and UART adapters through a dedicated Python application running on our testing workstation, which enables the automated, reproducible execution of applications on Occamy and the collection of current measurements across domains;
this application would later be extended to a generic solution for the rapid \mmp{bringup} and evaluation of RISC-V-based \glspl{soc} called \emph{DUTCTL}~\cite{benz2024dutctl}.

In the following compute workload evaluations, we use the %
measurement setup described above to determine execution times and chiplet power consumption at room temperature, repeating execution as necessary to allow for stable current measurements.
Externally unobservable workload characteristics, such as the number of \glsunset{flop}\glspl{flop} or comparisons performed, are obtained analytically or through cycle-accurate simulation.

\subsecocc{Regular Workloads}
\label{chap:occamy:eval:dense}

\begin{figure}
    \captionsetup[subfigure]{aboveskip=3.5pt, belowskip=4.6pt}%
    \begin{subcaptionblock}{0.3964110930\linewidth}%
        \centering%
        \vspace{-1.0pt}%
        \includegraphics[width=0.98\linewidth]{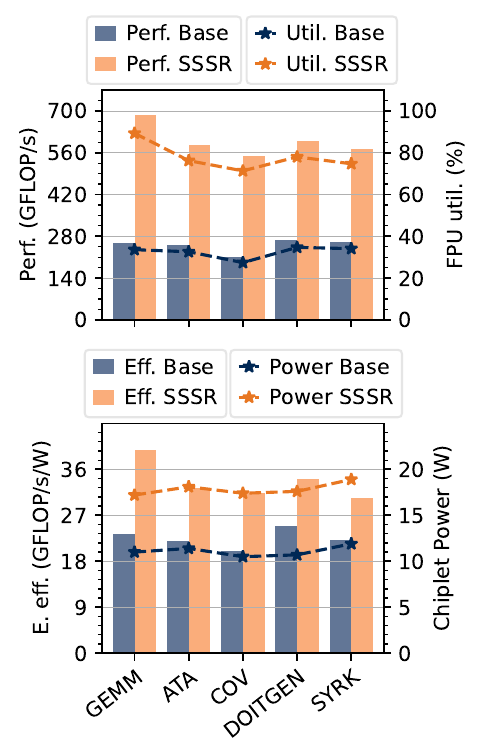}%
        \caption{Regular kernels}%
        \label{fig:occamy:eval:dense}%
        \vspace{-0.15em}%
    \end{subcaptionblock}\hfill%
    \begin{subcaptionblock}{0.6003262642\linewidth}%
        \centering%
        \vspace{-1.8pt}%
        \includegraphics[width=0.98\linewidth]{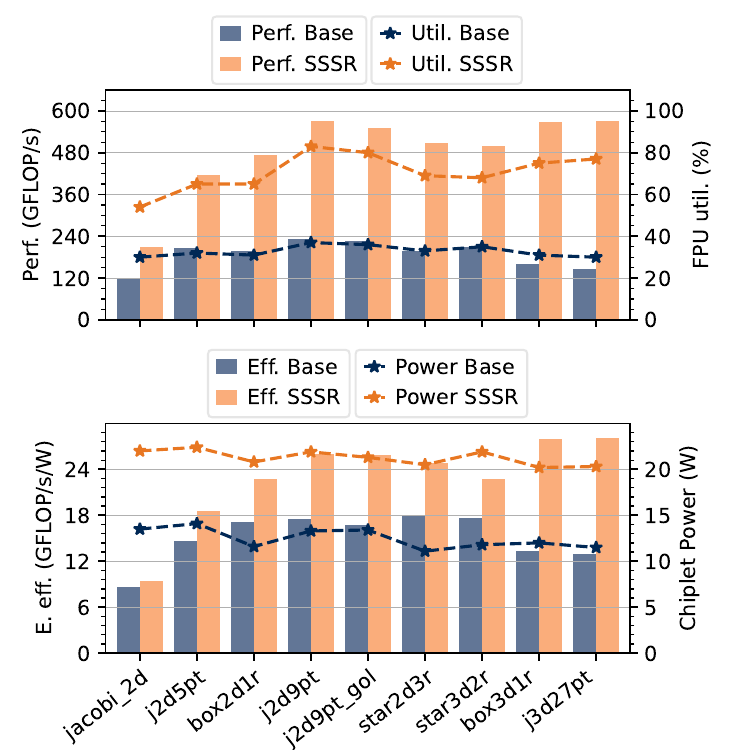}%
        \caption{Stencil codes}%
        \label{fig:occamy:eval:stc}%
        \vspace{-0.15em}%
    \end{subcaptionblock}\\%
    \begin{subcaptionblock}{0.497553018\linewidth}%
        \centering%
        \includegraphics[width=0.98\linewidth]{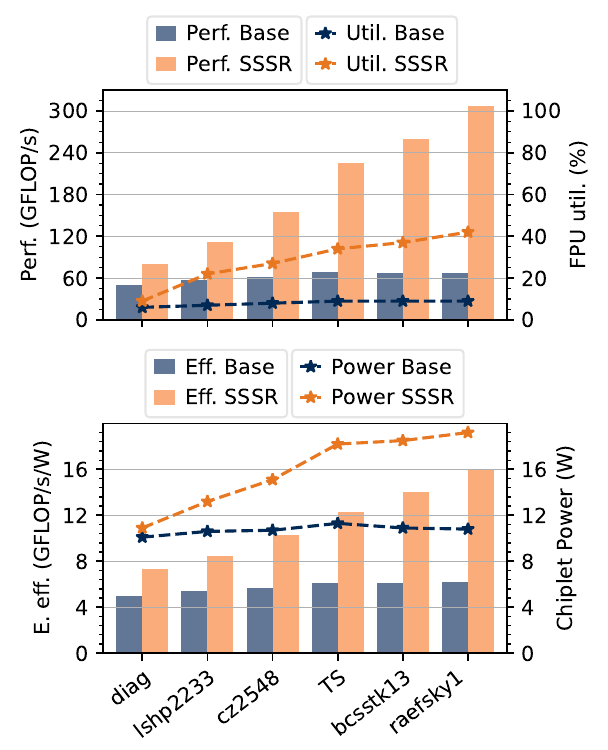}%
        \caption{SpMM}%
        \label{fig:occamy:eval:spmm}%
    \end{subcaptionblock}\hfill%
    \begin{subcaptionblock}{0.497553018\linewidth}%
        \centering%
        \includegraphics[width=0.98\linewidth]{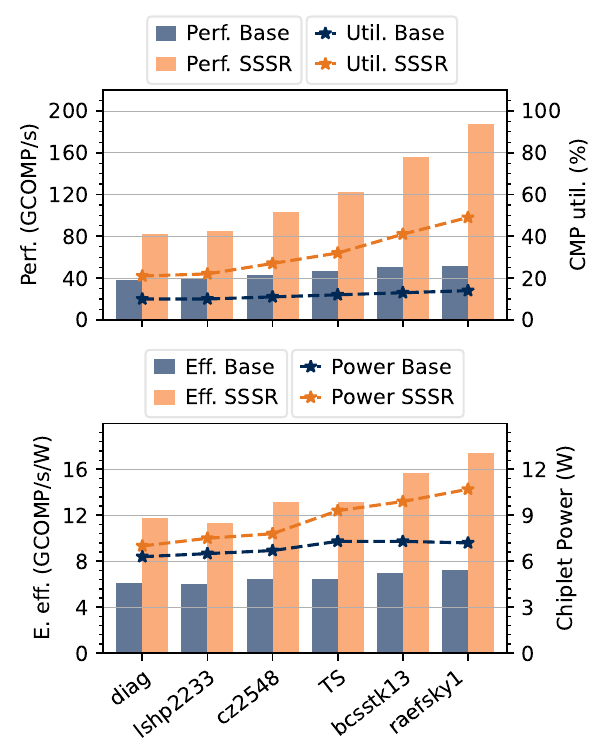}%
        \caption{SpMSpM}%
        \label{fig:occamy:eval:spmspm}%
    \end{subcaptionblock}%
    \vspace{-0.45em}%
    \caption{\fontsize{9.925pt}{12.1pt}\selectfont Performance, energy efficiency, and power for double-buffered regular and irregular FP64 workloads with and without \gls{sssr} acceleration. SpM(Sp)M left matrices (X axis) are real-world sparse matrices. SpMSpM right matrices are random with \cn{\SI{1}{\percent}} density.}%
    \label{fig:occamy:eval:workloads}%
\end{figure}

\Cref{fig:occamy:eval:dense} summarizes {\occamy}'s performance and chiplet energy efficiency at the nominal \SI{1}{\giga\hertz} compute clock on double-buffered regular (dense) FP64 workloads with and without \gls{sssr} acceleration.
\mmp{All initial input and final output data resides in HBM, and appropriately sized subtiles of the processed datasets are copied into and out of cluster \glspl{tcdm} for processing by the respective clusters.}

Through affine streams, \glspl{sssr} accelerate \gls{gemm} using \cn{$48^2$} data tiles by \cn{2.7\x} over baseline RV32G code, achieving a near-ideal \gls{fpu} utilization of \cn{\SI{89}{\percent}}, a performance of \cn{\SI{686}{\giga\flop\per\second}}, and an energy efficiency of \cn{\SI{39.8}{\giga\flop\per\second\per\watt}}.
On average, \gls{sssr}-accelerated dense workloads using the same \cn{$48^2$}
tile size achieve a somewhat lower FPU utilization of \cn{\SI{78}{\percent}}, as kernel and \gls{sssr} setup overheads become more pronounced and fewer FPU operations can be issued as fused (e.g. \gls{fma}) instructions.
Nevertheless, \glspl{sssr} enable average speedups and energy efficiency improvements over \fin{an RV32G baseline} of \cn{2.4\x}~and \cn{1.5\x}, respectively.

\subsecocc{Stencil Computations}
\label{chap:occamy:eval:stc}

\Cref{fig:occamy:eval:stc} shows our results on various double-buffered FP64 stencil codes with and without \gls{sssr} acceleration, which iterate the tiled cluster-level SARIS codes we developed in \Cref{chap:workloads:stc}.
We continue using \cn{$64^2$} and \cn{$16^3$} tiles for 2D and 3D grids, respectively.
\mmp{As done by Matsumura et al.}~\cite{Matsumura2020AN5DAS}, we use full grid sizes of \impval{$16384^2$} for 2D codes and \impval{$512^3$} for 3D codes.

Through indirect streams, \glspl{sssr} accelerate all stencil codes by up to \cn{3.9$\times$} compared to heavily optimized RV32G baseline code, achieving a peak FPU utilization of \cn{\SI{83}{\percent}} and up to \cn{\SI{571}{\giga\flop\per\second}} on the \emph{j3d27pt} stencil.
They also improve peak energy efficiency from \cn{\SIrange[]{17.9}{28.1}{\giga\flop\per\second\per\watt}}.
The remaining execution overheads preventing full \gls{fpu} utilization with \glspl{sssr} are due to kernel initialization, \gls{sssr} setup, and \gls{tcdm} access contentions between worker cores.

\subsecocc{Sparse Linear Algebra}
\label{chap:occamy:eval:sla}

\Cref{fig:occamy:eval:spmm,fig:occamy:eval:spmspm} show Occamy's performance and energy efficiency for double-buffered FP64 sparse-dense \glsunset{spmm}(SpMM) and \gls{spmspm}, respectively.
In both cases, we reuse our optimized matrix-vector cluster codes developed in \Cref{chap:sssrs}.
We multiply real-world left sparse matrices from \cite{Davis2011TheUO}~with random right matrices (\cn{\SI{1}{\percent}} density for \gls{spmspm}); the left matrices are dynamically tiled and reused within compute groups.
To demonstrate Occamy's flexibility in handling sparse operands, the real-world left matrices used are unstructured, cover different application domains, and span from \SI{0.12}{\percent} to \SI{2.8}{\percent} density (sorted left-to-right in figures).

Through indirection, \glspl{sssr} accelerate \gls{spmm} by up to \cn{4.6$\times$}, achieving up to \cn{\SI{307}{\giga\flop\per\second}}, \cn{\SI{16.0}{\giga\flop\per\second\per\watt}}, and \cn{\SI{42}{\percent}} \gls{fpu} utilization.
We introduce index comparison rate as a figure of merit for sparse-sparse matrix computation performance;
through index intersection, \glspl{sssr} accelerate \gls{spmspm} by up to \cn{3.6$\times$}, reaching up to \cn{\SI{187}{\giga\comp\per\second}}, \cn{\SI{17.4}{\giga\comp\per\second\per\watt}}, and index comparator utilizations of up to \cn{\SI{49}{\percent}}.
Normalizing our peak performance to the chiplet area dedicated to compute, we achieve up to \cn{\SI{3.63}{\giga\comp\per\second\per\milli\meter^2}}.
For both \gls{spmm} and \gls{spmspm}, performance improves with increasing left matrix density, which helps amortize \gls{sssr} setup costs as described in~\cref{chap:sssrs}.
Accordingly, while Occamy maintains consistently high utilization as operands become denser, its performance and energy efficiency on extremely sparse operands eventually becomes limited by \gls{sssr} setup costs and the cluster \glspl{tcdm}' finite indexing range.

\subsecocc{Multi-Precision Capabilities}
\label{chap:occamy:eval:mulprec}

\begin{figure}
    \begin{subcaptionblock}{0.49\linewidth}%
        \centering%
        \includegraphics[width=\linewidth]{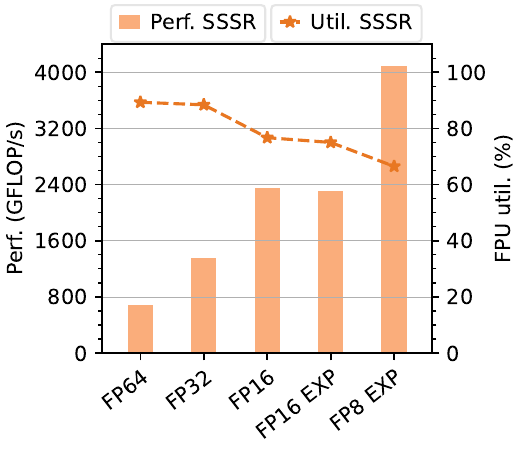}%
        \caption{Performance}%
        \label{fig:occamy:eval:mulprec_perf}%
    \end{subcaptionblock}\hfill%
    \begin{subcaptionblock}{0.49\linewidth}%
        \centering%
        \includegraphics[width=\linewidth]{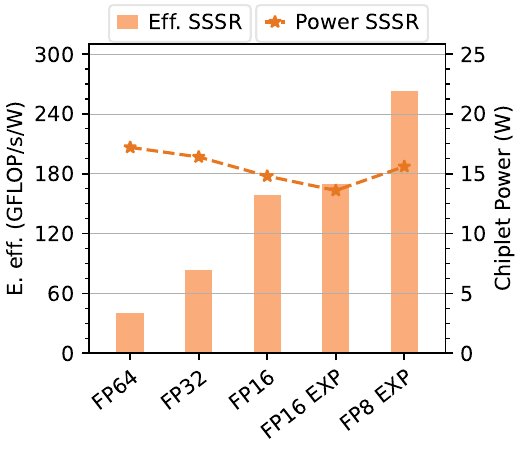}%
        \caption{Energy efficiency and power}%
        \label{fig:occamy:eval:mulprec_eff}%
    \end{subcaptionblock}\hfill%
\caption{Performance, energy efficiency, and power of GEMM at FP64 to FP8 precisions (EXP: expanding accumulation).}%
\label{fig:occamy:eval:mulprec}%
\end{figure}

\Cref{fig:occamy:eval:mulprec} shows how \gls{sssr}-accelerated \gls{gemm} performance and energy efficiency scale as we reduce floating-point precision from 64 to 8 bits, leveraging our FPU's \gls{simd} capabilities.
We consider FP16 GEMM without and with expanding FP32 accumulation (EXP suffix) and FP8 GEMM only with expanding FP16 accumulation.
As we reduce precision, the achieved FPU utilization drops from \cn{\SI{89}{\percent}} (FP64) to \cn{\SI{66}{\percent}} (FP8 EXP).
This is because the SIMD and mixed-precision kernels incur additional overheads: they must zero-initialize accumulators, reduce multiple accumulators packed in a $64$-bit register, and possibly convert the higher-precision result to the original precision~\cite{bertaccini2024fpu}.
The FP64 kernel does not require conversions and avoids accumulator initialization through the non-accumulating \texttt{fmul.d} instruction in the first loop iteration.
With these overheads in mind, the throughput and energy efficiency scale as expected;
for FP8 GEMM with expanding accumulation, we achieve a throughput of \cn{\SI{4.1}{QP\text{-}\tera\flop\per\second}} and an energy efficiency of \cn{\SI{263}{QP\text{-}\giga\flop\per\second\per\watt}}.
On FP16, expanding GEMM is \cn{\SI{6.5}{\percent}}~more energy-efficient than non-expanding GEMM thanks to dedicated expanding dot product units in Occamy's FPU design~\cite{bertaccini2024fpu}.

\subsecocc{Machine Learning Inference}
\label{chap:occamy:eval:mlinf}

We demonstrate Occamy's architectural benefits on end-to-end applications by evaluating its performance and energy efficiency on two \gls{ml} inference tasks%
: a \gls{gcn} layer, which requires sparse-dense graph aggregation, and a full \gls{gpt}-J \gls{llm}.

To evaluate \glspl{gcn}, we implement the layer computation described \mmp{by Kipf and Welling}~\cite{kipf2017semisupervised}, which is used in classification tasks and involves both dense (feature recombination) and sparse-dense (graph aggregation) computation steps.
We model the hidden layers of the \gls{gcn} benchmarked \mmp{by Dwivedi et al.}~\cite{dwivedi2024} on node classification, mapping an FP64 layer with 144 input and output features per graph node across both chiplets. We double-buffer incoming features and distribute the layer's output features among the 48 compute clusters.
We measure Occamy's performance and energy efficiency with and without \glspl{sssr} on the highly sparse citation graphs \emph{webkb}~\cite{lu2003linkbased}, \emph{cora}~\cite{sen2008collective}, and \emph{citeseer}~\cite{sen2008collective}  with average node degrees of 1.8, 2.0, and 1.4, respectively.

\begin{figure}
    \begin{subcaptionblock}{0.49\linewidth}%
        \centering%
        \includegraphics[width=\linewidth]{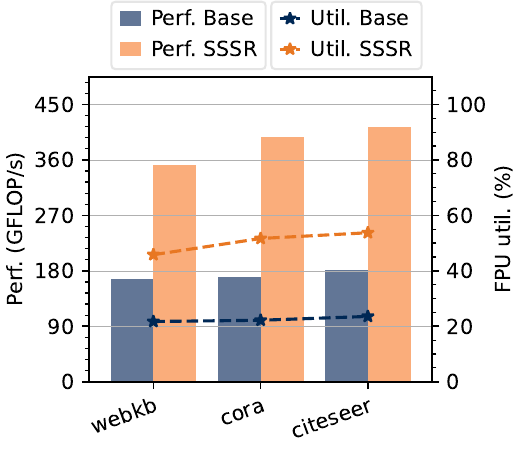}%
        \caption{\rev{Performance}}%
        \label{fig:occamy:eval:mlinf:gcn_perf}%
    \end{subcaptionblock}\hfill%
    \begin{subcaptionblock}{0.49\linewidth}%
        \centering%
        \includegraphics[width=\linewidth]{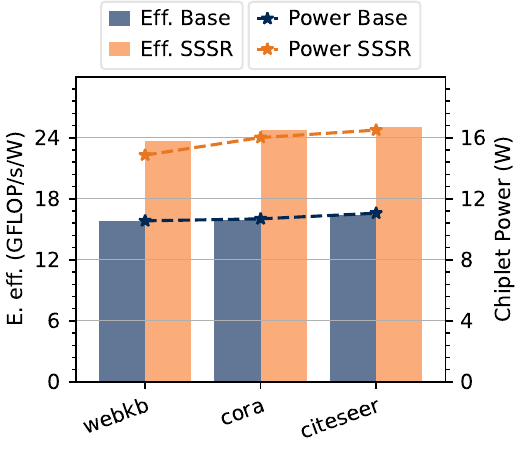}%
        \caption{\rev{Energy efficiency and power}}%
        \label{fig:occamy:eval:mlinf:gcn_pow}%
    \end{subcaptionblock}%
\caption{\rev{Performance, energy efficiency, and power on \gls{gcn} layer (144\x144, FP64) with and without \gls{sssr} acceleration.}}%
\label{fig:occamy:eval:mlinf:gcn}%
\end{figure}

\Cref{fig:occamy:eval:mlinf:gcn} shows the performance, energy efficiency, and power Occamy achieves on \gls{gcn} layer computation, with the evaluated graphs sorted left-to-right by increasing node count.
\glspl{sssr} enable an average speedup of 2.2\x~over an optimized RV32G baseline, raising the peak \gls{fpu} utilization from \SI{24}{\percent} to \SI{54}{\percent} and achieving up to \SI{413}{DP\text{-}\giga\flop\per\second} or \SI{8.01}{DP\text{-}\giga\flop\per\second\per\milli\meter^2} on \emph{citeseer}.
Peak energy efficiency improves from 16.4 to \SI{25.0}{DP\text{-}\giga\flop\per\second\per\watt}.
As would be expected, our results fall between those achieved for dense and sparse workloads, showcasing Occamy's ability to efficiently handle mixtures of regular and irregular compute.
Overall, we observe a slight increase in \gls{sssr} performance and energy efficiency as graphs grow \fin{larger: as was the case for \gls{spmm}, longer streams enable a} better amortization of \gls{sssr} setup costs.

\begin{figure}
    \centering
    \begin{subcaptionblock}{0.445\linewidth}%
        \centering%
        \includegraphics[width=\linewidth]{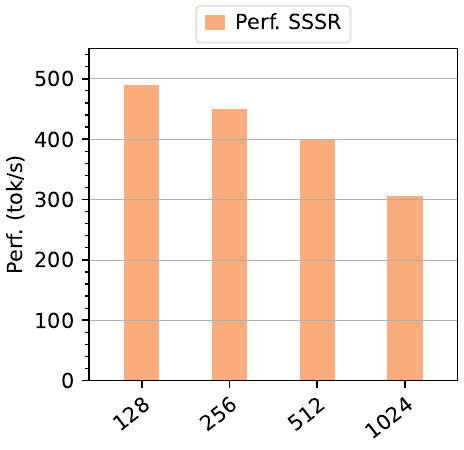}%
        \caption{\rev{Performance}}%
        \label{fig:occamy:eval:mlinf:llm_perf}%
    \end{subcaptionblock}\hspace{3mm}%
    \begin{subcaptionblock}{0.50\linewidth}%
        \centering%
        \includegraphics[width=\linewidth]{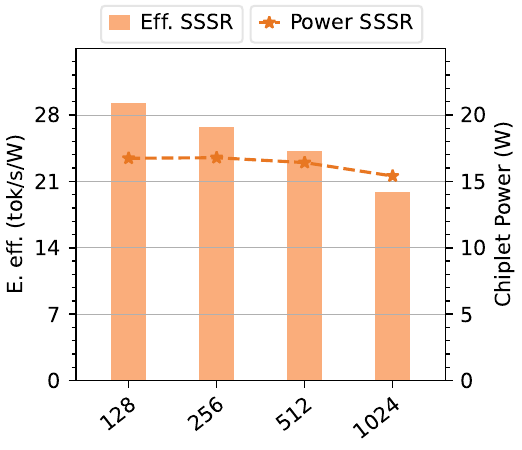}%
        \caption{\rev{Energy efficiency and power}}%
        \label{fig:occamy:eval:mlinf:llm_pow}%
    \end{subcaptionblock}%
\caption{\rev{Performance, energy efficiency, and power of \gls{sssr}-accelerated FP16 GPT-J LLM inference in non-autoregressive mode and for varying sequence lengths.}}%
\label{fig:occamy:eval:mlinf:llm}%
\end{figure}

\Cref{fig:occamy:eval:mlinf:llm} shows the performance, energy efficiency, and power of Occamy running the full FP16 \gls{gpt}-J model with \gls{sssr} acceleration in non-autoregressive mode for sequence lengths from 128 to 1024 tokens.
We use the standard measure of \emph{token rate} (\si{\tok\per\second}) to express \gls{llm} performance.
At the shortest sequence length, our implementation achieves a peak throughput of \SI{490}{\tok\per\second} and an energy efficiency of \SI{29.3}{\tok\per\second\per\watt}, \fin{reaching} a peak FPU utilization of \SI{75}{\percent}.
As can be seen in \Cref{fig:occamy:eval:mlinf:llm_perf}, increasing the sequence length leads to a gradual decrease in throughput,
which is due to the quadratic scaling of attention computation with sequence length.
Accordingly, we see a gradual decrease in energy efficiency with increasing sequence \fin{length:}
\fin{while} short sequences are dominated by matrix multiplication, as sequence length grows, more time is spent on the attention mechanism’s less \fin{energy-efficient} element-wise \emph{softmax} operation.

\subsecocc{Die-to-Die Link}
\label{chap:occamy:eval:d2d}

\begin{figure}
    \begin{subcaptionblock}{0.49\linewidth}
        \centering%
        \includegraphics[width=\linewidth]{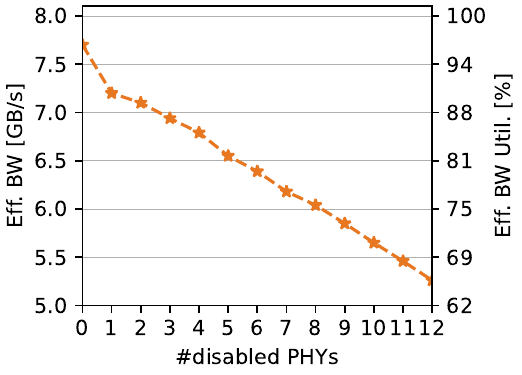}%
        \caption{}%
        \label{fig:occamy:eval:d2d:dis_phy}%
    \end{subcaptionblock}\hfill
    \begin{subcaptionblock}{0.49\linewidth}
        \centering%
        \includegraphics[width=\linewidth]{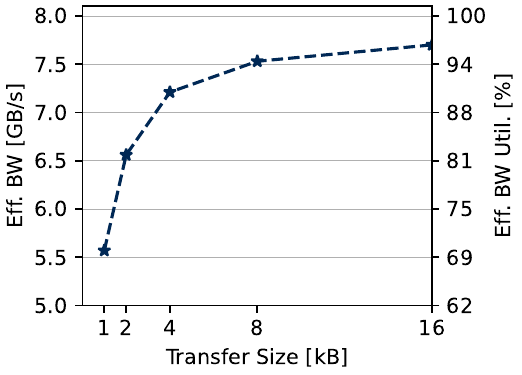}%
        \caption{}%
        \label{fig:occamy:eval:d2d:tf_size}%
    \end{subcaptionblock}\hfill
    \caption{(a) Linear effective bandwidth decrease for channel disabling on wide \gls{d2d} segment and (b) effective bandwidth utilization for varying transfer sizes on wide \gls{d2d} segment.}
    \label{fig:occamy:eval:d2d}
\end{figure}

\Cref{fig:occamy:eval:d2d} shows the effective bandwidth of the wide \gls{d2d} link under varying conditions analyzed in cycle-accurate \gls{rtl} simulation.
\mmp{We use simulations for these measurements to directly observe effective transfer durations and latencies through introspection.}
As expected, our bandwidth decreases linearly as we increase the number of disabled \glspl{phy} to mitigate faults.
Furthermore, a high link utilization of \cn{96\%} is achieved at a transfer size of \cn{\SI{16}{\kilo\byte}}.
We also determine the latency of communication between the two chips over the \gls{d2d} link:
accessing the other chiplet's narrow \gls{spm} through the narrow segment from the host core incurs a latency of \cn{27} cycles, while a \gls{dma} transfer over the wide segment experiences a latency of \cn{61} cycles.
On our silicon demonstrator, we measure the energy efficiency of the wide \gls{d2d} segment to be \cn{\SI{1.6}{\pico\joule\per\bit}}.

\secocc{Related Work}
\label{chap:occamy:relwork}

We will first review \gls{soa} hardware approaches to irregular compute acceleration, contrasting them with Occamy (\cref{chap:occamy:relwork:hwlit}), and then quantitatively compare Occamy's performance and energy efficiency to those achieved on \gls{soa}, silicon-proven CPUs and GPUs (\cref{chap:occamy:relwork:soacomp}).

\subsecocc{Hardware Targeting Irregular Workloads}
\label{chap:occamy:relwork:hwlit}

Large-scale \gls{ml} and \gls{hpc} applications primarily target general-purpose CPUs and GPUs.
These architectures efficiently handle regular compute~\cite{nvidia_a100_dense, siegmann2024first}, but achieve low FPU utilization on sparse and irregular workloads (\cn{$\leq$\SI{10}{\percent}} on sparse \gls{la}~\cite{alappat2020a64fx, Niu2021TileSpMVAT, siegmann2024first}) due to their data-dependent control flow and complex, indirect address computations.

To improve sparse compute efficiency, numerous accelerators have been proposed.
Most notably, sparse \gls{ml} accelerators achieving leading throughput and energy efficiency on \gls{ml} inference tasks have been demonstrated in silicon~\cite{huang2025hybr, yue2024cimblock, feng2024pntcloud, zhang2021snap};
however, they impose low-precision data formats ($\leq$\SI{16}{\bit}) and low ($\leq$\SI{80}{\percent}) or structured sparsity specific to \gls{ml} workloads and are thus unsuitable for \gls{hpc}.
Generic sparse~\gls{la}~\cite{Srivastava2020MatRaptorAS, Pal2018OuterSPACEAO, park2020output, Sadi2019EfficientSO} or \fin{\gls{ta}~\cite{Hegde2019ExTensorAA,zhong2024feasta}} accelerators are less domain-specific, with some \fin{having been} demonstrated in silicon~\cite{park2020output, Sadi2019EfficientSO}, but most are still hardwired for \fin{specific} operations~\cite{Srivastava2020MatRaptorAS, Pal2018OuterSPACEAO, park2020output, Sadi2019EfficientSO} or dataflows~\cite{Srivastava2020MatRaptorAS, Pal2018OuterSPACEAO}.

Discrete accelerators, while highly efficient, cannot match the flexibility and agility of general-purpose architectures.
Recently, \glspl{cgra} have been proposed that support streaming indirection, intersection, and union in hardware~\cite{Dadu2019TowardsGP,koul2024onyx,chen2025opal}. %
\mmp{Two among them, \emph{Onyx}~\cite{koul2024onyx} and \emph{Opal}~\cite{chen2025opal} were demonstrated in silicon, with Opal improving on Onyx specifically by increasing its sparse \gls{la} and \gls{ml} performance.}
However, \glspl{cgra} are still highly challenging to \mmp{program productively}~\cite{liu2019cgrasurvey}, and \mmp{both Onyx and Opal only support the} INT16 and BF16 data formats, \mmp{which are} of insufficient precision for most \gls{hpc} workloads.

Another common approach to maintain generality is through \gls{isa} extensions for CPUs and GPUs that accelerate sparse and irregular compute~\cite{siracusa2023tmu, Domingos2021UnlimitedVE, zhu2019sparsetensorcore, Wang2019StreambasedMA, Wang2021DualsideST, Rao2022SparseCoreSI};
however, most of these extensions are not demonstrated in silicon~\cite{siracusa2023tmu, Domingos2021UnlimitedVE, zhu2019sparsetensorcore, Wang2019StreambasedMA, Wang2021DualsideST, Rao2022SparseCoreSI} and only accelerate one-~\cite{siracusa2023tmu, Domingos2021UnlimitedVE, zhu2019sparsetensorcore, Wang2019StreambasedMA} or two-sided~\cite{Wang2021DualsideST, Rao2022SparseCoreSI} sparse operations.
Thus, Occamy is the first silicon-proven general-purpose system to efficiently and comprehensively handle sparse and irregular workloads in 8-to-64-bit precisions, all without compromising on regular compute efficiency.

\subsecocc{Comparison to SoA CPUs and GPUs}
\label{chap:occamy:relwork:soacomp}

\cref{tab:occamy:relwork:soa} compares Occamy to \gls{soa}, silicon-proven GPUs and CPUs supporting the dominant \glspl{isa} (x86, Arm, and \riscv).
To provide a holistic overview, we compare Occamy to both recently released processors (e.g.~AMD Genoa, Nvidia H100), which usually lack published stencil and sparse-dense \gls{la} performance evaluations, as well as established processors from prior generations (e.g. Fujitsu A64FX, Nvidia A100) whose real-world performance is well-studied in literature.

\renewcommand{\baselinestretch}{1.0}\selectfont

\newcommand{\bs}[1]{{#1}} %
\newcommand{\sh}[1]{{#1}} %
\newcommand{\vf}[1]{{#1}} %
\newcommand{\im}[1]{\item[#1]}
\newcommand{\dk}{-} %
\newcommand{\sep}{$\vert$}
\newcommand{\rto}[1]{\rotatebox[origin=c]{90}{#1}}
\newcommand{\uy}{\hspace{0.08em}}
\newcommand{\sqmmx}{\ensuremath{\,}mm$^2$}

\renewcommand\cellset{\renewcommand\arraystretch{1.0}%
    \setlength\extrarowheight{0.2ex}}

\def\hdls{1mm}

\newcommand{\tn}[1]{\uy\textcolor{black}{\textsuperscript{#1}}}

\newsavebox{\OccSaveBox}
\sbox{\OccSaveBox}{%
\renewcommand{\arraystretch}{1.0}
\begin{tabular}{@{}lllllllllll@{}}
\toprule
& %
& %
\makecell[cl]{\textbf{Schmidt} \\ \textbf{et al.}~\cite{schmidt2022rsoc}} &
\makecell[cl]{\textbf{Sophon}\\  \textbf{SG2042}~\cite{sophon-trm}} &
\makecell[cl]{\textbf{AMD Rome} \\ \cite{amd_zen_2_isscc20, amd_chiplet_isscc20}} & %
\makecell[cl]{\textbf{AMD Genoa} \\ \cite{munger2023zen4isscc, ravi2024zen4micro}} &
\makecell[cl]{\textbf{Fujitsu} \\ \textbf{A64FX}~\cite{a64fx_isscc22}} & %
\makecell[cl]{\textbf{Nvidia} \\ \textbf{A100}~\cite{a100_isscc21}} & %
\makecell[cl]{\textbf{Nvidia} \textbf{H100}\\\cite{hopper2022_hotchips, hopper2024_whitepaper}} &
\makecell[cl]{\textbf{Intel Sapphire} \\ \textbf{Rapids}~\cite{sapphire_rapids_isscc22, sapphire_rapids_hc33}} &
\makecell[cl]{\textbf{\textit{{\occamy}}} \\ \textbf{\textit{(Ours)}}} \\
\midrule\noalign{\vskip \hdls}
\multirow{10}{*}{\rto{\makecell[cc]{\\ General \\}}} &
Assembly &
Single Die &
Single Die &
Organic Substrate &
Organic Substrate &
\scalebox{0.9}[1.0]{2.5D Passive Interp.} &
2.5D CoWoS\tn{a} &
2.5D CoWoS\tn{a} &
2.5D EMIB\tn{b} &
\scalebox{0.95}[1.0]{2.5D Passive Interp.} \\
&
\makecell[cl]{2.5D Compute Cfg.} &
\makecell[cl]{\dk} &
\makecell[cl]{\dk} &
\makecell[cl]{8 Die} &
\makecell[cl]{12 Die} &
\makecell[cl]{1 Die} &
\makecell[cl]{1 Die} &
\makecell[cl]{1 Die} &
\makecell[cl]{4 Die} &
\makecell[cl]{2 Die} \\
&
\makecell[cl]{2.5D DRAM Cfg.} &
\makecell[cl]{\dk} &
\makecell[cl]{\dk} &
\makecell[cl]{1 IO Die} &
\makecell[cl]{1 IO Die} &
\makecell[cl]{4 HBM2} &
\makecell[cl]{6 HBM2} &
\makecell[cl]{5 HBM3} &
\makecell[cl]{4 HBM2E} &
\makecell[cl]{2 HBM2E} \\
&
\makecell[cl]{Die Technology} &
\makecell[cl]{16nm FinFET} &
\makecell[cl]{12nm FinFET\tn{c}} &
\makecell[cl]{7nm FinFET} &
\makecell[cl]{5nm FinFET} &
\makecell[cl]{7nm FinFET} &
\makecell[cl]{7nm FinFET} &
\makecell[cl]{TSMC 4N FinFET} &
\makecell[cl]{Intel 7 SuperFin} &
\makecell[cl]{12nm FinFET\tn{d}} \\

&
ISA &
RV64GXhwacha4 &
\makecell[cl]{RV64GCV\\(RVV v0.7.1)} &
x86-64, AVX2 &
x86-64, AVX-512 &
Armv8.2-A SVE\tn{e} &
PTX ISA, CC 8.0 &
PTX ISA, CC 9.0 &
\makecell[cl]{x86-64, AVX-512,\\DL Boost, AMX} &
RV32GXoc\tn{f,g} \\
&
Open-Source RTL &
Partially (Core) &
Partially (Core)\tn{h} &
No &
No &
No &
No &
No &
No &
Partially (Compute) \\
&
\makecell[cl]{%
Frequency~[\si{\giga\hertz}] \\ Voltage~[\si{V}]
} &
\makecell[cl]{%
1.44 \\ 1.0
} &
\makecell[cl]{%
2.00 \\ 0.8  
} &
\makecell[cl]{%
\vf{2.25}\tn{p} \\ \vf{0.8}  
} &
\makecell[cl]{%
\vf{2.40}\tn{p} \\ \dk  
} &
\makecell[cl]{%
\vf{2.20} \\ \vf{0.8}  
} &
\makecell[cl]{%
\vf{1.41} \\ \vf{0.8}
} &
\makecell[cl]{%
1.98 \\ \dk  
} &
\makecell[cl]{%
2.60 \\ \dk  
} &
\makecell[cl]{%
\vf{1.00} \\ \vf{0.8}  
} 
\vspace{\hdls} \\
\arrayrulecolor{ieee-dark-black-40}\hdashline\noalign{\vskip \hdls}
\multirow{3}{*}{\rto{\makecell[cc]{\vspace{-0.5em} \\ Circ. \\[-0.25em] Size \\}}} &
Transistors &
0.5B &
\dk &
\bs{39B} &
\bs{78B}\tn{k} &
\vf{8.8B} &
\vf{54B} &
\vf{80B} &
\dk &
\bs{4.8B} \\
&
Die Area [\sqmmx] &
24.0&
585\tn{i} &
\bs{592}\tn{k} &
\vf{840}\tn{k} &
\vf{420} &
\vf{826} &
\vf{814} &
1600 &
\vf{146} \\
&
\scalebox{.91}[1.0]{Norm. Comp. Area [\sqmmx]\tn{l}} &
3.86 (0.48/Core) &
292 (4.6/Core) &
\sh{442} (\sh{6.9}/Core) &
\sh{1183} (\sh{12.3}/Core) &
\sh{243} (\sh{5.1}/Core) &
\sh{632} (\sh{5.9}/SM) &
\sh{1170} (\sh{8.9}/SM) &
\sh{1024} (\sh{17}/Core) &
\sh{51.5} (\vf{1.0}/Cluster) 
\vspace{\hdls} \\
\arrayrulecolor{ieee-dark-black-40}\hdashline\noalign{\vskip \hdls}
\multirow{2}{*}{\rto{\makecell[cc]{\\ RAM \\}}} &
System DRAM & 
\dk &
$\leq$256GiB DDR4 &%
\vf{$\leq$4TiB} DDR4 &
\vf{$\leq$6TiB} DDR5 &
\vf{32GiB} HBM2 & 
\vf{40GiB} HBM2 & 
\vf{80GiB} HBM3 & 
\makecell[cl]{$\leq$4TiB DDR5 \\ 64GiB HBM2E} &
\vf{32GiB} HBM2E \\ 
&
On-Chip SRAM &
4.5 MiB &
64MiB L3\$ &
\vf{128MiB} L3\$ &
\vf{384MiB} L3\$ &
\vf{32MiB} L2\$ &
\vf{40MiB} L2\$ &
\vf{50MiB} L2\$ &
$\leq$112.5MiB L3\$ &
\bs{9MiB} SPM 
\vspace{\hdls} \\
\arrayrulecolor{ieee-dark-black-40}\hdashline\noalign{\vskip \hdls}
\multirow{1}{*}[0.12cm]{\rto{\makecell[cc]{\vspace{-0.5em} \\ D2D \\[-0.25em] Link}}} &
\makecell[cl]{%
Technology \\
\#Links \\
\#Wires/Link \\
R+W BW $[$GB/s$]$ \\
\scalebox{.90}[1.0]{R+W BW/Link $[$GB/s$]$} \\
E. Eff $[$pJ/bit$]$
} &
\dk &
\dk &
\makecell[cl]{%
IFOP\tn{m}\;~\cite{amd_chiplet_isscc20} \\
8 \\
72 \\
440 \\
55/Link \\
2.0
} &
\makecell[cl]{%
IFOP\tn{m}\;~\cite{ravi2024zen4micro} \\
12 \\
\dk \\
1044 \\
87/Link \\
\dk
} &
\dk &
\dk &
\dk &
\makecell[cl]{%
EMIB\tn{b}\;\uy/MDFIO\tn{n}\uy\cite{sapphire_rapids_isscc22} \\
20 \\
\dk \\
\textbf{10000} \\
\textbf{500/Link} \\
\textbf{0.5}
} &
\makecell[cl]{%
Fully digital DDR \\
38+1 \\
18 \\
16\tn{o} \\
0.42/Link \\
1.6\tn{o}
} 
\vspace{\hdls} \\
\arrayrulecolor{ieee-dark-black-40}\hdashline\noalign{\vskip \hdls}
\multirow{1}{*}[0.6cm]{\rto{\makecell[cc]{Peak \\[-0.25em] Compute \\[-0.25em] Throughp.}}} &
\makecell[cl]{%
FP Formats \\
$[$TFLOP/s$]$ \\
$[$GFLOP/s/\sqmmx$]$ \\
$[$GFLOP/s/\sqmmx$]$\tn{l}
} &
\makecell[cl]{%
\emph{FP64/32/16} \\
0.092/0.18/0.37 \\
20.9/41.8/83.7 \\
23.9/47.7/95.4
} &
\makecell[cl]{%
\emph{FP64/32/16} \\
1.00/2.00/4.00\hspace{0.1em}\cite{chen2020xuantie} \\
3.42/6.84/13.7 \\
3.42/6.84/13.7
} &
\makecell[cl]{%
\emph{FP64/32} \\
\sh{2.30\tn{p}\uy/4.61\tn{p}} \\
\sh{8.22\tn{p}\uy/16.4\tn{p}} \\
\sh{5.21\tn{p}\uy/10.4\tn{p}}
} &
\makecell[cl]{%
\emph{FP64/32} \\
\sh{5.53\tn{p}\uy/11.1\tn{p}} \\
\sh{11.8\tn{p}\uy/23.6\tn{p}} \\
\sh{4.67\tn{p}\uy/9.35\tn{p}}
} &
\makecell[cl]{%
\emph{FP64/32/16} \\
\vf{3.38/6.76/13.5} \\
\sh{21.9/43.9/87.8} \\
\sh{13.9/27.9/55.7}
} &
\makecell[cl]{%
\emph{FP64/32/16} \\
\vf{19.5\tn{q}\uy/156\tn{q}\uy/312\tn{q}} \\
\sh{48.6\tn{q}\uy/389\tn{q}\uy/777\tn{q}} \\
\sh{30.8\tn{q}\uy/247\tn{q}\uy/493\tn{q}}
} &
\makecell[cl]{%
\emph{FP64/32/16/8} \\
\scalebox{0.9}[1.0]{\textbf{66.9}\tn{q}\uy/\textbf{495}\tn{q}\uy/\textbf{989}\tn{q}\uy/\textbf{1979}\tn{q}\;\uy} \\ %
\scalebox{0.9}[1.0]{\textbf{172}\tn{q}\uy/\textbf{1270}\tn{q}\uy/\textbf{2541}\tn{q}\uy/\textbf{5082}\tn{q}\;\uy} \\
\scalebox{0.9}[1.0]{\textbf{57.2}\tn{q}\uy/\textbf{423}\tn{q}\uy/\textbf{845}\tn{q}\uy/\textbf{1691}\tn{q}\;\uy}
} &
\makecell[cl]{%
\emph{FP64/32/16/8} \\
\scalebox{0.9}[1.0]{2.85\tn{r}\uy\cite{intel-app-metrics}/5.71/75.3/151\uy\cite{intel-arch-day-2021}} \\
4.84/9.67/128/255 \\
2.78/5.57/73.5/147
} &
\makecell[cl]{%
\emph{FP64/32/16/8} \\
\vf{0.77/1.54/3.07/6.14} \\
\sh{14.9/29.8/59.6/119} \\
\sh{14.9/29.8/59.6/119}
} 
\vspace{\hdls} \\
\arrayrulecolor{ieee-dark-black-40}\hdashline\noalign{\vskip \hdls}
\multirow{1}{*}[0.45cm]{\rto{\makecell[cc]{Best \\[-0.25em] Dense \\[-0.25em] LA FP64}}} &
\makecell[cl]{%
$[$GFLOP/s$]$ \\
FPU util. \\
$[$GFLOP/s/W$]$ \\
$[$GFLOP/s/\sqmmx$]$ \\
$[$GFLOP/s/\sqmmx$]$\tn{l}
} &
\makecell[cl]{%
87.5\tn{s}\uy\uy\cite{schmidt2022rsoc, schmidt2018hwacha} \\
\textbf{\SI{95}{\percent}}\tn{t}\uy\uy\cite{schmidt2018hwacha} \\
20.0\tn{s}\uy\uy\cite{schmidt2022rsoc} \\
19.9\tn{s}\uy\uy\cite{schmidt2022rsoc} \\
22.7\tn{s}\uy\uy\cite{schmidt2022rsoc}
} &
\makecell[cl]{%
\emph{\dk} \\
\dk \\
\dk \\
\dk
} &
\makecell[cl]{%
1600\tn{p}\uy\uy\cite{a64fx_dense} \\
\SI{70}{\percent}\tn{p}\uy\uy\cite{a64fx_dense} \\
\dk \\
5.71\tn{p}\uy\uy\cite{a64fx_dense} \\
3.62\tn{p}\uy\uy\cite{a64fx_dense}
} &
\makecell[cl]{%
4428\uy\uy\cite{amd2024tune9004} \\
\SI{80}{\percent}\uy\uy\cite{amd2024tune9004} \\
\dk \\
9.43\uy\uy\cite{amd2024tune9004} \\
3.74\uy\uy\cite{amd2024tune9004}
} &
\makecell[cl]{%
1978\tn{u}\uy\uy\cite{siegmann2024first} \\
\SI{72}{\percent}\tn{u}\uy\uy\cite{siegmann2024first} \\
16.9\tn{u}\uy\uy\cite{a64fx_isscc22} \\
12.8\tn{u}\uy\uy\cite{siegmann2024first} \\
8.15\tn{u}\uy\uy\cite{siegmann2024first}
} &
\makecell[cl]{%
18500\uy\uy\cite{nvidia_a100_dense} \\
\SI{95}{\percent}\uy\uy\cite{nvidia_a100_dense} \\
41.4\tn{x,y}\uy\uy\cite{strohmaier2024green500} \\
46.1\uy\uy\cite{nvidia_a100_dense} \\
\textbf{29.3}\uy\uy\cite{nvidia_a100_dense}
} &
\makecell[cl]{%
\textbf{33348}\tn{x,y,z}\uy\uy\cite{strohmaier2024top500}\cite{mellor2024virganews} \\
\SI{81}{\percent}\tn{x,y,z}\uy\uy\cite{strohmaier2024top500} \\
\textbf{65.4}\tn{x,y}\uy\uy\cite{strohmaier2024green500} \\
\textbf{85.7}\tn{x,y,z}\uy\uy\cite{strohmaier2024top500} \\
28.5\tn{x,y,z}\uy\uy\cite{strohmaier2024top500}
} &
\makecell[cl]{%
2696\uy\uy\cite{siegmann2024first} \\
\SI{95}{\percent}\uy\uy\cite{siegmann2024first} \\
\dk \\
4.57\uy\uy\cite{siegmann2024first} \\
2.63\uy\uy\cite{siegmann2024first}
} &
\makecell[cl]{%
686 \\
\SI{89}{\percent} \\
39.8 \\
13.3 \\
13.3
} 
\vspace{\hdls} \\
\arrayrulecolor{ieee-dark-black-40}\hdashline\noalign{\vskip \hdls}
\multirow{1}{*}[0.300cm]{\rto{\makecell[cc]{Best \\[-0.25em] Stencil \\[-0.25em] FP64}}} &
\makecell[cl]{%
$[$GFLOP/s$]$ \\
FPU util. \\
$[$GFLOP/s/W$]$ \\
$[$GFLOP/s/\sqmmx$]$ \\
$[$GFLOP/s/\sqmmx$]$\tn{l}
} &
\makecell[cl]{%
\dk \\
\dk \\
\dk \\
\dk \\
\dk
} &
\makecell[cl]{%
\emph{\dk} \\
\dk \\
\dk \\
\dk \\
\dk
} &
\makecell[cl]{%
855\tn{p}\uy\uy\cite{cfd_amd_epyc_rome_tpds21} \\
\sh{\SI{37}{\percent}}\tn{p}\uy\uy\cite{cfd_amd_epyc_rome_tpds21} \\
{\dk} \\
\sh{3.05}\tn{p}\;\cite{cfd_amd_epyc_rome_tpds21} \\
\sh{1.93}\tn{p}\;\cite{cfd_amd_epyc_rome_tpds21}
} &
\makecell[cl]{%
\dk \\
\dk \\
\dk \\
\dk \\
\dk
} &
\makecell[cl]{%
323\tn{u}\uy\uy\cite{hirokawa2022large} \\
\sh{\SI{11}{\percent}}\tn{u}\uy\uy\cite{hirokawa2022large} \\
{\dk} \\
\sh{2.10}\tn{u}\;\cite{hirokawa2022large} \\
\sh{1.33}\tn{u}\;\cite{hirokawa2022large}
} &
\makecell[cl]{
\rev{\textbf{6063}\uy\uy\cite{chen2024convstencil}} \\
\sh{\SI{49}{\percent}}\tn{v}\uy\uy\cite{Zhang2023RevisitingTB} \\
{\dk} \\
\rev{\sh{\bf{15.1}}\uy\cite{chen2024convstencil}} \\
\rev{\sh{9.58}\uy\cite{chen2024convstencil}}
} &
\makecell[cl]{%
\dk \\
\dk \\
\dk \\
\dk \\
\dk
} &
\makecell[cl]{%
\emph{\dk} \\
\dk \\
\dk \\
\dk \\
\dk
} &
\makecell[cl]{%
571 \\
\sh{\textbf{\SI{83}{\percent}}} \\
\sh{\textbf{28.1}} \\
\sh{11.1} \\
\sh{\textbf{11.1}}
} 
\vspace{\hdls} \\
\arrayrulecolor{ieee-dark-grey-60}\hdashline\noalign{\vskip \hdls}
\multirow{1}{*}[0.5cm]{\rto{\makecell[cc]{Best \\[-0.25em] Sp-dense \\[-0.25em] LA FP64}}} &
\makecell[cl]{%
$[$GFLOP/s$]$ \\
FPU util. \\
$[$GFLOP/s/W$]$ \\
$[$GFLOP/s/\sqmmx$]$ \\
$[$GFLOP/s/\sqmmx$]$\tn{l}
} &
\makecell[cl]{%
\dk \\
\dk \\
\dk \\
\dk \\
\dk
} &
\makecell[cl]{%
\emph{\dk} \\
\dk \\
\dk \\
\dk \\
\dk
} &
\makecell[cl]{%
\sh{187}\tn{p}\uy\uy\cite{amd_sparse} \\
\sh{\SI{8.1}{\percent}}\tn{p}\uy\uy\cite{amd_sparse} \\
\dk \\
\sh{0.67}\tn{p}\;\cite{amd_sparse} \\
\sh{0.42}\tn{p}\;\cite{amd_sparse}
} &
\makecell[cl]{%
67.5\uy\cite{amd2024tune9004} \\
\SI{1.2}{\percent}\uy\cite{amd2024tune9004} \\
\dk \\
0.14\uy\cite{amd2024tune9004} \\
0.06\uy\cite{amd2024tune9004}
} &
\makecell[cl]{%
\sh{131}\tn{u}\uy\uy\cite{alappat2020a64fx} \\
\sh{\SI{4.7}{\percent}}\tn{u}\uy\uy\cite{alappat2020a64fx} \\
\dk \\
\sh{0.85}\tn{u}\uy\uy\cite{alappat2020a64fx} \\
\sh{0.54}\tn{u}\uy\uy\cite{alappat2020a64fx}
} &
\makecell[cl]{%
\sh{286}\tn{v}\uy\uy\cite{Niu2021TileSpMVAT} \\
\sh{\SI{2.9}{\percent}}\tn{v}\uy\uy\cite{Niu2021TileSpMVAT} \\
\dk \\
\sh{0.71}\tn{v}\uy\uy\cite{Niu2021TileSpMVAT} \\
\sh{0.45}\tn{v}\uy\uy\cite{Niu2021TileSpMVAT}
} &

\makecell[cl]{%
\emph{\dk} \\
\dk \\
\dk \\
\dk \\
\dk
} &
\makecell[cl]{%
\sh{198}\uy\uy\cite{siegmann2024first} \\
\SI{6.9}{\percent}\uy\uy\cite{siegmann2024first} \\
{\dk} \\ 
0.33\uy\uy\cite{siegmann2024first} \\
0.19\uy\uy\cite{siegmann2024first}
} &
\makecell[cl]{%
\sh{\textbf{307}} \\
\sh{\textbf{\SI{42}{\percent}}} \\
\sh{\textbf{16.0}} \\
\sh{\textbf{5.95}} \\
\sh{\textbf{5.95}}
} 
\vspace{\hdls}\\
\arrayrulecolor{black}\bottomrule
\end{tabular}
}

\afterpage{%
    \clearpage%
    \ifodd\value{page}
        \expandafter\afterpage%
    \fi
    {%

\begin{table}[p]

    \captionsetup{labelsep=contcols}
    \caption{Comparison of Occamy to \gls{soa} silicon-proven GPUs and CPUs supporting the dominant ISAs (x86, Arm, and RISC-V).}
    \label{tab:occamy:relwork:soa}

    \begin{threeparttable}
    \centering

    \adjustbox{clip,trim=0 0 {.5\width} 0,totalheight=0.73\textheight}{\usebox{\OccSaveBox}}

    \begin{adjustbox}{scale=0.6}
    \begin{minipage}{1.66667\linewidth}
        \begin{tablenotes}[para,flushleft]
            \im{a} Chip on Wafer on Substrate
            \im{b} Embedded Multi-Die Interconnect Bridges
            \im{c} Based on core presented in~\cite{chen2020xuantie}
            \im{d} Interposer: 65nm
            \im{e} 512 bit
            \im{f} Xoc := Xsf\_Xsssr
            \im{g} Host: RV64GC
            \im{h} RV64GC part of core as \emph{OpenC910}
            \im{i} Measured on released product
            \im{k} w/o IO die
            \im{l} Scaled to 12LP+ node using NAND2X1 gate areas (\si{GE\per\micro\metre\squared}) or transistor density
            \im{m} Infinity-Fabric On-Package
            \im{n} Multi-Die Fabric I/O
        \end{tablenotes}
    \end{minipage}
    \end{adjustbox}

    \end{threeparttable}
\end{table}

\clearpage

\begin{table}[p]

    \caption*{\Cref{tab:occamy:relwork:soa} (columns continued from previous page): Comparison of Occamy to \gls{soa} silicon-proven GPUs and CPUs supporting the dominant ISAs (x86, Arm, and RISC-V).}

    \begin{threeparttable}
    \centering

    \adjustbox{clip,trim={.5\width} 0 0 0,totalheight=0.73\textheight}{\usebox{\OccSaveBox}}

    \begin{adjustbox}{scale=0.6}
    \begin{minipage}{1.66667\linewidth}
        \begin{tablenotes}[para,flushleft]
            \im{o} Wide segment (38 links)
            \im{p} At base (non-boost) clock
            \im{q} Fully utilized tensor cores
            \im{r} At 2.6 \si{\giga\hertz}
            \im{s} At nominal 1V supply
            \im{t} Based on vector unit only
            \im{u} At below-peak clock \rev{(FPU util. relative to peak at reduced clock)}
            \im{v} Tensor cores unused (FPU util. relative to peak without tensor cores)
            \im{x} Unknown whether tensor cores used
            \im{y} Based on many-GPU supercomputers
            \im{z} Based on "Virga" system using 448 H100 GPUs~\cite{mellor2024virganews}
        \end{tablenotes}
    \end{minipage}
    \end{adjustbox}

    \end{threeparttable}
\end{table}

\clearpage

    }%
}

\renewcommand{\baselinestretch}{\occbsone}\selectfont

Occamy's peak performance is far ahead of other academic {\riscv} silicon prototypes such as~\cite{schmidt2022rsoc}, providing \cn{8.3\x}~more FP64 compute;
in fact, it is much closer in performance to the commercial Sophon SG2042~\cite{sophon-trm, chen2020xuantie} \riscv~manycore (\cn{0.77} vs \cn{\SI{1.0}{DP\text{-}\giga\flop\per\second}}).
Our \gls{d2d} link, while comparatively low in throughput due to its simple \gls{phy} design, achieves a competitive energy efficiency \cn{\SI{23}{\percent}} higher than that in AMD's Rome~\cite{amd_zen2_hc31}.
Unlike other designs in \Cref{tab:occamy:relwork:soa}, Occamy's entire digital compute architecture, including the groups, interconnect, host, and \gls{d2d} link, is available as an open-source \gls{rtl} description.

On FP64 dense \gls{la}, Occamy's \glspl{sssr} enable competitive FPU utilization through affine streams.
Our \cn{\SI{89}{\percent}} peak \gls{fpu} utilization is notably higher than that achieved on AMD's Rome~\cite{a64fx_dense} and Fujitsu's A64FX~\cite{siegmann2024first} and is comparable to the \cn{\SI{95}{\percent}} reached on Nvidia's A100 GPU~\cite{nvidia_a100_dense}.
At up to \cn{\SI{13.3}{DP\text{-}\giga\flop\per\second\per\milli\meter^2}}, dense \gls{la} on Occamy is highly area-efficient:
scaled to our \SI{12}{\nano\meter} node, we achieve a higher compute density than all competitors except Schmidt et al.~\cite{schmidt2022rsoc}, A100, and H100~\cite{hopper2022_hotchips, hopper2024_whitepaper}.
Occamy's peak dense \gls{la} energy efficiency of \cn{\SI{39.8}{DP\text{-}\giga\flop\per\second\per\watt}} is notably higher than that of the \glspl{cpu} we compare to, but cannot reach the efficiency of Nvidia's highly optimized A100 and H100 GPUs fabricated in more advanced nodes.

For FP64 stencil codes, Occamy's indirection-capable \glspl{sssr} enable significant utilization benefits over the compared CPU and GPU designs.
Our \cn{\SI{83}{\percent}} peak FPU utilization is \cn{1.7\x}~higher than the leading \cn{\SI{49}{\percent}} achieved on A100~\cite{Zhang2023RevisitingTB}
using CUDA compute only (to ensure a fair comparison, we normalized this FPU utilization to A100's reduced \SI{9.7}{DP\text{-}\tera\flop\per\second} peak \emph{excluding} tensor cores). While specialized transforms \emph{can} leverage A100's tensor cores to reach a higher peak throughput~\cite{chen2024convstencil} (albeit at lower FPU utilization of \SI{31}{\percent}), the corresponding node-normalized compute density is still 1.2\x~lower than the \cn{\SI{11.1}{DP\text{-}\giga\flop\per\second\per\milli\meter^2}} Occamy achieves.

On sparse-dense FP64 \gls{la}, Occamy's \glspl{sssr} and explicitly managed memory hierarchy enable even higher utilization gains.
Although all competitors have \gls{isa}-level support for scatter-gather, they reach at most \cn{\SI{8.1}{\percent}} FPU utilization on AMD Rome~\cite{amd_sparse}, \cn{5.2\x}~less than the \cn{\SI{42}{\percent}} we achieve;
this highlights the importance of our explicitly managed dataflow.
We also achieve the highest technology-node-normalized compute density of \cn{\SI{5.95}{DP\text{-}\giga\flop\per\second\per\milli\meter^2}}, \cn{11\x}~higher than the leading competitor A64FX~\cite{alappat2020a64fx}.
While A100 introduces hardware support for $2$:$4$ structured sparsity ($\leq$2 nonzeros in each block of four elements)~\cite{nvidia2020a100tc}, this feature cannot be used in~\cite{Niu2021TileSpMVAT} as it is specifically designed for deep learning on $\leq$\SI{32}{\bit} datatypes and ineffective in processing the highly sparse ($\gtrsim$\SI{95}{\percent}) data found in graph \gls{ml} and \gls{hpc} applications.
As a result, A100's inflexible GPU datapath achieves at most \cn{\SI{2.9}{\percent}} \gls{fpu} utilization, which is less than more agile CPU architectures we compare to.

To summarize, Occamy is the first 2.5D-integrated multi-chiplet \riscv~manycore demonstrated in silicon.
Compared to \gls{soa} CPUs and GPUs, it achieves competitive FPU utilization (\cn{\SI{89}{\percent}}) on dense \gls{la}, as well as leading FPU utilizations (\cn{1.7\x}, \cn{5.2\x}) and technology-node-normalized compute densities (\cn{1.2\x}, \cn{11\x}) on stencil codes and sparse-dense \gls{la}, respectively.
Unlike existing sparse compute accelerators and \gls{isa} extensions, its \glspl{sssr} are highly flexible and do not narrowly specialize to specific problem domains or data precisions.
As such, Occamy combines high efficiency on both regular and irregular workloads with high architectural flexibility in an unprecedented way.

\secocc{Conclusion}
\label{chap:occamy:conclusion}

In this chapter, we scaled out the performance and energy efficiency benefits of \glspl{sssr} to create \emph{\occamy}, the first open-source silicon-proven 2.5D-integrated \riscv~manycore featuring 432 cores across two chiplets, \SI{32}{\gibi\byte} of HBM2E DRAM, and a fully digital fault-tolerant \cn{\SI{1.6}{\pico\joule\per\bit}} \gls{d2d} link.
{\occamy}'s compute cores feature an 8-to-64-bit SIMD-capable FPU kept busy by \glspl{sssr}, which enable high FPU utilization on both regular and irregular workloads through $\leq$4D affine and indirect streams as well as streaming intersection and union.
Agile, latency-tolerant DMA engines in each compute cluster copy large data tiles to and from local \glspl{tcdm} to be processed by its compute cores.

We described the hierarchical implementation of our \SI{12}{\nano\meter} FinFET compute chiplets and our \SI{65}{\nano\meter} interposer \emph{Hedwig}.
On regular workloads, {\occamy} achieves a competitive FPU utilization of up to \cn{\SI{89}{\percent}} and an energy efficiency of up to \cn{\SI{39.8}{DP\text{-}\giga\flop\per\second\per\watt}}.
On stencil codes, we reach an FPU utilization of up to \cn{\SI{83}{\percent}} and a technology-node-normalized compute density of up to \cn{\SI{11.1}{DP\text{-}\giga\flop\per\second\per\milli\meter^2}}, improving by \cn{1.7\x}~and \rev{\cn{1.2\x}}, respectively, over \gls{soa} CPUs and GPUs.
On sparse-dense \gls{la}, we achieve an FPU utilization of up to \cn{\SI{42}{\percent}} \linebreak and a technology-node-normalized compute density of up to \linebreak \SI{5.95}{DP\text{-}\giga\flop\per\second\per\milli\meter^2}, resulting in substantial gains of \cn{5.2\x}~and \cn{11\x} over the \gls{soa}.
We \rev{further} achieve a sparse-sparse \gls{la} throughput of up to \cn{\SI{187}{\giga\comp\per\second}} at \cn{\SI{17.4}{\giga\comp\per\second\per\watt}}, \cn{\SI{3.63}{\giga\comp\per\second\per\milli\meter^2}}, and a comparator utilization of \cn{\SI{49}{\percent}}.
\rev{Finally, we reach FPU utilizations of up to \SI{75}{\percent} and \SI{54}{\percent} on
\gls{gpt}-J inference and a \gls{gcn} layer, respectively.}
The \gls{rtl} description of {\occamy}'s entire compute architecture is publicly available under a permissive open-source license.%

\renewcommand{\baselinestretch}{1.0}\selectfont
\setlength{\baselineskip}{12pt plus 0.33pt minus 0.12pt}
\setlength{\parskip}{0pt plus 3pt}

\clearpage
\newcommand{\ap}{AXI Pack}%
\newcommand{\strideSu}{5.4}%
\newcommand{\indirSu}{2.4}%
\newcommand{\strideBu}{87}%
\newcommand{\indirBu}{39}%
\newcommand{\areaOver}{6.2}%
\newcommand{\stridePowerInc}{19.8}%
\newcommand{\indirPowerInc}{15.7}%
\newcommand{\avgPowerInc}{17.8}%
\newcommand{\strideEEF}{5.3}%
\newcommand{\indirEEF}{2.1}%

\chapter{Chip-level Extensions}
\label{chap:chiplevel}

So far, our acceleration efforts leveraging \glspl{sssr} in the Snitch cluster are subject to one important constraint: irregular streams are confined to the cluster \gls{tcdm} and its addressable range.
While this ensures high energy efficiency on tiled workloads, some irregular workloads with large datasets cannot easily be partitioned without incurring significant performance overheads.

In this chapter, we present \emph{AXI Pack}, a backward-compatible extension to Arm's widespread \gls{axi4} on-chip protocol that provides bandwidth-efficient end-to-end irregular streams \emph{at the chip level}.
AXI Pack defines two new burst types that pack the non-contiguous elements of strided and indirect streams onto wide buses, increasing bandwidth utilization, performance, and energy efficiency.
Thanks to its chip-level scope, AXI Pack can access entire chip-level memory endpoints, overcoming the addressing limitations and partitioning requirements of \glspl{sssr} in the Snitch cluster.

\section{Introduction}
\label{chap:chiplevel:intro}

Increasingly large and sparse datasets in domains like fluid simulation~\cite{Georgescu2010ConjugateGO}, graph analytics~\cite{Min2020EMOGI}, and recommender systems~\cite{Li2017CuSNMFAS} push irregular compute toward ever larger physical memory footprints.
In addition to the large control overheads and non-contiguous memory access patterns we have encountered so far, these large-scale irregular workloads suffer from extremely low data locality~\cite{boman2013scalable,lin2021egdram}, making them difficult to partition among \glspl{pe}:
\gls{soa} solutions require complex graph partitioning approaches~\cite{boman2013scalable,Mukhopadhyay2024comms,mi2023balancing} or a high volume of global communication~\cite{Mukhopadhyay2024comms,mi2023balancing}, incurring significant performance overheads.
This motivates the efficient handling of irregular memory accesses at the \emph{chip level}, where all \glspl{pe} have random access to all memory endpoints without significant bandwidth or capacity limitations.

Existing works on accelerating chip-level irregular memory accesses focus on either the \emph{core side} or the \emph{memory side} of communication.
Core-side extensions (some of which we discussed in \cref{chap:sssrs}) often use \emph{stream} abstractions \cite{Talati2021ProdigyIT,fu2024dmp,Wang2019StreambasedMA,Domingos2021UnlimitedVE} to describe sequences of irregular accesses, freeing processors from address calculations and decoupling memory accesses from execution.
They focus on \emph{strided} and \emph{indirect} streams, which are most common in practice \cite{Wang2019StreambasedMA}, with some solutions mapping them directly to architectural registers \cite{Wang2019StreambasedMA, Scheffler2021IndirectionSS,Domingos2021UnlimitedVE}.
However, these works largely ignore downstream interconnects and memory systems: at most, they propose stream throttling~\cite{Wang2019StreambasedMA,Talati2021ProdigyIT,Domingos2021UnlimitedVE} and cache bypassing~\cite{fu2024dmp,Domingos2021UnlimitedVE} policies to avoid requestor interference and thrashing.
Thus, they do not address some of the fundamental challenges of chip-level irregular accesses, including the high index-fetching overhead of core-side indirection, the low bus utilization of narrow accesses on wide buses, or the lack of burstive transport for non-contiguous data.

In contrast, memory-side extensions focus on interconnects and memory endpoints, aiming to increase bus efficiency and improve access latency.
They include specialized memory controllers with scatter-gather features~\cite{Carter1999ImpulseBA,tanabe2011memctrl,Hussain2018MemoryCF}, \glsreset{dlt}\gls{dlt} accelerators~\cite{Barredo2021PLANARAP,Lloyd2015InMemoryDR}, and even in-memory extensions accelerating irregular accesses~\cite{lin2021egdram,Adhinarayanan2025folded}.
Unlike core-side extensions, these solutions reduce access times and optimize chip-level bus utilization.
Many of them avoid narrow bus accesses by \emph{packing} non-contiguous narrow elements onto bus-wide lines, which are then mapped to virtual addresses \cite{Carter1999ImpulseBA}, written to internal scratchpads \cite{Hussain2017ANH, Hussain2018MemoryCF}, or written back to dedicated memory locations \cite{Barredo2021PLANARAP,Lloyd2015InMemoryDR}.
However, these solutions have their own drawbacks: they occupy virtual or physical memory and lack the tight architectural integration of core-side extensions, limiting their acceleration potential and complicating programming.

Thus, while existing core-side and memory-side extensions tackle inefficiencies at their respective ends, they forego each other's benefits and do not integrate with established interconnect protocols, failing to provide an end-to-end solution for bandwidth-efficient irregular streams at the chip level.

To address these shortcomings, we propose \emph{\ap}, an extension to Arm's widespread \gls{axi4} on-chip protocol~\cite{Arm2021AMBAAXI} enabling end-to-end, tightly-packed strided and indirect memory streams.
\ap~transparently extends \gls{axi4}'s existing contiguous bursts, leveraging their decoupled, latency-resilient nature.
It remains compatible with all existing \gls{axi4} features and even existing interconnect blocks that do not reshape bursts. It encodes stream semantics (stride or index base and size) directly into burst requests, ensuring performance and flexibility even for short streams. Indirection is efficiently handled at memory endpoints, avoiding additional index traffic.
Like \gls{axi4} itself, \ap~supports both core and non-core requestors (e.g., accelerators), as well as systems with multiple requestors and endpoints\footnote{Since streams are resolved at endpoints, all elements accessed within one stream must hit one endpoint. We further discuss this caveat in \Cref{chap:conclusion:fwork_outlook}.}.

To demonstrate \ap~in an end-to-end full-system context, we extend the open-source RISC-V vector processor \emph{Ara}~\cite{Perotti2022Newara} for efficient strided and indexed accesses and design a banked memory controller efficiently handling irregular bursts.
On a system with a 256-bit-wide \ap~bus running various irregular FP32 workloads, we achieve bus utilizations of up to \SI{\strideBu}{\percent} on strided and \SI{\indirBu}{\percent} on indirect benchmarks, resulting in peak speedups of \strideSu\x~and \indirSu\x~over a baseline with a standard \gls{axi4} bus.
We implement our evaluation system in GlobalFoundries' \SI{22}{\nano\meter} FD-SOI technology and find that {\ap} improves energy efficiency by up to {\strideEEF}\x~and {\indirEEF}\x~in strided and indirect benchmarks while incurring only  \SI{\areaOver}{\percent} of our vector processor's area for our controller.
Finally, we analyze the impact of element and index size as well as bank count on \ap~performance and controller complexity.

To summarize, our contributions are as follows:
\begin{enumerate}
    \item We present \emph{\ap}, a backward-compatible extension to the widespread chip-level protocol \gls{axi4} that adds support for end-to-end bus-packed strided and indirect streams.

    \item We extend the open-source RISC-V vector processor \emph{Ara} with {\ap}, enabling high bus \mmp{utilization} and significant speedups on irregular workloads, and demonstrate a simple banked memory controller \mmp{that efficiently handles its irregular} bursts.

    \item We evaluate {\ap} by benchmarking various irregular workloads on our extended vector processor, achieving bus utilizations of up to \SI{\strideBu}{\percent} and \SI{\indirBu}{\percent} and speedups of up to \strideSu\x~and \indirSu\x~for strided and indirect workloads, respectively.

    \item We evaluate our {\ap} system and controller in terms of timing, area, and energy efficiency benefits, finding \fin{a low controller area impact and} energy efficiency improvements of up to \fin{{\strideEEF}\x~for strided and {\indirEEF}\x~for indirect workloads}.
\end{enumerate}

The remainder of this chapter will be structured as follows:
\cref{chap:chiplevel:protocol} \mmp{introduces} our {\ap} protocol extension.
In \cref{chap:chiplevel:integr}, we \mmpcut{will }discuss the integration of {\ap} into Ara and the implementation of an AXI-Pack-capable banked memory controller.
In \cref{chap:chiplevel:eval}, we \mmpcut{will }evaluate the performance and energy efficiency benefits of our work, as well as its area and timing impact.
\cref{chap:chiplevel:relwork} \mmpcut{will }discusses related work on core-side and memory-side extensions.
\mmp{Finally, in \cref{chap:chiplevel:conclusion}, we will provide a conclusion.}

\section{AXI Pack Extension}
\label{chap:chiplevel:protocol}

\ap~extends Arm's \gls{axi4} \cite{Arm2021AMBAAXI}, a widely used high-performance non-coherent on-chip memory protocol.
\gls{axi4} defines five independent channels: the \emph{AR} and \emph{AW} channels carry read and write requests, the \emph{R} and \emph{W} channels carry read and write data, and the \emph{B} channel carries write responses.
Without extensions, linear, fixed, and wrapping bursts are supported.
Each channel provisions a user field of parametric width that allows for the addition of new functionality without compromising compatibility with the baseline protocol.

\begin{figure}
  \centering
  \includegraphics[width=0.92\linewidth]{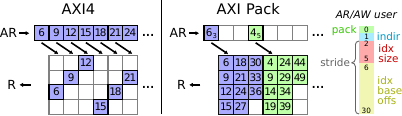}
  \caption{\ap~AR and AW channel user extensions and packed strided read example.}
  \label{fig:chiplevel:protocol:ap_user}
\end{figure}

\ap~extends the request channels AR and AW with user signals to communicate packed irregular bursts, as illustrated in \Cref{fig:chiplevel:protocol:ap_user}.
The \texttt{pack} bit indicates whether our extension is used, while the \texttt{indir} bit differentiates between strided and indirect bursts.
The remaining bits are shared between both burst types; they indicate either the element stride for strided bursts or the index size and base offset for indirect bursts.

While active, the new irregular burst types alter the semantics of existing channel fields. Most notably, data elements of the requested size, scattered in memory, are tightly packed onto the R and W data buses to fully utilize them. Additionally, the start of irregular bursts is aligned with the bus instead of the base address to simplify feeding packed data to and from vectorized functional units. Finally, the AR and AW \texttt{size} fields, previously only relevant for narrow beats, indicate the data element size.

In addition to performance and bus utilization, these semantics aim to maximize the \emph{transparency} and \emph{portability} of \ap: they ensure that any existing \gls{axi4} \gls{ip} blocks handling non-modifiable transactions without splitting, such as the routing blocks provided in \cite{kurth2021open}, are already compatible with \ap~without any modifications. \glspl{ip} that require burst splitting or reshaping, such as bus width converters, can \mmpcut{easily }be extended to support \ap~by re-packing bus-aligned data elements as is already done for other burst types.

\section{Implementation}
\label{chap:chiplevel:integr}

We \mmpcut{will }first extend the \emph{Ara} vector processor to support AXI Pack's strided and indirect bursts, and then present a simple banked memory controller as an example endpoint efficiently handling these bursts.

\subsection{Vector Processor Extension}

To demonstrate the benefits and flexibility of \ap, we extend the open-source \emph{Ara} \cite{Perotti2022Newara} RISC-V vector processor to leverage it for efficient irregular memory accesses.
Ara acts as a coprocessor to the application-class CVA6 core \cite{zaruba2019cva6}, which dispatches vector instructions to Ara. Both CVA6 and Ara access memory over \gls{axi4}.

As mandated by its \gls{rvv} instruction set, Ara supports three vector memory access types: \emph{contiguous}, \emph{strided}, and \emph{indexed}. Without extensions, only contiguous accesses can leverage bursts. For strided and indexed accesses, Ara must compute the address and issue \fin{an} individual narrow \fin{access} for each element, leaving the data channels severely underutilized as shown in \Cref{fig:chiplevel:protocol:ap_user}.

\Cref{fig:chiplevel:integr:ara_extension} shows our extensions to Ara. We modify its \gls{vlsu} to use \ap~for strided and indexed vector accesses. For strided accesses, we simply translate the existing \texttt{vlse} and \texttt{vsse} instructions to \ap~requests and exchange the read or written data directly with vector registers or functional lanes for chaining.
The existing indexed access instructions \texttt{vl(o|u)xei} and \texttt{vs(o|u)xei} in the RISC-V vector extension presume that indices are already loaded into vector registers, necessitating the prior move of indices into the core and precluding efficient memory-side indirection.
To remedy this, we extend Ara's decoder to introduce two new \emph{in-memory} indexed access instructions, \texttt{vlimxei} and \texttt{vsimxei}, which use index arrays in memory for indirection.
\mmp{The assembly syntax of these instructions is shown in \Cref{fig:chiplevel:integr:ara_extension}: instead of a vector register containing the indices themselves, they read a scalar register holding their base address in memory.}
They are directly translated into indirect \ap~bursts and allow for packed data to be exchanged with Ara's unmodified functional units and registers without data format changes.

\subsection{Banked Memory Controller}

To demonstrate the efficient handling of \ap~at banked memory endpoints, we design a proof-of-concept controller translating \ap~requests to sequences of parallel banked memory accesses. Our controller is fully backward-compatible with and efficiently handles regular \gls{axi4} bursts.

\begin{figure}
  \subfloat[Vector processor extensions]{
    \label{fig:chiplevel:integr:ara_extension}
    \includegraphics[width=0.46\linewidth, height=0.6\linewidth]{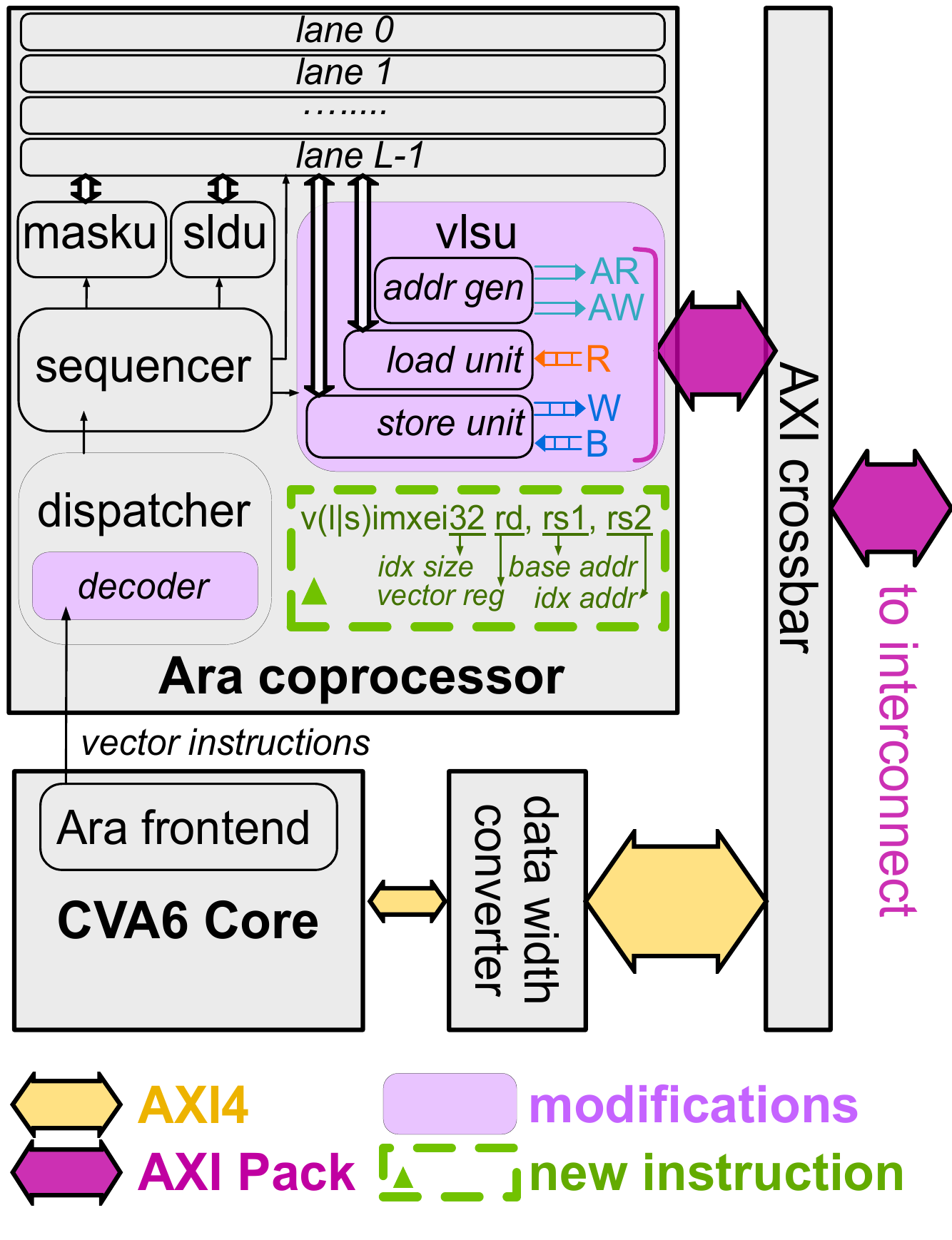}
  }
  \subfloat[Controller top level]{
    \label{fig:chiplevel:integr:arch_contr_top}
    \includegraphics[width=0.44\linewidth, height=0.6\linewidth]{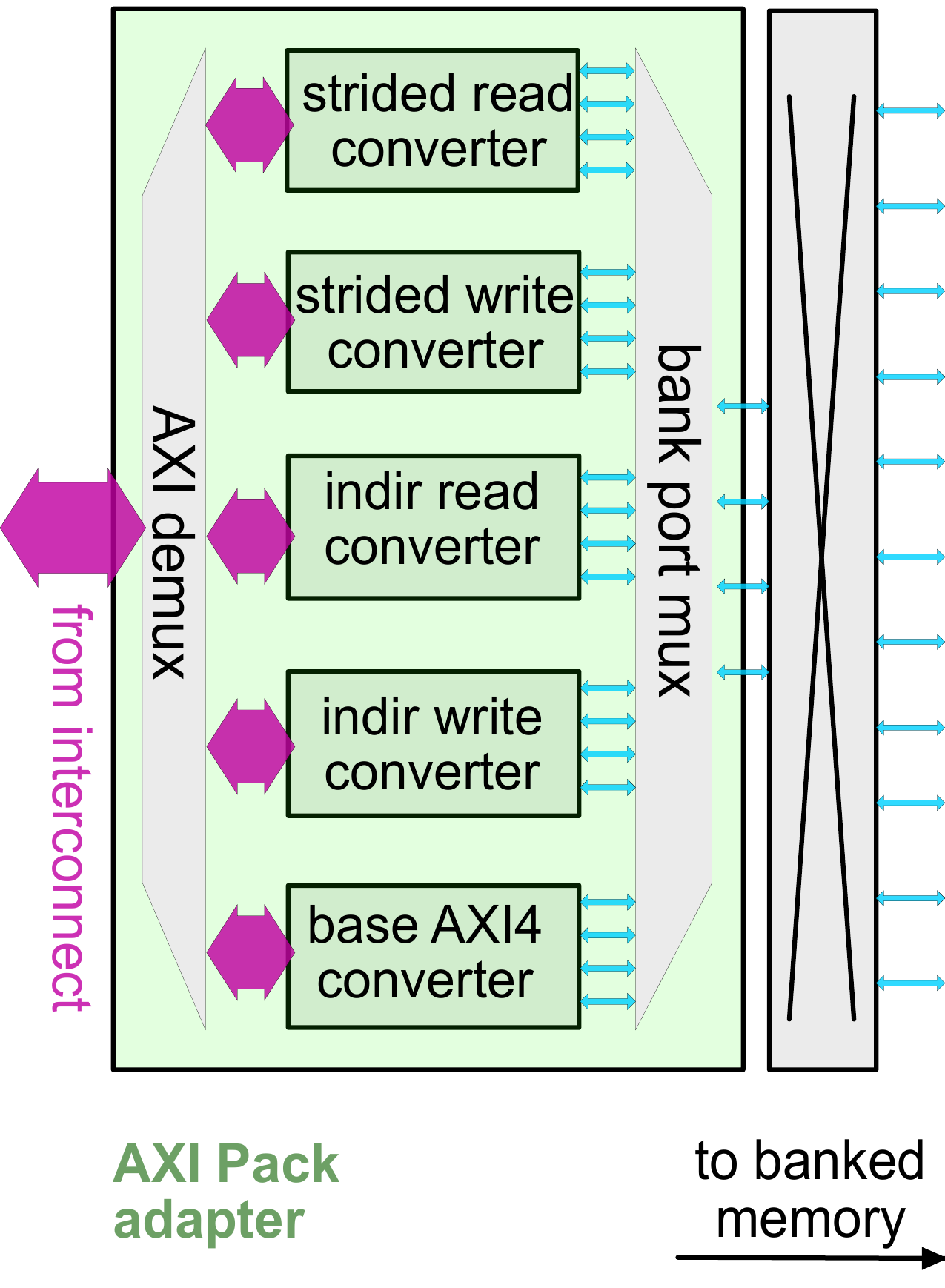}
  }

  \vspace{1em}

  \subfloat[Strided read converter]{
    \label{fig:chiplevel:integr:arch_contr_rstrided}
    \includegraphics[width=0.4446\linewidth, height=0.6\linewidth]{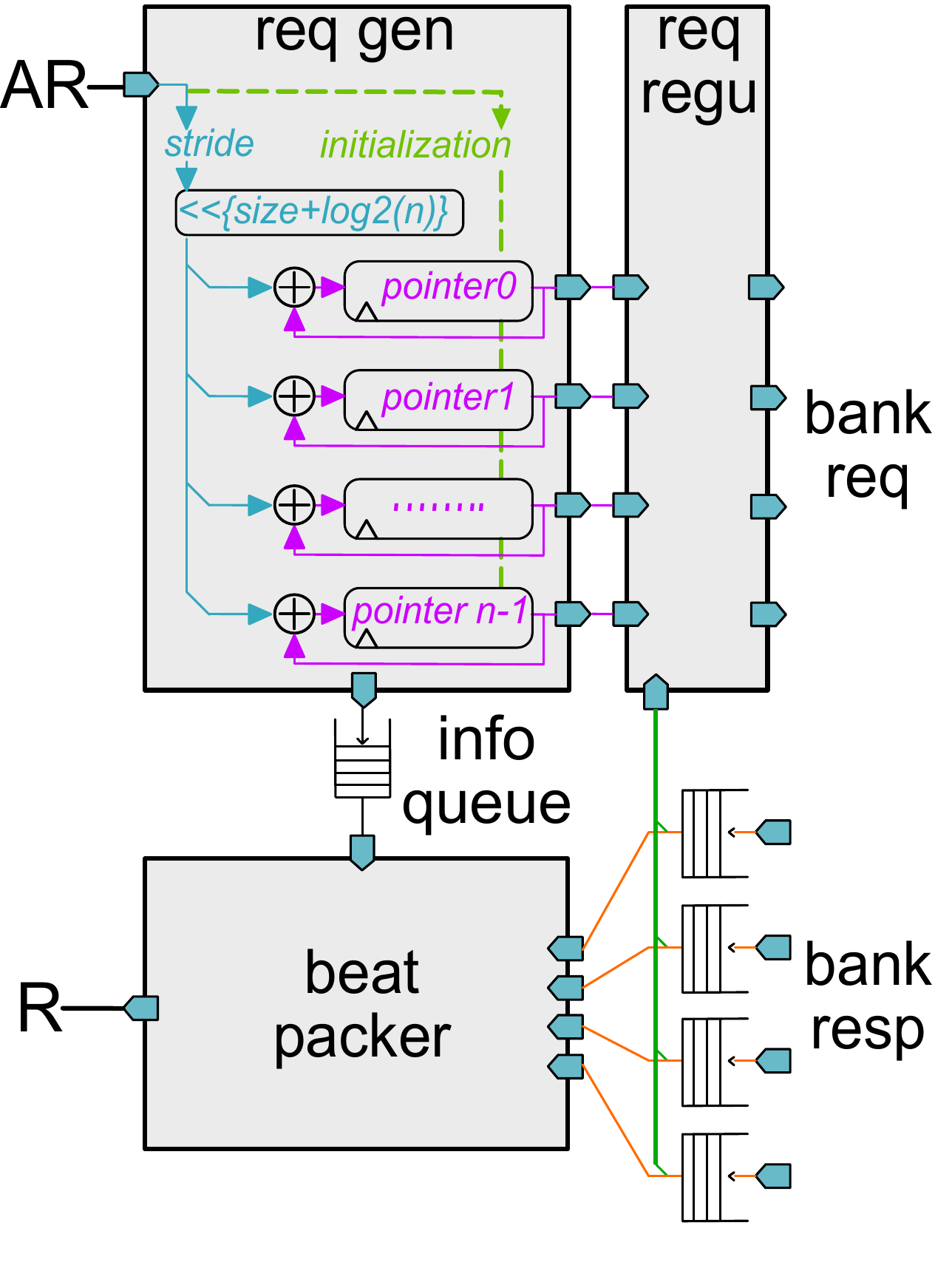}
  }
  \subfloat[Indirect read converter] {
    \label{fig:chiplevel:integr:arch_contr_rindir}
    \includegraphics[width=0.4892\linewidth, height=0.6\linewidth]{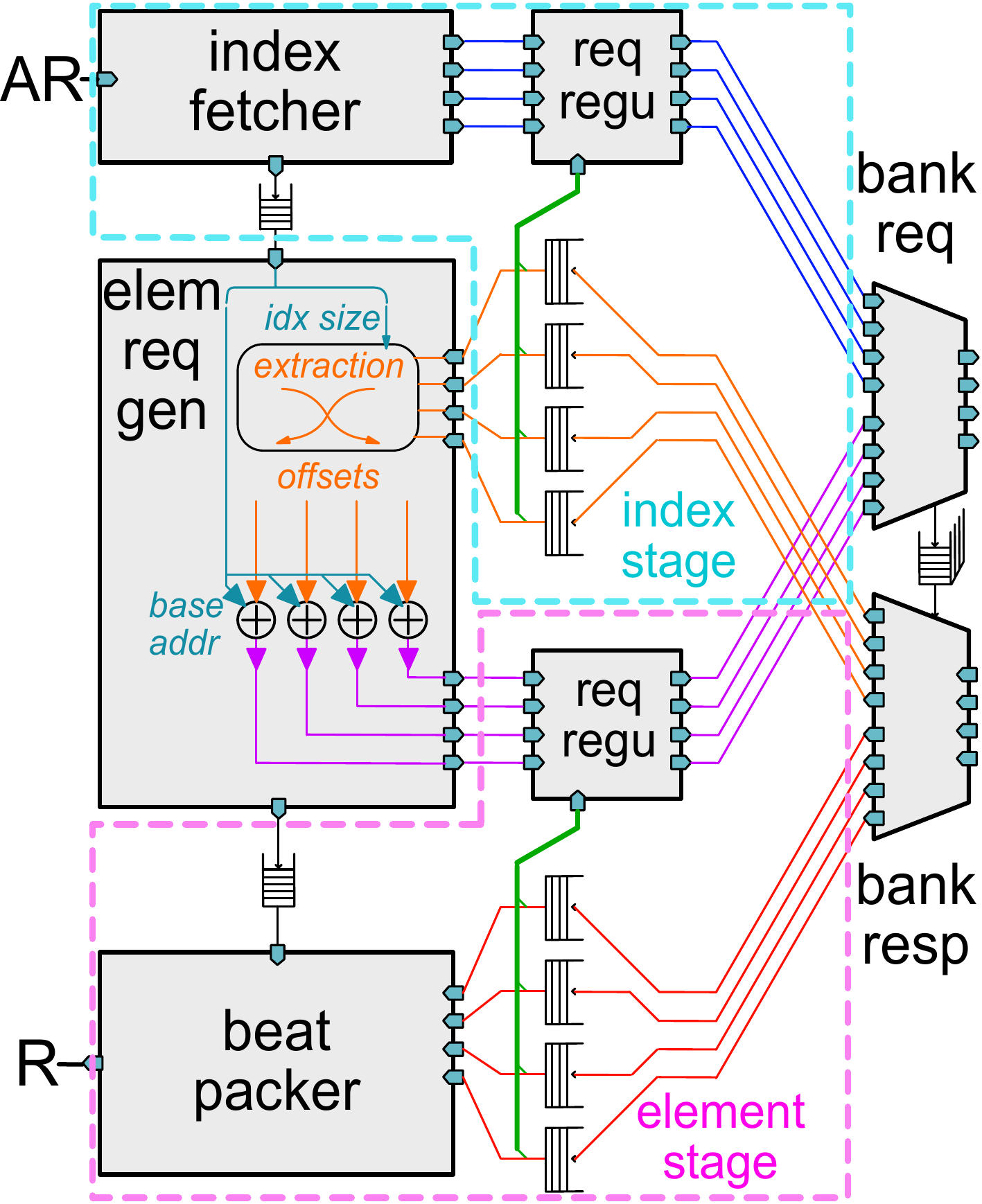}
    }
  \caption{AXI-Pack-extended vector processor, multi-banked memory controller, and read protocol converter architectures.}
  \label{fig:chiplevel:integr:ap_contr}
\end{figure}

\Cref{fig:chiplevel:integr:arch_contr_top} shows the controller architecture. The \emph{adapter} translates both regular and irregular bursts to sequences of $n$ parallel \emph{word} accesses, where a word has the same bitwidth $W$ as the used memory banks and determines the smallest efficiently-handled element size. For $D$-bit-wide \ap~data buses, $n=D/W$, since we must read or write $D/W$ words in parallel to saturate them. The adapter connects to an $n\times m$ fully connected crossbar mapping the $n$ word access ports to $m$ interleaved banks.

Internally, the adapter forwards requests to one of five protocol converters, which may concurrently handle bursts. The \emph{base} converter handles regular \gls{axi4} bursts, while the remaining converters are dedicated to strided and indirect read and write operations, respectively. Handling reads and writes individually leverages the inherent concurrency of the R and W channels.

\Cref{fig:chiplevel:integr:arch_contr_rstrided} details the strided read converter architecture. For each beat in a burst, the \emph{request generator} issues $n$ parallel word requests fetching the elements to be packed and pushes metadata needed for later packing into an \emph{info} queue. The words read from the banks are stored in decoupling queues and then passed to the \emph{beat packer}, which packs the words as specified by metadata popped from the \emph{info} queue to form the R beats. To prevent word queue overflows, a \emph{request regulator} limits the number of requests in-flight for each word lane.

\Cref{fig:chiplevel:integr:arch_contr_rindir} shows the indirect read converter architecture. It involves two stages sharing the $n$ word request ports through round-robin arbitration: the \emph{index stage} fetches indices from memory and the \emph{element stage} uses these indices to fetch indirect elements and pack them into R beats. The index stage is analogous to the strided read converter, but issues only contiguous word requests. The fetched indices are passed to the \emph{element request generator}, which shifts and adds them to the specified base address to generate word requests for the desired elements. Finally, the requested elements are packed by a beat packer as specified by metadata from the element request generator to form the desired R beats.

The corresponding strided and indirect \emph{write} converters are similar and differ only in the direction of the datapath: a \emph{beat unpacker} splits beats into individual words, which are then used as write data for the generated write requests. The memory responses are combined and forwarded to the B channel.

While the controller architecture we present here targets banked narrow endpoints, {\ap} streams can also be used effectively with wide single-port interfaces:
in \cite{zhang2024coal}, we explored the use of an AXI-Pack-compliant low-overhead coalescer to enable efficient irregular streams with wide DRAM interfaces like HBM and LPDDR, improving performance and energy efficiency on various irregular workloads.
In this chapter, we will focus on evaluating {\ap} using our banked controller with word-granularity banks.

\section{Evaluation}
\label{chap:chiplevel:eval}

We \mmpcut{will }first describe the system setup and irregular workloads we used for evaluation (\cref{chap:chiplevel:eval:suwl}).
We \mmpcut{will }then evaluate the performance (\cref{sec:chiplevel:eval:res_perf}) and energy efficiency benefits (\cref{sec:chiplevel:eval:enpow}) of {\ap}, followed by area and timing results on our protocol adapter (\cref{sec:chiplevel:eval:at}).
Finally, we \mmpcut{will }analyze the impact of bank count, element size, and index size on our results (\cref{sec:chiplevel:eval:parsens}).

\subsection{\mmp{Experimental} Setup and Workloads}
\label{chap:chiplevel:eval:suwl}

To evaluate \ap, we consider three RISC-V \glspl{soc} using CVA6 with Ara as a vector processor, \gls{axi4}-based interconnects, and a banked on-chip \glsunset{sram}\gls{sram} memory:

\begin{itemize}
    \item \textsc{base}: unmodified CVA6 and Ara connecting to a regular banked memory over a standard \gls{axi4} bus.

    \item \textsc{pack}: unmodified CVA6 and AXI-Pack-extended Ara connecting to a banked memory with an \ap~controller over an AXI-Pack-extended bus.

    \item \textsc{ideal}: same as \textsc{base}, but Ara connects directly to an exclusive, idealized memory with one port per lane, serving data with ideal packing, bandwidth, and latency.
\end{itemize}

\textsc{ideal} provides an upper bound for possible \ap~benefits by idealizing interactions between Ara's \gls{vlsu} and memory; however, it does not avoid  inefficiencies arising from Ara's internal microarchitecture or CVA6.
In all systems, Ara is parameterized to eight vector lanes and 256-bit-wide data buses. The banked memories provide eight 32-bit-wide word ports backed by 17 banks, which we determine in \cref{sec:chiplevel:eval:parsens} to provide a good performance-area tradeoff.

On each system, we evaluate a set of vectorized benchmarks benefiting from efficient strided and indirect memory accesses:

\begin{itemize}
    \item \texttt{ismt}: \emph{in-situ matrix transpose}. We transpose a square matrix in place by swapping and rotating elements above and below its diagonal using strided memory accesses.

    \item \texttt{gemv}: \emph{general matrix-vector multiply}. We investigate both \emph{row-} and \emph{column-wise} dataflows, with the latter trading vector reductions for strided accesses, and use the fastest approach on each system for fair comparisons.

    \item \texttt{trmv}: \emph{triangular matrix-vector multiply}, a \texttt{gemv} with an upper-triangular matrix. Only nonzero elements are streamed, incurring bursts of varying lengths. We again use the fastest dataflow on each system.

    \item \texttt{spmv}: \emph{sparse matrix-vector multiply}, a widely used sparse-dense \gls{la} kernel that incurs streaming indirections.

    \item \texttt{prank}: \emph{PageRank} \cite{Page1999ThePC}, which rates each node in a graph based on the edges inbound to it. The graph is represented as a sparse weighted adjacency matrix.

    \item \texttt{sssp}: \emph{single-source shortest path}, which calculates the shortest path from one node to all others in a weighted, directed graph represented as a sparse matrix.
\end{itemize}

We run the first three benchmarks leveraging strided streams on randomly-generated square matrices, and the latter three leveraging indirect streams on real-world sparse matrices from the SuiteSparse collection \cite{Davis2011TheUO} in the \gls{csr} format.
Elements are stored as 32-bit floats and indices as 32-bit integers. On indirect workloads, the \textsc{pack} system uses our extensions to handle indirection near memory, whereas \textsc{base} and \textsc{ideal} systems fetch indices into Ara.

\subsection{Performance}
\label{sec:chiplevel:eval:res_perf}

We determine the performance and read bus utilization for each benchmark in cycle-accurate \gls{rtl} simulation of our systems.
We initially assume a fixed matrix size of \mmp{256\x 256} for strided workloads and the sparse matrix \texttt{heart1} (390 average nonzeros per row) for indirect workloads.
\texttt{gemv} and \texttt{trmv} use a column-wise dataflow on \textsc{pack} and \textsc{ideal} and a row-wise dataflow on \textsc{base}, which we will later show to be optimal for each respective system.

\begin{figure}
\centering
  \subfloat[Speedups and bus utilizations across workloads.]{
    \label{fig:chiplevel:eval:all-bench-perf}
    \includegraphics[width=0.94\linewidth, height=0.4617\linewidth]{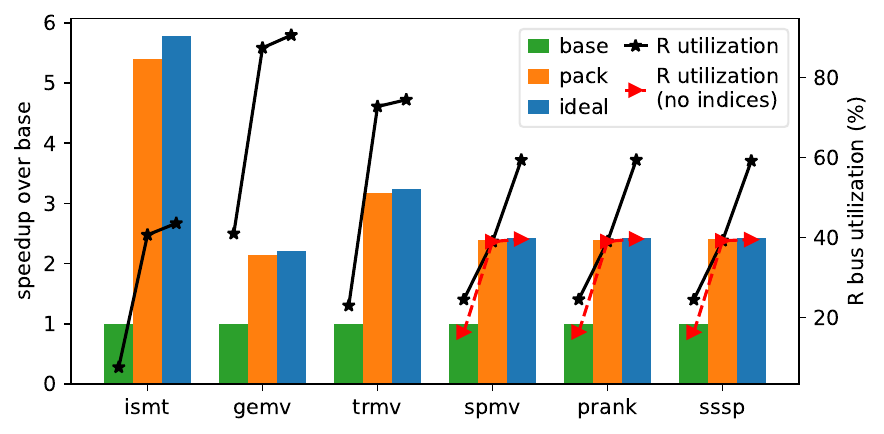}
    \vspace{-0.16em}
  } \\ \vspace{1em}
  \subfloat[\texttt{gemv} dataflows compared.]{
    \label{fig:chiplevel:eval:gemv-comp}
    \includegraphics[width=0.45\linewidth, height=0.40\linewidth]{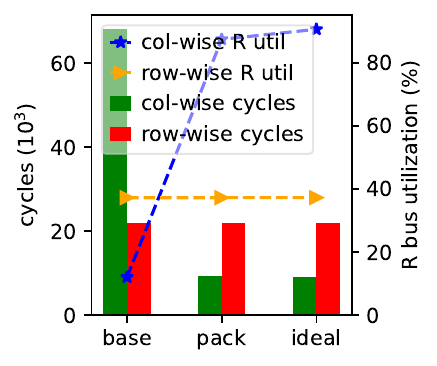}
    \vspace{-0.16em}
  }
  \subfloat[\texttt{trmv} dataflows compared.] {
    \label{fig:chiplevel:eval:trmv-comp}
    \includegraphics[width=0.45\linewidth, height=0.40\linewidth]{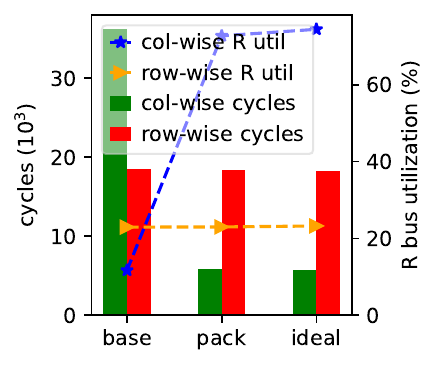}
    \vspace{-0.16em}
  } \\ \vspace{1em}
  \subfloat[\texttt{ismt} \textsc{pack} speedup scaling.]{
    \label{fig:chiplevel:eval:ismt-speedup}
    \includegraphics[width=0.45\linewidth, height=0.344\linewidth]{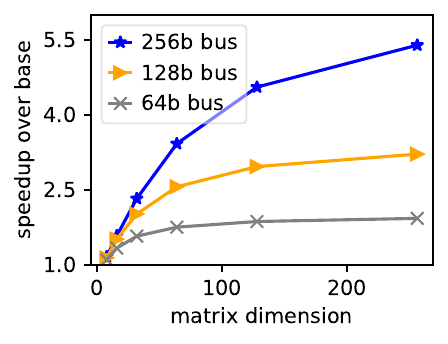}
    \vspace{-0.16em}
  }
  \subfloat[\texttt{spmv} \textsc{pack} speedup scaling.] {
    \label{fig:chiplevel:eval:spmv-speedup}
    \includegraphics[width=0.45\linewidth, height=0.344\linewidth]{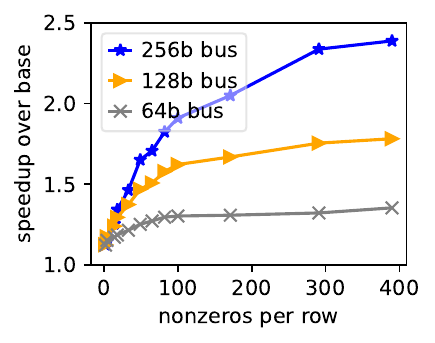}
    \vspace{-0.16em}
  }
  \caption{\ap~performance results.}
  \label{fig:chiplevel:eval:ap_perf}
\end{figure}

\cref{fig:chiplevel:eval:all-bench-perf} shows \textsc{pack} speedups over \textsc{base} and read bus utilizations with and without index transfers. \ap~significantly improves bus utilization and performance for all workloads, achieving {\SI{97}{\percent}} of the \textsc{ideal} performance on average.
On strided workloads, we measure peak speedups of \strideSu\x~(\texttt{ismt}) and bus utilizations of  \SI{\strideBu}{\percent} (\texttt{gemv}). We note that read bus utilizations on \texttt{ismt} are limited to \SI{50}{\percent} due to read-write ordering in Ara.
On indirect workloads, we achieve speedups of up to \indirSu\x~({\texttt{spmv}}) and bus utilizations of up to \SI{\indirBu}{\percent} ({\texttt{sssp}}). \textsc{pack} handles indirection directly in its \ap~controller, avoiding \textsc{ideal}'s waste of up to {\SI{20}{\percent}} (\texttt{spmv}) of bus time on index traffic and shifting indexed workloads further from the memory-bound toward the compute-bound regime.

To motivate our dataflow choices, \cref{fig:chiplevel:eval:gemv-comp,fig:chiplevel:eval:trmv-comp} compare the row- and column-wise dataflow performance for \texttt{gemv} and \texttt{trmv}.
Row-wise flows use only long contiguous accesses, so their performance is identical for \textsc{base} and \textsc{pack} and very close to \textsc{ideal}.
However, they require costly vector reductions, limiting \textsc{base} bus utilizations to \SI{37}{\percent} and \SI{23}{\percent}, respectively.
Column-wise flows avoid reduction by working on multiple results at once, providing higher \textsc{ideal} performance and \textsc{pack} utilizations of \SI{87}{\percent} and \SI{72}{\percent}, respectively.
However, for \textsc{base}, we stick to a row-wise flow, as the performance impact of strided accesses outweighs that of reductions without our extensions.

We also analyze the impact of input size and bus width on \ap~speedups for representative strided and indirect workloads.
\Cref{fig:chiplevel:eval:ismt-speedup} shows \texttt{ismt} speedups for matrix dimensions of 8 to 256 and bus widths of 64 to 256 bit, corresponding to 2 to 8 Ara lanes and thus a growing number of functional units.
As matrix size increases, speedups converge and reach up to 1.9\x, 3.2\x, and 5.4\x; as we widen the bus, the narrow accesses of \textsc{base} become less efficient, increasing peak \textsc{pack} speedups.
As matrix size decreases, streams and useful computation phases shorten and become bottlenecked by the overhead of row iteration, decreasing speedups.
\Cref{fig:chiplevel:eval:spmv-speedup} shows \texttt{spmv} speedups for sparse matrices with 2 to 390 average nonzeros per row and the same bus widths as before.
Speedups again converge and reach up to 1.4\x, 1.8\x, and 2.4\x. We see similar scaling trends as for \texttt{ismt} because in \texttt{spmv}, the nonzeros per row determine the computation phase \fin{duration} and stream \fin{length} in each row iteration. We note that thanks to its access-bundling nature, using \ap~never results in a slowdown no matter how short streams become.

\subsection{Energy and Power}
\label{sec:chiplevel:eval:enpow}

\begin{figure}
  \includegraphics[width=0.96\linewidth]{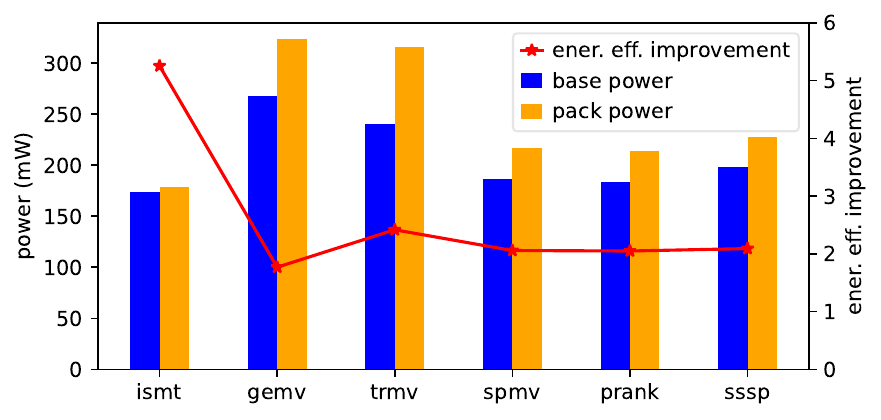}
  \caption{Estimated \ap~power consumption and energy efficiency improvements in GlobalFoundries' \SI{22}{\nano\meter} FD-SOI technology.}
  \label{fig:chiplevel:eval:power_analysis}
\end{figure}

We estimate the power consumption of \textsc{pack} and \textsc{base}, excluding the SRAM crossbar and banks, in the TT corner (\SI{0.8}{\volt}, \SI{25}{\celsius}) of GlobalFoundries' \SI{22}{\nano\meter} FD-SOI technology at \SI{1}{\giga\hertz}. We synthesize our systems using Synopsys \emph{Design Compiler} and estimate power on the benchmarks from \Cref{sec:chiplevel:eval:res_perf} using Synopsys \emph{PrimeTime}. \Cref{fig:chiplevel:eval:power_analysis} shows the average power and energy efficiency improvement of \textsc{pack} over \textsc{base} for each benchmark.
Despite small power consumption increases in \textsc{pack} of at most \SI{31}{\percent} (\texttt{trmv}) due to increased compute and bus utilization, all workloads see notable energy efficiency improvements, achieving peaks of {\strideEEF}\x~(\texttt{ismt}) and {\indirEEF}\x~(\texttt{sssp}) on strided and indirect benchmarks, respectively.

\subsection{Area and Timing}
\label{sec:chiplevel:eval:at}

\begin{figure}%
\centering%
  \subfloat[Adapter area versus clock period.]{
    \includegraphics[width=0.543\linewidth, height=0.375\linewidth]{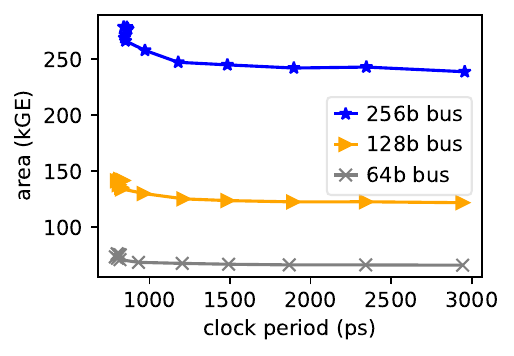}%
    \label{fig:chiplevel:eval:a_t}%
  }\hfill%
  \subfloat[Adapter area breakdown.] {
    \hspace{0.5em}%
    \includegraphics[width=0.356\linewidth, height=0.375\linewidth]{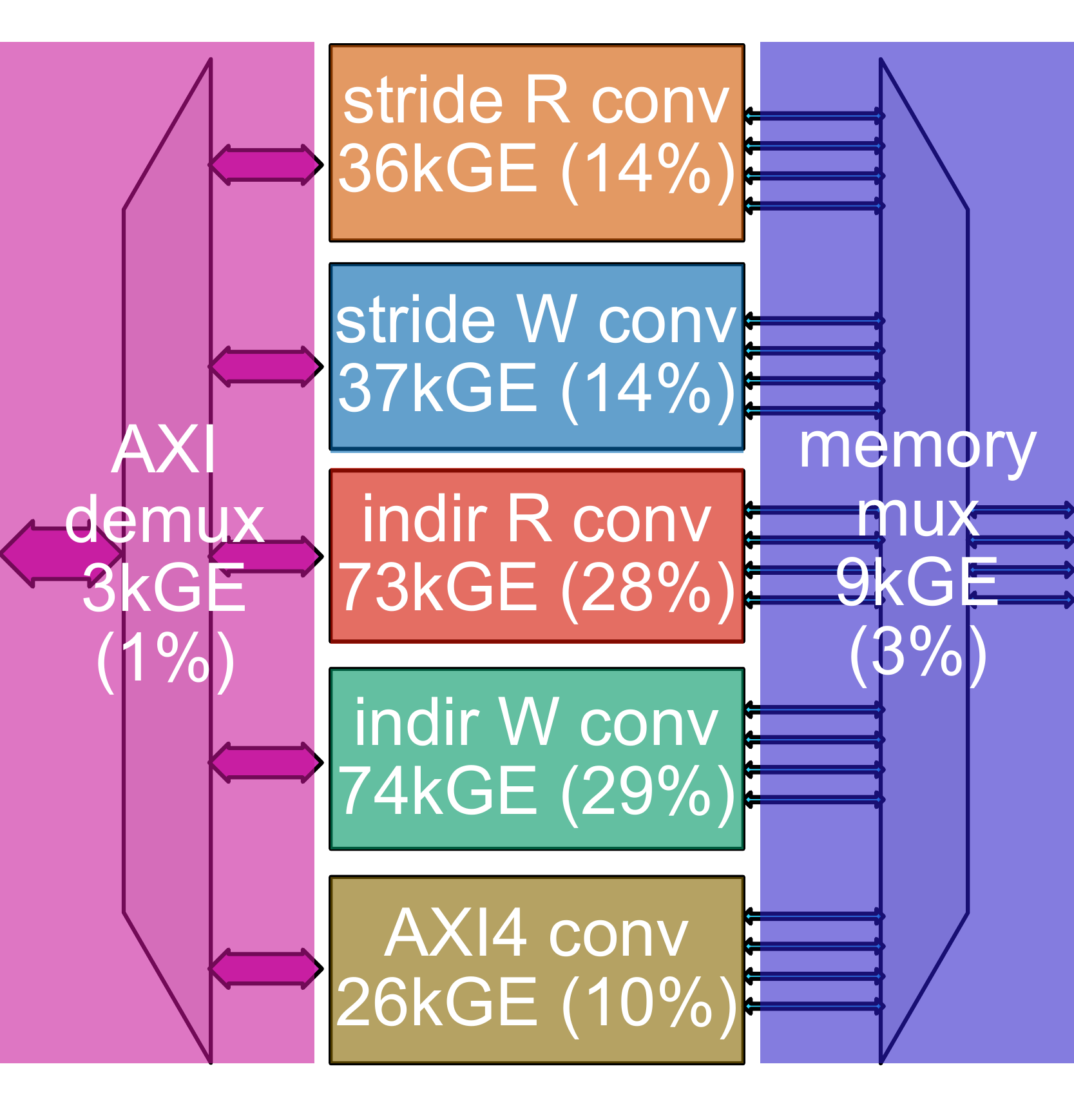}%
    \hspace{0.5em}%
    \label{fig:chiplevel:eval:area_break}%
  }%
  \caption{\ap~\fin{adapter} area and timing results for logic synthesis targeting GlobalFoundries' \SI{22}{\nano\meter} FD-SOI technology.}
  \label{fig:chiplevel:eval:ap_synthesis}
\end{figure}

We synthesize our {\ap} adapter with Synopsys \emph{Design Compiler} for GlobalFoundries' \SI{22}{\nano\meter} FD-SOI technology, targeting the SSG corner (\SI{0.72}{\volt}, \SI{-40}{\celsius}) with low-$V_t$ cells and no back-biasing.
Unless otherwise specified, we constrain a \SI{1}{\giga\hertz} clock and \SI{100}{\pico\second} IO delays and parameterize the decoupling queues to a depth of four.

\cref{fig:chiplevel:eval:a_t} shows the minimum achieved clock period and area for different clock constraints and bus widths of 64, 128, and 256 bit. Our adapter shows good scalability, increasing linearly in area with bus width and incurring 69, 130, and \SI{257}{\kGE} at \SI{1}{\giga\hertz}. Our full 256-bit controller incurs merely \SI{\areaOver}{\percent} of Ara's area, demonstrating that \ap~handling at banked endpoints is reasonably inexpensive.
As we decrease the constrained clock \fin{period}, we see that adapter area scales gracefully past Ara's \SI{1}{\giga\hertz} clock target and reaches minimum periods of 787, 800, and \SI{839}{\pico\second} with only small increases in area.

\cref{fig:chiplevel:eval:area_break} shows a hierarchical area breakdown of the adapter. As expected, the read and write converters are similar in size for both irregular burst types, since they simply reverse each other's datapaths. While the simpler strided converters are only up to \SI{42}{\percent} larger than the base \gls{axi4} converter, the indirect converters are nearly double this size due to their two stages.

\subsection{Parameter Sensitivity}
\label{sec:chiplevel:eval:parsens}

To gain deeper insight into the scaling of \ap~performance and hardware complexity, we investigate the impact of element and index size as well as bank count on read bus utilization and bank crossbar area.
For all performance measurements, we connect our controller to an ideal requestor issuing continuous read requests of length 256 and use random indices.
Unless otherwise specified, parameters default to their \textsc{pack} system configuration, but we increase decoupling queue depths to 32 to avoid bottlenecks unrelated to our analysis.
We consider power-of-two bank counts from 8 to 32, which result in minimal addressing logic, as well as prime counts in this range, which minimize bank conflicts across different strides. We also consider an ideal memory without bank conflicts for reference.

\begin{figure}
\captionsetup[subfigure]{aboveskip=3.3pt, belowskip=-1.8pt}
\centering
  \subfloat[Indirect read utilization.] {
    \label{fig:chiplevel:eval:indirect_snes}
    \includegraphics[width=0.96\linewidth, height=0.48\linewidth]{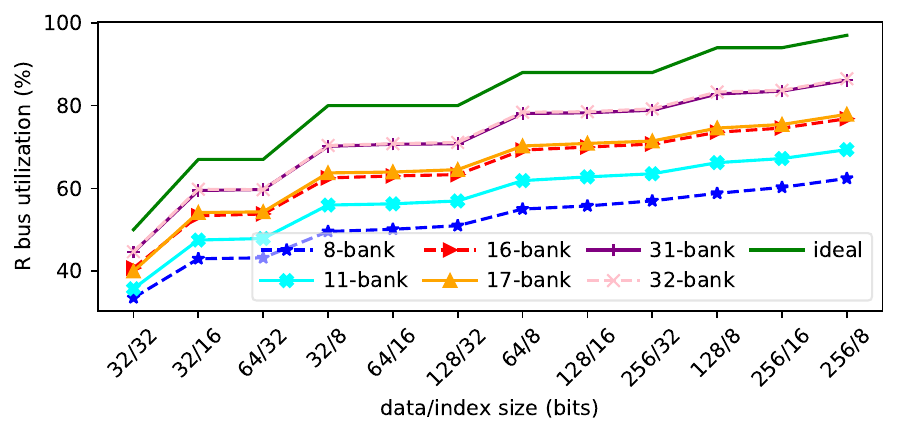}
    } \\ \vspace{1em}
  \subfloat[Strided read utilization.]{
    \label{fig:chiplevel:eval:stride_snes}
    \includegraphics[width=0.45\linewidth, height=0.3375\linewidth]{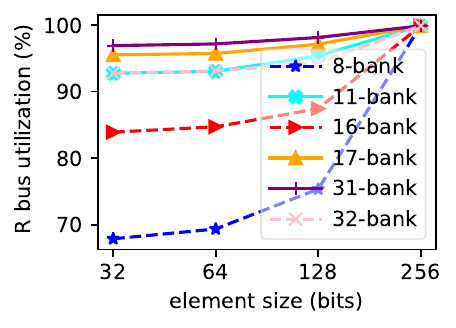}
  }
  \subfloat[Bank crossbar area.]{
    \label{fig:chiplevel:eval:area_tcdm}
    \includegraphics[width=0.45\linewidth, height=0.3375\linewidth]{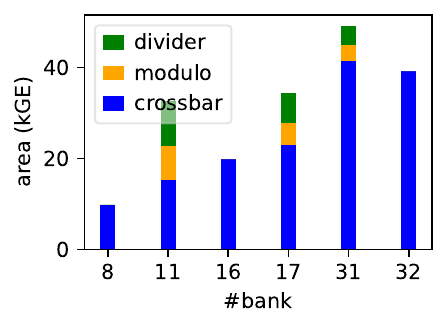}
  }
  \caption{\ap~parameter sensitivity results.}
  \label{fig:chiplevel:eval:ap_sens}
\end{figure}

\topar{Indirect accesses} \Cref{fig:chiplevel:eval:indirect_snes} shows the bus utilization achieved on indirect reads for different element-index size pairs and bank counts. For all size pairs, utilization increases monotonically with bank count as fewer bank conflicts occur. Since indirect bursts involve one contiguous and one random, but no strided bank access sequences, prime bank counts show no inherent advantage here.
Across all bank counts, utilization improves mainly with the \emph{ratio} $r$ of element size to index size,
similar to what we observed for \glspl{sssr} in \cref{chap:sssrs:arch:sssrs}: since we fetch indices as whole bus lines, we must fetch one index line for every $r$ data beats on average, limiting our ideal bus utilization to $\nicefrac{r}{r+1}$. For 32-bit elements and index sizes of 32, 16, and 8 bit, this corresponds to ideal utilizations of 50, 67, and \SI{80}{\percent}, respectively.
Thus, with larger elements or smaller indices, \ap~indirection bus utilizations can further exceed those shown in \cref{sec:chiplevel:eval:res_perf}.

\topar{Strided accesses} \Cref{fig:chiplevel:eval:stride_snes} shows the bus utilization of strided reads for different element sizes and bank counts, averaged across element strides of 0 to 63. As expected, prime bank counts (11, 17, 31) offer significant performance benefits on strided accesses by avoiding \pagebreak[3] repeated bank hit patterns, although more banks further improve performance for both power-of-two and prime bank counts. With increasing element size, conflicts become less likely for all bank counts as there are fewer aligned elements in each bus-wide line.

\topar{Bank crossbar area} \Cref{fig:chiplevel:eval:area_tcdm} shows the bank crossbar's total area for different bank counts, highlighting the overhead prime bank counts incur for modulo and division units to compute bank addresses. Power-of-two-banked crossbars are generally cheaper and prime-banked overheads decrease with increasing bank counts. Since 17 banks provide good area-overhead and area-performance tradeoffs (95\% and 81\% of ideal performance on strided and indirect reads on average), this is the bank count we chose for our evaluation systems.

\section{Related Work}
\label{chap:chiplevel:relwork}

Existing hardware works on efficient strided and indirect streams in chip-level interconnects focus on either end of the memory system.
\emph{Core-side extensions} integrate tightly with processors issuing streams, whereas \emph{memory-side extensions} focus on efficient stream handling in interconnects and at endpoints.

\subsection{Core-Side Extensions}

Core-side extensions relieve processors from load-store and bookkeeping overheads and decouple memory accesses from execution.
When connected to chip-level buses, stream register extensions like \glspl{sssr} fall under this category, as do some of the streaming \gls{isa} extensions we discussed in \cref{chap:sssrs:relwork:sisaext}.
However, the integration of core-side extensions with interconnects and memory endpoints is usually limited, and they do not address bus-level inefficiencies.

\glspl{sssr} (\cref{chap:sssrs}) map strided or indirect streams directly to registers, but they do not support burstive accesses or bus packing.
\mbox{Z. Wang} et al.~\cite{Wang2019StreambasedMA} also present a core-side extension for strided and indirect streams, but they do not explore bus packing or interconnect-level streaming; they merely propose reducing interference between requestors through stream throttling and cache bypassing.
Indirection-capable prefetchers like Prodigy~\cite{Talati2021ProdigyIT} and DMP~\cite{fu2024dmp} focus on improving prefetch accuracy rather than bus efficiency; they do not provide bus packing, although Prodigy does support request throttling and cache bypassing.
The Unlimited Vector Extension (UVE)~\cite{Domingos2021UnlimitedVE} maps strided and indirect streams directly to vector registers, similar to our extensions to Ara in principle; however, it does not provide packed transport or interconnect-level streams, at most dynamically bypassing the L1 and L2 caches.

Except for passive interference mitigations, none of the above extensions consider the interconnect or the remaining memory system:
they focus on the requesting cores and do not provide bus-packing, burstive transport, or near-memory handling of strided and indirect streams.
Luckily, \ap~is orthogonal to all of them: it may be used with any bus width, burst length, or mapping mechanism, providing a reusable end-to-end protocol that carries irregular streams through interconnects and to stream-aware endpoints with high bus efficiency.

\subsection{Memory-Side Extensions}

Memory-side extensions focus on interconnects and memory endpoints to increase bus utilization and hide access latency.
They come in the form of specialized memory controllers, \gls{dlt} accelerators, and \gls{pim} extensions.

\mmp{Corbal et al. propose command vector memory systems~\cite{corbal1998cvms}, in which SDRAM controllers can handle entire strided vector accesses as a single request (a \emph{command}). This improves application performance by effectively hiding precharge and row access times, but it does not accelerate indirect accesses.}
The Impulse memory controller~\cite{Carter1999ImpulseBA} maps irregular streams to virtual memory pages; it provides inherent, on-the-fly bus packing, but relies on carefully managed virtual addressing.
Tanabe et al.~\cite{tanabe2011memctrl} present a memory controller %
that can gather or scatter DRAM data to or from controller-internal vector registers; however, they do not propose a streaming interface to efficiently access these registers through an interconnect.
Hussain et al.~propose pattern-aware memory controllers~\cite{Hussain2018MemoryCF} and systems~\cite{Hussain2017ANH} prefetching irregular stream descriptors to dedicated scratchpads; this enables fast, packed core accesses, but incurs notable complexity overheads.

PLANAR~\cite{Barredo2021PLANARAP} accelerates layout transforms by writing packed, cacheable irregular data to memory ahead of use.
The data rearrangement engine~\cite{Lloyd2015InMemoryDR} is a \gls{dlt} solution integrated directly into a hybrid memory cube architecture.
While \gls{dlt} accelerators are highly bandwidth-efficient, they require the allocation of physical memory buffers and explicit, ahead-of-time invocation to be beneficial.

Lin et al.~\cite{lin2021egdram} present Elastic-Granularity DRAM, a \gls{pim} extension that serves packed narrow requests from memory controllers by individually accessing DRAM chips.
Folded Banks (FB) HBM~\cite{Adhinarayanan2025folded} leverages the 3D design of HBM to redistribute bank subarrays across dies, significantly improving the bandwidth of fine-grained irregular accesses.
\mmp{SparseP~\cite{giannoula2022sparsep} is the first open-source \glsunset{spmv}\gls{spmv} software package for real \gls{pim} architectures, achieving significant performance and energy efficiency gains on an existing UPMEM \gls{pim} system.}
While these in-memory works significantly accelerate irregular DRAM accesses, they do not propose an accompanying protocol to efficiently transport irregular data through the interconnect.

In general, \ap~\emph{supports} the above extensions by providing them with an efficient on-chip protocol:
bus packing can be done on the fly by a protocol adapter or ahead of time by an AXI-Pack-capable \gls{dma} engine,
and our lightweight irregular requests do not preclude the addition of more complex, memory-mapped descriptors.
However, unlike other proposals, \ap~builds on an established protocol and extends irregular streams \emph{throughout} interconnects, feeding directly into cores in an end-to-end fashion and closing the logistical gap between core-side and memory-side extensions.

\section{Conclusion}
\label{chap:chiplevel:conclusion}

In this chapter, we presented \emph{\ap}, an extension to the \gls{axi4} chip-level protocol that enables highly efficient end-to-end strided and indirect memory streams.
\ap~is backward-compatible with many existing \gls{axi4} \glspl{ip} and enables decoupled, bus-packed streams whose semantics are directly encoded into requests, ensuring high performance and flexibility even for short streams.
To demonstrate \ap~in a full system, we extended the Ara RISC-V vector processor to use it for strided and indexed memory accesses and designed a banked memory controller serving its irregular bursts.

We evaluated the performance and energy efficiency benefits of \ap~in the resulting vector processor system through benchmarks involving strided and indirect accesses.
On strided-access benchmarks, {\ap} enables bus utilizations of up to \SI{\strideBu}{\percent}, resulting in speedups of up to {\strideSu}\x~and energy efficiency gains of up to {\strideEEF}\x.
On indirect-access benchmarks, {\ap} reaches bus utilizations of up to \SI{\indirBu}{\percent} and enables speedups and energy efficiency gains of up to {\indirSu}\x~and {\indirEEF}\x, respectively; its memory-side indirection also avoids the loss of up to {\SI{20}{\percent}} of bus time on index traffic.
Our full 256-bit {\ap} protocol adapter incurs only {\areaOver}\% of Ara's area while exceeding its \SI{1}{\giga\hertz} clock target.
We find that {\ap} is complementary to many existing core-side and memory-side extensions, providing them with a chip-level protocol to communicate efficiently through bus-packed streams and combine their performance benefits.

\clearpage
\chapter{Conclusion}
\label{chap:conclusion}

The slowdown of technology scaling and the increasing importance of data-parallel compute-intensive workloads have motivated computer architects to specialize their designs for \emph{highly regular} computation, resulting in wide \gls{simd} datapaths, deep and wide cache hierarchies, and a focus on low control-to-compute ratios.
As this specialization is hitting diminishing returns and compute problems keep growing in complexity, attention is shifting towards \emph{sparse and irregular} workloads.
Unfortunately, these are difficult to accelerate uniformly and lack the regularity modern architectures are optimized for, leading to low performance and energy efficiency on current CPUs and GPUs.

In this thesis, we explored the design and application of \emph{energy-efficient hardware stream extensions} to overcome these shortcomings and accelerate sparse and irregular workloads in general-purpose systems.
We first created \emph{\glspl{sssr}}, a tightly core-coupled extension that accelerates irregular workloads through streaming indirection, intersection, and union while maintaining high regular-workload performance through affine streams.
We then integrated \glspl{sssr} into the energy-efficient Snitch compute cluster and used them to accelerate both general sparse \gls{la} and non-\gls{la} irregular workloads, showcasing the flexibility of our stream-based approach.
To leverage \glspl{sssr} at scale, we created \emph{Occamy}, the first silicon-proven 2.5D-integrated \riscv~manycore that features 48 \gls{sssr}-extended clusters\pagebreak[3] (432 cores) across two chiplets, \SI{32}{\gibi\byte} of HBM2E DRAM, and a fully digital \gls{d2d} link.
Finally, we brought irregular streams to chip-level interconnects with \emph{AXI Pack}, an extension to Arm's widespread \gls{axi4} bus protocol that adds support for end-to-end bus-packed indirect and strided streams and significantly improves irregular workload performance and bus utilization in a RISC-V vector processor system.
At every step, our stream extensions provide not only important performance benefits, but also significant increases in energy efficiency.

In the remainder of this chapter, we provide a structured summary of the main results of this thesis, followed by an outlook that discusses possible directions for future work.

\section{Summary of Main Results}
\label{chap:conclusion:main_results}

We revisit each of our contributions outlined in \cref{chap:introduction} to summarize the results achieved in this thesis.

\subsubsection*{Lightweight, Flexible Core-Coupled Streams}

We first focused our attention on sparse \gls{la}, which is performance-critical to numerous irregular applications, but exhibits low \gls{fpu} utilization and energy efficiency on \gls{soa} processors.
Sparse \gls{la} is difficult to accelerate uniformly due to significant algorithmic differences across operations, formats, sparsity regimes, and dataflows.
Existing CPU, GPU, and accelerator proposals either specialize for a limited subset of scenarios or incur significant architectural cost.

To address these shortcomings, we designed \emph{sparse stream semantic registers} (SSSRs), a lightweight, modular, core-coupled streaming \gls{isa} extension that accelerates \emph{general} sparse \gls{la}.
\glspl{sssr} map streams directly to architectural register accesses, enabling high \gls{fpu} utilization even on single-issue in-order processors.
Through \emph{affine} and \emph{indirect} streams, \glspl{sssr} can accelerate sparse-dense \gls{la} using \emph{any} fiber-based sparse tensor format, which includes the widespread \gls{csr}, \gls{csc}, \gls{csf}, and many of their variations.
Through a shared index comparison unit, two indirection-capable \glspl{sssr} can form the \emph{intersection} and \emph{union} of two sparse tensor axes, enabling streaming sparse-sparse\pagebreak[3] multiplication and addition;
a third \emph{egress} \gls{sssr} can write out the resulting index stream alongside computed data to enable element-wise sparse-sparse operations.
As explored in \cref{chap:workloads}, \glspl{sssr} are a \emph{general-purpose} extension: their flexible indirection and stream-join capabilities can significantly accelerate both sparse \gls{la} and non-\gls{la} irregular workloads.

We integrated \glspl{sssr} into the RISC-V Snitch compute cluster.
Each fully-featured streamer incurs a mere \SI{30}{\kilo\GE} in complexity, of which only \SI{37}{\percent} are dedicated to irregular streams.
Adding \gls{sssr} streamers to each worker core incurs a negligible \SI{1.8}{\percent} in additional cluster area over affine streamers and does not degrade cluster timing.

\subsubsection*{General Sparse Linear Algebra Acceleration}

To demonstrate the performance and energy efficiency benefits of \glspl{sssr} on sparse~\gls{la}, we evaluated them on various sparse-dense and sparse-sparse \gls{la} operations. %
We created a library of baseline RISC-V (RV32G) and \gls{sssr}-accelerated single-core kernels, as well as parallel %
cluster implementations of sparse-dense (\texttt{sM\x dV}) and sparse-sparse (\texttt{sM\x sV}) matrix-vector multiplication.
We ran our code in cycle-accurate \gls{rtl} simulation, with cluster-level simulations using realistic interconnect and HBM2E DRAM models.
We also estimated cluster power consumption in GlobalFoundries' \SI{12}{\nano\meter} LP+ technology.

For a single core using 16-bit indices, \gls{sssr} indirection enables \gls{fpu} utilizations of up to \SI{80}{\percent} on \verb|sV×dV| and speedups of up to 7.0\x~on \verb|sM×dV|.
Across a wide sparsity range, \gls{sssr} intersection and union enable speedups of $3.0$\,--\,$7.7$\x~on sparse-sparse dot products and $5.4$\,--\,$9.8$\x~on sparse-sparse vector addition, as well as \verb|sM×sV| speedups of up to 6.3\x.
On an eight-core cluster, \verb|sM×dV| is up to 4.9\x~faster and 2.9\x~more energy efficient, reaching overall \gls{fpu} utilizations of up to \SI{47}{\percent}.
Cluster-level \verb|sM×sV| with a vector density of \SI{1}{\percent} is up to 5.9\x~faster and 3.0\x~more energy efficient.

Unlike \mmp{many} recently proposed CPU, GPU, and accelerator solutions, \glspl{sssr} accelerate one- \emph{and} two-sided sparse \gls{la} across a wide sparsity range \mmp{\fin{at minimal area} cost and} without imposing a specific dataflow.
A Snitch cluster with \glspl{sssr} running \verb|sM×dV| achieves 9.9\x~and 1.7\x~higher peak floating-point utilizations than recent CPU and GPU software solutions.

\subsubsection*{Irregular \fin{Workload Acceleration} Beyond Linear Algebra}

To demonstrate the generality and flexibility of \glspl{sssr}, we used them to accelerate three irregular workloads on the Snitch cluster whose performance-critical compute goes \emph{beyond} sparse \gls{la}: %
\emph{stencil codes}, \emph{spiking neural networks} (SNNs), and \emph{model predictive control} (MPC).
For each, we used a cluster without \glspl{sssr} as a baseline.

For stencil codes, we created a generic methodology called \emph{SARIS} to map the memory access patterns of stencils to streaming indirections.
Using SARIS, \glspl{sssr} enable geomean speedups and energy efficiency improvements of 2.7\x~and 1.6\x, respectively, and increase geomean \gls{fpu} utilization from \SI{35}{\percent} to \SI{81}{\percent}.
On our Occamy manycore system, \glspl{sssr}-accelerated stencil codes achieve up to \SI{83}{\percent} of theoretical peak performance, \SI{20}{\percent} more than a leading code generator on GPUs.

To accelerate \glspl{snn}, we accelerated the sparse reduction of spiking weights with \glspl{sssr}.
This enables \gls{sssr} speedups and energy efficiency gains of \nchk{5.5\x} and \nchk{3.3\x} on FP16 S-VGG11 inference, respectively, increasing \gls{fpu} utilization from \nchk{\SI{9.3}{\percent}} to \nchk{\SI{52}{\percent}}.
FP8 inference further improves both runtime and energy efficiency by \nchk{1.7\x}.
We find that FP8 \gls{snn} inference on the \gls{sssr}-extended cluster is \nchk{4.4\x} more energy efficient than on the leading neuromorphic processor \emph{LSMCore}.

To accelerate \gls{mpc}, we sped up a performance-critical sparse solver step.
We designed \emph{ParSPL}, an optimized solver that reduces internal data dependencies and increases stream lengths.
With \emph{ParSPL}, \glspl{sssr} enable speedups and energy efficiency gains of up to 4.6\x~and 2.2\x, respectively, and raise the peak \gls{fpu} utilization from \SI{11}{\percent} to \SI{57}{\percent}.
On a 9\x9 manycore \gls{etm} problem, ParSPL and \glspl{sssr} together accelerate \gls{mpc} by 12.7\x, enabling online control;
the resulting \gls{etm} solution is more area- and energy-efficient than prior \gls{mpc}-based \gls{etm} proposals and competitive with \mmp{\gls{soa} \gls{mpc} approaches focused on performance.}

\subsubsection*{A Manycore for Efficient Mixed-Regularity Compute}

After establishing the performance benefits and flexibility of \glspl{sssr} across sparse and irregular workloads, we integrated them in a large-scale system:
we presented \emph{Occamy}, a silicon-proven 2.5D-integrated RISC-V manycore featuring 48 \gls{sssr}-extended Snitch clusters, or 432 cores, across two \SI{12}{\nano\meter} FinFET chiplets.
Each chiplet features a hierarchical \gls{axi4}-based interconnect and \SI{16}{\gibi\byte} of HBM2E DRAM.
A fully-digital \gls{d2d} link connects both chiplets through a \SI{65}{\nano\meter} passive interposer called \emph{Hedwig}.
We use a tiled programming model for Occamy: while the \gls{sssr}-extended worker cores process data in local TCDM, the DMA engines in each cluster simultaneously transfer double-buffered tiles between their TCDM and external memory.

Occamy tackles the shortcomings of existing CPU, GPU, and accelerator designs by efficiently handling general sparse and irregular workloads while maintaining high regular compute efficiency.
On regular workloads, affine \gls{sssr} streams enable competitive FPU utilizations of up to \cn{\SI{89}{\percent}} and an energy efficiency of up to \cn{\SI{39.8}{DP\text{-}\giga\flop\per\second\per\watt}}.
On stencil codes, Occamy reaches an FPU utilization of up to \cn{\SI{83}{\percent}} and a node-normalized compute density of up to \cn{\SI{11.1}{DP\text{-}\giga\flop\per\second\per\milli\meter^2}}, improving on \gls{soa} processors by \cn{1.7\x}~and \rev{\cn{1.2\x}}, respectively.
On sparse-dense \gls{la}, we achieve an FPU utilization of up to \cn{\SI{42}{\percent}} and a node-normalized compute density of up to \SI{5.95}{DP\text{-}\giga\flop\per\second\per\milli\meter^2}, resulting in substantial gains of \cn{5.2\x}~and \cn{11\x} over the \gls{soa}.
We \rev{further} achieve a sparse-sparse \gls{la} throughput of up to \cn{\SI{187}{\giga\comp\per\second}} at \cn{\SI{17.4}{\giga\comp\per\second\per\watt}}, \cn{\SI{3.63}{\giga\comp\per\second\per\milli\meter^2}}, and a comparator utilization of up to \cn{\SI{49}{\percent}}.
Finally, we reach FPU utilizations of up to \SI{75}{\percent} on \gls{gpt}-J inference and up to \SI{54}{\percent} on a \gls{gcn} layer.

To the best of our knowledge, Occamy is the first open-source multi-chiplet RISC-V manycore demonstrated in silicon.
The \gls{rtl} description of its digital core, including the compute clusters, hierarchical interconnect, and \gls{d2d} link, is freely available under a permissive open-source license.

\subsubsection*{\fin{Packed} Irregular Streams in Chip-Level Interconnects}

In the Snitch cluster with \glspl{sssr}, irregular streams are confined to the cluster \gls{tcdm}.
While this ensures high efficiency for tiled workloads, irregular applications with very large and sparse datasets are often difficult to partition efficiently, incurring significant overheads.
To address this, we turned our attention to end-to-end irregular streams in \emph{chip-level} interconnects, which can address global memory endpoints.
We created \emph{AXI Pack}, an extension to Arm's widespread \gls{axi4} on-chip protocol that provides new burst types for strided and indirect streams, packing narrow elements onto wide chip-level buses.

\ap~is backward-compatible with non-burst-reshaping \gls{axi4} \glspl{ip} and encodes stream semantics directly into burst requests, ensuring high performance and flexibility.
To evaluate the costs and benefits of \ap~in a full system, we extended a RISC-V vector processor to use it for strided and indexed memory accesses and created a banked memory controller capable of serving its packed streams.
On benchmarks involving strided accesses, {\ap} enables bus utilizations of up to \SI{\strideBu}{\percent}, providing speedups and energy efficiency gains of up to {\strideSu}\x~and {\strideEEF}\x, respectively.
On benchmarks involving indirect accesses, {\ap} reaches bus utilizations of up to \SI{\indirBu}{\percent} and enables speedups and energy efficiency gains of up to {\indirSu}\x~and {\indirEEF}\x, respectively.
By handling indirections at the memory endpoint, it further avoids using up to {\SI{20}{\percent}} of bus time on index traffic.
A 256-bit-wide {\ap} protocol adapter incurs only {\areaOver}\% of our vector processor's area while exceeding its \SI{1}{\giga\hertz} clock target.

Unlike existing approaches to chip-level irregular streams, which focus on either the core side or the memory side of the interconnect, we provide an \emph{end-to-end} solution ensuring high bus efficiency throughout.
In fact, {\ap} as a protocol is largely \emph{complementary} to existing core-side and memory-side extensions, providing them with a way to \fin{communicate efficiently} and ideally combine their performance benefits.

\section{Outlook and Future Work}
\label{chap:conclusion:fwork_outlook}

In this thesis, we focused on creating energy-efficient, \emph{general-purpose} stream extensions that provide a consistent approach to accelerating sparse and irregular workloads.
We embraced fiber-based formats with common 8-to-64-bit index types, prioritizing flexibility and scalability over leading performance in low- and extreme-sparsity settings.
We also made few prescriptions about target applications or host systems to keep our proposed extensions portable.

However, this leaves ample room for future work, especially on combining our solutions with specialized approaches \mmp{and on scaling chip-level irregular streams to larger systems.}
In the following, we outline some possible directions for further research.

\paragraph*{Improved \gls{sssr} Handling of Narrow Datatypes}
Unless a contiguous or \gls{simd}-compatible blocked format is streamed, \glspl{sssr} can only stream one data element each cycle, even if they are narrower than the memory bus.
However, narrow datatypes are becoming increasingly important, especially for \gls{ml} inference.
Thus, one could improve bandwidth utilization on narrow datatype streaming through bank parallelism, coalescing, or a two-level fiber-based format wherein each element is a structured-sparse block.

\paragraph*{Orthogonal Structured Sparsity Support}
Structured $N$:$M$ formats, particularly 2:4, are finding mainstream adoption for \gls{ml} inference on GPUs as they are easy to handle.
Since the indexing range of $N$:$M$ formats is extremely limited, they can be handled at the \emph{register} level inside \gls{simd} or vector functional units, remaining orthogonal to memory streaming extensions.
In fact, combining register-level $N$:$M$ support with indexed memory streams of $N$:$M$ blocks could be one approach to support the two-level sparse format described above.

\paragraph*{Cache-Aware Chip-Level Streams}
The AXI Pack chip-level streams we presented could make use of cacheability flags to communicate where and how they should be cached.
For instance, at each cache hierarchy level, they could be cached in packed or scattered form, only as their indices, or they could bypass specific cache levels altogether as explored by other chip-level extensions.
Devising appropriate caching policies and mechanisms for specific target architectures could significantly benefit the effectiveness of these streams.

\paragraph*{DRAM Coalescing and Processing in Memory}
While we shortly touched on using chip-level streams with wide, coalescing DRAM endpoints, there is much left to explore on this topic.
For instance, it is unclear whether narrow-access coalescing is better performed at memory controllers or at last-level caches, and how it interacts with caching.
One could also explore whether \gls{pim} solutions that leverage DRAM bank parallelism, like those we discussed in \cref{chap:chiplevel:relwork}, could help reduce or eliminate the need for coalescing.

\paragraph*{Multi-Endpoint Chip-Level Streams} Our proposed chip-level irregular streams seamlessly support multiple requestors, but can only fetch elements from one endpoint per stream.
This limitation can be avoided through multi-port memory controllers and careful stream sanitization, but such solutions do not scale endlessly.
To support packed irregular streams across disaggregated endpoints, narrow accesses must be resolved and combined per endpoint at the requestor, increasing design complexity and bringing new efficiency challenges.

\paragraph*{Irregular Streams in \glspl{noc} and Across Chiplets} We demonstrated packed irregular streams in noncoherent crossbar-based \gls{axi4} interconnects, but these exhibit limited scalability in the number of requestors and endpoints.
To scale further, we could bring irregular streams to the coherent \glspl{noc} of application-class manycores by extending Arm's \fin{\glsunset{chi}\emph{Coherent Hub Interface} (CHI)}~\cite{Arm2021AMBACHI}.
As a coherent protocol, \gls{chi} brings extensive cache management capabilities we could leverage for the targeted caching of streams we outlined earlier.
It also defines its own \gls{c2c} architecture~\cite{Arm2021AMBACHIC2C}, which would allow us to explore system-level irregular streams across chiplets in large 2.5D manycores.

\appendix
\crefalias{chapter}{appendix}
\crefalias{section}{subappendix}
\crefalias{subsection}{subsubappendix}

\clearpage
\chapter{Chip Gallery}
\label{chap:chip_gallery}

This appendix lists all fabricated chips the author
has contributed to in the form of technical work or supervision, \mmp{in chronological order of their tape-out}.
A complete, up-to-date list can be found online at:
\url{http://asic.ethz.ch/authors/Paul_Scheffler.html}.

Where available, we provide both a render of the chip layout and a photograph of the chip die. The renders are generated using the \emph{ArtistIC} toolchain~\cite{benz2025artistic}, which the author has also contributed to.

\clearpage
\section{Xavier}
\label{apdx:cg:xavier}

\begin{figure}[H]
    \centering
    \includegraphics[width=147.545pt,height=147.54861pt]{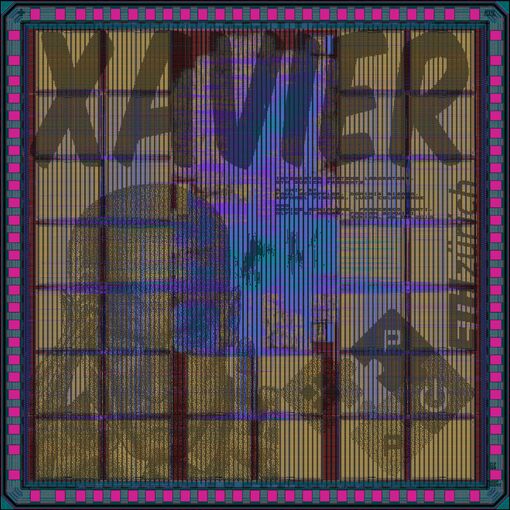}\hspace{0.5em}%
    \includegraphics[width=147.52956pt,height=147.54861pt]{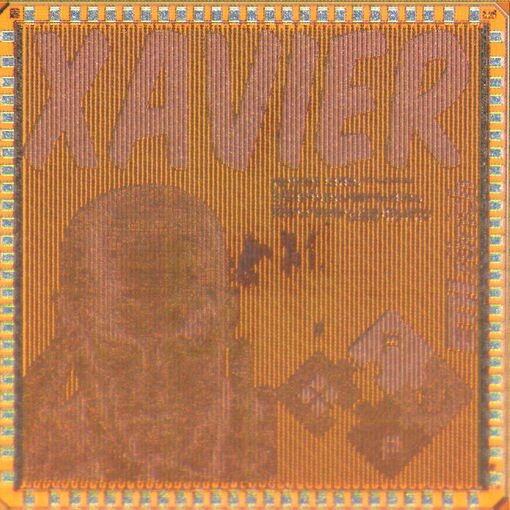}
\end{figure}

\begin{table}[H]
    \centering
    \small
    \begin{tabularx}{\linewidth}{@{}>{\bfseries}lX@{}}
        \toprule
        Designers                 & Paul Scheffler, Dominik Bisang, \\
                                  & Raphael Fischer, Luca Colagrande, \\
                                  & Davide Schiavone, Francesco Conti,\\
                                  & Alfio Di Mauro, Fabian Schuiki \\
        \midrule
        Application               & Neural recording \& processing \\
        Technology \& Package     & UMC \SI{65}{\nano\meter}, QFN64 \\
        Size \& Complexity        & 2.63~\x~\SI{2.63}{\milli\meter}, \SI{3.3}{\mega\GE} \\
        Clock Speed \& Voltage    & \SI{250}{\mega\hertz}, \SI{1.2}{\volt} \\
        \bottomrule
    \end{tabularx}
\end{table}

Xavier is a PULPissimo-based system extended for neural recording and subsequent quantized \fin{\gls{ml}} inference.
It features a RI5CY (RV32IMFC) core, \SI{512}{\kibi\byte} of \gls{spm}, a uDMA subsystem with 8 SPI ports, and a hardware accelerator for quantized neural networks (\SI{4}{\bit} weights, \SI{4}{\bit} inputs) that reaches up to \SI{4.81}{\giga MAC\per\second}.
The uDMA subsystem provides integrated filtering during data transfers for IIR, template matching, and spike detection.
The HWPE-based accelerator uses a programmable micro-riscy core (RV32EMC) as its controller.

\clearpage
\section{Thestral}
\label{apdx:cg:thestral}

\begin{figure}[H]
    \centering
    \includegraphics[width=143.02336pt,height=143.01111pt]{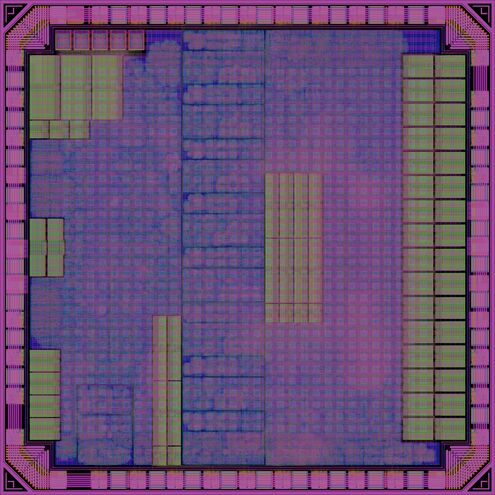}\hspace{0.5em}%
    \includegraphics[width=151.83351pt,height=143.01111pt]{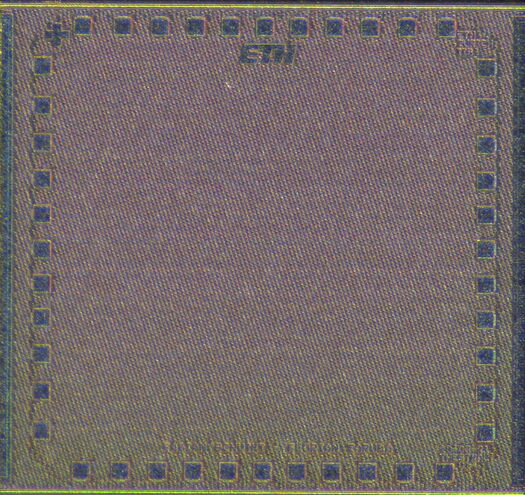}
\end{figure}

\begin{table}[H]
    \centering
    \small
    \begin{tabularx}{\linewidth}{@{}>{\bfseries}lX@{}}
        \toprule
        Designers                 & Fabian Schuiki, Florian Zaruba, \\
                                  & Thomas Benz, Paul Scheffler, \\
                                  & Wolfgang R\"{o}nninger~\cite{benz2021thestral} \\
        \midrule
        Application               & Fine-grain power gating \\
        Technology \& Package     & GF \SI{22}{\nano\meter} (22FDX), QFN40 \\
        Size \& Complexity        & 1.25~\x~\SI{1.25}{\milli\meter}, \SI{6}{\mega\GE} \\
        Clock Speed \& Voltage    & \SI{650}{\mega\hertz}, \SI{0.8}{\volt} \\
        \bottomrule
    \end{tabularx}
\end{table}

Thestral is a Snitch cluster test chip exploring fine-grain power gating.
It demonstrates a number of new Snitch cluster features in silicon including improved prefetching, \glspl{ipu}, SV32 virtual memory support, and \glspl{fpu} clocked twice as fast as the surrounding core logic.
The eight-core FP32 cluster with \SI{64}{\kibi\byte} of \gls{spm} is managed by an always-on \emph{governor} core with its own \gls{fpu}, \gls{ipu}, and private instruction caches.
Each \gls{fpu}, \gls{ipu}, and the remaining compute cluster are in their own power domains, resulting in a total \mmp{of} 20 fine-grain domains capable of ultra-fast (sub-\SI{10}{\nano\second}) power gating.
Peripherals include a Hyperbus interface, a digital \gls{c2c} link, a USB device interface, and an I2C host.

\clearpage
\section{Dogeram}
\label{apdx:cg:dogeram}

\begin{figure}[H]
    \centering
    \includegraphics[width=213.40034pt,height=106.69028pt]{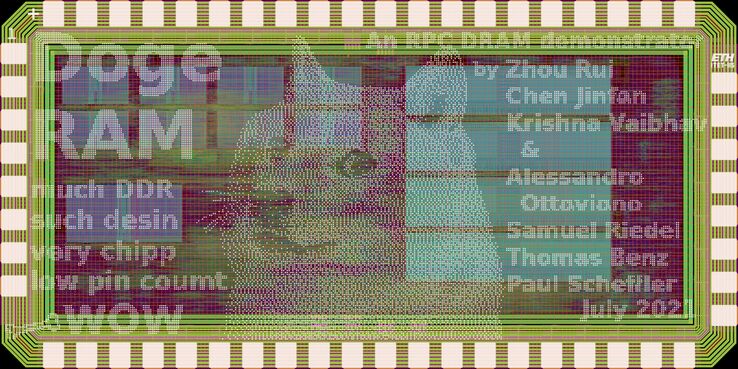}\\\vspace{0.3em}
    \includegraphics[width=213.49147pt,height=108.9625pt]{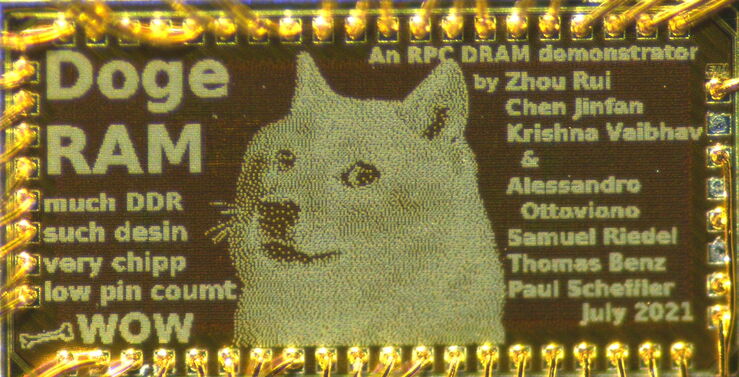}
\end{figure}

\begin{table}[H]
    \centering
    \small
    \begin{tabularx}{\linewidth}{@{}>{\bfseries}lX@{}}
        \toprule
        Designers                 & Rui Zhou, Jinfan Chan, Vaibhav Krishna, \\                          & Alessandro Ottaviano, Samuel Riedel, \\
                                  & Thomas Benz, Paul Scheffler \\
        \midrule
        Application               & Low-pin-count DRAM interface \\
        Technology \& Package     & TSMC \SI{65}{\nano\meter}, QFN56 \\
        Size \& Complexity        & 2.0~\x~\SI{1.0}{\milli\meter}, \SI{700}{\kilo\GE} \\
        Clock Speed \& Voltage    & \SI{200}{\mega\hertz}, \SI{1.2}{\volt} \\
        \bottomrule
    \end{tabularx}
\end{table}

Dogeram was created to test a new, fully digital \gls{rpc} \gls{dram} interface implementation.
The interface is connected to a minimal Snitch-based \gls{soc} over an \gls{axi4} crossbar.
At \SI{200}{\mega\hertz} under worst case conditions, the chip \mmp{achieves} up to \SI{752}{\mega\byte\per\second} throughput over the \gls{rpc} \gls{dram} interface.
The name references the \emph{Doge} meme, with the logo depicting the image it is based on.

\pagebreak
\section{Zest}
\label{apdx:cg:zest}

\begin{figure}[H]
    \centering
    \includegraphics[width=149.5139pt,height=112.13528pt]{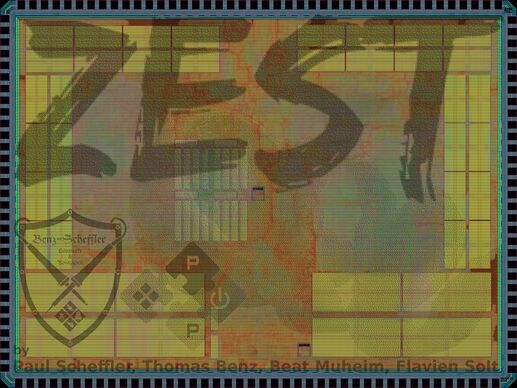}\hspace{0.5em}%
    \includegraphics[width=146.95694pt,height=112.13528pt]{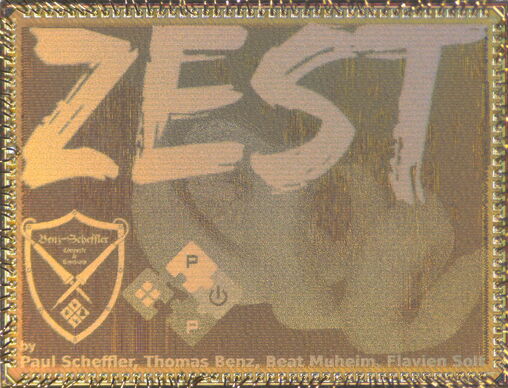}
\end{figure}

\begin{table}[H]
    \centering
    \small
    \begin{tabularx}{\linewidth}{@{}>{\bfseries}lX@{}}
        \toprule
        Designers                 & Flavien Solt, Thomas Benz, \\
                                  & Paul Scheffler \\
        \midrule
        Application               & Scalable peripheral subsystem \\
        Technology \& Package     & TSMC \SI{65}{\nano\meter}, QFN88 \\
        Size \& Complexity        & 4.0~\x~\SI{3.0}{\milli\meter}, \SI{6}{\mega\GE} \\
        Clock Speed \& Voltage    & \SI{230}{\mega\hertz}, \SI{1.2}{\volt} \\
        \bottomrule
    \end{tabularx}
\end{table}

Zest is the first PULP SoC that features both Snitch and PULP clusters in a common architecture to accelerate both integer and floating-point compute efficiently.
The four-core Snitch cluster features the first silicon implementation of \glspl{issr}.
Zest demonstrates the Fenrir peripheral system, which provides a standard specification for simple peripherals and ensures their scalability through flexible interconnect pipelining.
IO interfaces include UART, I2C, QSPI, CPI, and DVSI \fin{for event cameras}, as well as a custom \gls{c2c} interface and Hyperbus.

\clearpage
\section{Neo}
\label{apdx:cg:neo}

\begin{figure}[H]
    \centering
    \includegraphics[width=146.7275pt,height=91.70612pt]{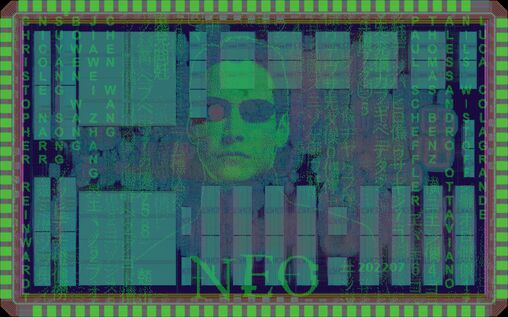}\hspace{0.5em}%
    \includegraphics[width=145.53964pt,height=91.70612pt]{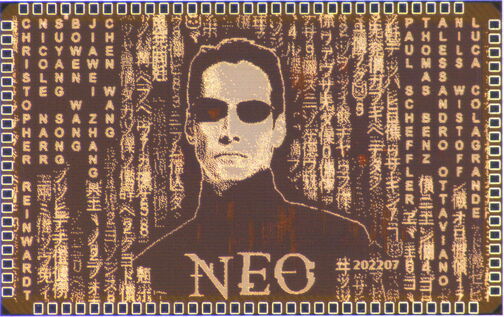}
\end{figure}

\begin{table}[H]
    \centering
    \small
    \begin{tabularx}{\linewidth}{@{}>{\bfseries}lX@{}}
        \toprule
        Designers                 & Christopher Reinwardt, Nicole Narr, \\
                                  & Suyong Song, Bowen Wang, Jiawei Zhang, \\
                                  & Chen Wang, Paul Scheffler, Thomas Benz, \\
                                  & Alessandro Ottaviano, Nils Wistoff, \\
                                  & Luca Colagrande~\cite{ottaviano2023cheshire} \\
        \midrule
        Application               & Linux-capable RISC-V \gls{soc} \\
        Technology \& Package     & TSMC \SI{65}{\nano\meter}, QFN64 \\
        Size \& Complexity        & 3.2~\x~\SI{2.0}{\milli\meter}, \SI{2.5}{\mega\GE} \\
        Clock Speed \& Voltage    & \SI{325}{\mega\hertz}, \SI{1.2}{\volt} \\
        \bottomrule
    \end{tabularx}
\end{table}

Neo is the original silicon implementation of our \emph{Cheshire} \gls{soc} platform, which is built around the 64-bit CVA6 core (RV64GC) and provides an autonomously \mmp{booting,} Linux-capable host to systems that need one.
This includes minimal Linux-capable \glspl{soc} like \emph{Basilisk} (\cref{apdx:cg:basilisk}) and larger heterogeneous systems like \emph{Carfield} (\cref{apdx:cg:carfield}), both of which are based on Cheshire.
Neo couples Cheshire to an improved version of the \gls{rpc} DRAM interface from Dogeram (\cref{apdx:cg:dogeram}).
It is configured with \SI{128}{\kibi\byte} of \gls{spm} and numerous peripherals including UART, I2C, QSPI, a VGA video output, and a fully digital \gls{c2c} link.

\clearpage
\section{Occamy}
\label{apdx:cg:occamy}

\begin{figure}[H]
    \centering
    \begin{minipage}{0.38\textwidth}
        \includegraphics[width=113.46368pt,height=75.11844pt]{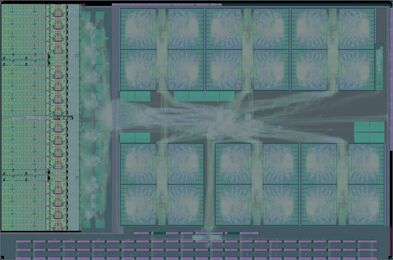}\\%
        \includegraphics[width=113.46368pt,height=74.94351pt]{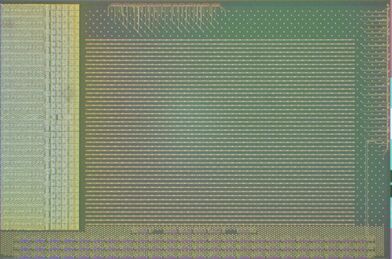}
    \end{minipage}
    \begin{minipage}{0.579\textwidth}
        \includegraphics[width=172.87906pt,height=151.19897pt]{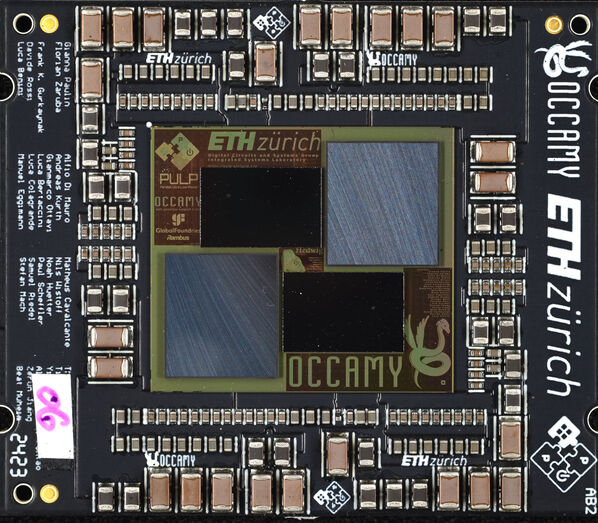}
    \end{minipage}
    \vspace{-0.25em}
\end{figure}

\begin{table}[H]
    \centering
    \small
    \begin{tabularx}{\linewidth}{@{}>{\bfseries}lX@{}}
        \toprule
        Designers           & \scalebox{.97}[1.0]{Gianna Paulin, Paul Scheffler, Thomas Benz,} \\
                            & Florian Zaruba, Stefan Mach, Manuel \\
                            & \scalebox{.93}[1.0]{Eggimann, Matheus Cavalcante, Yichao Zhang,} \\
                            & Tim Fischer, Nils Wistoff, Luca Bertaccini, \\
                            & Luca Colagrande, Alfio Di Mauro, Andreas \\
                            & Kurth, Samuel Riedel, Noah H\"{u}tter, \\
                            & Gianmarco Ottavi, Zerun Jiang, Beat \\
                            & \scalebox{.95}[1.0]{Muheim, Frank K. Gurkaynak, Davide Rossi,} \\
                            & Luca Benini~\cite{scheffler2025occamy} \\
        \midrule
        Application               & Efficient general-purpose acceleration \\
                                  & of both regular \& irregular workloads \\
        Technology \& Package     & GF \SI{12}{\nano\meter} (12LP+), custom \\
        Size \& Complexity        & 10.5~\x~\SI{6.95}{\milli\meter}, \SI{600}{\mega\GE} (chiplet)\\
        Clock Speed \& Voltage    & \SI{1}{\giga\hertz}, \SI{0.8}{\volt} \\
        \bottomrule
    \end{tabularx}
\end{table}

Occamy is a dual-chiplet 2.5D RISC-V system with 432 cores, \SI{32}{\gibi\byte} of HBM2E DRAM, and a peak performance of \SI{768}{DP\text{-}\giga\flop\per\second}. It scales up our \gls{sssr}-extended Snitch cluster and is designed to efficiently handle both regular and general irregular compute. We present and evaluate Occamy in detail in \cref{chap:occamy} of this thesis.

\clearpage
\section{Carfield}
\label{apdx:cg:carfield}

\begin{figure}[H]
    \centering
    \includegraphics[width=116.76881pt,height=120.30972pt]{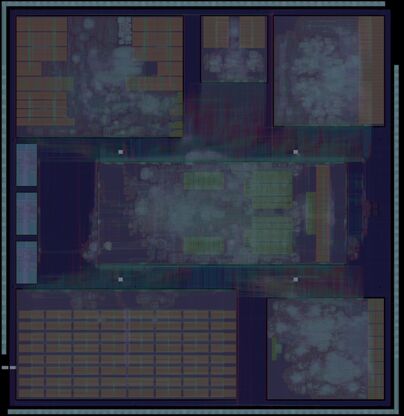}\hspace{0.5em}%
    \includegraphics[width=154.95259pt,height=120.30972pt]{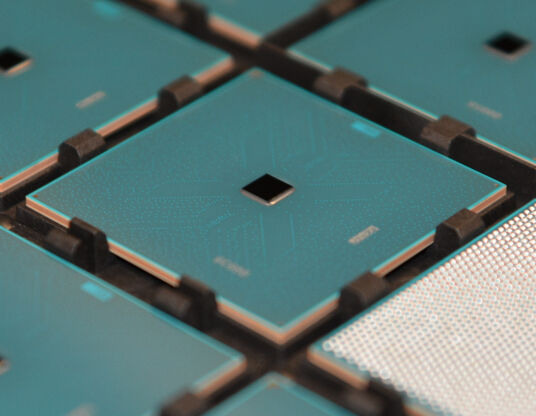}
\end{figure}
\vspace{-0.5em}
\begin{table}[H]
    \centering
    \small
    \begin{tabularx}{\linewidth}{@{}>{\bfseries}lX@{}}
        \toprule
        Designers               & Angelo Garofalo, Alessandro Ottaviano, \\
                                & Matteo Perotti, Robert Balas, Yvan \\
                                & \scalebox{.97}[1.0]{Tortorella, Michael Rogenmoser, Chi Zhang,}\\
                                & Luca Bertaccini, Nils Wistoff, Maicol \\
                                & Ciani, Thomas Benz, Cyril Koenig, Luca \\
                                & Valente, Mattia Sinigaglia, Paul Scheffler,\\ & Manuel Eggimann, Matheus Cavalcante, \\
                                & Beat Muheim, Zerun Jiang, Davide Rossi, \\
                                & Frank K. Gurkaynak, Luca Benini~\cite{garofalo2025carfield} \\
        \midrule
        Application               & Mixed-criticality systems \\
        Technology \& Package     & Intel \SI{16}{\nano\meter}, BGA1733 \\
        Size \& Complexity        & 4.09~\x~\SI{3.93}{\milli\meter}, \SI{70}{\mega\GE}\\
        Clock Speed \& Voltage    & \SI{600}{\mega\hertz}, \SI{0.8}{\volt} \\
        \bottomrule
    \end{tabularx}
\end{table}

Carfield is a heterogeneous mixed-criticality research \gls{soc} for safe, resilient, and time-predictable systems.
It features hardware solutions for time-predictable communication, fault recovery, secure boot, cryptographic acceleration, virtualization, and two acceleration clusters for floating-point and (quantized) integer workloads, respectively.
Its Cheshire-based host domain features two memory-consistent CVA6 cores and hypervisor support, while its MCU-like safety island employs triple-redundant cores running in lockstep.
A secure domain based on the \emph{OpenTitan} project provides a hardware root of trust.

\clearpage
\section{Iguana}
\label{apdx:cg:iguana}

\begin{figure}[H]
    \centering
    \includegraphics[width=142.98991pt,height=143.01111pt]{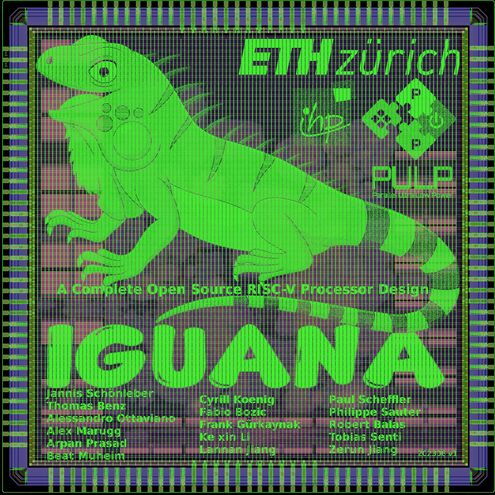}\hspace{0.5em}%
    \includegraphics[width=143.22252pt,height=143.01111pt]{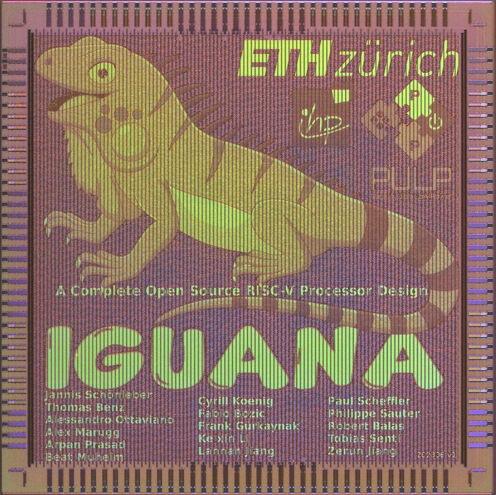}
\end{figure}

\begin{table}[H]
    \centering
    \small
    \begin{tabularx}{\linewidth}{@{}>{\bfseries}lX@{}}
        \toprule
        Designers                 & Jannis Sch\"{o}nleber, Thomas Benz, \\
                                  & Alessandro Ottaviano, Alex Marugg, \\
                                  & Arpan Prasad, Beat Muheim, Cyril Koenig, \\
                                  & \scalebox{.97}[1.0]{Fabio Bozic, Frank K. Gurkaynak, Kexin Li,} \\
                                  & Lannan Jiang, Paul Scheffler, Philippe \\
                                  & Sauter, Robert Balas, Tobias Senti, \\
                                  & Zerun Jiang~\cite{benz2023iguana} \\
        \midrule
        Application               & Linux-capable \gls{soc} in open PDK \\
        Technology \& Package     & IHP \SI{130}{\nano\meter}, QFN88 \\
        Size \& Complexity        & 6.26~\x~\SI{6.26}{\milli\meter}, \SI{3}{\mega\GE}\\
        Clock Speed \& Voltage    & \SI{60}{\mega\hertz}, \SI{1.2}{\volt} \\
        \bottomrule
    \end{tabularx}
\end{table}

Iguana is a minimal Linux-capable RISC-V \gls{soc} based on Cheshire and implemented in IHP's open \SI{130}{\nano\meter} BiCMOS technology.
It features Cheshire's default set of peripherals including UART, I2C, QSPI, VGA, and a fully-digital \gls{c2c} link, as well as a HyperRAM DRAM interface.
Iguana's open-source \gls{rtl} description was implemented using an open PDK and open-source standard cells.
We planned to use only open-source EDA tools for implementation, but could not resolve all tool-related issues in time, falling back to commercial tools.

\pagebreak
\section{Basilisk}
\label{apdx:cg:basilisk}

\begin{figure}[H]
    \centering
    \includegraphics[width=146.97086pt,height=129.39166pt]{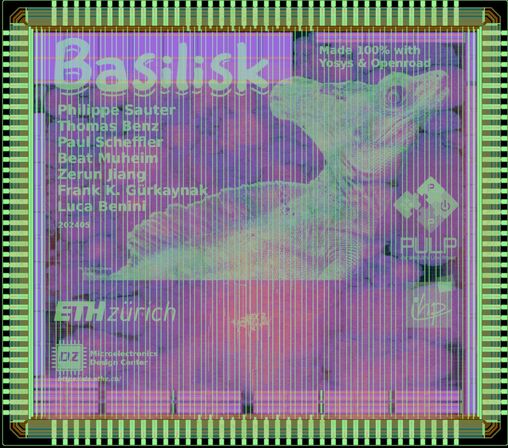}\hspace{0.5em}%
    \includegraphics[width=147.58206pt,height=129.39166pt]{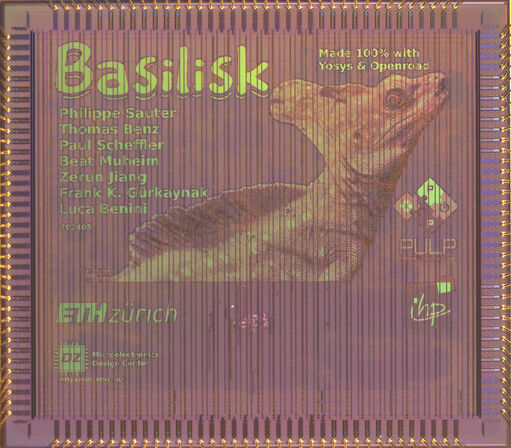}
\end{figure}

\begin{table}[H]
    \centering
    \small
    \begin{tabularx}{\linewidth}{@{}>{\bfseries}lX@{}}
        \toprule
        Designers                 & Philippe Sauter, Thomas Benz, \\
                                  & Paul Scheffler, Beat Muheim, \\
                                  & Zerun Jiang, Frank K. Gurkaynak, \\
                                  & Luca Benini~\cite{sauter2024insbas} \\
        \midrule
        Application               & End-to-end open-source Linux-capable \gls{soc} \\
        Technology \& Package     & IHP \SI{130}{\nano\meter}, QFN88 \\
        Size \& Complexity        & 6.26~\x~\SI{5.50}{\milli\meter}, \SI{3}{\mega\GE}\\
        Clock Speed \& Voltage    & \SI{62}{\mega\hertz}, \SI{1.2}{\volt} \\
        \bottomrule
    \end{tabularx}
\end{table}

Basilisk is an iteration on Iguana's design (\cref{apdx:cg:iguana}) that was created using only open-source EDA tools.
Its \gls{rtl} description, PDK, standard cells, and EDA flow are all open-source;
thus, it is the first \emph{end-to-end} open-source Linux-capable \gls{soc} and the largest end-to-end open-source design implemented to date at \SI{34}{\milli\meter^2}.
Like Iguana, Basilisk features a default set of Cheshire peripherals and a HyperRAM interface, enabling the autonomous boot and execution of a full Linux OS.
While implementing Basilisk, we made significant improvements to the existing open-source tool flow using \emph{Yosys} and \emph{OpenROAD}, improving the quality of results and shrinking the die area while significantly reducing the synthesis runtime and memory footprint.

\clearpage
\section{Mlem}
\label{apdx:cg:mlem}

\begin{figure}[H]
    \centering
    \includegraphics[width=145.28894pt,height=145.28333pt]{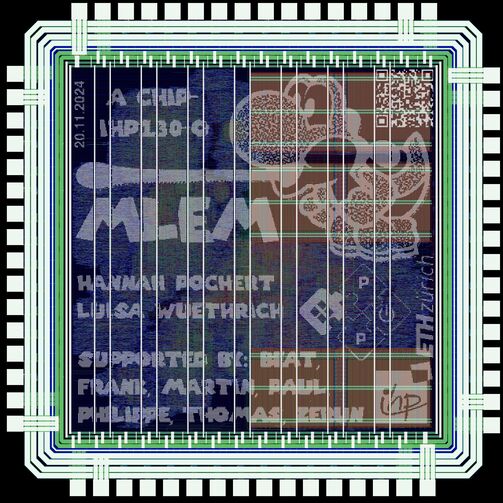}
    \includegraphics[width=147.31422pt,height=145.28333pt]{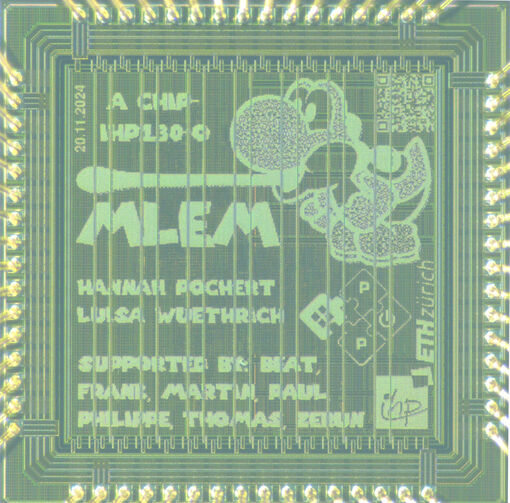}
\end{figure}

\begin{table}[H]
    \centering
    \small
    \begin{tabularx}{\linewidth}{@{}>{\bfseries}lX@{}}
        \toprule
        Designers                 & Hannah Pochert, Luisa W\"{u}thrich, \\
                                  & Philippe Sauter, Paul Scheffler, \\
                                  & Thomas Benz~\cite{sauter2025croc} \\
        \midrule
        Application               & End-to-end open-source educational MCU \\
        Technology \& Package     & IHP \SI{130}{\nano\meter}, QFN56 \\
        Size \& Complexity        & 2.24~\x~\SI{2.24}{\milli\meter}, \SI{350}{\kilo\GE}\\
        Clock Speed \& Voltage    & \SI{80}{\mega\hertz}, \SI{1.2}{\volt} \\
        \bottomrule
    \end{tabularx}
\end{table}

Mlem is the first silicon implementation of \emph{Croc}, a simple RISC-V-based MCU-class \gls{soc} for educational purposes using PULP IPs.
Mlem served as a pilot design for the new Open VLSI Lecture started at ETH Zurich in spring 2025, where students \mmpcut{}prepare an end-to-end open-source Croc tape-out with any custom addition they may choose.
Croc's end-to-end open-source tool flow is a streamlined version of Basilisk's flow (\cref{apdx:cg:basilisk}), also targeting IHP's open \SI{130}{\nano\meter} PDK.
For Mlem, students contributed a SystemVerilog UART peripheral to replace the existing VHDL implementation, a GPIO peripheral, and a Neopixel hardware unit for controlling RGB LEDs.

\clearpage
\section{Koopa}
\label{apdx:cg:koopa}

\begin{figure}[H]
    \centering
    \includegraphics[width=163.42834pt,height=163.44028pt]{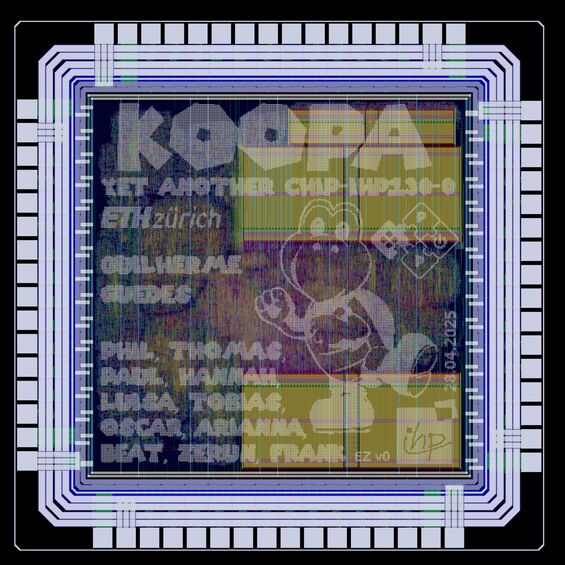}
\end{figure}

\begin{table}[H]
    \centering
    \small
    \begin{tabularx}{\linewidth}{@{}>{\bfseries}lX@{}}
        \toprule
        Designers                 & Guilherme Guedes, Philippe Sauter, \\
                                  & Thomas Benz, Paul Scheffler, Hannah \\
                                  & Pochert, Luisa W\"{u}thrich, Tobias Senti, \\
                                  & Oscar Castaneda, Arianna Rubino, Beat \\
                                  & Muheim, Zerun Jiang, Frank K. Gurkaynak \\
        \midrule
        Application               & Evaluating new standard cells \\
        Technology \& Package     & IHP \SI{130}{\nano\meter}, QFN56 \\
        Size \& Complexity        & 2.24~\x~\SI{2.24}{\milli\meter}, \SI{250}{\kilo\GE}\\
        Clock Speed \& Voltage    & \SI{74}{\mega\hertz}, \SI{1.2}{\volt} \\
        \bottomrule
    \end{tabularx}
\end{table}

Following up on Mlem (\Cref{apdx:cg:mlem}), Koopa is an implementation of Croc in IHP's open \SI{130}{\nano\meter} technology that uses the initial version of the new \emph{EZ} standard cell library.
These cells were designed for use in the VLSI 5 course on standard cell design at ETH Zurich, led by the Integrated Information Processing (IIP) group.
Additionally, Koopa brings various minor IP updates and a new QSPI interface.

\chapter{Acronyms}
\label{chap:acronyms}

\vspace{-0.8em}

\renewcommand{\glossarysection}[2][]{}

\setglossarystyle{long}
\setlength{\glsdescwidth}{8.5cm}
\renewcommand{\glsnamefont}[1]{\textbf{#1}}
\renewcommand{\glspostdescription}{}         %
\renewcommand{\glsnumberformat}[1]{\hspace{1.5mm}(p.$\,$\glshypernumber{#1})}  %

\renewcommand{\glossarymark}[1]{\markright{\MakeUppercase{#1}}}  %

\printglossary[nonumberlist=true]

\backmatter
\def\UrlBigBreaks{\do\/\do-\do:\do.}

\renewcommand{\thesauthor}{P.~Scheffler}


%
\clearpage
\chapter{Curriculum Vitae}
\label{chap:cv}

\begin{wrapfigure}{l}{0.31\textwidth}
    \vspace{-0.5cm}
    \includegraphics[width=80.94484pt,height=101.18105pt]{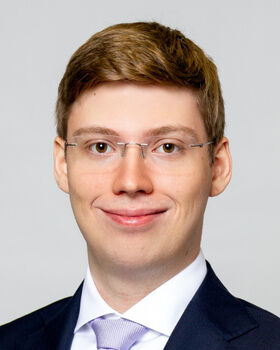}
    \vspace{0.1cm}
\end{wrapfigure}

Paul Scheffler was born in 1996 in Luxembourg City, Luxembourg.
He received his BSc and MSc degrees in electrical engineering and information technology from ETH Zurich in 2018 and 2020, respectively.
He completed his PhD in the Digital Circuits and Systems group of \mbox{Prof. Luca Benini} in 2026.
His research interests include sparse and irregular workload acceleration, on-chip interconnects, manycore architectures, and energy-efficient SoC design.
He has published more than 14 research articles in peer-reviewed journals and conference proceedings and contributed to over 11 SoC tapeouts.
He has also advised over 20 projects and theses by bachelor’s and master’s students at ETH Zurich.

\end{document}